\documentclass[%
preprint
 amsmath, amssymb, %
 aip,cha,%
]{revtex4-1}

\usepackage{CJKutf8}
\usepackage{xcolor}
\usepackage{amsmath}
\usepackage{amssymb}
\usepackage{docs}%
\usepackage{bm}%
\usepackage{xcolor}
\usepackage{dcolumn}% Align table columns on decimal point
\usepackage{graphicx}% Include figure files
\usepackage[export]{adjustbox}
\usepackage[colorlinks=true,linkcolor=blue]{hyperref}%
\usepackage{algorithm}
\usepackage{algpseudocode}
\usepackage{subcaption}
\usepackage{color}
\usepackage{scalerel,stackengine}
\usepackage[normalem]{ulem}

\stackMath % \widecheck symbol required
\newcommand\widecheck[1]{%
	\savestack{\tmpbox}{\stretchto{%
			\scaleto{%
				\scalerel*[\widthof{\ensuremath{#1}}]{\kern-.6pt\bigwedge\kern-.6pt}%
				{\rule[-\textheight/2]{1ex}{\textheight}}%WIDTH-LIMITED BIG WEDGE
			}{\textheight}% 
		}{0.5ex}}%
	\stackon[1pt]{#1}{\scalebox{-1}{\tmpbox}}%
}

\expandafter\ifx\csname package@font\endcsname\relax\else
 \expandafter\expandafter
 \expandafter\usepackage
 \expandafter\expandafter
 \expandafter{\csname package@font\endcsname}%
\fi
\begin{document}
		
\begin{CJK*}{UTF8}{gbsn}
\title{Adjoint-based variational optimal mixed models for large-eddy simulation of turbulence}

\author{Zelong Yuan (袁泽龙)}%
\author{Yunpeng Wang (王云朋)}%
\author{Xiaoning Wang (王小宁)}%
\author{Jianchun Wang (王建春)}%
\email[Email address for correspondence:\;]{wangjc@sustech.edu.cn}
\affiliation{\small Department of Mechanics and Aerospace Engineering, Southern University of Science and Technology, Shenzhen 518055, People's Republic of China}
\affiliation{\small Guangdong–Hong Kong–Macao Joint Laboratory for Data-Driven Fluid Mechanics and Engineering Applications, Southern University of Science and Technology, Shenzhen 518055, People's Republic of China}

%\author{}
%\affiliation{}
%\author{}
%\affiliation{}
%\author{}
%\affiliation{}

\date{\today}% It is always \today, today,

\begin{abstract}
An adjoint-based variational optimal mixed model (VOMM) is proposed for subgrid-scale (SGS) closure in large-eddy simulation (LES) of turbulence. The stabilized adjoint LES equations are formulated by introducing a minimal regularization to address the numerical instabilities of the long-term gradient evaluations in chaotic turbulent flows. The VOMM model parameters are optimized by minimizing the discrepancy of energy dissipation spectra between LES calculations and \emph{a priori} knowledge of direct numerical simulation (DNS) using the gradient-based optimization. The \emph{a posteriori} performance of the VOMM model is comprehensively examined in LES of three turbulent flows, including the forced homogeneous isotropic turbulence, decaying homogenous isotropic turbulence, and temporally evolving turbulent mixing layer. The VOMM model outperforms the dynamic Smagorinsky model (DSM), dynamic mixed model (DMM) and approximate deconvolution model (ADM) in predictions of various turbulence statistics, including the velocity spectrum, structure functions, statistics of velocity increments and vorticity, temporal evolutions of the turbulent kinetic energy, dissipation rate, momentum thickness and Reynolds stress, as well as the instantaneous vortex structures at different grid resolutions and times. In addition, the VOMM model only takes up 30\% time of the DMM model for all flow scenarios. These results demonstrate that the proposed VOMM model improves the numerical stability of LES and has high \emph{a posteriori} accuracy and computational efficiency by incorporating the \emph{a priori} information of turbulence statistics, highlighting that the VOMM model has a great potential to develop advanced SGS models in the LES of turbulence.       
\end{abstract}

%\pacs{47.20.Qr,  47.65.-d}% PACS, the Physics and Astronomy
\maketitle
\end{CJK*}

\section{Introduction}\label{sec:level1}

Large-eddy simulation (LES) has become an effective tool for the investigation of turbulent flows, and has been widely applied to many industrial problems including the aeroacoustics, combustions, meteorological physics, interfacial mixing and turbomachinery \citep{sagaut2006,garnier2009}. The dominant large-scale motions of turbulence are directly resolved by the LES, leaving the effects of residual subgrid scales (SGS) on the resolved large scales modeled by the SGS models \citep{lesieur1996,meneveau2000}. In contrast, direct numerical simulation (DNS) of turbulence requires a sufficiently high mesh resolution to fully resolve all flow scales down to the size of the Kolmogorov eddies, whose computational cost is prohibitively expensive at a high Reynolds number \citep{pope2000}. Therefore, LES is much more computationally efficient than the DNS by significantly reducing the degrees of freedom of turbulence, meanwhile accurately reconstructing large-scale flow structures \citep{pope2000,sagaut2006,durbin2018}.

The modeling of the unclosed SGS stress is crucial for the accuracy of predictions in LES. SGS models can be generally categorized into functional models, structural models and mixed models \citep{sagaut2006,garnier2009}. The functional SGS models utilize the explicit dissipative terms to correctly reconstruct the forward kinetic energy cascade from large scales to small scales \citep{rozema2015,abkar2016}. The Smagorinsky model is one of the most popular functional SGS models and is favored for its substantial numerical stability and excellent robustness of LES calculations \citep{smagorinsky1963,lilly1967}. However, the functional SGS models generally exhibit excessive dissipation and fail to predict the sophisticated small-scale flow structures. In contrast, the structural SGS models recover the unclosed SGS stress with high \emph{a priori} accuracy by exactly truncating the Taylor series expansions or the assumption of scale similarity. These structural models include the approximate deconvolution method \citep{stolz1999,stolz2001}, scale-similarity model \citep{bardina1980,liu1994}, velocity gradient model \citep{clark1979}, one-equation nonviscosity model \citep{pomraning2002dynamic} and algebraic structural model \citep{lu2010modulated}. The structural SGS models can accurately capture the spatial distribution of SGS energy flux and backscatter of the kinetic energy, but suffer from the numerical instability without sufficient SGS dissipation in the \emph{a posteriori} studies of LES. 

The mixed models consist of the structural models and functional eddy-viscosity models to balance the numerical stability and accuracy of LES and compensate their inherent model deficiencies. The Clark model combines the velocity gradient model with the Smagorinsky eddy viscosity \citep{clark1979}. \citet{erlebacher1992} proposed a mixed model which consists of the scale-similarity model and the dissipative Smagorinsky term. In the early stage, the SGS model parameters were either theoretically derived from the isotropic turbulent flows \citep{lilly1967} or estimated by the \emph{a priori} analysis of DNS and experimental observations \citep{deardorff1970,clark1979}, yielding poor predictions in the \emph{a posteriori} LES \citep{lesieur1996,meneveau2000}. A pioneering dynamical procedure with the Germano identity was developed to determine the Smagorinsky coefficient adaptively by the least-squares algorithm \citep{germano1991,lilly1992}. Subsequently, the dynamic versions of mixed models were successively proposed, including the one-parameter \citep{zang1992} and two-parameter dynamical mixed models \citep{liu1994,shi2008}, the dynamic Clark model \citep{vreman1994} and dynamic ADM model \citep{habisreutinger2007}. The coefficients of a general multi-parameter dynamic mixed model (DMM)  can be conveniently determined by the Germano-identity-based dynamic approach \citep{sagaut2000a}. However, extensive previous studies have shown that these DMM models are excessively dissipative in the transitional regions, but underestimate the SGS dissipation in situations of coarse mesh resolutions and grid anisotropy \citep{meneveau2000,moser2021}. In addition, the dissipative Smagorinsky part in the DMM models  is usually dominant over the structural part, leading to little advantage in the high \emph{a priori} accuracy of structural models. The basis tensors of the DMM model, comprising the functional eddy-viscosity and the accurate structural part, give a complete representation of the SGS stress and SGS energy flux (SGS dissipation), which is essential for the SGS modeling of LES. \citet{yuan2022} preliminarily explored a scale-similarity dynamic procedure (SSD) with a dynamic nonlinear algebraic model, yielding more accurate predictions of various turbulence statistics and instantaneous vortex structures for both \emph{a priori} and \emph{a posteriori} analyses of LES than the Germano-identity-based dynamic (GID) approach in the homogeneous isotropic turbulence. However, the SSD procedure still suffers from the numerical instability at coarse-grid-resolution cases, where the spatial discretization error dominates the SGS modeling error. It might be challenging to develop a general dynamic framework for the model coefficient determination at various grid resolutions applicable to different types of turbulence problems. These results demonstrate that the adjustment of SGS model parameters can effectively improve the accuracy of SGS modeling and enhance the predictions of LES.

Besides, additional artificial viscous or penalized regularization terms have been also introduced to enhance the \emph{a posteriori} stability of structural models. A secondary filtering regularization technique was proposed by \citet{stolz2001} and \citet{adams2004} to maintain the numerical stability of ADM models. \citet{vollant2016} efficiently regularized the velocity gradient model by dynamically clipping the SGS backscatter. A spectral-vanishing-viscosity method \citep{tadmor1989} was proposed to effectively suppress the Gibbs oscillations at high wavenumbers \citep{cerutti2000} and has been successfully applied to the prediction of turbulent channel flows \citep{karamanos2000}. \citet{xie2020a} used a hyperviscosity term to address the stability issue of the spatial-artificial-neural-network models. The effective hyperviscosity term was further applied to other data-driven SGS models \citep{yuan2020, wang2021}. \citet{yuan2021} developed a small-scale eddy-viscosity model to enhance the \emph{a posteriori} stability of dynamic iterative approximate deconvolution models, without affecting the accurate predictions of large-scale flow structures. A kinetic-energy-flux constrained SGS model proposed by \citet{yu2022a} regularizes the DSM model by the correct kinetic energy flux approximated by the tensor-diffusivity model and accurately predicts the transition to turbulence of a compressible flat-plate boundary layer. Additional numerical parameters would be introduced for most regularization techniques, which are sensitive to the grid resolution of LES, requiring multiple tedious testings for different turbulence scenarios. To our knowledge, there might not be a unified adaptive regularization framework proposed for the stability of structural SGS models that can be universally applied to various types of turbulence with different grid resolutions of LES calculations. The dependence of SGS model parameters on grid resolutions of LES might be effectively addressed by incorporating the \emph{a priori} knowledge of DNS or experimental observations.

In recent years, many data-driven closure approaches \citep{tracey2015,ling2016a,xiao2016,maulik2017,wang2018,zhou2019,yang2019,park2021,guan2022} have been extensively developed to improve the modeling of unclosed terms in turbulence, as more high-fidelity DNS or experimental data become available \citep{kutz2017,duraisamy2019}. \citet{ling2016} proposed a representative tensor-basis-neural-network (TBNN) model with the multiplicative layer that predicts coefficients of the basis tensors for the modeled Reynolds stress by taking velocity invariants as input to preserve Galilean invariance. The TBNN architecture can accurately reconstruct the anisotropy of Reynolds stress and predict the flow separation better than the baseline linear or nonlinear eddy-viscosity model. \citet{xie2020d} further developed the artificial-neural-network-based nonlinear algebraic models yielding better predictions of LES statistics than classical dynamic SGS models. The gene-expression-programming technique was proposed to acquire the explicit mathematical expression of the unclosed SGS stress modeled by basis functions for LES using an evolutionary algorithm \citep{schoepplein2018,li2021e,wu2022b}. The multi-agent reinforcement-learning framework was developed to discover Smagorinsky model coefficients using the control policy rewarded by the statistical discrepancy of energy spectrum \citep{novati2021,kurz2023}, and further applied to modeling the near-wall dynamics \citep{bae2022b}.

Although the machine-learning-based closure models can improve the \emph{a priori} accuracy of turbulence models fairly well, they have been reported to suffer from the ill-conditioned issues in the \emph{a posteriori} studies. The small \emph{a priori} errors of the modeled Reynolds stress can be significantly amplified and then propagated into the mean velocity field in the \emph{a posteriori} testings \citep{wu2019}. \citet{gamahara2017} established an artificial-neural-network framework for the SGS closures of turbulent channel flows, which accurately predicts the unclosed SGS stress in \emph{a priori} studies, but shows no obvious advantages over the Smagorinsky model in the reconstruction of the mean velocity profiles. The recurrent neural network was employed to learn the coarse-grained discretization errors of LES and expected to construct the perfect LES formulation \citep{beck2019}.  However, these perfect SGS closure terms also encounter serious \emph{a posteriori} instability issues, even though the \emph{a priori} predictions show high correlations with the exact unclosed terms \citep{beck2019}. These results indicate that most current data-driven closure approaches can acquire sufficiently high \emph{a priori} accuracy after being trained by the high-fidelity DNS or experimental data, but still lack indispensable extrapolation capabilities and are difficult to be applied to the \emph{a posteriori} testings of out-of-sampling flow scenarios. 

The data-assimilation techniques can effectively remedy the deficiencies of insufficient \emph{a posteriori} accuracy of closure models by iteratively evaluating and minimizing the discrepancies between coarse-grained \emph{a posteriori} calculations and benchmark high-fidelity DNS or experimental observations. The data-assimilation approaches can be generally classified into three categories: ensemble-based statistical methods \citep{colburn2011,zhang2022b,deng2021,wang2023}, adjoint-based variational approaches \citep{bewley2001,delport2009a,badreddine2014,he2022a,li2022h} and their mixed variants \citep{mons2021}. The ensemble-based statistical techniques use ensemble statistics to approximately measure the model uncertainty and continuously correct the measurement errors of observations by the classical Kalman-filtering strategies or nudging methods \citep{clarkdileoni2020,pawar2020f,li2022c}. These statistical assimilation methods allow the convenient inference of flow states and statistics, without any detailed information of dynamical systems, facilitating their wide application in complex practical scenarios. However, the state estimations of these ensemble-based approaches frequently evaluate the matrix multiplication and inverse operations, resulting in the massive computation expense and large memory usage for the high degree-of-freedom turbulence problems at a high Reynolds number.  In contrast, the adjoint-based variational techniques employ the optimal control strategy to efficiently optimize the model parameters or state variables by minimizing the discrepancies between the benchmark observations and \emph{a posteriori} predictions. \citet{singh2016} proposed a field-inversion procedure to infer model discrepancies in the source terms of Reynolds-averaged Navier–Stokes (RANS) transport equations using Bayesian posterior estimation. \citet{he2018} simplified the field-inversion strategy and employed the continuous adjoint formulation to optimize a spatially varying turbulence production term in the Spalart–Allmaras model of RANS equations. 

In comparison with the extensive studies of data-assimilation-based RANS models \citep{kato2013,kato2015,xiao2016,li2017,xiao2019}, investigations on SGS models of LES assimilated with high-fidelity simulation data are still preliminary. A spatially-varying parameter in a local uncertainty model and initial conditions were optimized based on experimental observations of the cylindrical wake flow using the discrete adjoint algorithm \citep{chandramouli2020}. \citet{mons2021} developed a non-intrusive ensemble-variational approach (EnVar) to enhance the predictions of the mean flow and Reynolds stresses by adjusting the wall-normal distribution of the Smagorinsky coefficient or injecting an artificial steady force in the LES momentum equations. The SGS force modeled by the artificial neural network was optimized by the point-to-point errors of the filtered velocity field using the discrete adjoint method for the decaying isotropic turbulence and plane jet flows \citep{sirignano2020,macart2021}. However, these discrete adjoint or ensemble-based variational methods require massive matrix operations with significant memory usage. 

In this paper, a variational optimal mixed model (VOMM) is proposed to reconstruct the unclosed SGS stress by assimilating the turbulence statistics of high-fidelity filtered DNS data using the continuous adjoint approach. The main difference from the previous work is that we derive adjoint LES equations with the general SGS model and conduct the energy budget analysis of adjoint equations. The continuous adjoint algorithm can enhance the physical understanding of the adjoint-based sensitivities and provide flexibility in selecting the discretization scheme for the adjoint equations. The quadratic terms of shear strain rate in adjoint LES equations turn out to be responsible for the exponential temporal growth of the adjoint-based gradients, giving rise to the numerical divergence in a long time horizon for the chaotic turbulent flows. Hence, the stabilized adjoint LES equations are correspondingly formulated to enhance the numerical stability of the adjoint LES calculations. To the extent of the authors’ knowledge, few previous studies have given detailed derivations of the adjoint LES equations with general SGS mixed models and formulated the stabilized version for long-term gradient evaluations. In addition, the selected cost functional is essential for the convergence and performance of adjoint-based gradient optimizations. Compared to the previous studies, turbulence statistical discrepancies rather than the chaotic point-to-point prediction errors are adopted to quantify the multiscale statistical behaviours of turbulence. The \emph{a priori} information about statistics of turbulence acquired from experimental data or DNS results, including energy spectra, structure functions, and probability density functions of physical quantities, can be used to determine or correct SGS model parameters to improve the \emph{a posteriori} accuracy of LES greatly. Turbulent statistical assimilation can effectively alleviate the impact of chaotic field observations on the performance of data assimilation. Furthermore, the \emph{a posteriori} performance of VOMM model is comprehensively investigated and compared to classical SGS models at multiple grid resolutions in different turbulence scenarios, including the forced and decaying homogeneous isotropic turbulence, as well as the temporally evolving turbulent mixing layer.

The remainder of this paper is structured as follows. Sec.~\ref{sec:level2} describes the governing equations of the large-eddy simulation. The conventional subgrid-scale models, including DSM, DMM and ADM models, are briefly introduced in Sec.~\ref{sec:level3}. In Sec.~\ref{sec:level4}, we first derive the adjoint LES equations with a general form of mixed SGS models, then conduct the energy budget analysis of adjoint equations, and correspondingly propose the stabilized adjoint LES equations. Afterwards, the adjoint-based variational optimal mixed model is developed. Sec.~\ref{sec:level5} further investigates the \emph{a posteriori} performance of the VOMM model in comparison to the classical SGS models for three turbulent flow scenarios, including the forced homogeneous isotropic turbulence, decaying homogeneous isotropic turbulence, and temporally evolving turbulent mixing layer. The generalization ability of the VOMM model is further discussed in Sec.~\ref{sec:level6}. Conclusions are finally drawn in Sec.~\ref{sec:level7}.

\section{Governing equations of the large-eddy simulation}\label{sec:level2}
The three dimensional incompressible turbulence is governed by the Navier-Stokes equations \citep{pope2000}, namely
\begin{equation}
	\frac{{\partial {u_i}}}{{\partial {x_i}}} = 0,
	\label{ns1}
\end{equation}

\begin{equation}
	\frac{{\partial {u_i}}}{{\partial t}} + \frac{{\partial \left( {{u_i}{u_j}} \right)}}{{\partial {x_j}}} =  - \frac{{\partial p}}{{\partial {x_i}}} + \nu \frac{{{\partial ^2}{u_i}}}{{\partial {x_j}\partial {x_j}}} + {{ \mathcal F}_i}, 
	\label{ns2}
\end{equation}
where $u_i$ is the $i$-th component of velocity, $p$ denotes the pressure divided by the constant density, $\nu$ is the kinematic viscosity, and $\mathcal {F}_i$ represents the large-scale forcing on the fluid momentum in the $i$-th coordinate direction. The summation convection for the repeated indices is adopted by default for simplicity in this paper. Besides, the dimensionless governing parameter for the incompressible turbulence, namely, the Taylor microscale Reynolds number ${\rm{Re}}_\lambda$ is defined as \citep{pope2000}
\begin{equation}
	{{\rm Re}_{\lambda }}=\frac{{{u}^{\rm{rms}}}\lambda }{\sqrt{3} \nu },
	\label{Re_l}
\end{equation}
where ${{u}^{\rm{rms}}}=\sqrt{\left\langle {{u}_{i}}{{u}_{i}} \right\rangle}$ represents the root-mean-square (rms) value of the velocity magnitude, and $\left\langle {\cdot} \right\rangle $ represents a spatial average along the homogeneous direction (\emph{i.e.}, average over the entire domain for the isotropic turbulence and the horizontal average for the temporally evolving mixing layer). Here, $\lambda ={{u}^{\rm {rms}}}\sqrt{5\nu /\varepsilon }$  is the Taylor microscale, where $\varepsilon =2\nu \left\langle {{S}_{ij}}{{S}_{ij}} \right\rangle $ represents the average dissipation rate and  ${{S}_{ij}}=\frac{1}{2}\left( \partial {{u}_{i}}/\partial {{x}_{j}}+\partial {{u}_{j}}/\partial {{x}_{i}} \right)$ denotes the strain-rate tensor. 

To obtain the governing equations of the large-eddy simulation, a spatial filtering operation, $\bar{f}\left( \mathbf{x} \right)=\int\limits_{\Omega}{f\left( {\mathbf{{x}'}} \right)G\left( \mathbf{x}-\mathbf{{x}'}; \bar \Delta  \right)d\mathbf{{x}'}}$ is applied to the Navier-Stokes equations. Here, an overbar denotes the spatial filtering, $\Omega$ is the entire domain. $G$ and $\bar \Delta$ are the filter kernel and filter width, respectively. The governing equations for the LES can be correspondingly derived as \citep{sagaut2006}
\begin{equation}
	\frac{\partial {{{\bar{u}}}_{i}}}{\partial {{x}_{i}}}=0,
	\label{fns1}
\end{equation}
\begin{equation}
	\frac{\partial {{{\bar{u}}}_{i}}}{\partial t}+\frac{\partial \left( {{{\bar{u}}}_{i}}{{{\bar{u}}}_{j}} \right)}{\partial {{x}_{j}}}=-\frac{\partial \bar{p}}{\partial {{x}_{i}}}-\frac{\partial {{\tau }_{ij}}}{\partial {{x}_{j}}}+\nu\frac{{{\partial }^{2}}{{{\bar{u}}}_{i}}}{\partial {{x}_{j}}\partial {{x}_{j}}}+{{\bar{\mathcal{F}}}_{i}}.
	\label{fns2}
\end{equation}
Here, the unclosed SGS stress tensor $\tau_{ij}=\overline{{{u}_{i}}{{u}_{j}}}-{{\bar{u}}_{i}}{{\bar{u}}_{j}}$ cannot be directly calculated using the resolved variables $\bar u_i$, and additional SGS stress modeling is required to make the LES equations solvable.  

\section {Conventional subgrid-scale models for LES}\label{sec:level3}
The SGS models aim to establish the approximate constitutive equation for SGS unclosed terms using the known resolved variables, and reconstruct the nonlinear interactions between the resolved large scales and unsolved small scales as accurately as possible \citep{moser2021,johnson2022}. The explicit SGS models consist of the functional and structural models. The functional modeling adopts the eddy-viscosity forms to mimic the forward kinetic energy transfer from the resolved large scales to the residual small scales, while the structural models can accurately recover the unclosed SGS stress by the hypothesis of scale similarity or using the truncated series expansions with high \emph{a priori} accuracy \citep{sagaut2006,fowler2022}. One of the most widely-used functional models is the Smagorinsky model \citep{smagorinsky1963,lilly1967}, expressed as
\begin{equation}
	\tau_{ij}^A=\tau_{ij}-\frac{\delta_{ij}}{3}\tau_{kk}=-2C_S^2{\bar \Delta}^2|\bar{S}|\bar{S}_{ij},
	\label{tau_sm}
\end{equation}
where $\delta_{ij}$ denotes the Kronecker delta operator, ${\bar S_{ij}} = \frac{1}{2}\left( {\partial {{\bar u}_i}}/{{\partial {x_j}}} + {\partial {{\bar u}_j}}/{{\partial {x_i}}} \right)$ is the filtered strain-rate tensor and $|\bar{S}|=(2\bar{S}_{ij}\bar{S}_{ij})^{1/2}$ represents the characteristic filtered strain rate. The superscript ``A'' represents the trace-free anisotropic part of the arbitrary variables, namely, $\left(  \bullet  \right)_{ij}^A = {\left(  \bullet  \right)_{ij}} - {\left(  \bullet  \right)_{kk}}{\delta _{ij}}/3$. The isotropic SGS stress $\tau_{kk}$ is absorbed into the pressure term. $C_S^2$ is the Smagorinsky coefficient and can be determined empirically or by a theoretical analysis. The most common approach is based on the least-squares dynamic procedure using the Germano identity, giving rise to the dynamic Smagorinsky model (DSM), whose coefficient is given by \citep{germano1991,lilly1992}
\begin{equation}
	C_S^2=\frac{\langle{L_{ij}^A \mathcal M_{ij}}\rangle}{\langle{\mathcal M_{kl} \mathcal M_{kl}}\rangle},
	\label{dsm_cs}
\end{equation} 
where the Leonard stress $ L_{ij}=\widetilde {{{\bar u}_i}{{\bar u}_j}} - {{\tilde {\bar u}}_i}{{\tilde {\bar u}}_j}$, $ L_{ij}^A = { L_{ij}} - \frac{1}{3}{\delta _{ij}}{ L_{kk}}$ and $\mathcal M_{ij}=\tilde{\alpha}_{ij}-\beta_{ij}$. Here, a tilde stands for the test filtering operation at the double-filtering scale $\tilde{{\Delta}}=2 \bar \Delta$,  the variables $\alpha_{ij}= 2 \bar \Delta^2 |\bar{S}|\bar{S}_{ij}$ and $\beta_{ij}= 2 \tilde{ \Delta}^2 |\tilde{\bar{S}}|\tilde{\bar{S}}_{ij}$. The scale-similarity model ${\tau_{ij}} = \widetilde {{{\bar u}_i}{{\bar u}_j}} - {\tilde {\bar u}_i}{\tilde {\bar u}_j}$ is a typical structural model and can correctly reconstruct the SGS stress with high \emph{a priori} accuracy. However, these structural models often exhibit insufficient dissipation and numerical instability in the \emph{a posteriori} testings of LES due to the underestimation of the forward kinetic energy cascade. The dynamic mixed model (DMM)  combines the scale-similarity model with the dissipative Smagorinsky term, and is given by \citep{liu1994,shi2008}
\begin{equation}
	{\tau _{ij}} = {C_1}{\bar \Delta ^2}\left| {\bar S} \right|{\bar S_{ij}} + {C_2}\left( {\widetilde {{{\bar u}_i}{{\bar u}_j}} - {{\tilde {\bar u}}_i}{{\tilde {\bar u}_j}}} \right).
	\label{dmm1}
\end{equation}
Similar to the DSM model, model coefficients of the DMM model $C_1$ and $C_2$ are dynamically determined by the least-squares algorithm using the Germano identity, expressed respectively as \citep{xie2020d,yuan2020}
\begin{equation}
	{C_1} = \frac{{\left\langle {N_{ij}^2} \right\rangle \left\langle {{L}_{ij}{M_{ij}}} \right\rangle  - \left\langle {{M_{ij}}{N_{ij}}} \right\rangle \left\langle {{L}_{ij}{N_{ij}}} \right\rangle }}{{\left\langle {N_{ij}^2} \right\rangle \left\langle {M_{ij}^2} \right\rangle  - {{\left\langle {{M_{ij}}{N_{ij}}} \right\rangle }^2}}},
	\label{dmm_c1}
\end{equation}
\begin{equation}
	\quad{C_2} = \frac{{\left\langle {M_{ij}^2} \right\rangle \left\langle {{L}_{ij}{N_{ij}}} \right\rangle  - \left\langle {{M_{ij}}{N_{ij}}} \right\rangle \left\langle {{L}_{ij}{M_{ij}}} \right\rangle }}{{\left\langle {N_{ij}^2} \right\rangle \left\langle {M_{ij}^2} \right\rangle  - {{\left\langle {{M_{ij}}{N_{ij}}} \right\rangle }^2}}},
	\label{dmm_c2}
\end{equation}
where ${M_{ij}} = H_{1,ij} - \tilde{h}_{1,ij}$, and ${N_{ij}} = H_{2,ij} - \tilde{h}_{2,ij}$. Here, $h_{1,ij} = -2{\bar \Delta ^2}\left| {\bar S} \right|{\bar S_{ij}}$, ${h_{2,ij}} = \widetilde {{{\bar u}_i}{{\bar u}_j}} - {\tilde {\bar u}_i}{\tilde {\bar u}_j}$, $H_{1,ij} =-2{\tilde { \Delta}^2}\left| {\tilde {\bar S}} \right|{\tilde {\bar S}_{ij}}$, and ${H_{2,ij}} = \widehat {{{\tilde {\bar u}}_i}{{\tilde {\bar u}}_j}} - {\hat {\tilde {\bar u}}_i}{\hat {\tilde {\bar u}}_j}$. The hat stands for the test filtering at scale $\hat{\Delta}=4\bar \Delta$.

The unfiltered variables can be accurately recovered by the resolved filtered field using the iterative approximate deconvolution procedure, namely \citep{stolz1999,stolz2001}
\begin{equation}
	u_i^* = A{D^N}\left( {{{\bar u}_i}} \right) = \sum\limits_{n = 1}^N {{{(I - G)}^{n - 1}}} \otimes  {{{\bar u}_i}},
	\label{u_AD}
\end{equation}
where the asterisk represents the approximately unfiltered variables, $A{D^N}$ is the abbreviation of the $N$-th order approximate deconvolution, $I$ is the identity, and the symbol ``$\otimes$'' stands for the spatial convolution operator. For any two functions $f$  and $g$, $f \otimes g=\int_{-\infty}^{+\infty} f\left(\mathbf{x}^{\prime}\right) g\left(\mathbf{x}-\mathbf{x}^{\prime}\right) d \mathbf{x}^{\prime}$. The unclosed SGS stress then can be recovered with the scale-similarity form by the approximate deconvolution method (ADM), given by \citep{bardina1980} 
\begin{equation}
	\tau_{i j}=\overline{u_i^* u_j^*}-\bar{u_i^*} \bar{u_j^*}.
	\label{ADM}
\end{equation}
The number of iterations for the ADM model is recommended to be $N \!=\! 3 \sim 5$ \citep{stolz2001}. The accuracy of the ADM model becomes higher, while the numerical stability drops, as the number of iterations increases. Hence, $N \!=\! 5$ is selected in this paper. To maintain the numerical stability of the \emph{a posteriori} testings of LES [${\partial {{\bar u}_i}}/{{\partial t}} = {{\bar R}_i}\left( {{{\bar u}_i},t } \right)$], \citet{stolz2001} and \citet{adams2004} introduced a secondary filtering relaxation term [${\partial {{\bar u}_i}}/{{\partial t}} = {{\bar R}_i}\left( {{{\bar u}_i},t} \right) + {{\bar S}_i}\left( {{{\bar u}_i} } \right)$], yielding
\begin{equation}
	{{\bar S}_i}\left( {{{\bar u}_i} } \right) =  - \chi \left[ {I - G \otimes \sum\limits_{n = 1}^N {{{(I - G)}^{n - 1}}} } \right] \otimes {{\bar u}_i},
	\label{ADM_relax}
\end{equation}
where $\chi$ is the empirical regularization coefficient, which is approximately insensitive to the LES results in previous studies, and we choose  $\chi=$ 0 and 1 for comparisons in our paper.  

\section {Adjoint-based variational optimal mixed models (VOMM)}\label{sec:level4}
The mixed model is composed of the structural parts and the dissipative functional terms, and its general form can be written as \citep{sagaut2000a}
\begin{equation}
	{\tau _{ij}}\left( {{u_i};\bar \Delta } \right) = \sum\limits_{n = 1}^N {{C_n}T_{ij}^{\left( n \right)}\left( {{{\bar u}_i};\bar \Delta } \right)} ,
	\label{tau_mm}
\end{equation}
where ${T_{ij}^{\left( n \right)}\left( {{{\bar u}_i};\bar \Delta } \right)}$ represents the $n$-th basis stress tensor. ${C_n}\left( {n = 1,2,...,N} \right)$ denotes the corresponding model coefficient and $N$ is the number of basis stress tensors. The model coefficients are generally respectively determined by the multivariate least-squares algorithm proposed by \citet{germano1991} and \citet{lilly1992}. Many previous studies have shown that the dynamic mixed models give rise to an excessive dissipation of energy in the transitional regions and dissipation underestimation if the filter scales are sufficiently large, especially in situations of grid anisotropy \citep{meneveau2000,moser2021}.

In recent years, data-driven based high-accuracy SGS models are successively proposed \citep{kutz2017,duraisamy2019}. \citet{xie2019a} proposed an artificial-neural-network-based mixed model which accurately recovers the unclosed SGS terms by estimating mixed model coefficients with local flow characteristics as inputs of the machine-learning strategy, yielding better predictions of LES statistics than the classical dynamic mixed model. The input features of the data-driven closure models are crucial for the accuracy of SGS models \citep{gamahara2017,beck2019,xie2019,park2021}. Incorporating the accurate structural parts, \emph{i.e.}, filtered velocity gradients at the neighboring stencil turn out to improve the performance of data-driven SGS models effectively \citep{xie2019,xie2020a}.  Moreover, the spatial flow structures at scales between $\bar \Delta/2$ and $2 \bar\Delta$ are found to be essential for the SGS modeling of LES at the filter scale $\bar \Delta$ \citep{xie2020b}. The strategy of the blind deconvolution with the artificial neural network was proposed to recover the unknown original unfiltered variables from the known filtered quantities with high accuracy \citep{maulik2017,maulik2019a}. A deconvolutional-artificial-neural-network (DANN) framework was further proposed to accurate reconstruct the SGS unclosed terms both in \emph{a priori} and \emph{a posteriori} analyses of isotropic turbulence \citep{yuan2020,yuan2021a}, and successfully applied to the chemically reacting compressible turbulence \citep{teng2022}. It was demonstrated that the DANN models embed the properties of symmetry and realizability conditions, which preserve the physical reliability of the DANN framework \citep{yuan2020}. To enhance the interpretability of black-box machine-learning SGS models, a semi-explicit ANN-based spatial gradient model and constant-coefficient spatial gradient models are successively proposed by the elaborate Taylor expansions of velocity gradients in the neighboring stencil locations \citep{wang2021,wang2022a}. The machine-learning-based SGS models trained by high-fidelity simulation data can be regarded as the structural models with high \emph{a priori} accuracy, requiring additional indispensable dissipation to account for the spatial discretization effect and ensure the numerical stability in the \emph{a posteriori} studies of LES.

In addition to the machine-learning-assisted SGS models, some \emph{a priori} information about statistics of turbulence acquired from experimental data or DNS results like energy spectra, structure functions, and probability density functions of physical quantities can be used to determine or correct the model coefficients of SGS models to improve the model accuracy greatly. These \emph{a priori} knowledge of turbulent statistical quantities can be dynamically assimilated into the closure models via the data-assimilation based approaches. Among these data-assimilation techniques, adjoint-based variational methods adopt the optimal control strategy to efficiently calculate all the gradients of cost functionals for the model coefficients by solving the forward governing equations and the backward adjoint equations \citep{bewley2001}. Then, the model coefficients of SGS models are iteratively updated using the gradient-based optimization algorithm until the optimal values are obtained. The cost functionals measure the discrepancies of statistical quantities in turbulence between the LES results and measurements from the experimental or DNS data, which can greatly alleviate the impact of chaotic field observations on the performance of data assimilation. In this work, we resort to the state-of-art adjoint-based data-assimilation approaches to establish a general optimal SGS framework to determine model parameters adaptively for various grid resolutions of LES in different turbulence scenarios.   
\subsection {Adjoint LES equations and gradient evaluations with the mixed model}
We optimize the model coefficients of the SGS closure model to minimize the statistical discrepancies between the LES calculations and the reference values acquired from the experimental or DNS data, which can be defined as the minimal optimization problem constrained by the governing equations (see Eqs.~\ref{fns1} and \ref{fns2}). The constrained optimization problem for the turbulent closure modeling is expressed as
\begin{equation}
	\begin{array}{*{20}{c}}
		{\mathop {{\rm{min}}}\limits_{{C_n}} }&{{\mathcal J}\left[ {\phi \left( {{{\bar u}_i};{C_n}} \right),\phi \left( {\bar u_i^{{\rm{ref}}}} \right)} \right]},\\
		{{\rm{s}}.{\rm{t}}.}&{\begin{array}{*{20}{c}}
				{{R_0}\left( {{{\bar u}_i}} \right) = \frac{{\partial {{\bar u}_i}}}{{\partial {x_i}}} = 0,}\\
				{{R_i}\left( {{{\bar u}_i},\bar p} \right) = \frac{{\partial {{\bar u}_i}}}{{\partial t}} + \frac{{\partial \left( {{{\bar u}_i}{{\bar u}_j}} \right)}}{{\partial {x_j}}} + \frac{{\partial \bar p}}{{\partial {x_i}}} - \nu \frac{{{\partial ^2}{{\bar u}_i}}}{{\partial {x_j}\partial {x_j}}} - {{\overline {\mathcal F} }_i} + \frac{{\partial {\tau _{ij}}}}{{\partial {x_j}}} = 0,}
		\end{array}}
	\end{array}
	\label{opt_problem}
\end{equation}
where $\mathcal{J}\left[ {\phi \left( {{{\bar u}_i{;{C_n}}}} \right),\phi \left( {\bar u_i^{\rm{ref}}} \right)} \right]=\int \limits_0^T \;{\int\limits_\Omega{J\left[ {\phi \left( {{{\bar u}_i};{C_n},{\bf{x}},t} \right),\phi \left( {\bar u_i^{\rm{ref}};{\bf{x}},t} \right)} \right]d{\bf{x}}dt} }$ denotes the total cost functions, ${J\left[ {\phi \left( {{{\bar u}_i};{C_n},{\bf{x}},t} \right),\phi \left( {\bar u_i^{\rm{ref}};{\bf{x}},t} \right)} \right]}$ is the discrepancy of statistical quantities $\phi $ (\emph{e.g.} kinetic energy spectra, structure functions and probability density functions of the targeted variables) between the LES results ${{{\bar u}_i}}$ and reference values ${\bar u_i^{\rm{ref}}}$ (experimental or DNS data) at a certain state $\left( {{C_n},{\bf{x}},t} \right)$. ${{C_n} \left({n = 1,2,...,N}\right)}$ denotes model coefficients of the SGS mixed model ${\tau _{ij}} = \sum\limits_{n = 1}^N {{C_n}T_{ij}^{\left( n \right)}}$, and $t \in \left[ {0,T} \right]$ is the time horizon. Here, ``s.t.'' stands for the abbreviation of ``subject to''. ${R_0}$ and ${R_i} \left( {i = 1,2,3} \right)$ represent the LES continuity equation and momentum equations, respectively. 

The Lagrangian functional $\mathcal{L}$ is introduced to take the dynamics of LES variables ${\bf{\bar v}} = {\left[ {\bar p,{{\bar u}_1},{{\bar u}_2},{{\bar u}_3}} \right]^T}$ into account and convert the constrained optimization (Eq.~\ref{opt_problem}) into the unconstrained optimization problem, namely \citep{lewis2006}
\begin{equation}
	\mathop {{\rm{min}}}\limits_{{C_n}} \;\mathcal{L}\left( {{\bf{\bar v}};{C_n}} \right),\;{\rm{where}}\;\mathcal{L} = {\mathcal J}\left[ {\phi \left( {{\bf{\bar v}};{C_n}} \right),\phi \left( {{{{\bf{\bar v}}}^{{\rm{ref}}}}} \right)} \right] -  \sum\limits_{k = 0}^3 {\int\limits_0^T {\int\limits_\Omega  {{R_k}\left( {{\bf{\bar v}};{C_n}} \right) \cdot {{{{\bar v_k}}}^{{\dag}}}d{\bf{x}}dt} } }.
	\label{opt_problemL}
\end{equation}
Here, ${{{\bf{\bar v}}}^\dag } = {\left[ {{{\bar p}^\dag },\bar u_1^\dag,\bar u_2^\dag,\bar u_3^\dag } \right]^T}$ are the adjoint LES variables of ${\bf{\bar v}}$, where ${{{\bar p}^\dag }}$ and ${\bar u_i^\dag }$ are the adjoint pressure and adjoint velocity, respectively. For the sake of brevity, the inner product of time and space is defined by ${\left\langle {{\bf{f}},{\bf{g}}} \right\rangle _{{\bf{x}},t}} = \int\limits_0^T {\int\limits_\Omega  {{\bf{f}}\left( {{\bf{x}},t} \right) \cdot {\bf{g}}\left( {{\bf{x}},t} \right)d{\bf{x}}dt} }, $ where ${{\bf{f}}\left( {{\bf{x}},t} \right)}$ and ${{\bf{g}}\left( {{\bf{x}},t} \right)}$ denote the arbitrary physical variables. The Lagrangian functional $\mathcal{L}$ can be simplified as $\mathcal{L}\left( {{\bf{\bar v}};{C_n}} \right) = {\mathcal J}\left( {{\bf{\bar v}};{C_n}} \right) - \sum\limits_{k = 0}^3 {{{\left\langle {{R_k}\left( {{\bf{\bar v}};{C_n}} \right),{{{\bf{\bar v}}}^\dag }} \right\rangle }_{{\bf{x}},t}}}$.
The sensitivity of the Lagrangian functional $\mathcal{L}$ can be derived by
\begin{equation}
	\begin{aligned}
		\delta \mathcal{L}\left( {{\bf{\bar v}};{C_n}} \right) &= \delta {\mathcal J}\left( {{\bf{\bar v}};{C_n}} \right) - \sum\limits_{k = 0}^3 {{{\left\langle {{R_k}\left( {\delta {\bf{\bar v}};{C_n}} \right),{{{\bf{\bar v}}}^{\dag}}} \right\rangle }_{{\bf{x}},t}}}  - \sum\limits_{k = 0}^3 {{{\left\langle {{R_k}\left( {{\bf{\bar v}};\delta {C_n}} \right),{{{\bf{\bar v}}}^{\dag}}} \right\rangle }_{{\bf{x}},t}}} ,\\
		&= \delta {\mathcal J}\left( {{\bf{\bar v}};{C_n}} \right) - {\sum\limits_{k = 0}^3 \left\langle { {\frac{{\partial {R_k}\left( {{\bf{\bar v}};{C_n}} \right)}}{{\partial {\bf{\bar v}}}} \cdot \delta {\bf{\bar v}}} ,{{{\bf{\bar v}}}^{\dag}}} \right\rangle _{{\bf{x}},t}}  - \sum\limits_{k = 0}^3 {{{\left\langle {\frac{{\partial {R_k}\left( {{\bf{\bar v}};{C_n}} \right)}}{{\partial {C_n}}} \cdot \delta {C_n},{{{\bf{\bar v}}}^{\dag}}} \right\rangle }_{{\bf{x}},t}}} ,
	\end{aligned}
	\label{delta_L}
\end{equation}
where $\partial {R_k}/\partial {\bf{\bar v}}$ and $\partial {R_k}/\partial {C_n}$ are the tangent operators of the governing equations ${R_k}\;\left( {k = 0,1,2,3} \right)$ for the variables ${\bf{\bar v}}$ and parameters ${C_n}$ with the perturbation field $\delta {\bf{\bar v}} = {\bf{\bar v}}\left( {{C_n} + \delta {C_n}} \right) - {\bf{\bar v}}\left( {{C_n}} \right),\;n \in \left\{ {1,2,...,N} \right\}$. 
The first term in Eq.~\ref{delta_L} is the sensitivity of the cost functional ${\mathcal J}$ and calculated as the Gâteaux-Fréchet derivative \citep{bewley2001} of  ${\mathcal J}$ at $C_n$ in the direction $\delta C_n$, namely
\begin{equation}
	\delta {\mathcal J}\left( {{\bf{\bar v}};\delta {C_n}} \right) = \mathop {\lim }\limits_{\varepsilon  \to 0} \frac{d}{{d\varepsilon }}{\mathcal J}\left( {{\bf{\bar v}}\left( {{C_n} + \varepsilon \delta {C_n}} \right)} \right) = {\left\langle {\frac{{\partial { J}}}{{\partial {\bf{\bar v}}}},\delta {\bf{\bar v}}} \right\rangle _{{\bf{x}},t}}.   
	\label{delta_J}
\end{equation}
The adjoint identity \citep{bewley2001} can be obtained via the integral by part, given by
\begin{equation}
	{\left\langle {{\bf{R}}\left( {{\bf{\bar v}}} \right),{{{\bf{\bar v}}}^{\dag}}} \right\rangle _{{\bf{x}},t}} = {\left\langle {{\bf{\bar v}},{{\bf{R}}^{\dag}}\left( {{{{\bf{\bar v}}}^{\dag}}} \right)} \right\rangle _{{\bf{x}},t}} + {\left. {{{\left\langle {{\bf{\bar F}},{{{\bf{\bar v}}}^{\dag}}} \right\rangle }_t}} \right|_\Gamma } + \left. {{{\left\langle {{\bf{\bar v}},{{{\bf{\bar v}}}^{\dag}}} \right\rangle }_{\bf{x}}}} \right|_0^T = {\left\langle {{\bf{\bar v}},{{\bf{R}}^{\dag}}\left( {{{{\bf{\bar v}}}^{\dag}}} \right)} \right\rangle _{{\bf{x}},t}} + BT,
	\label{adj_identity}
\end{equation}
where the partial differential equations ${\bf{R}}\left( {{\bf{\bar v}}} \right) = \partial {\bf{\bar v}}/\partial t + \partial {\bf{\bar F}}/\partial {\bf{x}} = 0$ with the associated adjoint operator ${{{\bf{R}}^\dag }\left( {{{{\bf{\bar v}}}^\dag }} \right)}$, ${{\bf{\bar F}}}$ denotes the fluxes and $\Gamma $ is the boundary of the domain $\Omega$. Here, $BT = {\left. {{{\left\langle {{\bf{\bar F}},{{{\bf{\bar v}}}^{\dag}}} \right\rangle }_t}} \right|_\Gamma } + \left. {{{\left\langle {{\bf{\bar v}},{{{\bf{\bar v}}}^{\dag}}} \right\rangle }_{\bf{x}}}} \right|_0^T $ represents the boundary and temporal integral terms, which determines the boundary and terminal conditions of the adjoint equations to give $BT=0$. ${\left\langle {{\bf{f}},{\bf{g}}} \right\rangle _t} = \int\limits_0^T {{\bf{f}}\left( {{\bf{x}},t} \right) \cdot {\bf{g}}\left( {{\bf{x}},t} \right)dt}$ and  ${\left\langle {{\bf{f}},{\bf{g}}} \right\rangle _{\bf{x}}} = \int\limits_\Omega  {{\bf{f}}\left( {{\bf{x}},t} \right) \cdot {\bf{g}}\left( {{\bf{x}},t} \right)d{\bf{x}}}$ denote the temporal and spatial inner products, respectively.
The second term in Eq.~\ref{delta_L} can be expressed with the adjoint identity, namely \citep{bewley2001,delport2009a,delport2011a}
\begin{equation}
	{\left\langle {\frac{{\partial {R_k}}}{{\partial {\bf{\bar v}}}}\cdot\delta {\bf{\bar v}},{{{\bf{\bar v}}}^\dag }} \right\rangle _{{\bf{x}},t}} = {\left\langle {\delta {\bf{\bar v}}, {{\left( {\frac{{\partial {R_k}}}{{\partial {\bf{\bar v}}}}} \right)}^\dag } \cdot {{{\bf{\bar v}}}^\dag }} \right\rangle _{{\bf{x}},t}} + BT,
	\label{adj_identity1}
\end{equation}
where ${\left( {\partial {R_k}/\partial {\bf{\bar v}}} \right)^{\dag}}$ is the adjoint operator of the LES tangent Jacobian tensor ${\partial {R_k}/\partial {\bf{\bar v}}}$, $\left( {k = 0,1,2,3} \right)$.
Substitute the Fréchet derivative $\delta {\mathcal J}$ (Eq.~\ref{delta_J}) and the adjoint identity (Eq.~\ref{adj_identity1}) into the sensitivity of the Lagrangian functional $\mathcal{L}$ (Eq.~\ref{delta_L}), and we get
\begin{equation}
	\delta \mathcal{L}\left( {{\bf{\bar v}};{C_n}} \right) = {\left\langle {\frac{{\partial J}}{{\partial {\bf{\bar v}}}} - \sum\limits_{k = 0}^3 {{{\left( {\frac{{\partial {R_k}}}{{\partial {\bf{\bar v}}}}} \right)}^{\dag}} \cdot {{{\bf{\bar v}}}^{\dag}}} ,\delta {\bf{\bar v}}} \right\rangle _{{\bf{x}},t}} - \sum\limits_{k = 0}^3 {{{\left\langle {\frac{{\partial {R_k}\left( {{\bf{\bar v}};{C_n}} \right)}}{{\partial {C_n}}} \cdot \delta {C_n},{{{\bf{\bar v}}}^{\dag}}} \right\rangle }_{{\bf{x}},t}}}  - BT,
	\label{delta_L1}
\end{equation}
To avoid calculating the perturbation field $\delta {\bf{\bar v}}$ in the first term of Eq.~\ref{delta_L1}, the inner product should be equal to 0 and the corresponding adjoint LES equations can be derived by
\begin{equation}
	\sum\limits_{k = 0}^3 {{{\left( {\frac{{\partial {R_k}}}{{\partial {\bf{\bar v}}}}} \right)}^{\dag}} \cdot {{{\bf{\bar v}}}^{\dag}}} -\frac{{\partial J}}{{\partial {\bf{\bar v}}}} = 0.
	\label{adj_Eqn}
\end{equation}
Substitute the specific forms of the LES equations  ${R_k}\;\left( {k = 0,1,2,3} \right)$ (see Eq.~\ref{opt_problem}), and the adjoint LES equations can be written as
\begin{equation}
	\frac{{\partial \bar u_i^\dag }}{{\partial {x_i}}} = 0,
	\label{adj_LES1}
\end{equation}
\begin{equation}
	\frac{{\partial \bar u_i^{\dag}}}{{\partial t}} + \left( {\frac{{\partial \bar u_i^{\dag}}}{{\partial {x_j}}} + \frac{{\partial \bar u_j^{\dag}}}{{\partial {x_i}}}} \right){{\bar u}_j} + \frac{{\partial {{\bar p}^{\dag}}}}{{\partial {x_i}}} + \nu \frac{{{\partial ^2}\bar u_i^{\dag}}}{{\partial {x_j}\partial {x_j}}} + \frac{{\partial \tau _{ij}^{\dag}}}{{\partial {x_j}}} + \frac{{\partial J}}{{\partial {{\bar u}_i}}} = 0,
	\label{adj_LES2}
\end{equation}
where $\tau _{ij}^\dag = \sum\limits_{n = 1}^N {{C_n}T_{ij}^{\left( n \right),\dag }}$ denotes the adjoint SGS mixed model and ${T_{ij}^{\left( n \right),\dag }}$ is the $n$-th adjoint basis stress tensor. The detailed derivation of the adjoint LES equations can refer to the Appendix~\ref{AppendixA}. 
The terminal conditions of the adjoint LES equations is determined by the last term of adjoint identity (Eq.~\ref{adj_identity}), namely
\begin{equation}
	\left. {{{\left\langle {{{{\bf{\bar v}}}^{\dag}}, \delta {\bf{\bar v}}} \right\rangle }_{\bf{x}}}} \right|_0^T = {\left\langle {{{{\bf{\bar v}}}^\dag }\left( T \right) , \delta {\bf{\bar v}}\left( T \right)} \right\rangle _{\bf{x}}} - {\left\langle {{{{\bf{\bar v}}}^\dag }\left( 0 \right) , \delta {\bf{\bar v}}\left( 0 \right)} \right\rangle _{\bf{x}}} = {\left\langle {{{{\bf{\bar v}}}^\dag }\left( T \right), \delta {\bf{\bar v}}\left( T \right)} \right\rangle _{\bf{x}}},
	\label{adj_IC}
\end{equation}
where ${\delta {\bf{\bar v}}\left( 0 \right)}=0$, since the unperturbed initial LES field is exactly given by the filtered DNS (fDNS) data. The terminal conditions ${{{\bf{\bar v}}}^\dag }\left( T \right) = {\left[ {\bar u_i^\dag \left( T \right),{{\bar p}^\dag }\left( T \right)} \right]^T} = \bf{0}$ make the temporal integral terms $\left[ {{{\left\langle {\delta {\bf{\bar v}},{{{\bf{\bar v}}}^\dag }} \right\rangle }_{\bf{x}}}} \right]_0^T$ equal to zero and the calculation of the terminal perturbation ${\delta {\bf{\bar v}}\left( T \right)}$ is obviated.  
The terminal conditions ($\bar u_i^{\dag}\left( T \right) = 0$, ${{\bar p}^{\dag}}\left( T \right) = 0$) and boundary conditions of the adjoint LES equations are identified by setting $BT=0$ in Eq.~\ref{delta_L1}. The sensitivity of the Lagrangian functional $\mathcal{L}$ can be further expressed as
\begin{equation}
	\delta {\mathcal L}\left( {{\bf{\bar v}};{C_n}} \right) =  - \sum\limits_{k = 0}^3 {{{\left\langle {\frac{{\partial {R_k}\left( {{\bf{\bar v}};{C_n}} \right)}}{{\partial {C_n}}} \cdot \delta {C_n},{{{\bf{\bar v}}}^{\dag}}} \right\rangle }_{{\bf{x}},t}}} ,
	\label{delta_L2}
\end{equation}
where $\partial {R_0}/\partial {C_n} = 0$, and $\partial {R_i}/\partial {C_n} = \frac{\partial }{{\partial {C_n}}}\left( {\frac{{\partial {\tau _{ij}}}}{{\partial {x_j}}}} \right) = {\rm{ }}\partial T_{ij}^{\left( n \right)}/\partial {x_j} \left( {n = 1,2,...,N} \right)$ denotes the $n$-th SGS basis force.
Once the LES equations (Eqs.~\ref{fns1} and \ref{fns2}) temporally advances forward in the time horizon $t \in \left[ {0,T} \right]$ and the adjoint LES equations (Eqs.~\ref{adj_LES1} and \ref{adj_LES2}) are integrated backward with zero terminal conditions, the gradients of Lagrangian functional for the SGS model coefficients can be calculated efficiently by
\begin{equation}
	\frac{{\partial {\mathcal L}}}{{\partial {C_n}}} = \frac{{\delta {\mathcal L}\left( {{\bf{\bar v}};{C_n}} \right)}}{{\delta {C_n}}} =  - {\left\langle {\frac{{\partial T_{ij}^{\left( n \right)}}}{{\partial {x_j}}},\bar u_i^{\dag}} \right\rangle _{{\bf{x}},t}},\;\left( {n = 1,2,...,N} \right).
	\label{grad_L}
\end{equation}

The adjoint-based gradient evaluations are independent of the parameter perturbations ${\delta {C_n}} \left( {n = 1,2,...,N} \right)$, which are very efficient compared to the finite difference algorithm and forward sensitivity analysis with at least $N$ parameter perturbations and $N+1$ LES equation calculations for each optimization iteration \citep{chandramouli2020,sirignano2020,macart2021}.           

\subsection {Energy budget analysis of the adjoint LES equations}
Before proceeding to the introduction of the variational optimal mixed models, it is essential to analyze the energy budget of the adjoint LES equations. The adjoint LES kinetic energy (${{\bar {\mathcal E}}^\dag } = \bar u_i^\dag \bar u_i^\dag /2$) equation is derived through multiplying the adjoint velocity $\bar u_i^\dag$ on both sides of the adjoint LES momentum equations (Eq.~\ref{adj_LES2}), namely
\begin{equation}
	\frac{{\partial {{\bar {\mathcal E}}^\dag }}}{{\partial t}} + \frac{{\partial {{\bar {\mathcal P}}_j}}}{{\partial {x_j}}} = \bar u_i^\dag {{\bar S}_{ij}}\bar u_j^\dag  + {{\bar D}^\dag } - {{\bar \Pi }^\dag } - {{\bar J}^\dag },
	\label{EK_eqn}
\end{equation}
where ${\bar u_i^\dag {{\bar S}_{ij}}\bar u_j^\dag }$ denotes the adjoint energy production term due to the shear strain rate  $\bar S_{ij} $. Here, ${{\bar {\mathcal P}}_j}$ is the adjoint spatial transport flux, ${\bar D}$ is the adjoint viscous dissipation term, ${{\bar \Pi }^\dag }$ is the adjoint variable of the SGS energy flux $\bar \Pi  =  - {\tau _{ij}}{{\bar S}_{ij}}$  and ${{\bar J}^\dag }$ is the energy injected from the discrepancy between LES results and reference data. These terms are respectively defined by
\begin{equation}
	{{\bar {\mathcal P}}_j} = {{\bar {\mathcal E}}^\dag }{{\bar u}_j} + \left( {{{\bar p}^\dag } + {{\bar u}_i}\bar u_i^\dag } \right)\bar u_j^\dag  + \left( {\nu \frac{{\partial \bar u_i^\dag }}{{\partial {x_j}}} + \tau _{ij}^\dag } \right)\bar u_i^\dag,
	\label{EK_Pj}
\end{equation}
\begin{equation}
	\bar D = \nu \frac{{\partial \bar u_i^\dag }}{{\partial {x_j}}}\frac{{\partial \bar u_i^\dag }}{{\partial {x_j}}},
	\label{EK_D}
\end{equation}
\begin{equation}
	{{\bar \Pi }^\dag } =  - \tau _{ij}^\dag \bar S_{ij}^\dag,
	\label{EK_Pi}
\end{equation}
\begin{equation}
	{{\bar J}^\dag } = \bar u_i^\dag \frac{{\partial J}}{{\partial {{\bar u}_i}}},
	\label{EK_J}
\end{equation}
where $\bar S_{ij}^\dag = \left( {\partial \bar u_i^\dag /\partial {x_j} + \partial \bar u_j^\dag /\partial {x_i}} \right)/2 $ represents the adjoint strain-rate tensor.
The backward evolution of the adjoint volume-averaged kinetic energy can be written as
\begin{equation}
	- \frac{{\partial \left\langle {{{\bar {\mathcal E}}^\dag }} \right\rangle }}{{\partial t}} =  - \left\langle {\bar u_i^\dag {{\bar S}_{ij}}\bar u_j^\dag } \right\rangle  - \left\langle {{{\bar D}^\dag }} \right\rangle  + \left\langle {{{\bar \Pi }^\dag }} \right\rangle  + \left\langle {{{\bar J}^\dag }} \right\rangle ,
	\label{avgEK_eqn}
\end{equation}
where $\left\langle {{{\bar D}^\dag }} \right\rangle $ is pure dissipation term that drains out the adjoint energy. $\left\langle {{{\bar \Pi }^\dag }} \right\rangle$ denotes the adjoint SGS energy transport term which represents the forward adjoint energy transfer from large scales to unsolved residual scales if $\left\langle {{{\bar \Pi }^{\dag}}} \right\rangle  > 0$, otherwise stands for the adjoint SGS energy backscatter. The accurate reconstruction of $\left\langle {{{\bar \Pi }^\dag }} \right\rangle $ is crucial for the SGS modeling of LES and gradient evaluations with respect to the SGS model coefficients. $\left\langle {{{\bar J}^\dag }} \right\rangle $ is the loss-induced adjoint energy injection term. $\left\langle {{{\bar D}^\dag }} \right\rangle $ is the viscous dissipation which enhances the numerical stability of the adjoint LES field. $\left\langle {{{\bar J}^\dag }} \right\rangle $ is the adjoint energy production due to the discrepancy between LES evaluation and reference data, which dominates the accuracy of the sensitivity calculations.
The large-scale strain-rate tensor $\bar S_{ij}$ can be decomposed into its principal components using the eigendecomposition approach, such that \citep{wang2013d}
\begin{equation}
	{{\bar S}_{ij}}{\rm{ = }}{\lambda _1}q_i^{\left( 1 \right)}q_j^{\left( 1 \right)} + {\lambda _2}q_i^{\left( 2 \right)}q_j^{\left( 2 \right)} + {\lambda _3}q_i^{\left( 3 \right)}q_j^{\left( 3 \right)}=\sum\limits_{k = 1}^3 {{\lambda _k}q_i^{\left( k \right)}q_j^{\left( k \right)}},
	\label{Sij_decomp}
\end{equation}
where $\lambda_1$, $\lambda_2$ and $\lambda_3$ are the eigenvalues of the shear strain rate, with $q_i^{\left( 1 \right)}$, $q_i^{\left( 2 \right)}$ and $q_i^{\left( 3 \right)}$ being the associated eigenvectors. Here, ${\lambda _1} + {\lambda _2} + {\lambda _3} = 0$ for the trace-free strain rate ${{\bar S}_{ij}}$ in the incompressible turbulent flows. Hence, the quadratic term $-\left\langle {\bar u_i^\dag {{\bar S}_{ij}}\bar u_j^\dag } \right\rangle $ in Eq.~\ref{avgEK_eqn} is further expressed as 
\begin{equation}
	- \left\langle {\bar u_i^\dag {{\bar S}_{ij}}\bar u_j^\dag } \right\rangle  =  - \sum\limits_{k = 1}^3 {\left\langle {{\lambda _k}\left( {q_i^{\left( k \right)}\bar u_i^\dag } \right)\left( {q_j^{\left( k \right)}\bar u_j^\dag } \right)} \right\rangle }  =  - \sum\limits_{k = 1}^3 {\left\langle {{\lambda _k}{{\left( {q_i^{\left( k \right)}\bar u_i^\dag } \right)}^2}} \right\rangle } . 
	\label{uSu}
\end{equation}
The sign of the eigenvalues ${{\lambda _k}},\;\left( {k = 1,2,3} \right)$ determines the contribution of the adjoint energy from the quadratic term $-\left\langle {\bar u_i^\dag {{\bar S}_{ij}}\bar u_j^\dag } \right\rangle $ is productive or dissipative. The quadratic terms with negative eigenvalues of the shear strain rate produce the positive adjoint energy production, while those with positive eigenvalues drain out the adjoint energy. In previous studies of chaotic adjoint methods, the adjoint-based gradients are found to grow exponentially with time and finally numerically diverge in a long time horizon for the chaotic flows \citep{wang2013d,ashley2019,garai2021}. The terms $\left\langle {{{\bar D}^\dag }} \right\rangle $,  $\left\langle {{{\bar \Pi }^\dag }} \right\rangle$  and $\left\langle {{{\bar J}^\dag }} \right\rangle $ in the volume-averaged adjoint energy equation (Eq.~\ref{avgEK_eqn}) are less likely to cause the exponential growth of the adjoint energy, since the adjoint energy term $\left\langle {{{\bar {\mathcal E}}^\dag }} \right\rangle$ does not appears explicitly in these terms. It can be further shown that the quadratic term $-\left\langle {\bar u_i^\dag {{\bar S}_{ij}}\bar u_j^\dag } \right\rangle $ plays the dominant role in the exponential growth of the adjoint variables. We apply the Cauchy-Schwarz inequality to the inner product terms in Eq.~\ref{uSu}, such that \citep{talnikar2017}
\begin{equation}
	{\left( {q_i^{\left( k \right)}\bar u_i^\dag } \right)^2} \le \left[ {q_i^{\left( k \right)}q_i^{\left( k \right)}} \right]\left( {\bar u_i^\dag \bar u_i^\dag } \right) = 2\left\| {{{\bf{q}}^{\left( k \right)}}} \right\|\overline {\mathcal E}^\dag \;\;\left( {k = 1,2,3} \right),
	\label{Cauchy_ieq}
\end{equation}
where ``$\left\|  \cdot  \right\|$'' denotes the L2 norm of the vectors. For the quadratic terms with negative eigenvalues (adjoint energy production), the evolution of the adjoint energy can be approximated using the leading principal vectors as
\begin{equation}
	- \frac{{\partial \left\langle {{{\overline {\mathcal E} }^\dag}} \right\rangle }}{{\partial t}} \approx 2{\left| \lambda  \right|_\infty } {\left\| {\bf{q}} \right\|_\infty }\left\langle {{{\overline {\mathcal E} }^\dag}} \right\rangle,
	\label{dadjE_dt}
\end{equation}
where ${\left| \lambda  \right|_\infty } = \mathop {\max }\limits_\Omega  \left\{ { - {\lambda _1}, - {\lambda _2}, - {\lambda _3}} \right\}$ denotes the magnitude of the leading negative eigenvalue in the entire domain $\Omega$ and ${{{\left\| {\bf{q}} \right\|}_\infty }}$ represents the corresponding eigenvector magnitude. The adjoint energy is then calculated by the backward time interval, namely 
\begin{equation}
	\left\langle {{{\overline {\mathcal E} }^\dag }} \right\rangle \left( t \right) \approx \left\langle {{{\overline {\mathcal E} }^\dag }} \right\rangle \left( T \right)\exp \left[ {2{{\left| \lambda  \right|}_\infty } {{\left\| {\bf{q}} \right\|}_\infty }\left( {T - t} \right)} \right],\;\;t \in \left[ {0,T} \right].
	\label{adjE_t}
\end{equation}
The quadratic term $-\left\langle {\bar u_i^\dag {{\bar S}_{ij}}\bar u_j^\dag } \right\rangle $ with negative eigenvalues makes the adjoint energy grow exponentially over time and numerically unstable if it cannot be suppressed by the adjoint dissipation in a long time horizon $t \in \left[ {0,T} \right]$.
 To stabilize the adjoint equations during every iteration, an additional symmetric tensor $\bar S_{ij}^a$ \citep{ashley2019,garai2021} is introduced to maintain the numerical stability of the adjoint momentum (Eq.~\ref{adj_LES2}), and the stabilized adjoint momentum equations are then expressed as
\begin{equation}
	\frac{{\partial \bar u_i^{\dag}}}{{\partial t}} + \left( {\frac{{\partial \bar u_i^{\dag}}}{{\partial {x_j}}} + \frac{{\partial \bar u_j^{\dag}}}{{\partial {x_i}}}} \right){{\bar u}_j} + \bar S_{ij}^a\bar u_j^{\dag} + \frac{{\partial {{\bar p}^{\dag}}}}{{\partial {x_i}}} + \nu \frac{{{\partial ^2}\bar u_i^{\dag}}}{{\partial {x_j}\partial {x_j}}} + \frac{{\partial \tau _{ij}^{\dag}}}{{\partial {x_j}}} + \frac{{\partial J}}{{\partial {{\bar u}_i}}} = 0.
	\label{adj_MLES2}
\end{equation}
Consequently, the stabilized adjoint kinetic energy equation is written by
\begin{equation}
	\frac{{\partial {{\overline {\mathcal E} }^{\dag}}}}{{\partial t}} + \frac{{\partial {{\overline {\mathcal P} }_j}}}{{\partial {x_j}}} = \bar u_i^{\dag}\left( {{{\bar S}_{ij}} - \bar S_{ij}^a} \right)\bar u_j^{\dag} + {{\bar D}^{\dag}} - {{\bar \Pi }^{\dag}} - {{\bar J}^{\dag}}.
	\label{adj_MEK}
\end{equation}
Here, the quadratic term $\bar u_i^\dag {{\bar S}_{ij}}\bar u_j^\dag  < 0\;\left( { - \bar u_i^\dag {{\bar S}_{ij}}\bar u_j^\dag  > 0} \right)$ is responsible for the exponential growth of the adjoint energy, and the minimal artificial symmetric tensor is added to keep the adjoint variables numerically stable in advancing backward of the adjoint LES equations. The artificial symmetric tensor $\bar S_{ij}^a$ can be optimized by the suboptimal minimization problem \citep{ashley2019,garai2021}, such that
\begin{equation}
	\begin{array}{*{20}{c}}
		{\mathop {\min }\limits_{\bar S_{ij}^a} }&{\frac{1}{2}\bar S_{ij}^a\bar S_{ij}^a,}\\
		{s.t.}&{\bar u_i^\dag \left( {{{\bar S}_{ij}} - \bar S_{ij}^a} \right)\bar u_j^\dag  \ge 0.}
	\end{array}
	\label{Sija_SQP}
\end{equation}
We use the sequential quadratic programming (SQP) approach \citep{boggs1995,chung2022a} to efficiently solve the suboptimal problem, and the augmented Lagrangian functional $\mathbb{L}$ is applied to the constrained minimization problem, namely
\begin{equation}
	\mathbb{L}= \frac{1}{2}\bar S_{ij}^a\bar S_{ij}^a + \lambda \left[ {\bar u_i^\dag \left( {{{\bar S}_{ij}} - \bar S_{ij}^a} \right)\bar u_j^\dag } \right],
	\label{L_SQP}
\end{equation}
where $\lambda$ is the Lagrangian multiplier. The Karush–Kuhn–Tucker (KKT) optimal conditions \citep{kuhn1951,blonigan2018a} are obtained by taking the derivatives of the cost functional with respect to the augmented optimal variables ( $\bar S_{ij}^a$ and $\lambda$), derived by
\begin{equation}
	\frac{{\partial \mathbb{L}}}{{\partial \bar S_{ij}^a}} = \bar S_{ij}^a - \lambda \left( {\bar u_i^\dag \bar u_j^\dag } \right) = 0\; \Rightarrow \; \bar S_{ij}^a = \lambda \left( {\bar u_i^\dag \bar u_j^\dag } \right),\
	\label{KKT1}
\end{equation}
\begin{equation}
	\frac{{\partial \mathbb{L}}}{{\partial \lambda }} = \bar u_i^\dag \left( {{{\bar S}_{ij}} - \bar S_{ij}^a} \right)\bar u_j^\dag  = 0\; \Rightarrow \; \bar u_i^\dag {{\bar S}_{ij}}\bar u_j^\dag  = \bar u_i^\dag \bar S_{ij}^a\bar u_j^\dag.
	\label{KKT2}
\end{equation}
By multiplying Eq.~\ref{KKT1} by $\bar u_i^\dag$ from the left and right by $\bar u_j^\dag$ , and then substituting it into Eq.~\ref{KKT2}, the Lagrangian multiplier $\lambda$ is calculated by
\begin{equation}
	\lambda  = \frac{{\bar u_i^\dag {{\bar S}_{ij}}\bar u_j^\dag }}{{{{\left( {\bar u_k^\dag \bar u_k^\dag } \right)}^2}}}= \frac{{\bar u_i^\dag {{\bar S}_{ij}}\bar u_j^\dag }}{{4{{\overline {\mathcal E} }^\dag }^2}}.
	\label{lambda}
\end{equation}
The minimal artificial symmetric tensor $\bar S_{ij}^a$ can be obtained by substituting  Eq.~\ref{lambda} into Eq.~\ref{KKT1}, yielding
\begin{equation}
	\bar S_{ij}^a = \left\{ {\begin{array}{*{20}{c}}
			{\frac{{\bar u_m^\dag {{\bar S}_{mn}}\bar u_n^\dag }}{{4{{\overline {\mathcal E} }^\dag }^2}}\bar u_i^\dag \bar u_j^\dag ,}&{\text{if}\;\bar u_i^\dag {{\bar S}_{ij}}\bar u_j^\dag  < 0,}\\
			{0\;\;\;\;\;\;\;\;\;\;\;\;\;\;\;\;,}&{\text{if}\;\bar u_i^\dag {{\bar S}_{ij}}\bar u_j^\dag  \ge 0.}
	\end{array}} \right.
	\label{Sija}
\end{equation}
The artificial momentum term $\bar S_{ij}^a\bar u_j^\dag $ is thus additionally calculated in the stabilized adjoint momentum equations (Eq.~\ref{adj_MLES2}), namely
\begin{equation}
	\bar S_{ij}^a\bar u_j^\dag  = \left\{ {\begin{array}{*{20}{c}}
			{\frac{{\bar u_m^\dag {{\bar S}_{mn}}\bar u_n^\dag }}{{2{{\overline {\cal E} }^\dag }}}\bar u_i^\dag ,}&{\text{if}\;\bar u_i^\dag {{\bar S}_{ij}}\bar u_j^\dag  < 0,}\\
			{0\;\;\;\;\;\;\;\;\;\;\;\;\;\;\;,}&{\text{if}\;\bar u_i^\dag {{\bar S}_{ij}}\bar u_j^\dag  \ge 0.}
	\end{array}} \right.
	\label{SaijUj}
\end{equation}
The minimal stabilization term $\bar S_{ij}^a\bar u_j^\dag $ can efficiently maintain the numerical stability of LES adjoint variables in the long-term chaotic turbulent calculations as much as possible, without deteriorating the correct evaluations of the adjoint-based gradient.

\begin{figure}\centering
	\includegraphics[width=1.0\textwidth]{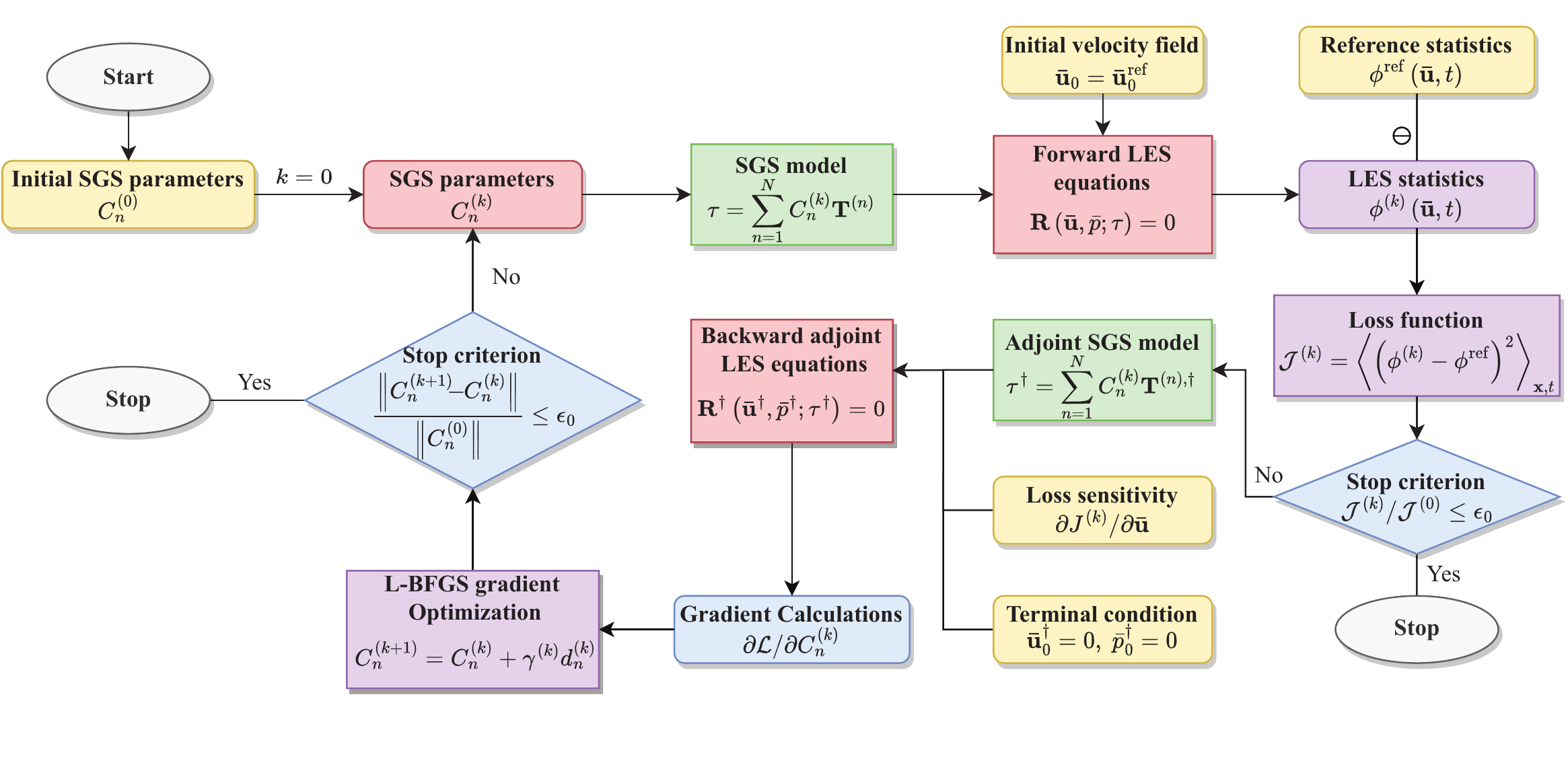}
	\caption{Schematic diagram of the adjoint-based variational optimal mixed models.}\label{fig:1}
\end{figure}

\subsection {Adjoint-based variational optimal mixed models (VOMM)}
In this research, we select the mixed model comprised of the Smagorinsky dissipative term (Eq.~\ref{tau_sm}) and approximate deconvolution model (ADM, Eq.~\ref{u_AD}) in the scale-similarity form, expressed as \citep{sagaut2006}
\begin{equation}
	{\tau _{ij}} = {C_1}\left( {{{\bar \Delta }^2}|\bar S|{{\bar S}_{ij}}} \right) + {C_2}\left( {\overline {u_i^*u_j^*}  - \overline {u_i^*} \;\overline {u_j^*}} \right),
	\label{VOMM}
\end{equation}
where ${u_i^*}$ denotes the approximate unfiltered velocity recovered by the iterative van Cittert procedure (Eq.~\ref{u_AD}). In previous studies \citep{yuan2020}, we have conducted error analyses to validate that deconvolutional-type SGS models with scale-similarity form (Eq.~\ref{ADM}) perform better than those with the conventional direct-modeling form ($\tau_{i j}=\overline{u_i^* u_j^*}-\bar{u}_i \bar{u}_j$), satisfying the properties of symmetry and realizability conditions. The model coefficients $C_1$ and $C_2$ are optimally identified by minimizing the discrepancy between statistical quantities calculated by the LES results and those measured by the filtered DNS (fDNS) data. The selected statistics should be able to sufficiently quantify the multiscale transport behaviours of turbulence, meanwhile facilitating the practical measurements. The SGS stress $\tau_{ij}$ and SGS force $\partial {\tau _{ij}}/\partial {x_j}$ are intermediate variables, and their statistics are relatively difficult to be obtained through the actual observations. In contrast, the statistics of velocity are more convenient to measure and the velocity spectrum clearly quantifies the turbulent kinetic energy distributions at different scales. The SGS modeling is especially concerned with the accurate reconstruction of small scales near the filter width, therefore we select the dissipation spectrum as the optimization statistical quantities $\phi \left( {{{\bar u}_i}} \right)$ to increase the weights of small scales, namely \citep{pope2000}
\begin{equation}
	\phi \left( {{{\bar u}_i},k,t} \right) = {D}\left( {k,t} \right) =  \int\limits_{\bf{k}} {\nu {k ^2}\bar v_i^*\left( {{\bf{k}},t} \right){{\bar v}_i}\left( {{\bf{k}},t} \right)\delta \left( {\left| {\bf{k}} \right| - k} \right)d{\bf{k}}} ,
	\label{DK}
\end{equation}
where $\delta \left(  \cdot  \right)$ denotes the Dirac delta function and the star symbol represents complex conjugate. $k $ and ${\bf{k}}$ stand for the wavenumber magnitude and wavenumber vectors, respectively. Here, ${{\bar v}_j}\left( {{\bf{\kappa }},t} \right)=\mathbb{F}\left\{ {{{\bar u}_j}\left( {{\bf{x}},t} \right)} \right\} = \sum\limits_{\bf{k}} {{{\bar u}_j}\left( {{\bf{x}},t} \right){e^{ - i{\bf{k}} \cdot {\bf{x}}}}}$ is the $j$-th velocity component in Fourier space, where $\mathbb{F}\left\{ \cdot \right\}$ represents the 3D Fourier transform, and $i$ is the imaginary unit with ${i^2} =  - 1$. 

\begin{table}
	\begin{center}	
		\normalsize
		\caption{ One-point statistics for the DNS of forced homogeneous isotropic turbulence with grid resolution of $1024^3$.}\label{tab:1}
		\small 
		\begin{tabular*}{0.95\textwidth}{@{\extracolsep{\fill}}lccccccccccc}
			\hline\hline
			${ {\rm Re}_{\lambda }}$ & $E_{k}$ & $k_{\rm{max}} \eta$ &  $\eta /{{h }_{\rm{DNS}}}$ & ${{L}_{I}}/\eta $ & $\lambda /\eta$ & ${{u}^{\rm {rms}}}$ & ${{\omega}^{\rm {rms}}}$ &  $\varepsilon $  \\  \hline
			252   & 2.63 & 2.11 & 1.01  & 235.2 & 31.2 & 2.30 & 26.90 & 0.73  \\	\hline\hline
		\end{tabular*}
	\end{center}
\end{table}

The optimization problem constrained by the governing equations for the SGS parameters $C_1$ and $C_2$ is defined in Eq.~\ref{opt_problem}, where the cost functional for the dissipation spectrum ${D}\left( {k,t} \right)$ is given by 
\begin{equation}
	{\cal J}\left( {\phi ,{\phi ^{{\rm{fDNS}}}}} \right) = \int\limits_0^T {\sum\limits_{k = 1}^{{k_{\max }}} {J\left[ {{D}\left( {k,t} \right),D^{{\rm{fDNS}}}\left( {k,t} \right)} \right]} dt} ,
	\label{J_VOMM}
\end{equation}
where ${{k_{\max }}=N_{\rm{LES}}/3}$ is the effective maximum wavenumber, $N_{\rm{LES}}$ is the number of LES grids, and the discrepancy function $J\left[ {{{D}}\left( {k,t} \right),{{D^{{\rm{fDNS}}}}}\left( {k,t} \right)} \right] = {\left[ {{{D}} \left( {k,t} \right) - {{{D^{{\rm{fDNS}}}} }}\left( {k,t} \right)} \right]^2}$ takes the L2 norm of the prediction error.

The gradients of the loss function with respect to the model coefficients $C_1$ and $C_2$ are evaluated by Eq.~\ref{grad_L}, where the adjoint variables $\bar u_i^{\dag}$ are calculated by backward advancing the stabilized adjoint LES equations (Eqs.~\ref{adj_LES1} and \ref{adj_MLES2}) with zero terminal conditions. The sensitivity term $\partial J/\partial {{\bar u}_i}$ is calculated by the chain rule, namely    
\begin{equation}
	\frac{{\partial J}}{{\partial {{\bar u}_i}}} = \frac{{\partial J}}{{\partial {D}}}\frac{{\partial {D}}}{{\partial {{\bar u}_i}}} = 2\left( {{D} - D^{{\rm{fDNS}}}} \right){\mathbb{F}^{ - 1}}\left\{ 2\nu{{k^2}{{\bar v}_i}\left( {{\bf{k}},t} \right)\delta \left( {\left| {\bf{k}} \right| - k} \right)} \right\},
	\label{dJ_du}
\end{equation}
where $\mathbb{F}^{-1}\left\{ \cdot \right\}$ denotes the 3D inverse Fourier transform. 
In the stabilized adjoint momentum equations (Eq.~\ref{adj_MLES2}), the adjoint SGS stress is given by $\tau _{ij}^\dag  = {C_1}T_{ij}^{\left( 1 \right),\dag } + {C_2}T_{ij}^{\left( 2 \right),\dag }$ , where the associated adjoint basis stress tensors $T_{ij}^{\left( 1 \right),\dag }$ and $T_{ij}^{\left( 2 \right),\dag }$ are expressed in detail as
\begin{equation}
	T_{ij}^{\left( 1 \right),\dag } = -{{\bar \Delta }^2}\left( {\left| {\bar S} \right|\bar S_{ij}^\dag  + 2\frac{{{{\bar S}_{kl}}\bar S_{kl}^\dag }}{{\left| {\bar S} \right|}}{{\bar S}_{ij}}} \right),
	\label{Tij1}
\end{equation}
\begin{equation}
	T_{ij}^{\left( 2 \right),\dag } = \sum\limits_{n = 1}^N {{{\left( {I - G} \right)}^{n - 1}} \otimes \left( {\overline {\bar u_i^\dag \overline {u_j^*} }  - \overline {\bar u_i^\dag } u_j^*} \right)},
	\label{Tij2}
\end{equation}
where $N=5$ denotes the number of iterations for the AD procedure. The detailed derivation of the adjoint SGS stress tensors for the VOMM model can refer to the Appendix~\ref{AppendixB}. To our knowledge, few previous works have studied the mixed SGS models and given the detailed derivations of the adjoint SGS models. 

\begin{figure}\centering
	\begin{subfigure}{0.5\textwidth}
		\centering
		{($a$)}
		\includegraphics[width=0.9\linewidth,valign=t]{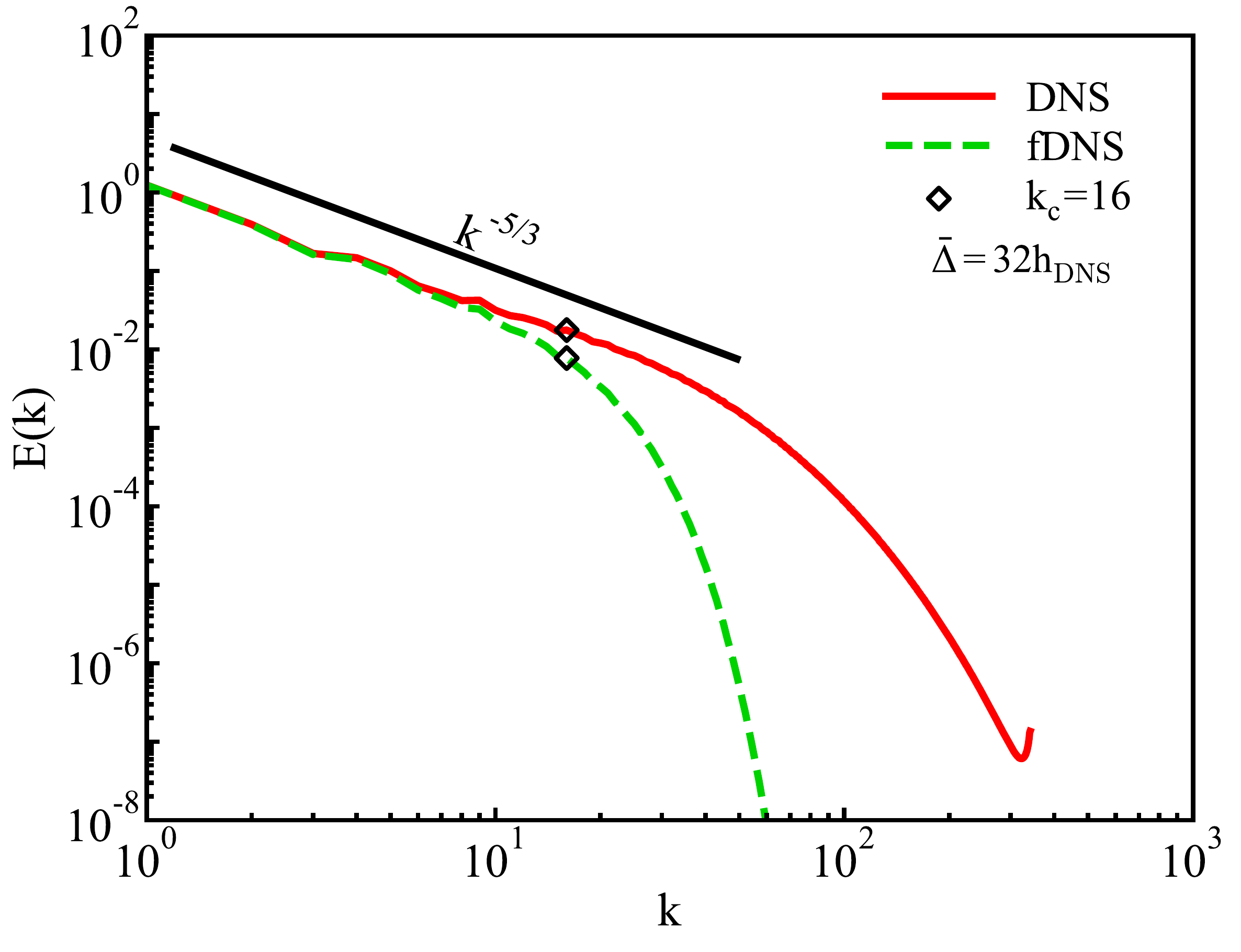}
	\end{subfigure}%
	\begin{subfigure}{0.5\textwidth}
		\centering
		{($b$)}
		\includegraphics[width=0.9\linewidth,valign=t]{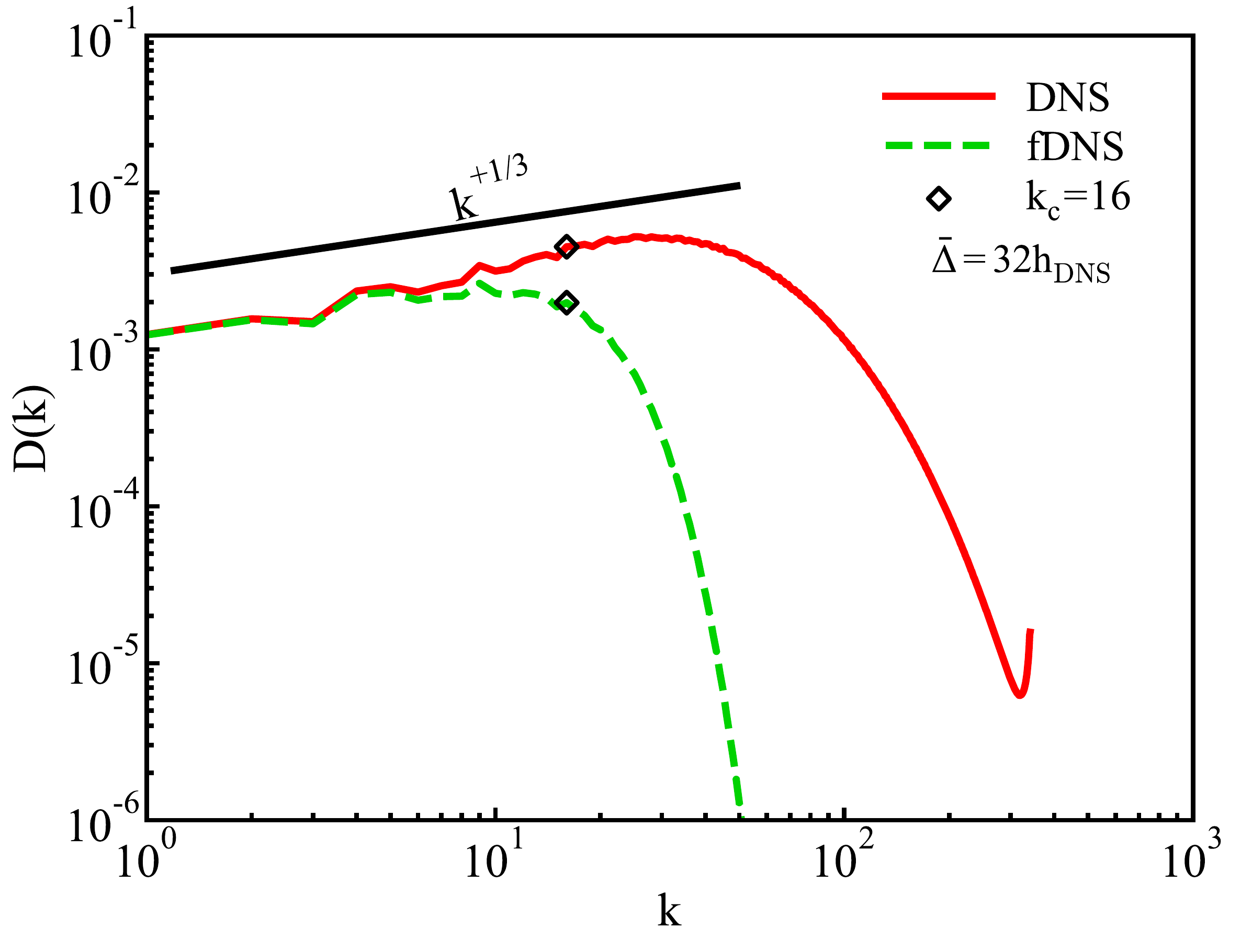}
	\end{subfigure}
	\caption{Velocity and dissipation spectra of DNS and filtered DNS in forced homogeneous isotropic turbulence with grid resolution of $1024^3$: ($a$) velocity spectra, and ($b$) dissipation spectra. Diamond represent the cutoff wavenumber $k_c$=16 ($\bar \Delta=32h_{\rm{DNS}}$).}\label{fig:2}
\end{figure}

Once the gradients of the cost functional for the model coefficients are obtained by successively solving the forward LES equations and backward stabilized adjoint LES equations, a gradient-based iterative optimization procedure can be established, namely \citep{liu1989,badreddine2014}
\begin{equation}
	C_n^{\left( {k + 1} \right)} = C_n^{\left( k \right)} + {\gamma ^{\left( k \right)}}d_n^{\left( k \right)},\;\;\left( {n = 1,2, \cdots ,N} \right),
	\label{opt_iterative}
\end{equation}
where $C_n^{\left( k \right)}$ is the $n$-th model coefficient during the $k$-th gradient-based optimal iteration, $d_n^{\left( k \right)}$ denotes the updated direction of the $n$-th model coefficient and ${\gamma ^{\left( k \right)}}$ represents the step size. We use a popular quasi-Newton method named limited-memory Broyden–Fletcher–Goldfarb–Shanno (L-BFGS) algorithm to update the directions $d_n^{\left( k \right)}$ \citep{liu1989}. The step size  ${\gamma ^{\left( k \right)}}$ is calculated by the backtracking-Armijo line search method in the L-BFGS algorithm \citep{armijo1966}. 

\begin{figure}\centering
	\includegraphics[width=0.7\textwidth]{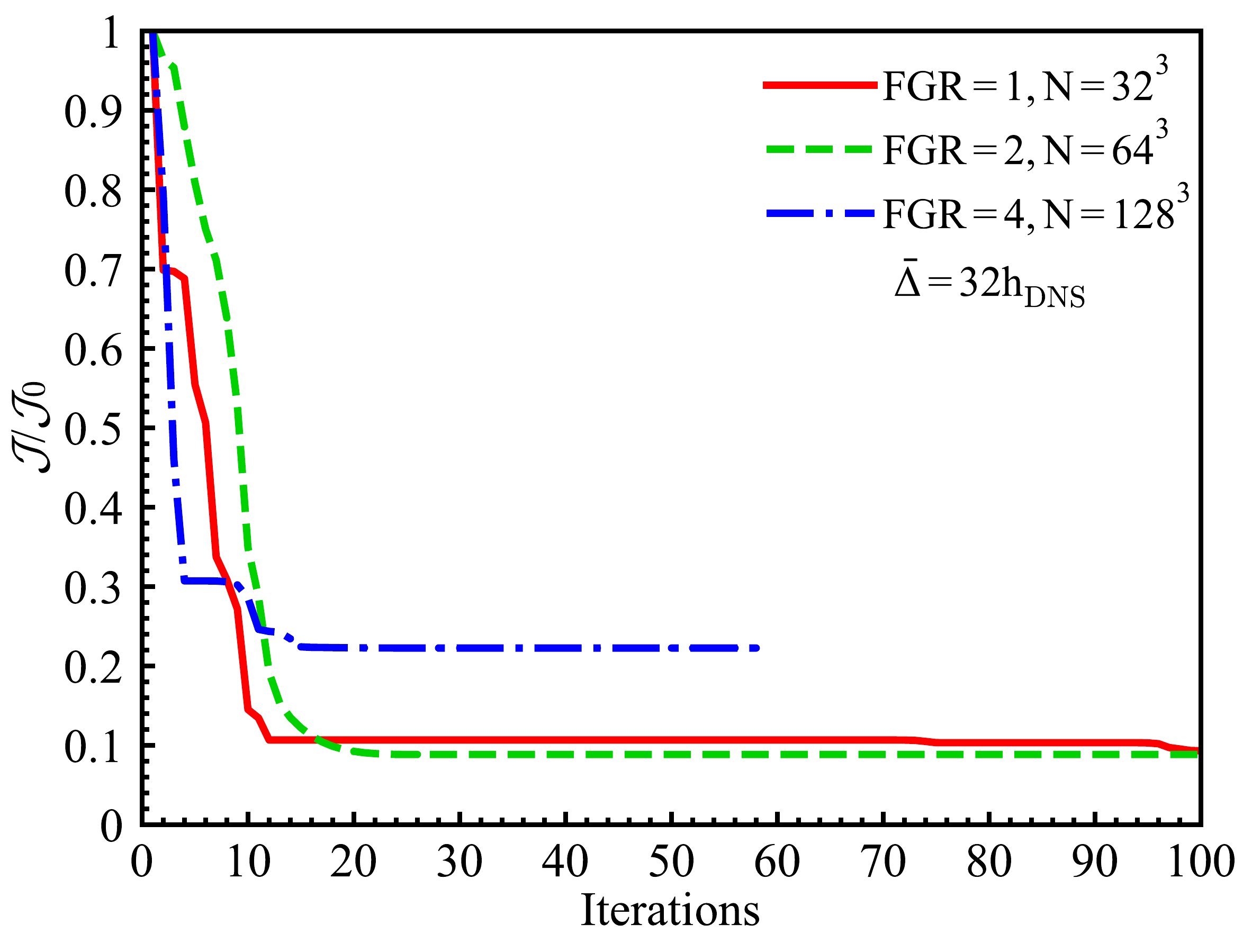}
	\caption{The evolution of the normalized cost function in forced homogeneous isotropic turbulence.}\label{fig:3}
\end{figure}

In summary, the diagram of the VOMM model is illustrated in Fig.~\ref{fig:1}, and the calculation steps are listed as follows.

(1) We first select the pure structural ADM model without the dissipative Smagorinsky term as the initial SGS model (Eq.~\ref{VOMM}) with model coefficients $C_1^{\left( 0 \right)}=0$ and  $C_2^{\left( 0 \right)}=1$.

(2) The LES transient statistics (\emph{e.g.} the dissipation spectrum shown in Eq.~\ref{DK}) is then evaluated by forward calculating the LES equations (Eqs.~\ref{fns1} and \ref{fns2}) initialized by the filtered DNS velocity field. The statistical discrepancy (Eq.~\ref{J_VOMM}) between the LES statistics and the \emph{a priori} measurable benchmark data (fDNS data) is measured to evaluate the performance of the SGS model with current parameters.

(3) Afterwards, the stabilized adjoint LES equations (Eqs.~\ref{adj_LES1} and \ref{adj_MLES2}) are integrated backward with zero terminal conditions, driven by the loss sensitivity (Eq.~\ref{dJ_du}) and corresponding adjoint SGS model (Eqs.~\ref{Tij1} and \ref{Tij2}). The adjoint-based gradients of augmented functional with respect to the model coefficients (Eq.~\ref{grad_L}) are sequentially evaluated using the adjoint variables and the SGS basis forces.

(4) The L-BFGS gradient-based optimization algorithm (Eq.~\ref{opt_iterative}) is adopted to iteratively update the SGS model parameters by repeating the above calculations until the stopping criteria are satisfied.

The stop criteria for the VOMM model for the optimization iterations are summarized as follows: 

(a) the number of iterations reaches the maximum number of iterations;

(b) the ratio of the current loss to the initial loss is smaller than a given error threshold $\epsilon_0$ (\emph{e.g.}, $\epsilon_0=1 \%$) , namely, ${{\mathcal J}^{\left( k \right)}}/{{\mathcal J}^{\left( 0 \right)}} \le \epsilon_0 $; 

(c) the difference of model coefficients between two successive iterations is negligible, namely, $\left\| {C_n^{\left( {k + 1} \right)} - C_n^{\left( k \right)}} \right\| / \left\| C_n^{\left( {0} \right)} \right\| \le \epsilon_0 $.

Eventually, the optimal parameters of the VOMM model are automatically obtained after reaching the given stopping optimization criteria.  

\begin{table}
	\begin{center}	
		\caption{The initial and optimal parameters of the VOMM model for LES computations with filter width $\bar \Delta = 32 h_{\rm{DNS}}$ in forced homogeneous isotropic turbulence.} 
		\small 
		\begin{tabular*}{0.95\textwidth}{@{\extracolsep{\fill}} lccccc}
			\hline\hline	
			FGR & LES Resolution & $C_1^{\left( 0 \right)}$ & $C_2^{\left( 0 \right)}$ & $C_1^{\rm{opt}}$ & $C_2^{\rm{opt}}$ \\ \hline
			1 & $32^3$  & 0 & 1 & -0.0529 & 1.229 \\ 
			2 & $64^3$  & 0 & 1 & -0.0101 & 1.027 \\ 
			4 & $128^3$ & 0 & 1 & -0.0030 & 1.000 \\ \hline\hline
		\end{tabular*}%	
		\label{tab:2}%	
	\end{center}
\end{table}% 

\begin{table}
	\begin{center}	
		\caption{The average computational cost of SGS stress modeling $\tau_{ij}$ for LES computations with filter width $\bar \Delta=32 h_{\rm{DNS}}$ in forced homogeneous isotropic turbulence.} 
		\small 	
		\begin{tabular*}{0.95\textwidth}{@{\extracolsep{\fill}} lccccc}
		\hline\hline
			Model(FGR=1,$N=32^3$)       & DSM   & DMM    & ADM($\chi$=0) & ADM($\chi$=1) & VOMM \\ \hline
			t(CPU$\cdot$s) & 0.142 & 0.243 & 0.056 & 0.056 & 0.066 \\
			t/t$_{\rm{DMM}}$ & 0.584 & 1 & 0.231 & 0.230 & 0.273 \\ 
			\hline
			Model(FGR=2,$N=64^3$)       & DSM   & DMM    & ADM($\chi$=0) & ADM($\chi$=1) & VOMM \\ 
			t(CPU$\cdot$s) & 0.870 & 1.465 & 0.368 & 0.361 & 0.418 \\
			t/t$_{\rm{DMM}}$ & 0.594 & 1 & 0.251 & 0.246 & 0.285 \\ 
			\hline
			Model(FGR=4,$N=128^3$)       & DSM   & DMM    & ADM($\chi$=0) & ADM($\chi$=1) & VOMM \\ 
			t(CPU$\cdot$s) & 6.512 & 10.103 & 2.517 & 2.588 & 3.240 \\
			t/t$_{\rm{DMM}}$ & 0.645 & 1 & 0.249 & 0.256 & 0.321 \\ \hline\hline
		\end{tabular*}%	
		\label{tab:3}%	
	\end{center}
\end{table}% 

\section {\emph{A posteriori} studies of the VOMM models}\label{sec:level5}
To examine the performance of the proposed VOMM model, the \emph{a posteriori} evaluations are respectively carried out for the forced, decaying homogeneous isotropic turbulence and temporally evolving turbulent mixing layer in this paper.
The results of the filtered direct numerical simulation (DNS) are the benchmark for the performance evaluations of the large-eddy simulation (LES). We first introduce the detailed settings of DNS for these three turbulent problems. The DNS data are then explicitly filtered by the commonly-used Gaussian filter, which is expressed as \citep{pope2000,sagaut2006}
\begin{equation}
	G\left( \mathbf{r}; \bar \Delta \right)={{\left( \frac{6}{\pi {{\bar \Delta }^{2}}} \right)}^{1/2}}\exp \left( -\frac{6{{\mathbf{r}}^{2}}}{{{\bar \Delta }^{2}}} \right).
	\label{G}
\end{equation}
The filter scale $\bar \Delta  = 32{h_{{\rm{DNS}}}}$ is selected for both the forced and decaying homogeneous isotropic turbulence, while $\bar \Delta  = 8{h_{{\rm{DNS}}}}$ for the temporally evolving turbulent mixing layer, where $h_{\rm{DNS}}$ denotes the grid spacing of DNS. Three conventional SGS models, \emph{i.e.}, the dynamic Smagorinsky model (DSM, Eq.~\ref{tau_sm}), the dynamic mixed model (DMM, Eq.~\ref{dmm1}) and the approximate deconvolution model with standard secondary filtering regularization (ADM, Eqs.~\ref{u_AD} $\sim $ \ref{ADM_relax}) are adopted to compare against the VOMM model. The consistent instantaneous snapshots of the filtered DNS data are used to initialize the LES calculations for different SGS models. Both the turbulent statistics and transient contours are evaluated and compared with different SGS models for the \emph{a posteriori} testings of the three canonical turbulent flows.

\begin{figure}\centering
	\begin{subfigure}{0.5\textwidth}
		\centering
		{($a$)}
		\includegraphics[width=0.9\linewidth,valign=t]{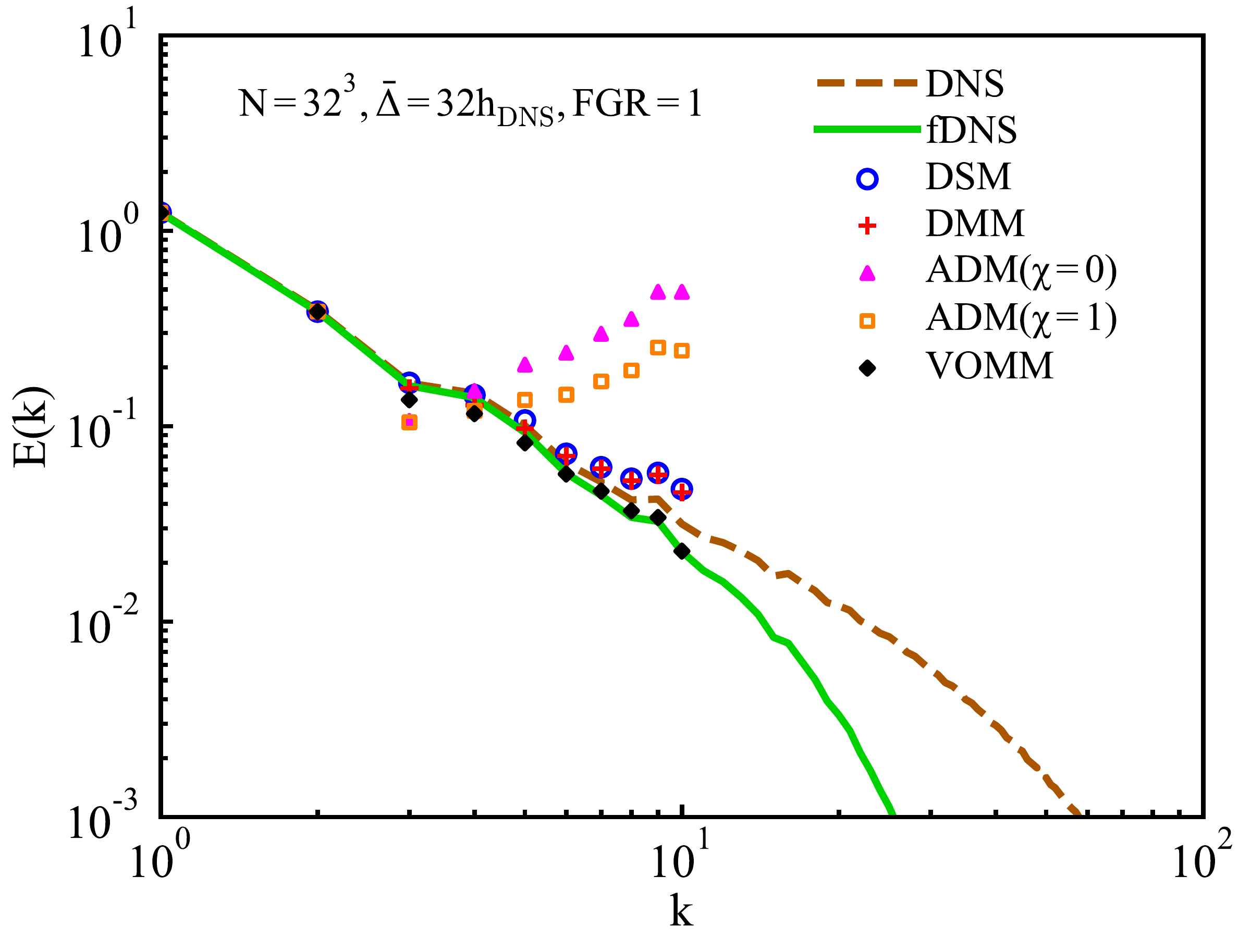}
	\end{subfigure}%
	\begin{subfigure}{0.5\textwidth}
		\centering
		{($b$)}
		\includegraphics[width=0.9\linewidth,valign=t]{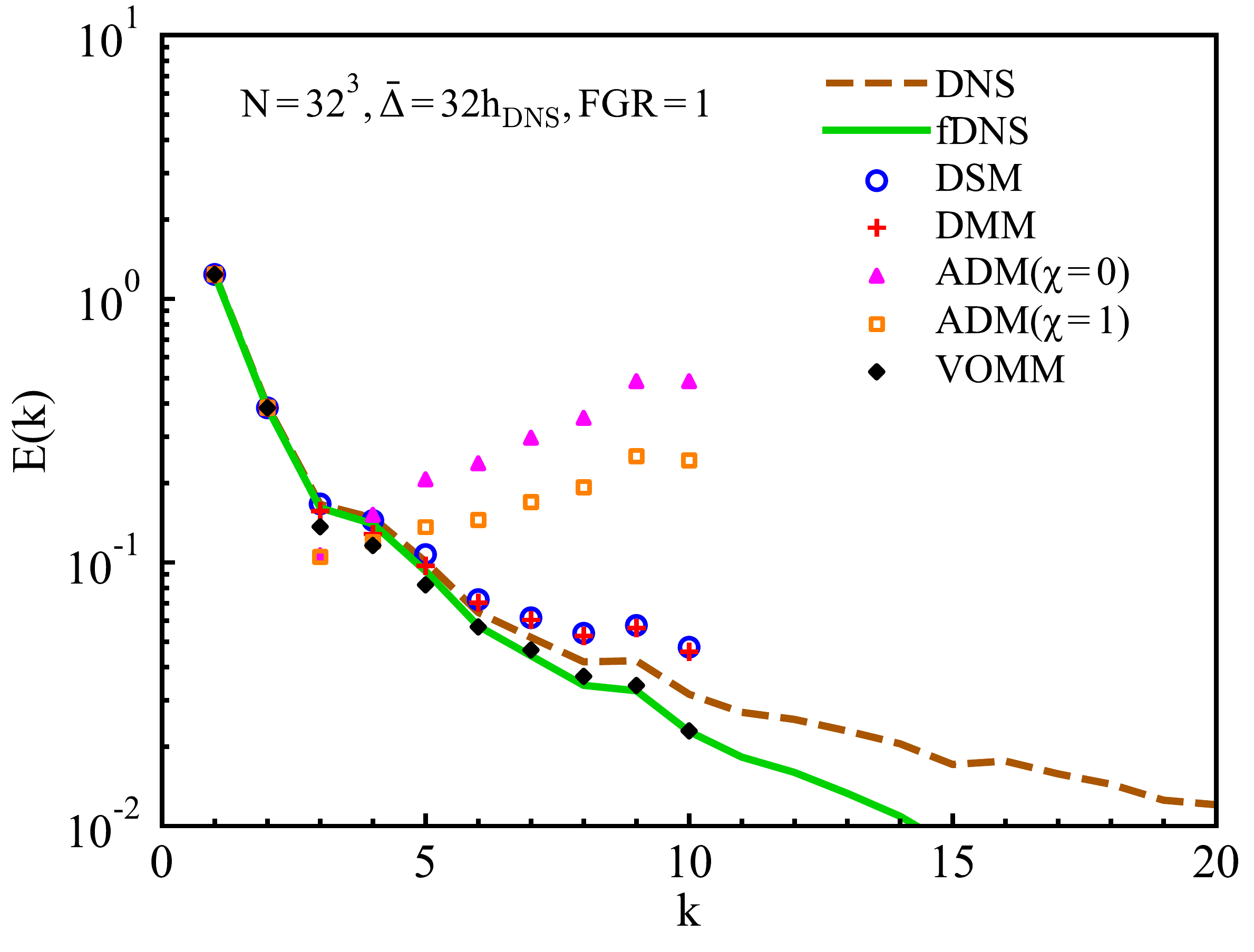}
	\end{subfigure}\\
	\begin{subfigure}{0.5\textwidth}
		\centering
		{($c$)}
		\includegraphics[width=0.9\linewidth,valign=t]{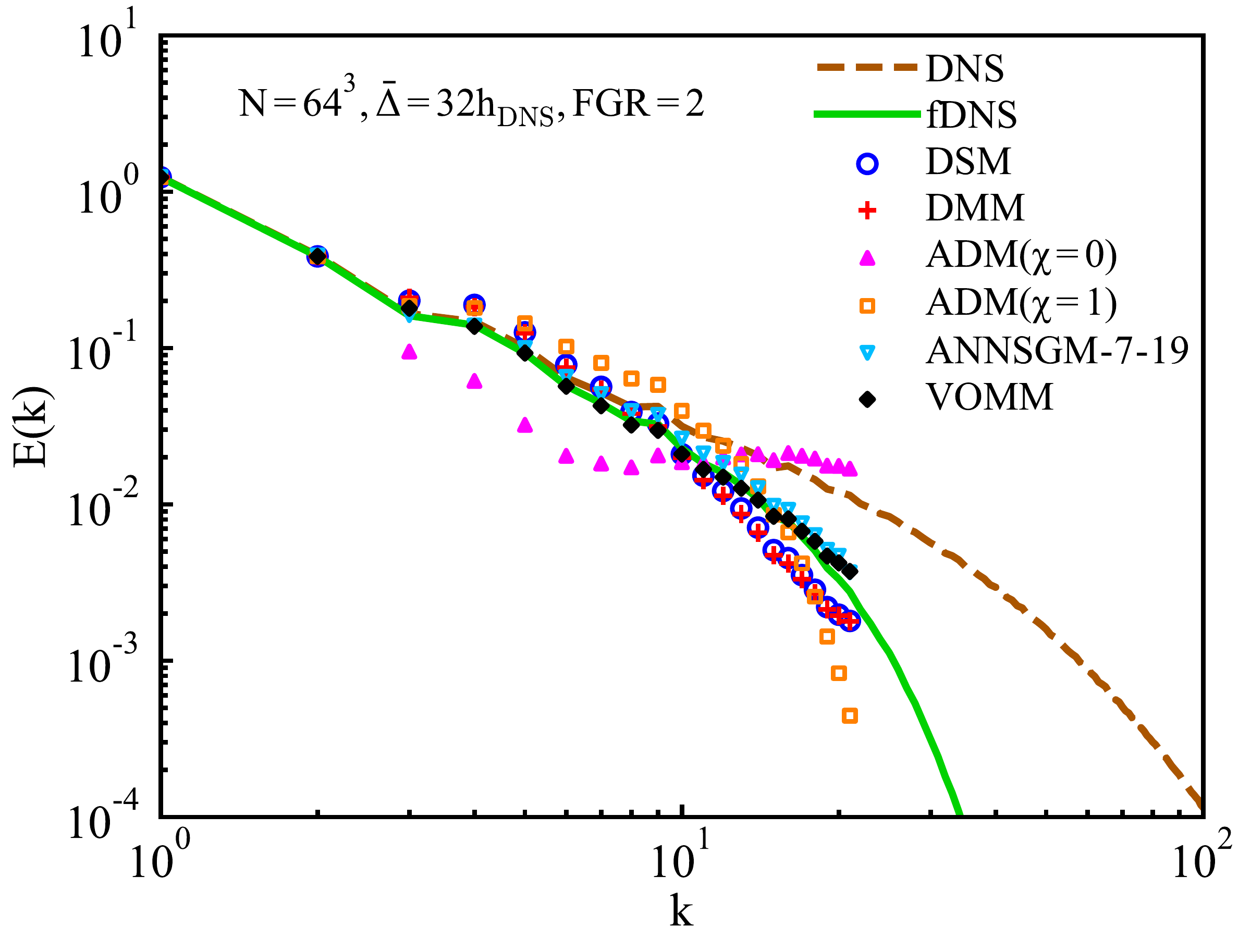}
	\end{subfigure}%
	\begin{subfigure}{0.5\textwidth}
		\centering
		{($d$)}
		\includegraphics[width=0.9\linewidth,valign=t]{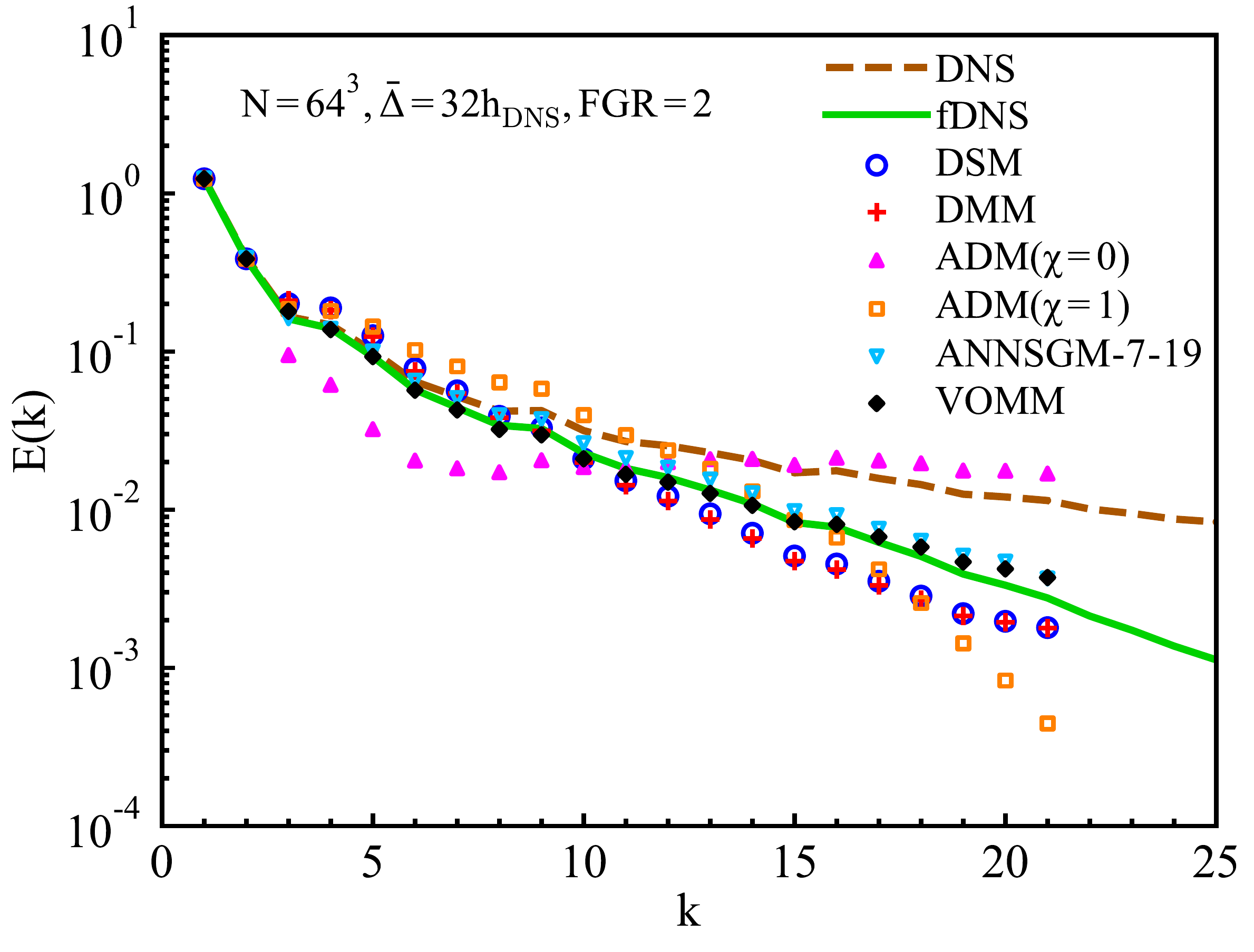}
	\end{subfigure}\\
	\begin{subfigure}{0.5\textwidth}
		\centering
		{($e$)}
		\includegraphics[width=0.9\linewidth,valign=t]{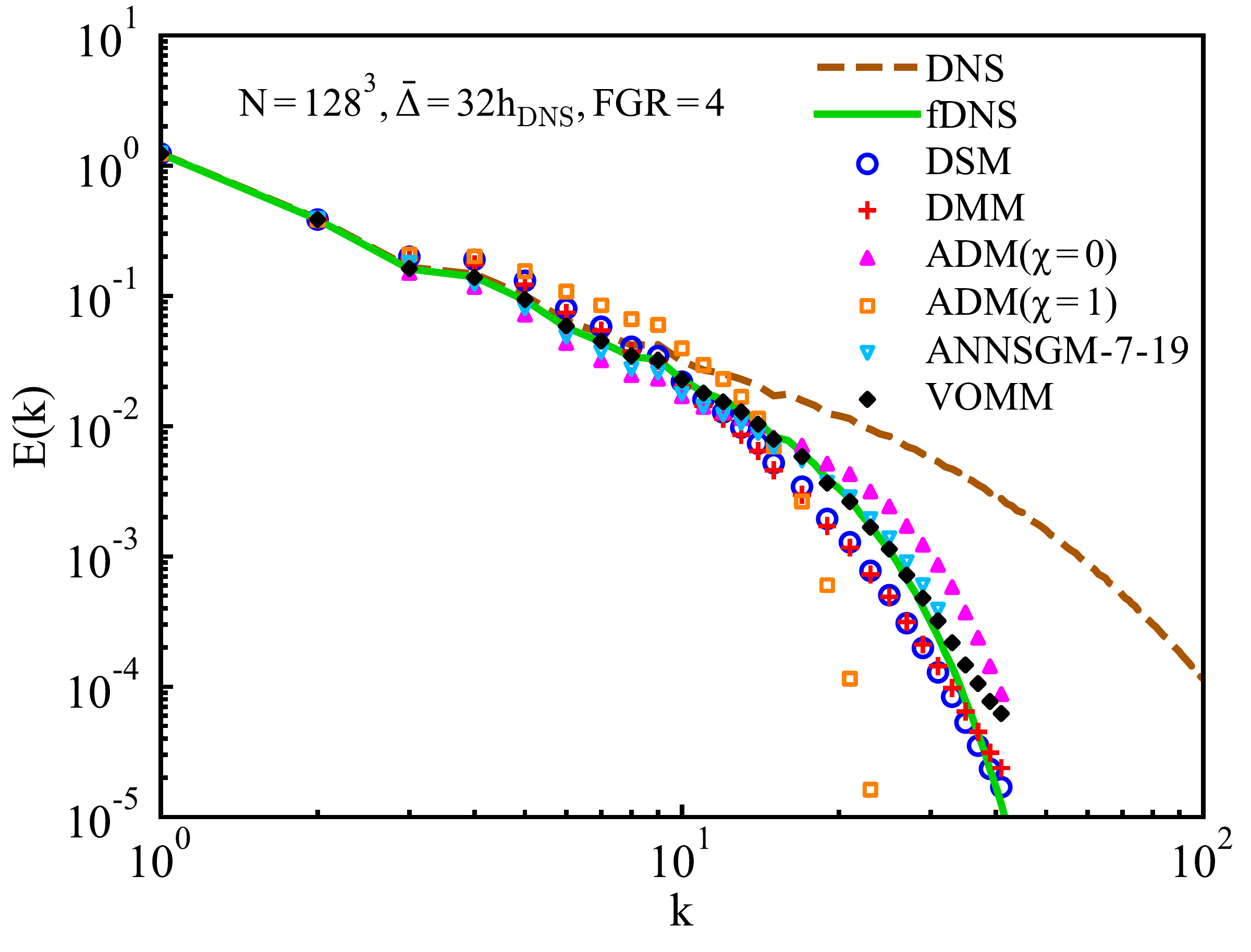}
	\end{subfigure}%
	\begin{subfigure}{0.5\textwidth}
		\centering
		{($f$)}
		\includegraphics[width=0.9\linewidth,valign=t]{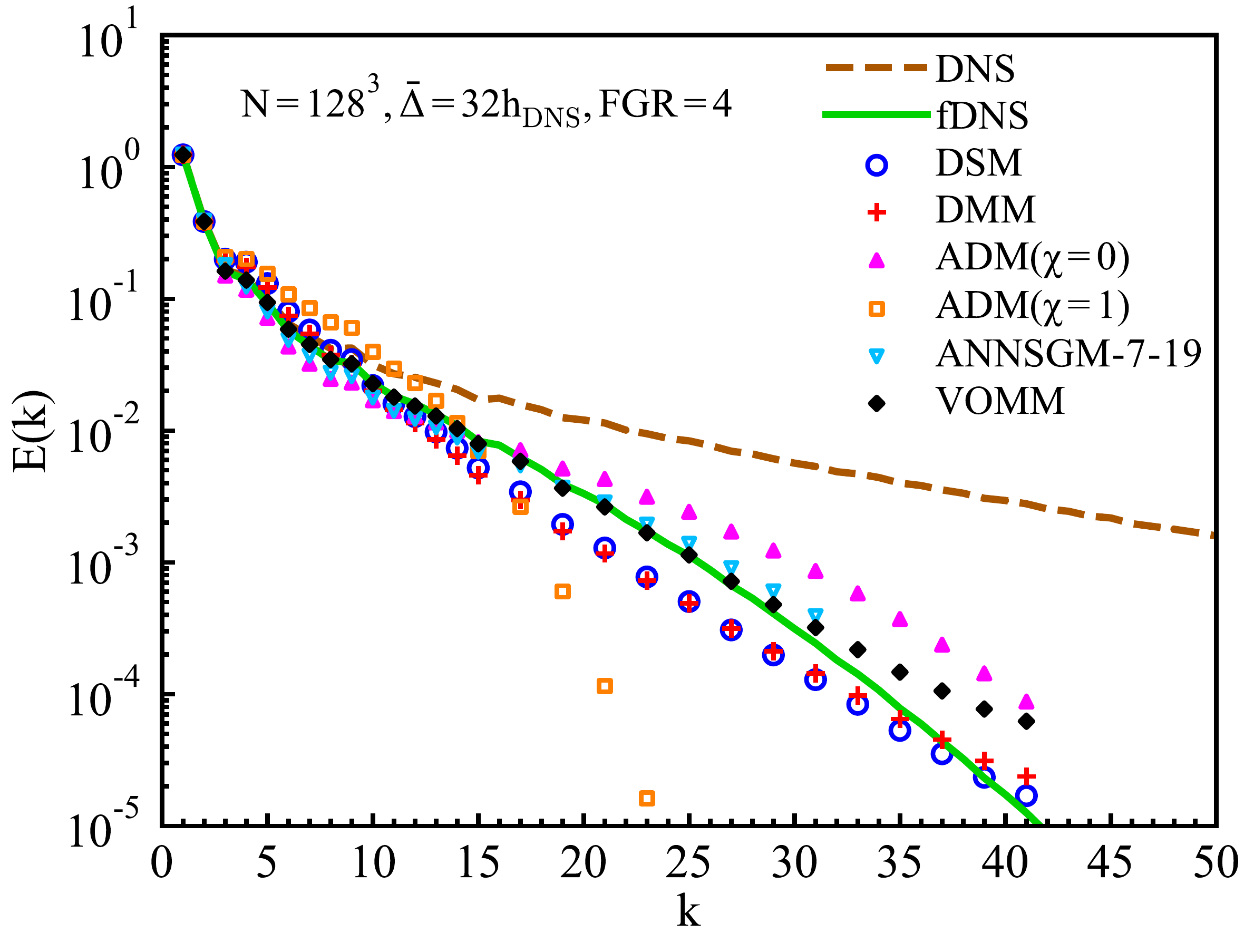}
	\end{subfigure}
	\caption{Velocity spectra for different SGS models in the \emph{a posteriori} analysis of forced homogeneous isotropic turbulence with the same filter scale $\bar \Delta  = 32{h_{{\rm{DNS}}}}$: (a) log-log for FGR=1, $N=32^3$; (b) semi-log for FGR=1, $N=32^3$; (c) log-log for FGR=2, $N=64^3$; (d) semi-log for FGR=2, $N=64^3$; (e) log-log for FGR=4, $N=128^3$; and (f) semi-log for FGR=4, $N=128^3$.}
	\label{fig:4}
\end{figure}

\subsection {Forced homogeneous isotropic turbulence}
We perform the direct numerical simulation of forced incompressible isotropic turbulence using the uniform grid resolution $N=1024^3$ in a cubic box of ${\left( {2\pi } \right)^3}$ with periodic boundary conditions $\left( {{h_{{\rm{DNS}}}} = 2\pi /1024} \right)$ \citep{xie2020a,xie2020d,yuan2020}. The pseudo-spectral method is used for the spatial discretization of the governing equations \citep{canuto1988,peyret2002}. The nonlinear advection terms are fully dealiased by the two-thirds dealiasing rule \citep{canuto1988}. A second-order two-step Adams-Bashforth explicit scheme is used for time integration \citep{chen1993}. 

The kinematic viscosity is chosen as $\nu = 0.001$, and large-scale forcing is applied to the two lowest wavenumber shells to maintain the turbulence in statistical equilibrium, giving rise to  the Taylor Reynolds number ${{\mathop{\rm Re}\nolimits} _\lambda } \approx 250$ \citep{wang2010,yuan2020}. The detailed one-point statistics of DNS data for the forced isotropic turbulence are summarized in Table~\ref{tab:1} \citep{yuan2022}. Here, ${k_{\rm{max}}} = \frac{{2\pi }}{{3{h_{{\rm{DNS}}}}}}$ denotes the largest effective wavenumber after the fully dealiasing, and $\omega^{\rm{rms}}= \sqrt{\left\langle {{\omega_i}{\omega_i}} \right\rangle }$ represents the root-mean-square value of the vorticity magnitude, where ${\bf{\omega }} = \nabla  \times {\bf{u}}$ stands for the vorticity which is the curl of the velocity field.
The Kolmogorov length scale $\eta$ and the integral length scale $L_I$ stand for the smallest resolved scale and the largest characteristic scale of turbulence, and are defined respectively by
\begin{equation}
	\eta ={{\left( \frac{{{\nu }^{3}}}{\varepsilon } \right)}^{1/4}},
	\label{eta}
\end{equation}
\begin{equation}
	{L_I}=\frac{3\pi}{2{{\left( {{u}^{\rm {rms}}} \right)}^{2}}}\int_{0}^{+\infty }{\frac{E\left( k \right)}{k}dk},
	\label{LI}
\end{equation}
where $\varepsilon$ is the spatial average dissipation rate of kinetic energy. The total turbulent kinetic energy $E_k = \left\langle {{u_i}{u_i}} \right\rangle /2 = \int_0^{ + \infty } {E\left( k \right)dk}$, and ${E\left( k \right)}$ represents the velocity spectrum. The resolution parameters ${k_{\max }}\eta  \ge 2.1$ and $\eta /{h_{{\rm{DNS}}}} \ge 1$ indicate that the grid resolution is sufficient to capture the smallest turbulent eddy scales and ensure the convergence of turbulent kinetic energy at all scales \citep{ishihara2007,ishihara2009}. To alleviate the impact of initial conditions, the forced homogeneous isotropic turbulence is run for a long period after the flow gradually reaches a statistically steady state (more than 50 large-eddy turnover times $\tau = {L_I}/{u^{{\rm{rms}}}}$). We select data of the last ten large-eddy turnover times as a benchmark for LES comparisons (total forty flow-field snapshots of DNS data).

\begin{figure}\centering
	\begin{subfigure}{0.33\textwidth}
		\centering
		{($a$)}
		\includegraphics[width=0.88\linewidth,valign=t]{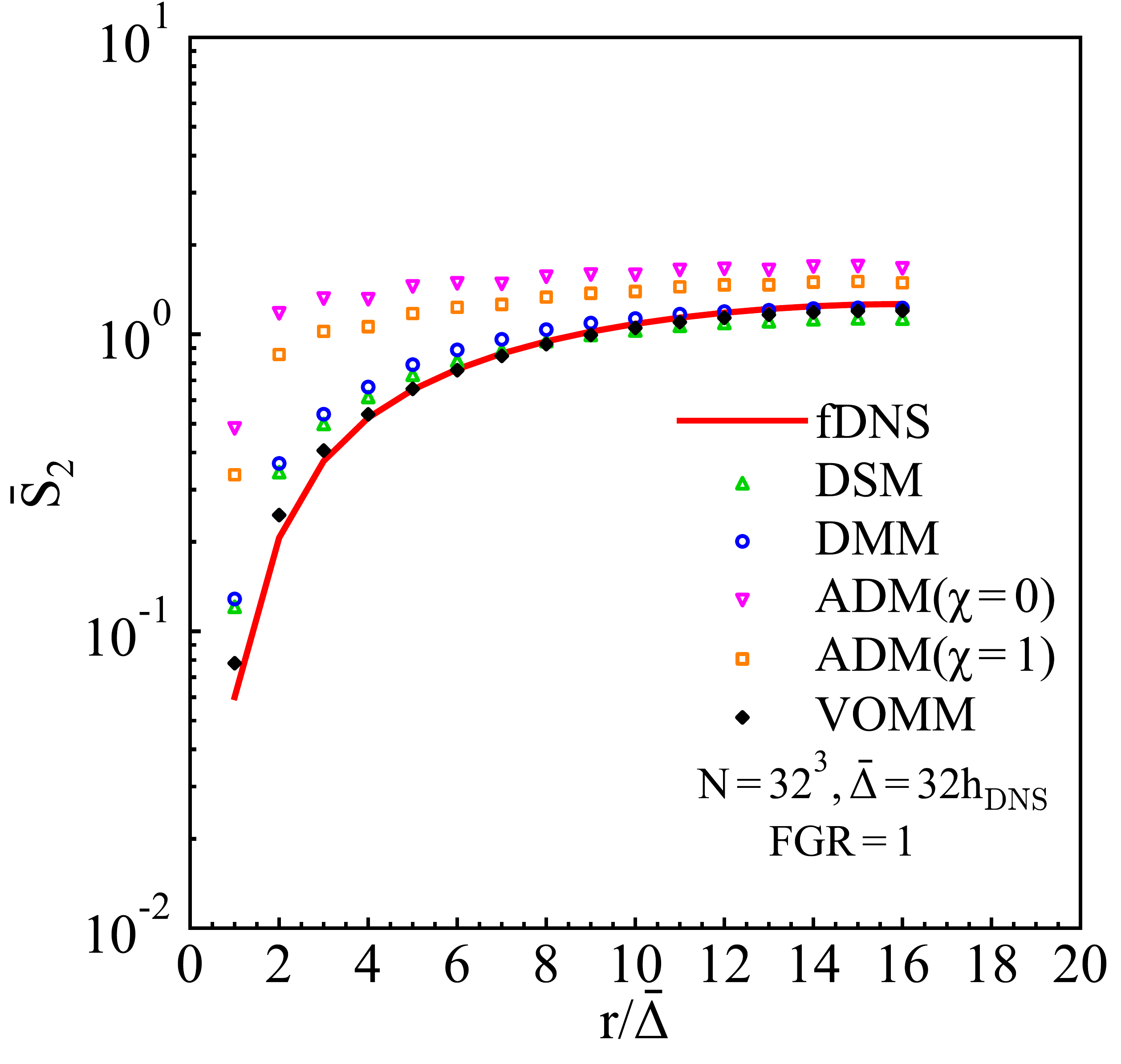}
	\end{subfigure}%
	\begin{subfigure}{0.33\textwidth}
		\centering
		{($b$)}
		\includegraphics[width=0.88\linewidth,valign=t]{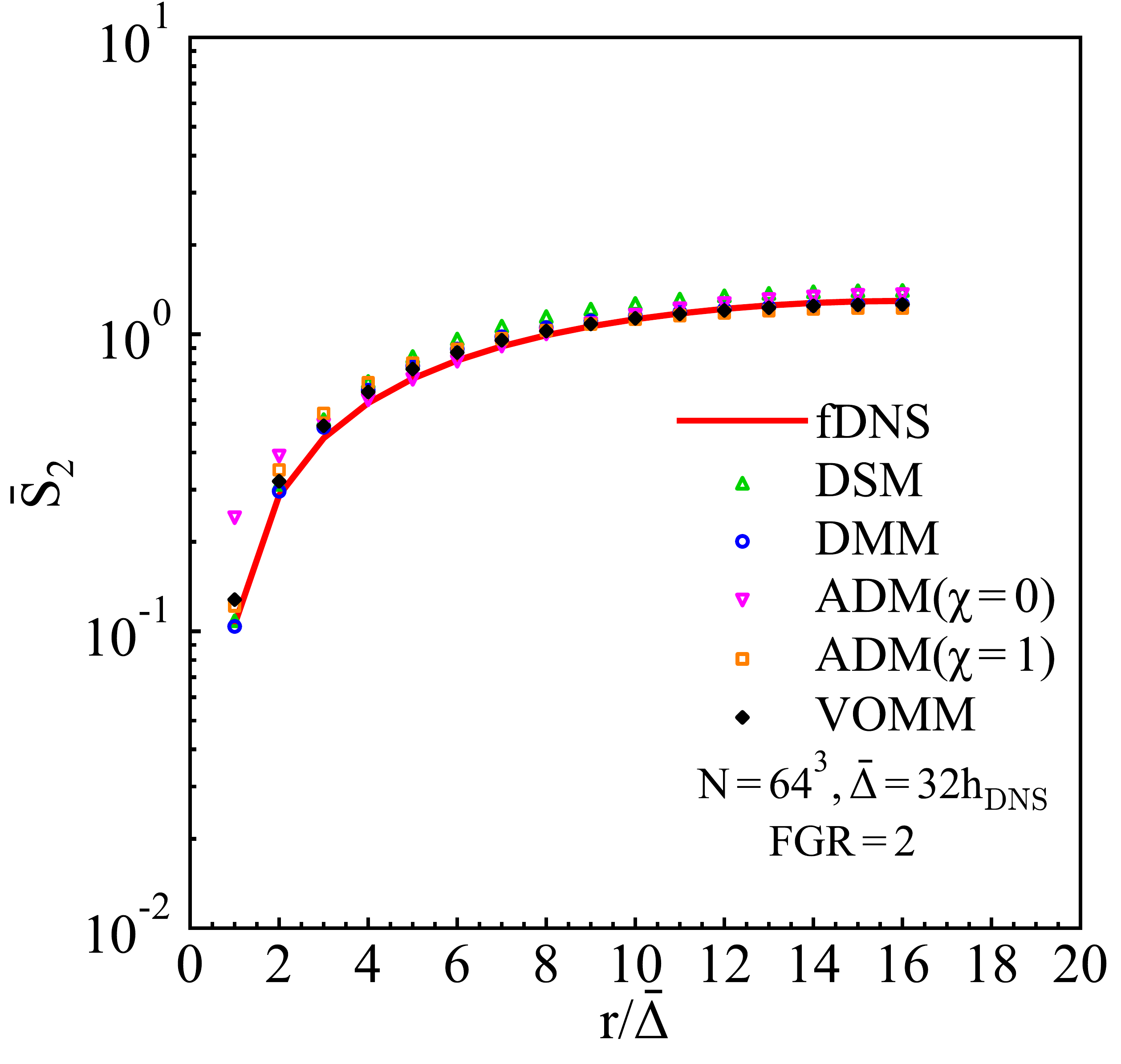}
	\end{subfigure}
	\begin{subfigure}{0.33\textwidth}
		\centering
		{($c$)}
		\includegraphics[width=0.88\linewidth,valign=t]{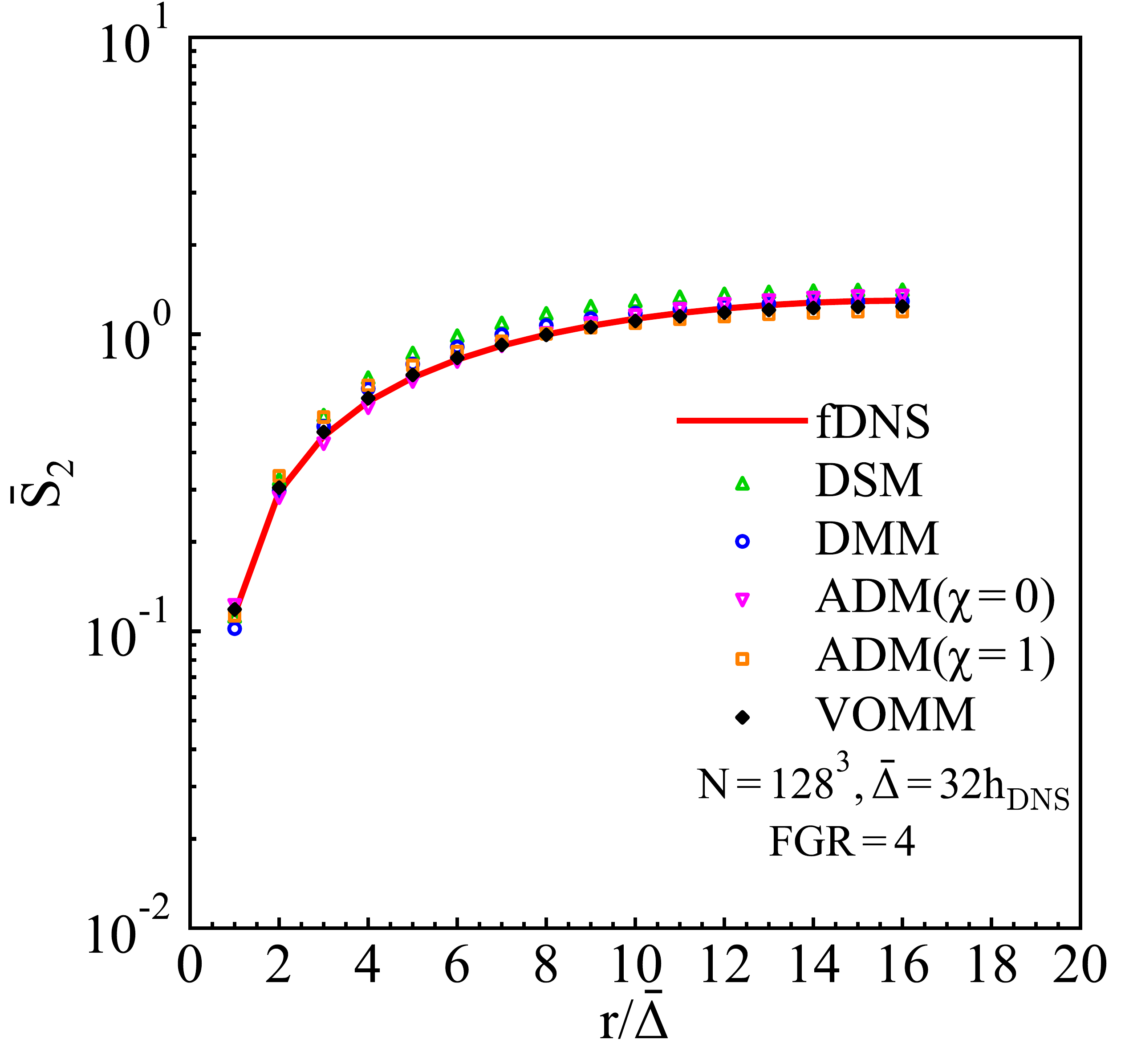}
	\end{subfigure}%
	\caption{Second-order structure functions of the filtered velocity for LES in the \emph{a posteriori} analysis of forced homogeneous isotropic turbulence with the same filter scale $\bar \Delta  = 32{h_{{\rm{DNS}}}}$: (a) FGR=1, $N=32^3$; (b) FGR=2, $N=64^3$; and (c) FGR=4, $N=128^3$.}
	\label{fig:5}
\end{figure}

In this paper, the Gaussian filter (Eq.~\ref{G}) is used as the explicit filter to calculate the filtered physical variables. The selected filter width $\bar \Delta  = 32{h_{{\rm{DNS}}}}$ and the corresponding cutoff wavenumber is ${k_c} = \pi /\bar \Delta  = 16$. The velocity and dissipation spectra of the DNS and filtered DNS at $\bar \Delta  = 32{h_{{\rm{DNS}}}}$ are illustrated in Fig.~\ref{fig:2}. The filtered velocity spectrum nearly overlaps with the DNS data in a Kolmogorov scaling law of $k^{-5/3}$ at the low wavenumber region, while it drops significantly at the region larger than the truncated wavenumber $k_c$. Overall 12\% of the turbulent kinetic energy is filtered out in the residual velocity field at the filter scale  $\bar \Delta  = 32{h_{{\rm{DNS}}}}$.  In contrast, the filtered dissipation spectrum gradually grows with the power of law scaling $k^{1/3}$ at the low-wavenumber inertial region, and drops sharply where the cutoff wavenumber exceeds. The small scales near the truncated wavenumbers are essential for the reconstruction of the filtered dissipation spectrum and also very important for the residual SGS modeling. However, these small scales account for a very small proportion of the turbulent kinetic energy, almost several orders of magnitude smaller than the large scales. Thus, the dissipation spectrum instead of the kinetic energy spectrum is chosen as the optimization objective function of the proposed VOMM model in the paper.  

The \emph{a posteriori} testings of LES are essential to validate the practical performance of the SGS models. LES calculations use the same kinematic viscosity ($\nu=0.001$) with the DNS. The filter width is fixed to $\bar \Delta = 32h_{\rm{DNS}}$ and the impact of the spatial discretization errors on the SGS models is investigated by changing the grid resolution of LES. Three different filter-to-grid ratios FGR=$\bar \Delta /h_{\rm{LES}}$=1, 2 and 4 are chosen to study the influence of spatial discretization on the SGS modeling, and the corresponding grid points of LES are $N=32^3$, $64^3$ and $128^3$, respectively. The proposed VOMM model (Eq.~\ref{VOMM}) is compared against the classical SGS models, including the dynamic Smagorinsky model (DSM, Eq.~\ref{tau_sm}), the dynamic mixed model (DMM, Eq.~\ref{dmm1}) and the standard approximate deconvolution model with secondary filtering regularization (ADM, Eqs.~\ref{ADM} and \ref{ADM_relax}). The relaxation factors of ADM model $\chi$=0 and 1 are chosen for comparisons. The ratios of the time steps for LES and DNS are $\Delta {t_{{\rm{LES}}}}/\Delta {t_{{\rm{DNS}}}} = \left\{ {10,10,5} \right\}$ for different grids (FGR=1, 2 and 4 with $N=32^3$, $64^3$ and $128^3$). Among the filtered DNS data of the ten large-eddy turnover periods, the data of the first two large-eddy turnover times are used for the adjoint optimization of the VOMM model (only the dissipation spectrum is used, stored once every 0.1$\tau$, twenty sets in total), and the remaining data of the last eight large-eddy turnover times are used for the \emph{a posteriori} accuracy validation of the LES models. 

\begin{figure}\centering
	\begin{subfigure}{0.33\textwidth}
		\centering
		{($a$)}
		\includegraphics[width=0.88\linewidth,valign=t]{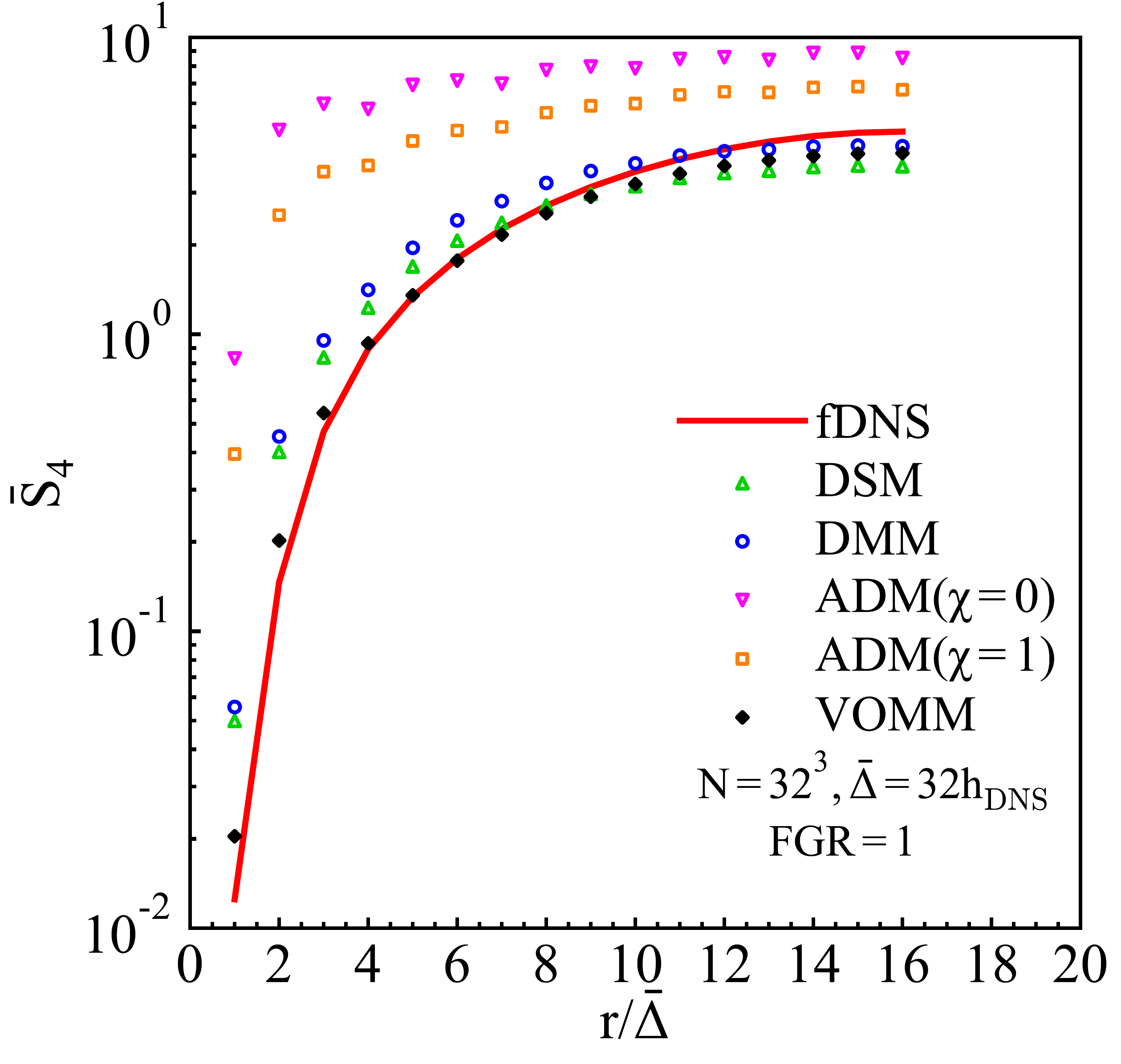}
	\end{subfigure}%
	\begin{subfigure}{0.33\textwidth}
		\centering
		{($b$)}
		\includegraphics[width=0.88\linewidth,valign=t]{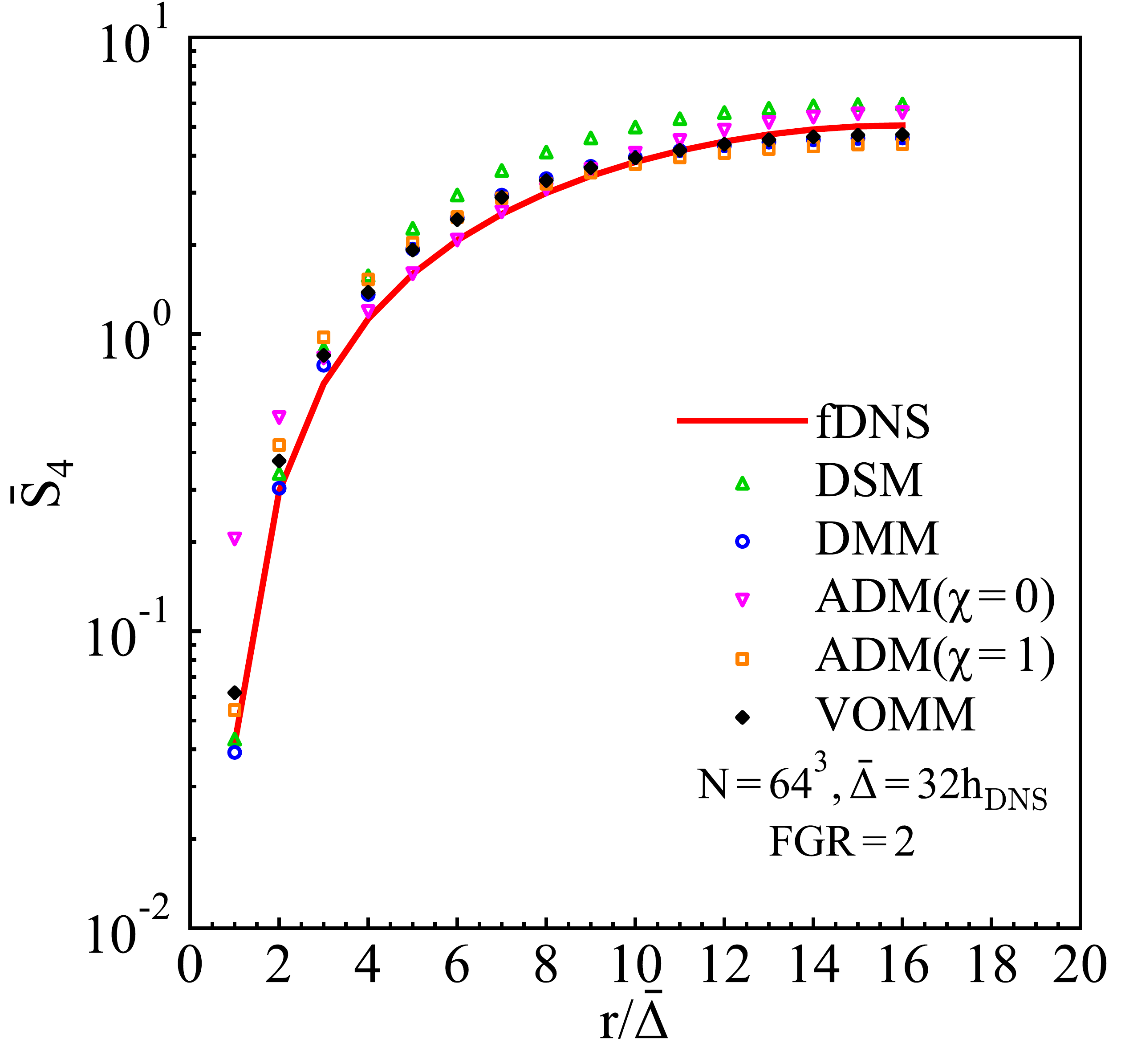}
	\end{subfigure}
	\begin{subfigure}{0.33\textwidth}
		\centering
		{($c$)}
		\includegraphics[width=0.88\linewidth,valign=t]{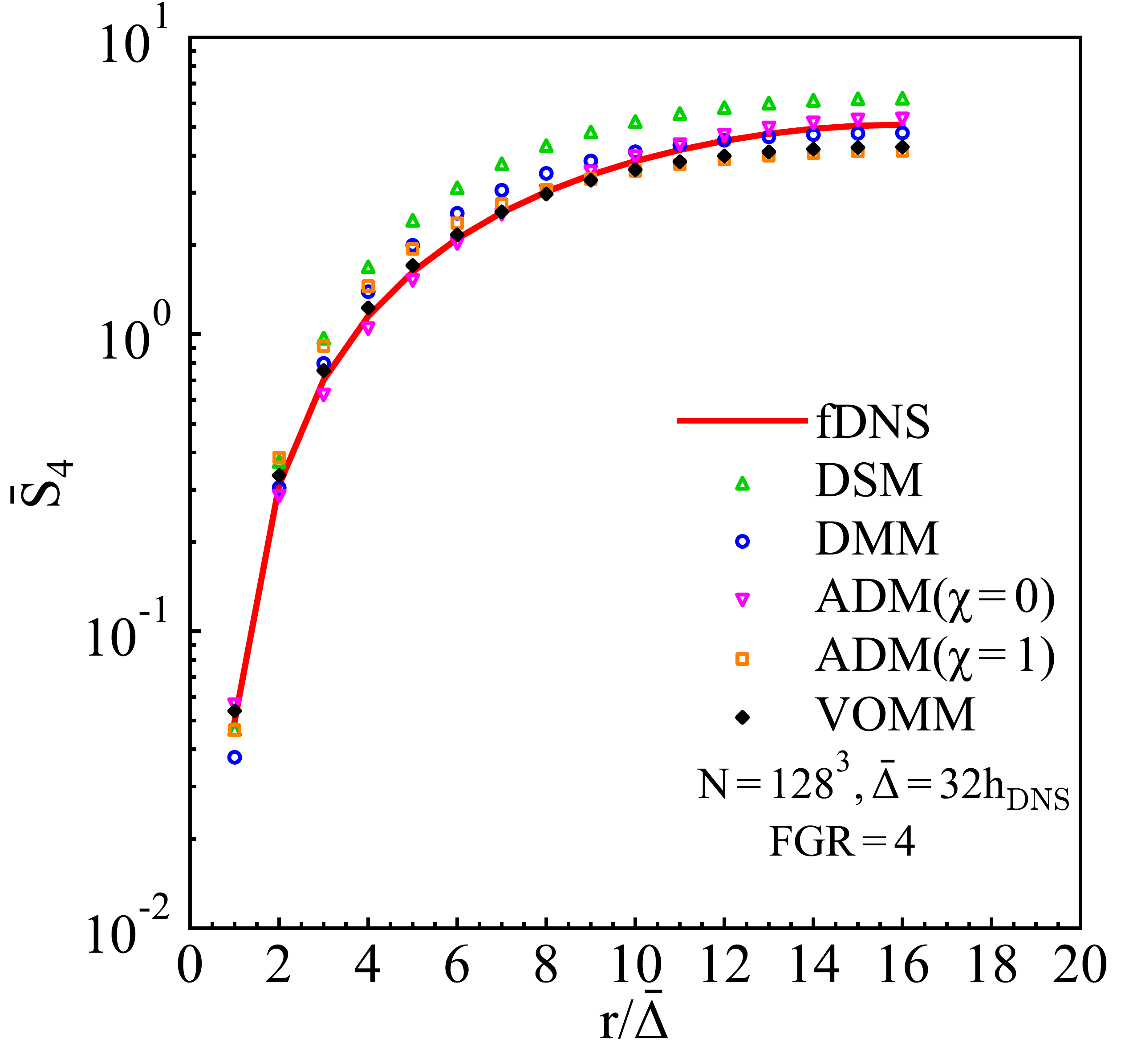}
	\end{subfigure}%
	\caption{Fourth-order structure functions of the filtered velocity for LES in the \emph{a posteriori} analysis of forced homogeneous isotropic turbulence with the same filter scale $\bar \Delta  = 32{h_{{\rm{DNS}}}}$: (a) FGR=1, $N=32^3$; (b) FGR=2, $N=64^3$; and (c) FGR=4, $N=128^3$.}
	\label{fig:6}
\end{figure}

At the adjoint-based optimization stage of the VOMM model, the calculations of the adjoint equations are consistent with the primary LES equations. We adopt the same pseudo-spectral numerical scheme to spatially discrete the stabilized adjoint momentum equations (Eq.~\ref{adj_MLES2}). A second-order two-step Adams-Bashforth explicit scheme is applied for the time backward integration with zero terminal conditions. Since the large-scale forcing is assumed to be nearly independent of the filtered velocity, the large-scale forcing term does not appear in the adjoint momentum equations. During the adjoint optimization stage (see Fig.~\ref{fig:1}) of the VOMM model, the pure structural ADM model without the dissipative Smagorinsky term is selected as the initial SGS model with model coefficients $C_1^{\left( 0 \right)}=0$ and  $C_2^{\left( 0 \right)}=1$. The LES forward evolution is initialized by the filtered DNS velocity field and the dissipation spectrum is calculated when the filtered DNS data are available (every  0.1$\tau$). The statistical discrepancy of the dissipation spectrum between the LES and fDNS data is evaluated and recorded as the cost functional. The adjoint-based gradients of the cost functional with respect to the model coefficients are calculated through backward integrating the stabilized adjoint LES equations (Eqs.~\ref{adj_LES1} and \ref{adj_MLES2}) with zero terminal conditions. The SGS model coefficients are then iteratively updated by the gradient-based L-BFGS optimization algorithm (Eq.~\ref{opt_iterative}) until reaching the stopping criteria.

Figure~\ref{fig:3} shows the evolution of the cost function normalized by the initial discrepancy during the adjoint-based optimization in forced homogeneous isotropic turbulence. The loss functions (prediction errors of dissipation spectra between LES and fDNS data) for all three different filter-to-grid ratio cases (FGR=1,2 and 4) gradually converge and become stationary within less than twenty iterations. The error is significantly reduced by nearly an order of magnitude for the cases of FGR=1 and 2 within about ten iterations, and is drastically reduced to 20\% of the initial state at FGR=4. These results indicate that the adjoint-based L-BFGS gradient optimization is very efficient and effectively obtains the optimal model coefficients within several iterations. The optimal parameters  of the VOMM model are summarized in Table~\ref{tab:2}. The magnitude of the eddy-viscosity coefficient (($\left| {C_1^{\rm{opt}}} \right|$) ) dramatically reduces from 0.0529 to 0.003 with the increasing of FGR and LES resolutions, while the coefficient of the ADM part ($C_2^{\rm{opt}}$) gradually approaches unity, which is identical to the theoretical value derived from the Taylor series expansions. Once the optimal model coefficients are obtained, we further examine the \emph{a posteriori} performance of the VOMM model using the filtered DNS data of the last eight large-eddy turnover periods. 

Table~\ref{tab:3} gives the average computational cost for the SGS stress modeling at the same filter width $\bar \Delta=32 h_{\rm{DNS}}$. For all three different grid resolutions, the computation time of the VOMM model is only about 30\% of that of the DMM model, without significantly increasing the computational cost in comparison to the ADM models ($\chi = 0$ and 1).

\begin{figure}\centering
	\begin{subfigure}{0.33\textwidth}
		\centering
		{($a$)}
		\includegraphics[width=0.88\linewidth,valign=t]{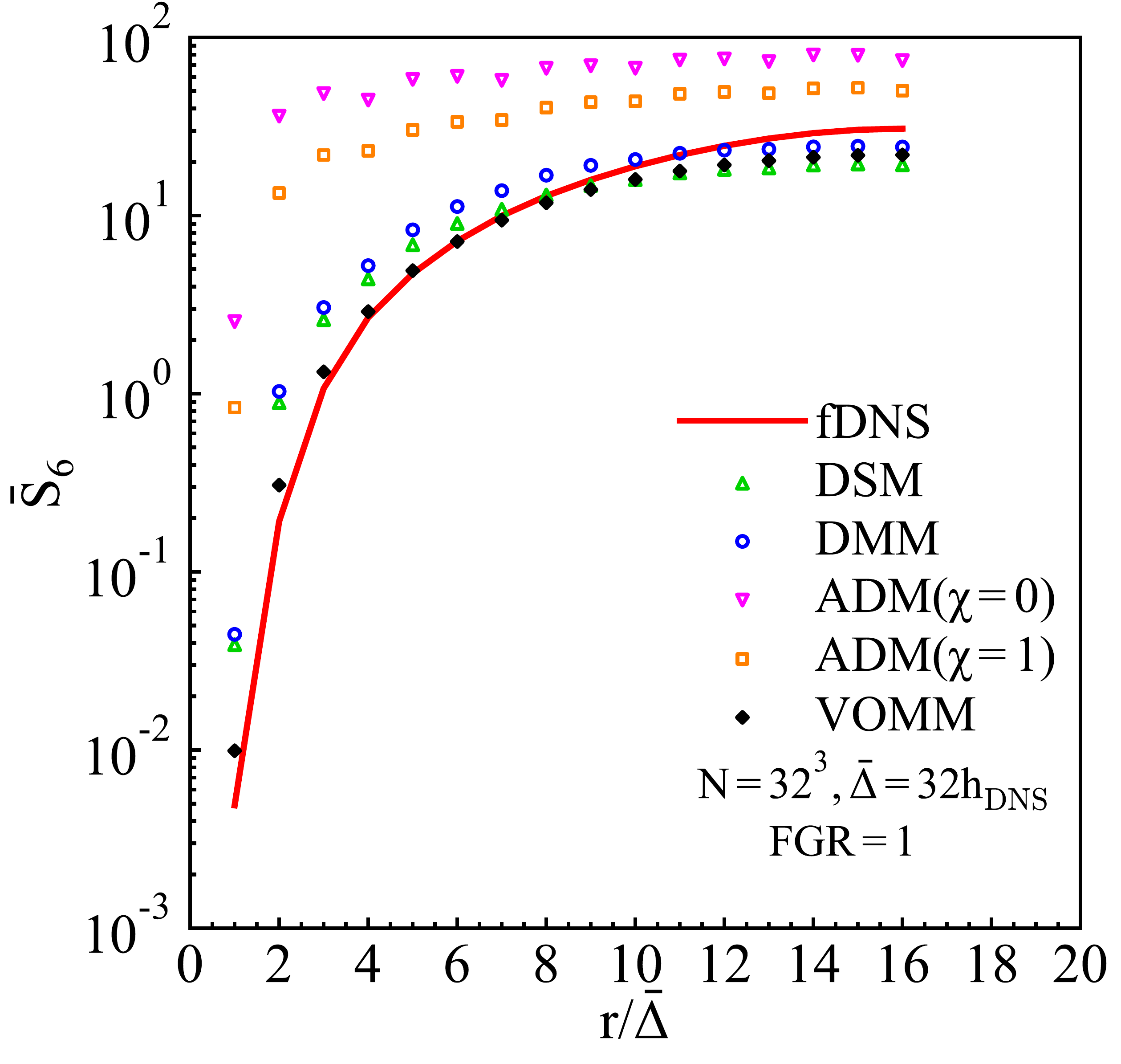}
	\end{subfigure}%
	\begin{subfigure}{0.33\textwidth}
		\centering
		{($b$)}
		\includegraphics[width=0.88\linewidth,valign=t]{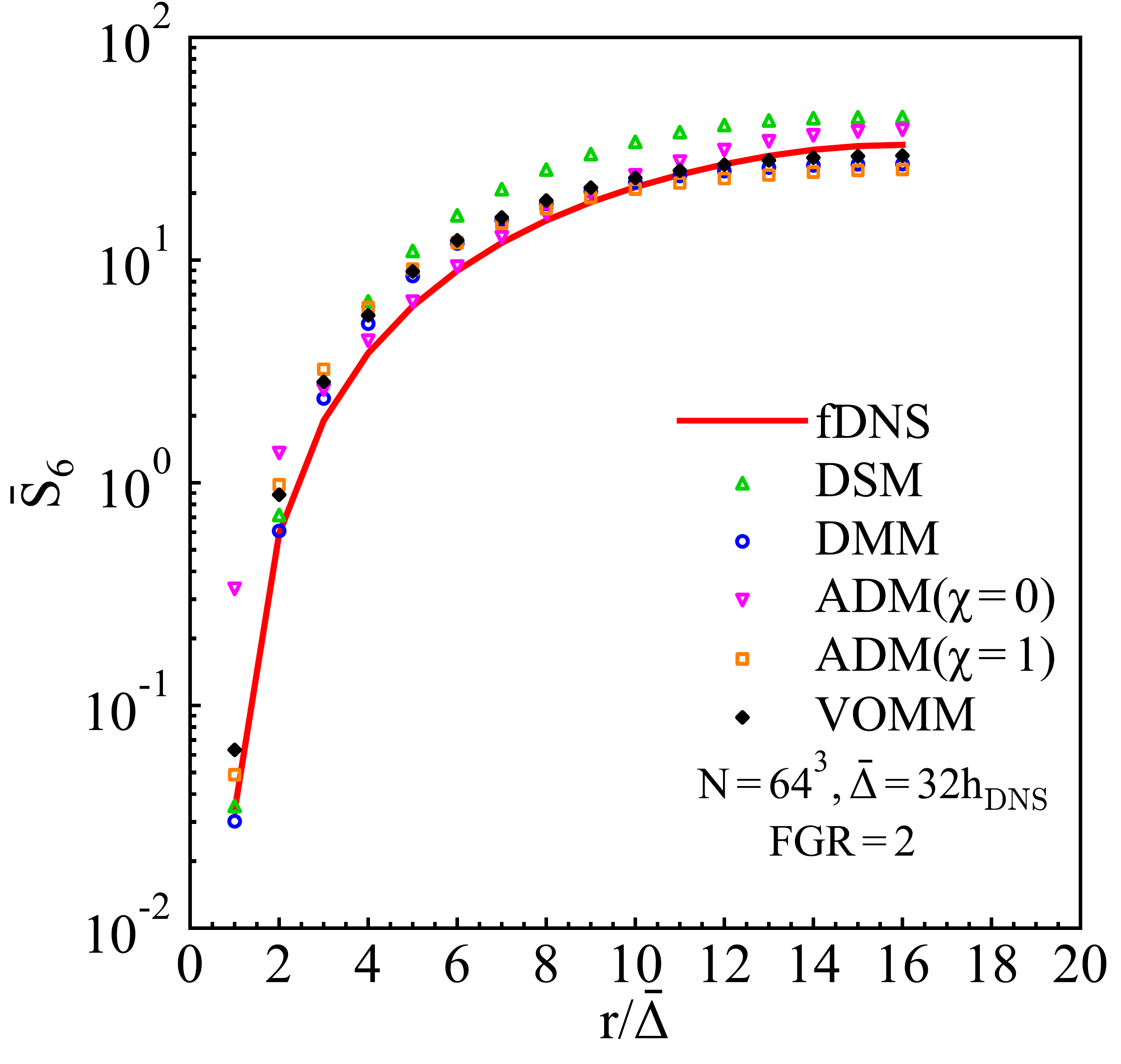}
	\end{subfigure}
	\begin{subfigure}{0.33\textwidth}
		\centering
		{($c$)}
		\includegraphics[width=0.88\linewidth,valign=t]{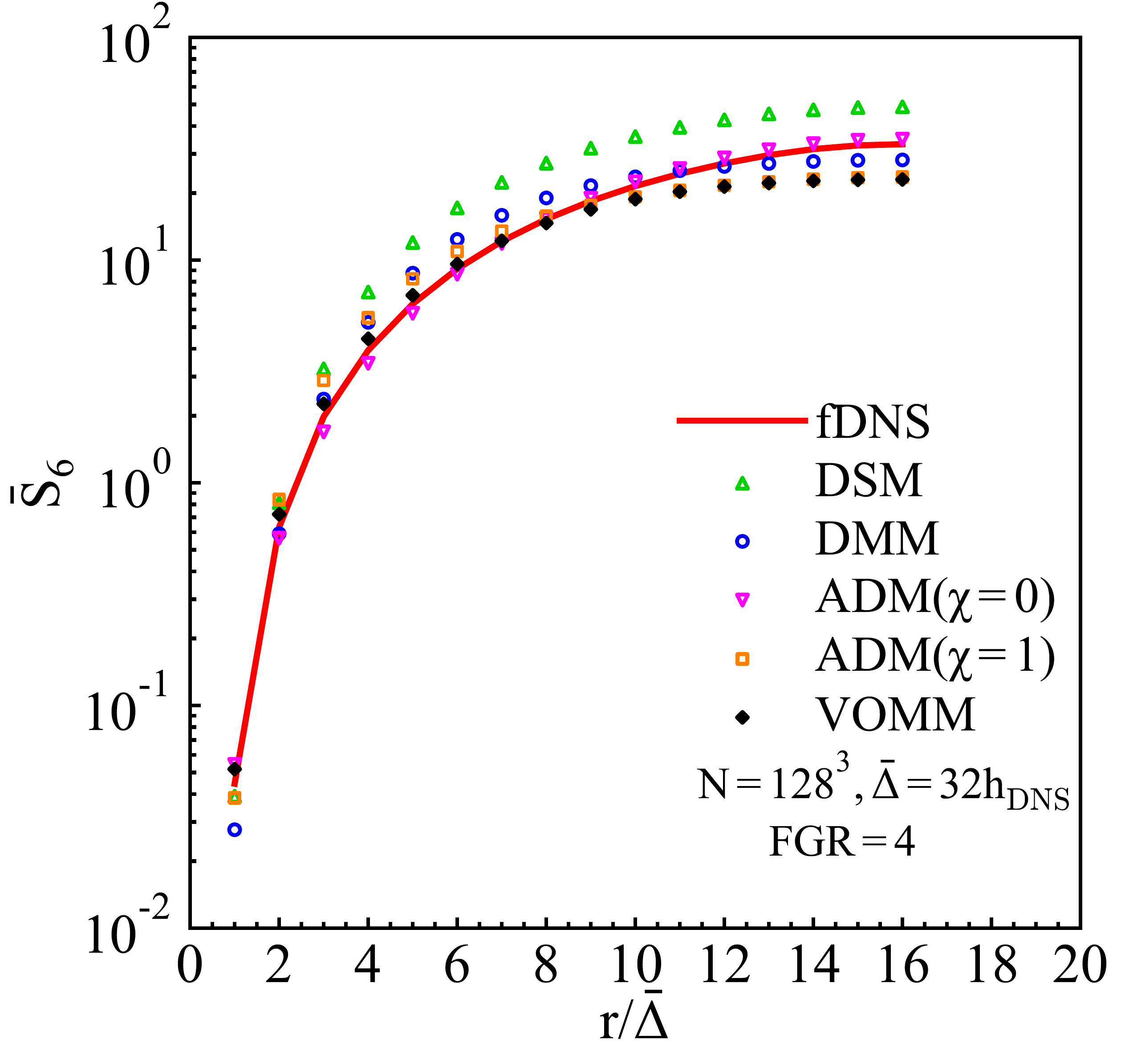}
	\end{subfigure}%
	\caption{Sixth-order structure functions of the filtered velocity for LES in the \emph{a posteriori} analysis of forced homogeneous isotropic turbulence with the same filter scale $\bar \Delta  = 32{h_{{\rm{DNS}}}}$: (a) FGR=1, $N=32^3$; (b) FGR=2, $N=64^3$; and (c) FGR=4, $N=128^3$.}
	\label{fig:7}
\end{figure}

The velocity spectra of different SGS models with the filter scale $\bar \Delta  = 32{h_{{\rm{DNS}}}}$ in comparison to those of the DNS and filtered DNS (fDNS) data are shown in Fig.~\ref{fig:4}. The velocity spectrum of DNS data exhibits a sufficiently long inertial range with a typical $k^{-5/3}$ scaling.  The spectrum of fDNS almost overlaps with that of DNS at the low-wavenumber region, but is obviously lower than that of DNS near the truncated wavenumber since the small-scale kinetic energy at high wavenumbers is filtered out. LES only solves the large-scale variables with the filtered Navier-Stokes equations (Eqs.~\ref{fns1} and \ref{fns2}), leaving the effect of residual small scales to be approximately reconstructed by the SGS model. Therefore the statistics of an ideal LES would overlap with that of the fDNS data as closely as possible. When the grid resolution of LES is sufficiently coarse and the grid spacing of LES is equal to the  filter scale (FGR=1, cf. Figs~\ref{fig:4}a and \ref{fig:4}b), the spatial discretization error is significant and deteriorates the accuracy of the SGS stress modeling. LES calculations with traditional SGS models are very difficult to obtain accurate predictions of the turbulent kinetic energy cascade at FGR=1. The velocity spectra predicted by the ADM models with $\chi = 0$ and 1 exhibit numerically unstable, and kinetic energy at high wavenumbers is obviously overestimated due to the insufficient dissipation. DSM and DMM models also have dramatic overestimations at high-wavenumber regions, with predictions even larger than that of the DNS data. In contrast, VOMM model predicts the velocity spectra most accurately among these SGS models whose results nearly coincide with that of fDNS. 

For the cases of fine grid resolutions (FGR=2 and 4), the pure ADM model ($\chi = 0$) is still numerically unstable since the pure structural model itself cannot produce sufficient SGS dissipation. The ADM model with the standard secondary-filtering regularization ($\chi = 1$) exhibits excessively dissipative, and the small-scale kinetic energy at high wavenumbers is extremely exhausted and much lower than that of fDNS. The predictions of DSM and DMM models illustrate the obviously tilted distribution, where kinetic energy at low wavenumbers is accumulated, while that near the truncated wavenumber is  diminished. The dynamic least-square procedure for both DSM and DMM models would overestimate the eddy-viscosity coefficient for the cases of fine grid resolutions (FGR=2 and 4), and small-scale flow structures near the truncated wavenumbers are exhausted by the excessive dissipation. The turbulent kinetic energy is transferred from large scales to small scales through the forward energy cascade process of the nonlinear advection. The lack of the sufficient flow structures near the cutoff wavenumber leads to the energy accumulation in the intermediate wavenumber region. In contrast, the VOMM model is superior to the other SGS models and can accurately predict the velocity spectra at all different grid resolutions of LES, with the predictions very close to the fDNS data. 

We also compare the VOMM model with the artificial neural network-based spatial gradient models (ANN-SGM) proposed by Wang \emph{et al.}\cite{wang2021} The ANNSGM-7-19 model consists of four fully-connected layers of neurons (4:20:20:19) and takes the integrity invariants as input to learn the model coefficients of the velocity gradient products for the neighboring seven-point stencil.\cite{wang2021} Figure~\ref{fig:4} compares the velocity spectra of the ANNSGM-7-19 model with the VOMM models in the \emph{a posteriori} testings of LES for different grid resolutions (FGR=2 with $N=64^3$ and FGR=4 with $N=128^3$). For the cases of coarse grid resolution (FGR=1 with $N=32^3$), the ANNSGM-7-19 model exhibits numerical instability and eventually diverges after several time steps of LES calculations. The VOMM model gives satisfactory predictions of the velocity spectra, which are very close to those predicted by the ANNSGM-7-19 model. Furthermore, the VOMM model can accurately predict the turbulence statistics of the coarse-grained LES calculations for FGR=1. In contrast, the machine-learning-assisted SGS models might require finer grid resolutions of LES, making SGS modeling dominate the grid discretization errors.\cite{wang2021} In general, the performance of the proposed VOMM model is very close to that of the ANNSGM-7-19 model. Compared to the ANN-based SGS model, the VOMM model has a better generalization ability (discussed in Sec.~\ref{sec:level6}) and model interpretability, which has the potential to be applied to the complex turbulence problems.

\begin{figure}\centering
	\begin{subfigure}{0.5\textwidth}
		\centering
		{($a$)}
		\includegraphics[width=0.9\linewidth,valign=t]{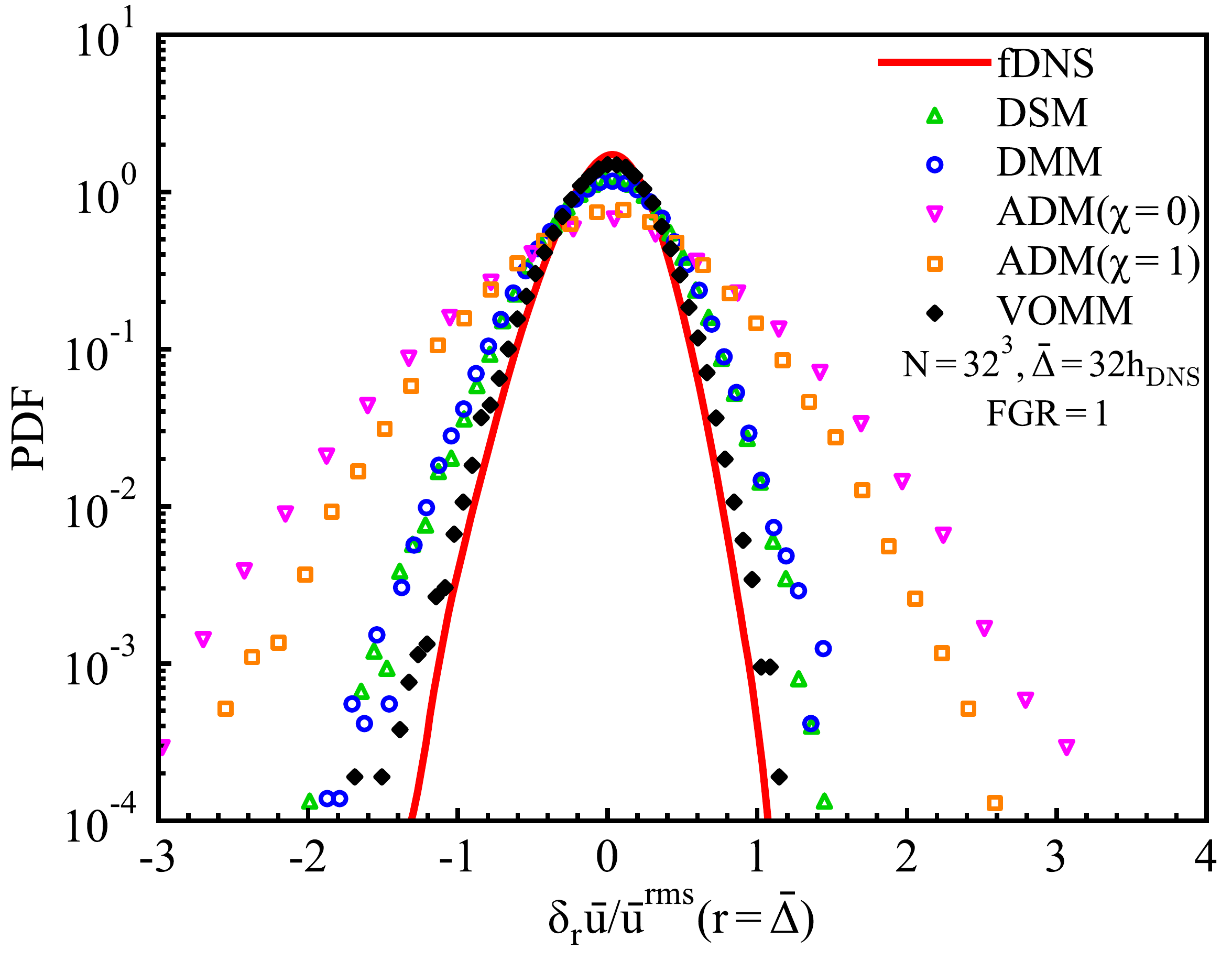}
	\end{subfigure}%
	\begin{subfigure}{0.5\textwidth}
		\centering
		{($b$)}
		\includegraphics[width=0.9\linewidth,valign=t]{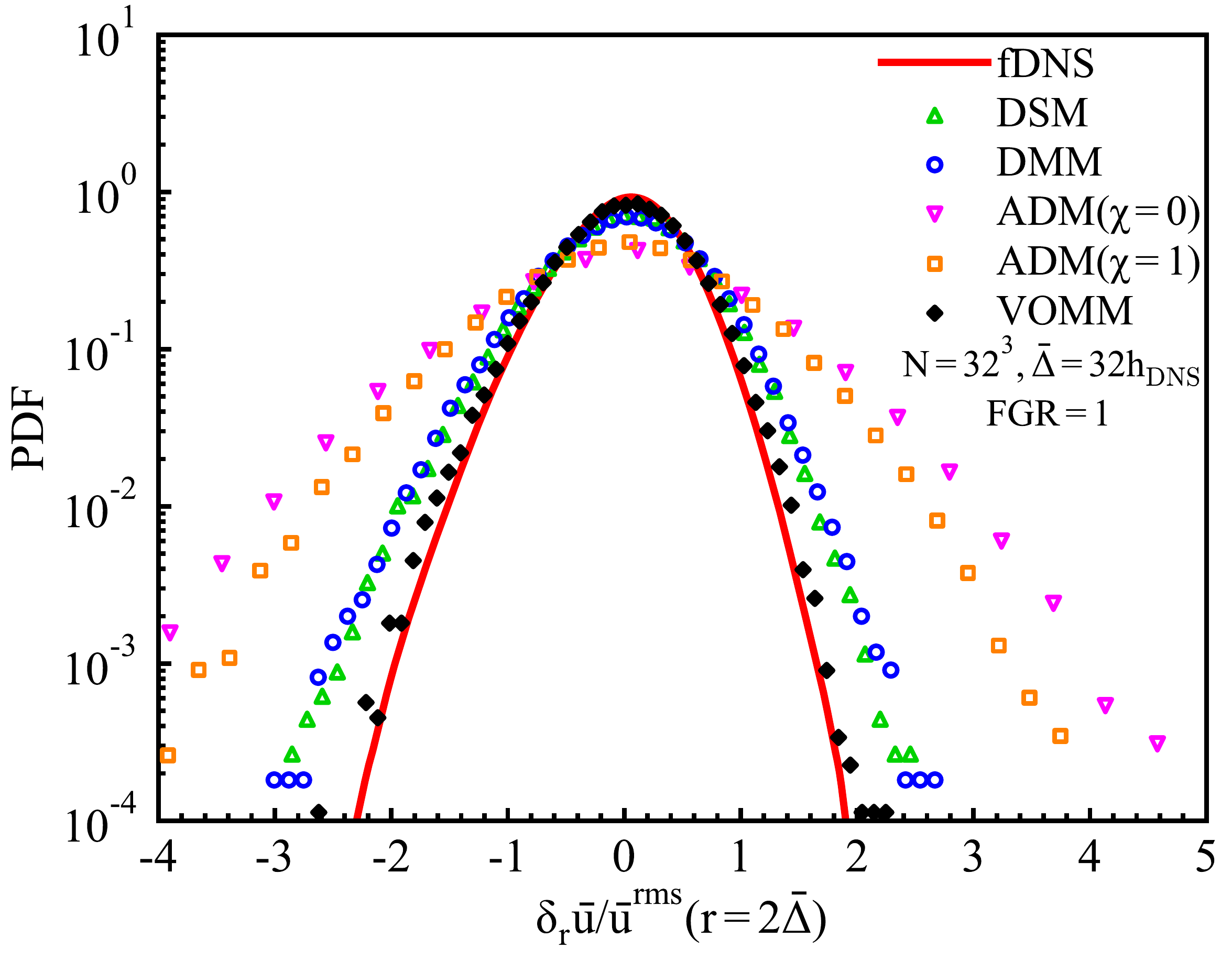}
	\end{subfigure}\\
	\begin{subfigure}{0.5\textwidth}
		\centering
		{($c$)}
		\includegraphics[width=0.9\linewidth,valign=t]{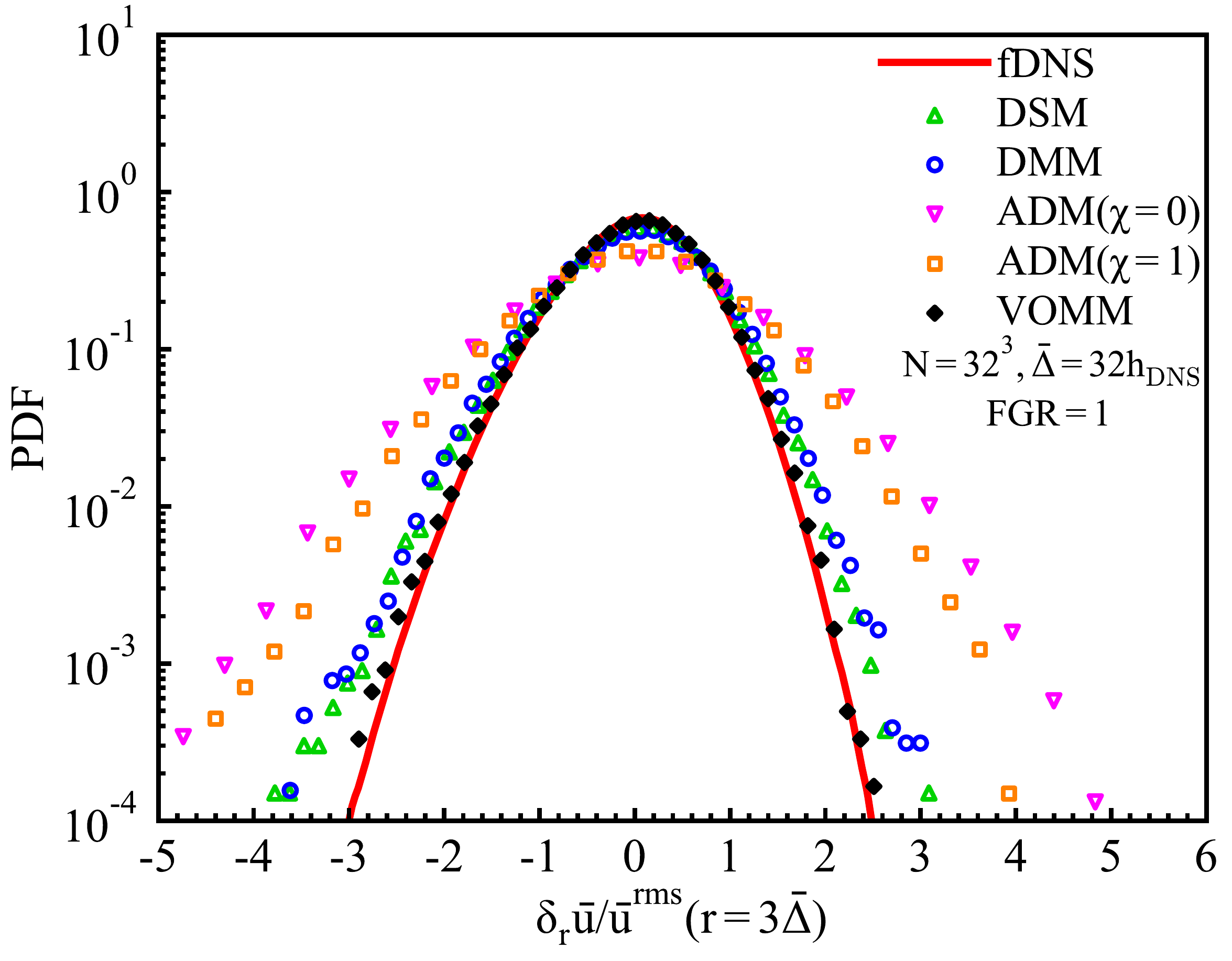}
	\end{subfigure}%
	\begin{subfigure}{0.5\textwidth}
		\centering
		{($d$)}
		\includegraphics[width=0.9\linewidth,valign=t]{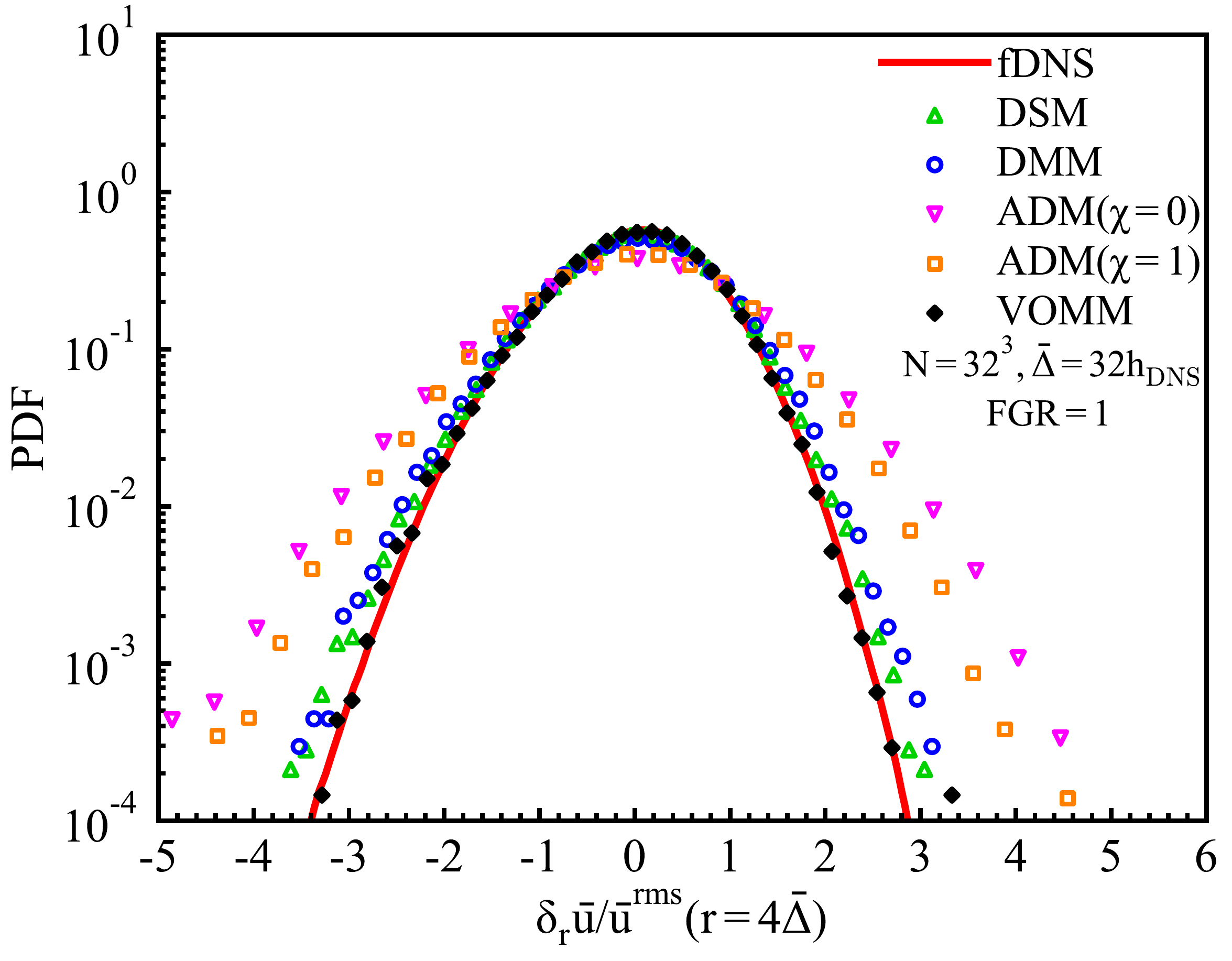}
	\end{subfigure}
	\caption{PDFs of the normalized velocity increments ${\delta _{\rm{r}}}\bar u/{{\bar u}^{{\rm{rms}}}}$ for LES at grid resolution of $32^3$ in the \emph{a posteriori} analysis of forced homogeneous isotropic turbulence with the same filter scale $\bar \Delta  = 32{h_{{\rm{DNS}}}}$: (a) ${\rm{r}} = \bar \Delta $; (b) ${\rm{r}} = 2\bar \Delta$; (c) ${\rm{r}} = 3\bar \Delta$; (d) ${\rm{r}} = 4\bar \Delta$.}
	\label{fig:8}
\end{figure}

To further examine the reconstruction of multiscale properties of turbulence by the SGS models, we calculate the longitudinal structure functions of the filtered velocity, namely \citep{xie2018,xie2019a}
\begin{equation}
	{\bar S_n}(r) = \left\langle {{{\left| {\frac{{{\delta _r}\bar u}}{{{{\bar u}^{{\rm{rms}}}}}}} \right|}^n}} \right\rangle,
	\label{Struct_n}
\end{equation}
where $n$ represents the order of structure function and ${\delta _r}\bar u = \left[ {{\bf{\bar u}}\left( {{\bf{x}} + {\bf{r}}} \right) - {\bf{\bar u}}\left( {\bf{x}} \right)} \right] \cdot {\bf{\hat r}}$ denotes the longitudinal velocity increment at the separation $\bf{r}$ with the unit distance vector ${\bf{\hat r}} = {\bf{r}}/\left| {\bf{r}} \right|$. Figures~\ref{fig:5}, \ref{fig:6} and \ref{fig:7} respectively compare the second-order, fourth-order and sixth-order structure functions of the filtered velocity for different SGS models with the filtered DNS data. For all three grid resolutions of LES (FGR=1, 2 and 4), all SGS models predict the lower-order structure functions (Fig.~\ref{fig:5}) much better than the higher-order structure functions (Figs.~\ref{fig:6} and \ref{fig:7}). Besides, the predictions of structure functions are improved greatly with the increasing of the grid resolution, and those of all SGS models almost coincide with each other at large separations. The ADM models (both $\chi = 0$ and 1) give the worst predictions and obviously overestimate the structure function at small distances $\bf{r}$. DSM and DMM models also predict the structure functions greater than the fDNS data at small separations but underestimate the structure functions at large distances. In contrast, the VOMM model can accurately reconstruct the structure functions with different orders at both small and large separations, almost overlapping with those of the filtered DNS. 

\begin{figure}\centering
	\begin{subfigure}{0.40\textwidth}
		\centering
		{($a$)}
		%\caption{fDNS}
		\includegraphics[width=0.9\linewidth,valign=t]{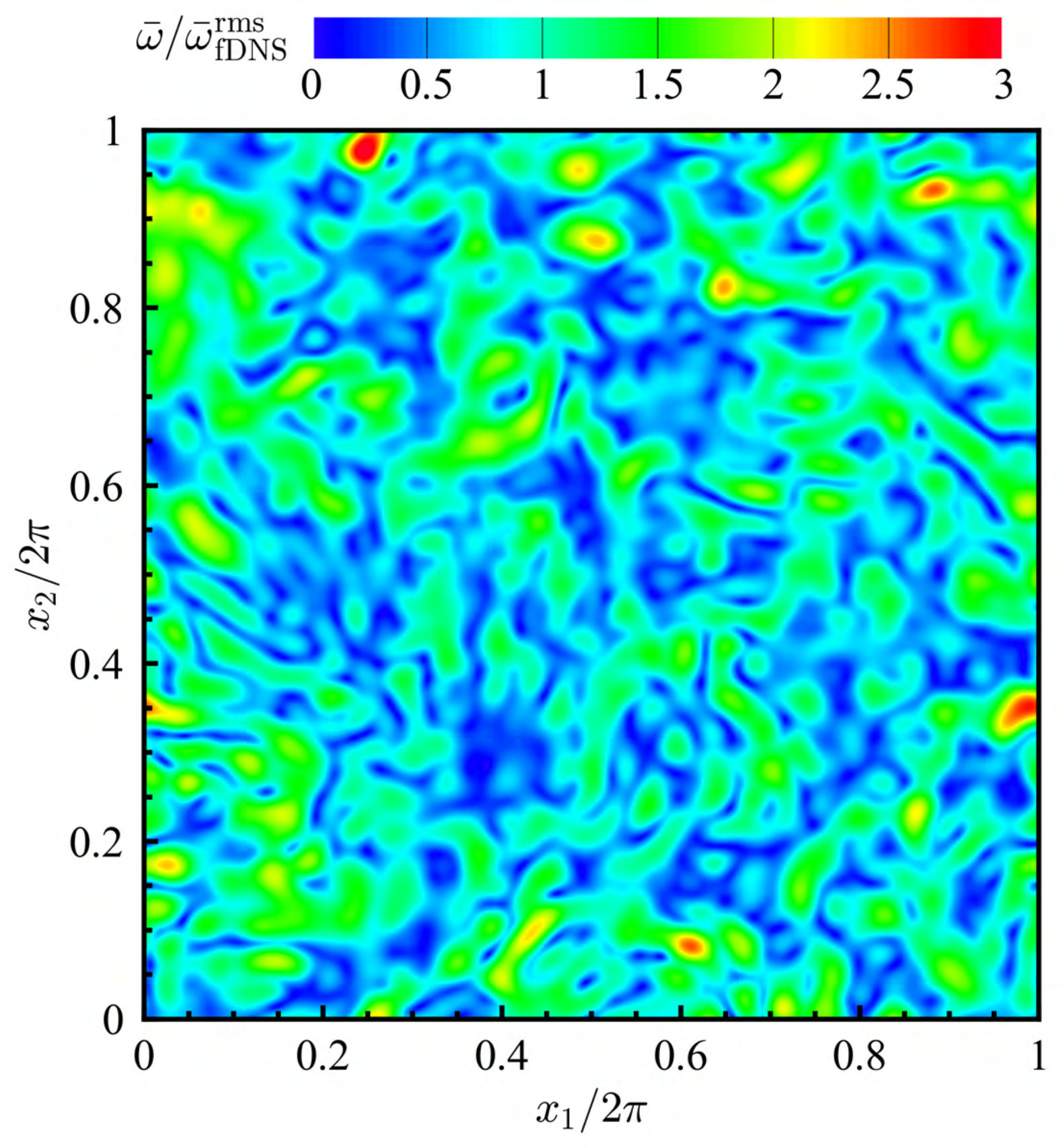}
		
	\end{subfigure}%
	\begin{subfigure}{0.40\textwidth}
		\centering
		{($b$)}
		%\caption{DMM}
		\includegraphics[width=0.9\linewidth,valign=t]{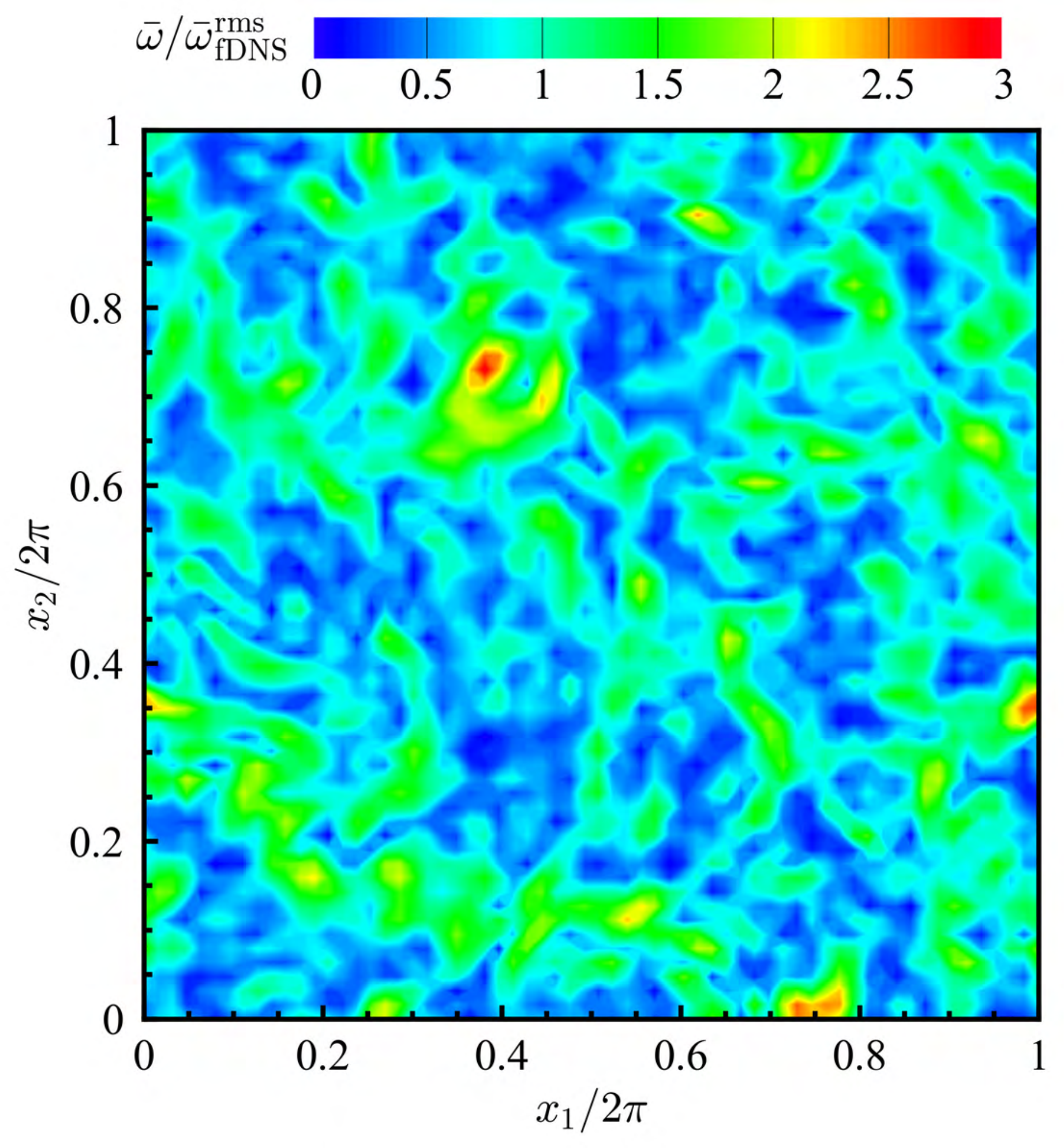}
	\end{subfigure}\\
	\begin{subfigure}{0.40\textwidth}
		\centering
		{($c$)}
		%\caption{ADM (${\chi \!=\! 0}$)}
		\includegraphics[width=0.9\linewidth,valign=t]{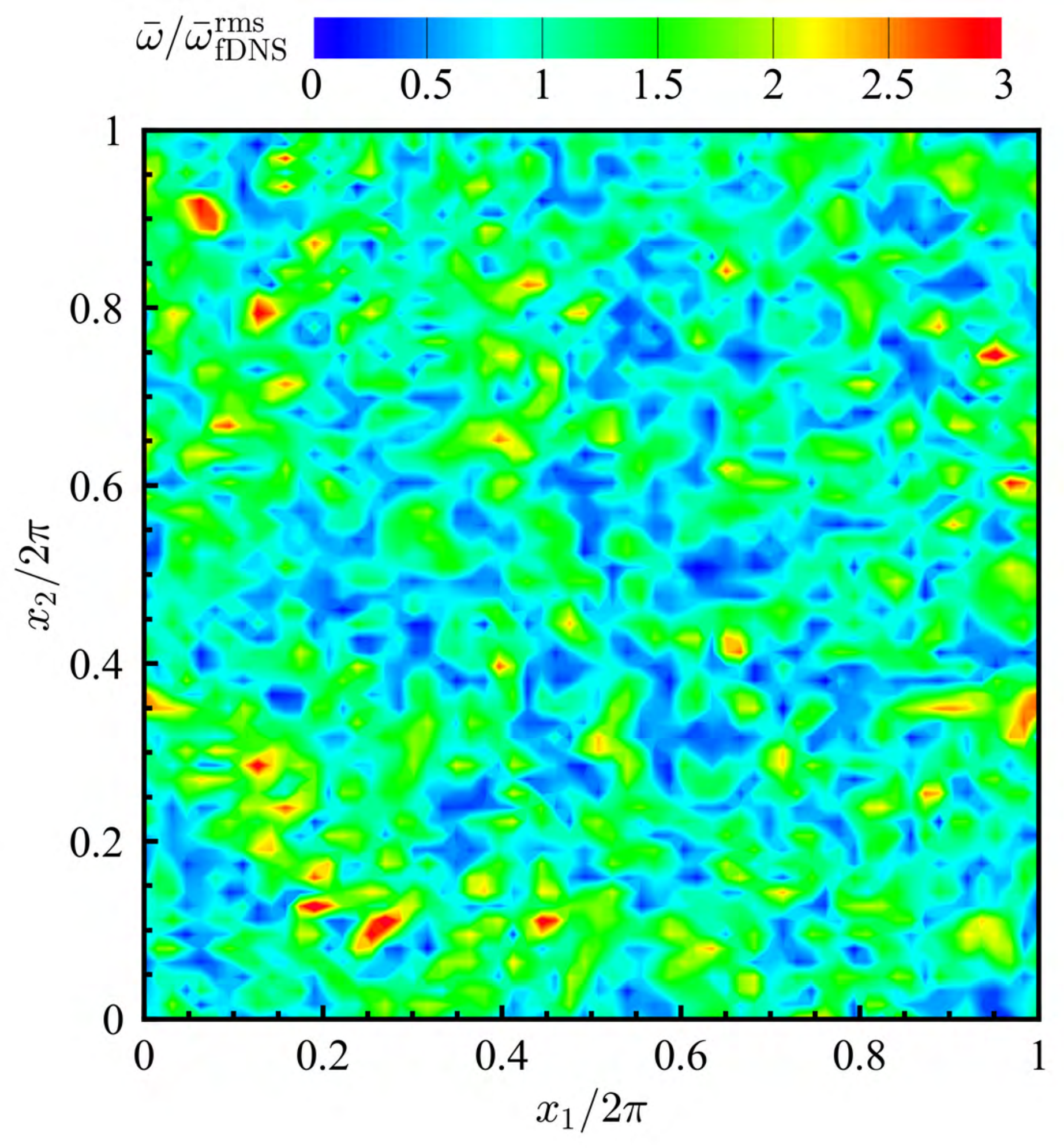}
	\end{subfigure}%
	\begin{subfigure}{0.40\textwidth}
		\centering
		{($d$)}
		%\caption{ADM (${\chi \!=\! 1}$)}
		\includegraphics[width=0.9\linewidth,valign=t]{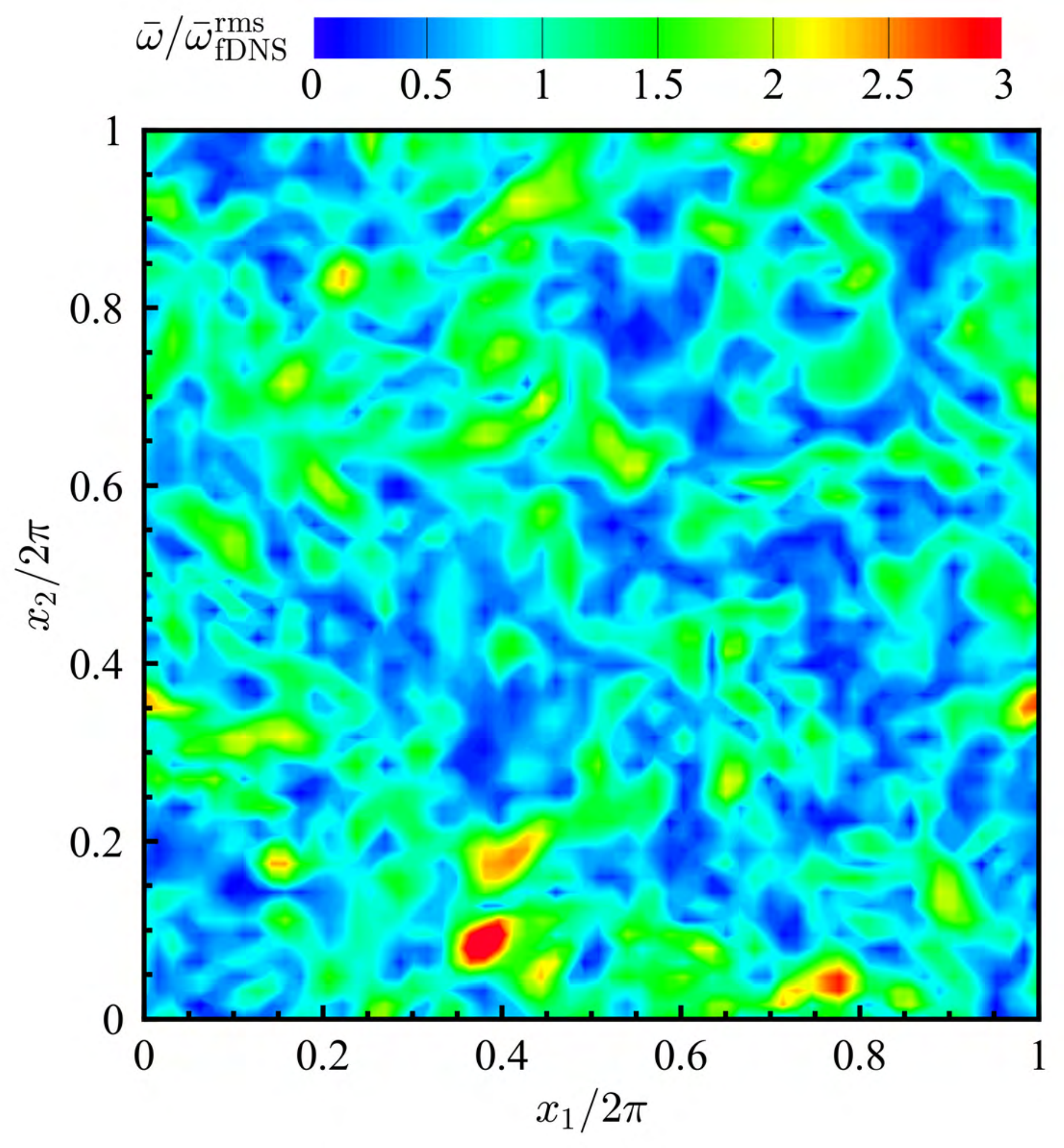}
	\end{subfigure}\\
	\begin{subfigure}{0.40\textwidth}
		\centering
		{($e$)}
		%\caption{DSM}
		\includegraphics[width=0.9\linewidth,valign=t]{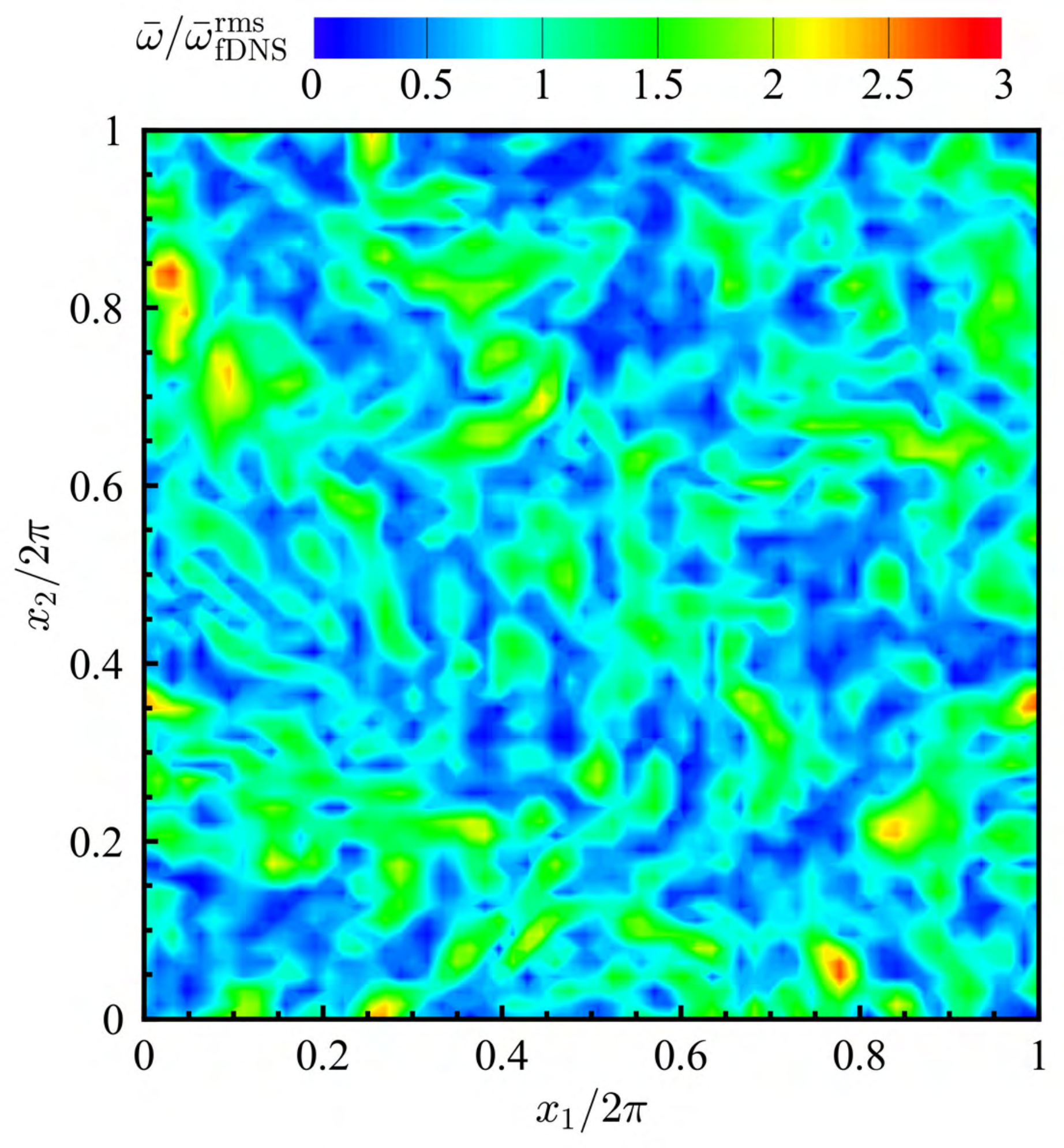}
	\end{subfigure}%
	\begin{subfigure}{0.40\textwidth}
		\centering
		{($f$)}
		%\caption{VOMM}
		\includegraphics[width=0.9\linewidth,valign=t]{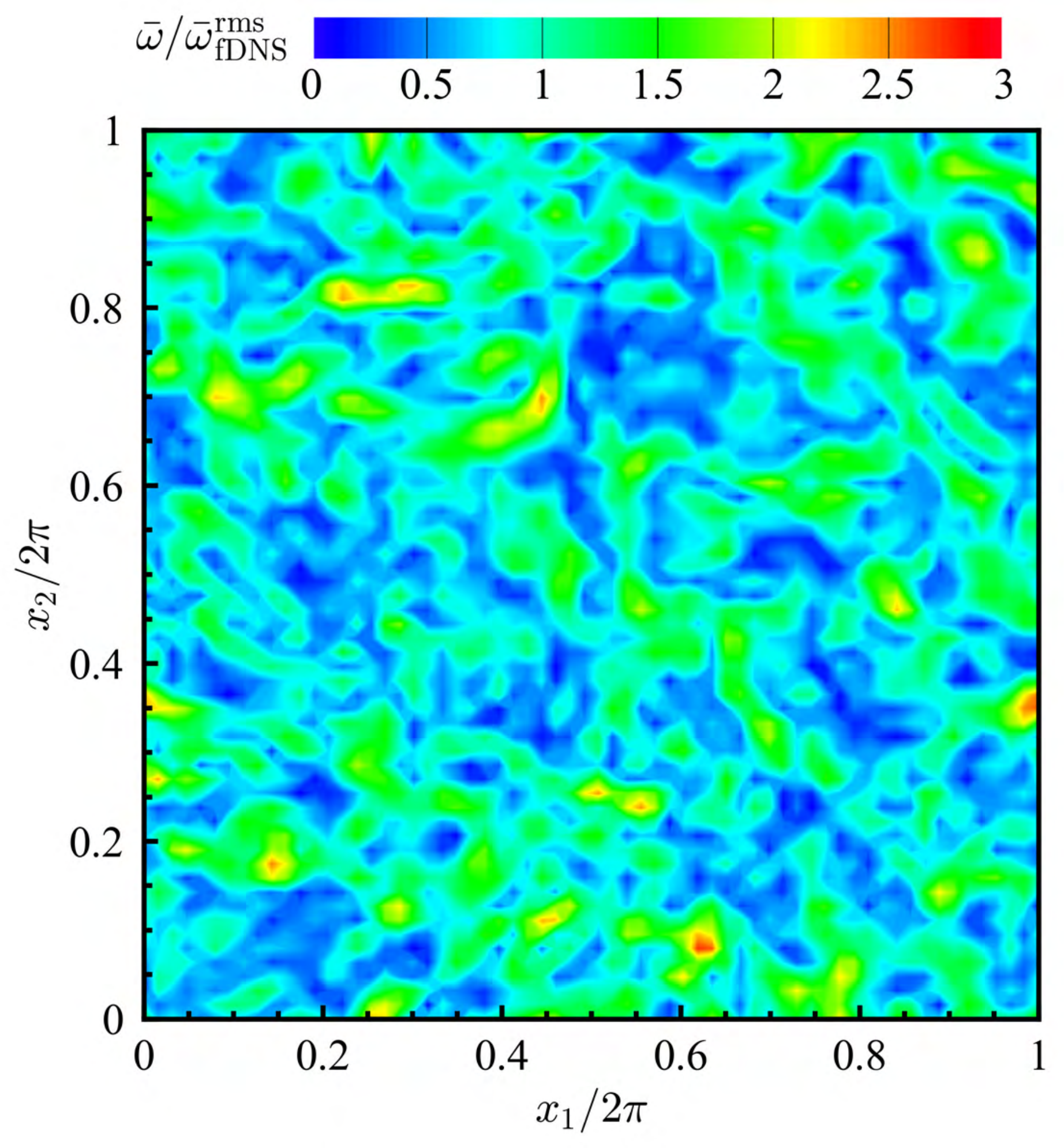}
	\end{subfigure}
	\caption{Contours of the normalized vorticity $\bar \omega /\bar \omega _{{\rm{fDNS}}}^{{\rm{rms}}}$ at an arbitrary $x_1$-$x_2$ plane at $t/\tau \approx$ 4 for LES at a grid resolution of $64^3$ (FGR=2) in forced homogeneous isotropic turbulence with the filter width $\bar \Delta  = 32{h_{{\rm{DNS}}}}$: (a) fDNS, (b) DMM, (c) ADM($\chi$=0), (d) ADM($\chi$=1), (e) DMM, and (f) VOMM.}
	\label{fig:9}
\end{figure}

We then evaluate the probability density functions (PDFs) of the filtered velocity increments to measure the spatial correlations of turbulence, as shown in Fig.~\ref{fig:8}, where the velocity increments ${\delta _r}\bar u/{{\bar u}^{{\rm{rms}}}}$ are normalized by the root-mean-square value of velocity. The cases of fine grid resolutions (FGR=2 and 4) are very similar to that of FGR=1 and not shown in the paper. The PDFs of the velocity increments exhibit approximately symmetrical distribution, relatively concentrated at small distances while gradually becoming wider as the distance increases. The PDFs predicted by the ADM, DSM and DMM models are significantly wider than those of the fDNS. In comparison with these traditional SGS models, the VOMM model gives the most accurate prediction of the velocity increments for different distances, which are in reasonable agreement with the fDNS data.    

We finally examine the reconstruction of instantaneous spatial flow structures by plotting the contours of the normalized vorticity magnitude as shown in Fig.~\ref{fig:9}. The vorticity contours are consistently extracted on an arbitrary $x_1$-$x_2$ plane for the isotropic turbulence at the same time with approximately four large-eddy turnover periods ( $t/\tau \approx 4$) at a grid resolution of $64^3$. The exact point-to-point correlations are difficult to achieve under the long-term forecasting of LES due to the chaotic nature of the turbulence and extreme sensitivity to perturbations \citep{pope2000,wang2022,wang2023}. The pure ADM model overpredicts some unrealistic small-scale structures, which are obviously different from the band-like or strip-like spatial structures of the fDNS data. The DSM, DMM and ADM ($\chi = 1$)  models only predict the large-scale vorticity structures and some small scales are excessively dissipated. Compared to these traditional SGS models, the VOMM model predicts the vortex structures very similar to the fDNS data.  

\begin{figure}\centering
	\includegraphics[width=0.7\textwidth]{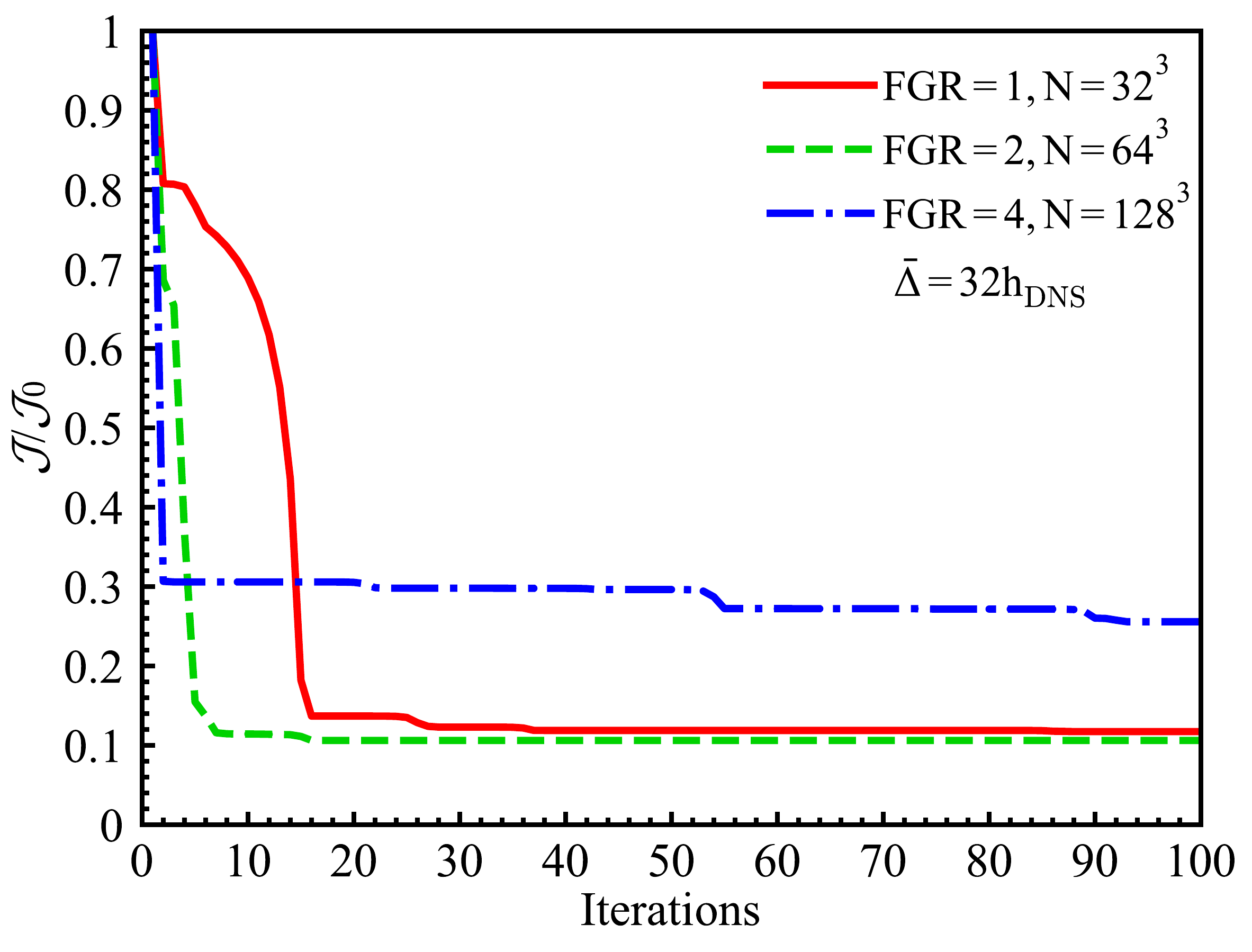}
	\caption{The evolution of the normalized cost function in decaying homogeneous isotropic turbulence.}\label{fig:10}
\end{figure}

\begin{table}
	\begin{center}
		\caption{The initial and optimal parameters of the VOMM model for LES computations with the filter width $\bar \Delta = 32 h_{\rm{DNS}}$ in decaying homogeneous isotropic turbulence.} 	
		\small
		\begin{tabular*}{0.95\textwidth}{@{\extracolsep{\fill}} lccccc} 
			\hline\hline	
			FGR & LES Resolution & $C_1^{\left( 0 \right)}$ & $C_2^{\left( 0 \right)}$ & $C_1^{\rm{opt}}$ & $C_2^{\rm{opt}}$ \\ \hline
			1 & $32^3$  & 0 & 1 & -0.0398 & 3.150 \\ 
			2 & $64^3$  & 0 & 1 & -0.0094 & 1.326 \\ 
			4 & $128^3$ & 0 & 1 & -0.0020 & 1.101 \\ \hline\hline
		\end{tabular*}%	
		\label{tab:4}%	
	\end{center}
\end{table}% 

\begin{table}
	\begin{center}	
		\caption{The average computational cost of SGS stress modeling $\tau_{ij}$ for LES computations with filter width $\bar \Delta=32 h_{\rm{DNS}}$ in decaying homogeneous isotropic turbulence.} 	
		\label{tab:5}%	
		\small 	
		\begin{tabular*}{0.95\textwidth}{@{\extracolsep{\fill}} lccccc}
		\hline\hline
			Model(FGR=1,$N=32^3$)       & DSM   & DMM    & ADM($\chi$=0) & ADM($\chi$=1) & VOMM \\ \hline
			t(CPU$\cdot$s) & 0.153 & 0.259 & 0.065 & 0.062 & 0.070 \\
			t/t$_{\rm{DMM}}$ & 0.590 & 1 & 0.249 & 0.239 & 0.269 \\ 
			\hline
			Model(FGR=2,$N=64^3$)       & DSM   & DMM    & ADM($\chi$=0) & ADM($\chi$=1) & VOMM \\ 
			t(CPU$\cdot$s) & 1.026 & 1.857 & 0.567 & 0.563 & 0.589 \\
			t/t$_{\rm{DMM}}$ & 0.553 & 1 & 0.306 & 0.303 & 0.317 \\ 
			\hline
			Model(FGR=4,$N=128^3$)       & DSM   & DMM    & ADM($\chi$=0) & ADM($\chi$=1) & VOMM \\ 
			t(CPU$\cdot$s) & 6.026 & 10.287 & 2.521 & 2.531 & 3.393 \\
			t/t$_{\rm{DMM}}$ & 0.586 & 1 & 0.245 & 0.246 & 0.330 \\ \hline\hline
		\end{tabular*}%
	\end{center}
\end{table}% 

\subsection {Decaying homogeneous isotropic turbulence}
To investigate the impact of turbulent unsteady evolution on SGS stress modeling, the numerical simulation of decaying homogeneous isotropic turbulence in a cubic box of ${\left( {2\pi } \right)^3}$ with periodic boundary conditions is investigated in this subsection. The numerical simulation method is consistent with the forced homogeneous isotropic turbulence. We spatially discretize the governing equations using the pseudo-spectral method with the two-thirds dealiasing rule at a uniform grid resolution of $N=1024^3$. The temporal discretization scheme adopts the second-order two-step Adams-Bashforth explicit method. The statistically steady isotropic turbulence data of the forced isotropic turbulence (detailed statistics see Fig.~\ref{fig:1}) is used as the initial field for DNS decaying turbulence without the large-scale forcing. The kinematic viscosity is set to $\nu=0.001$ and the initial Taylor Reynolds number is  ${{\mathop{\rm Re}\nolimits} _\lambda } \approx 250$. We calculate the DNS data of decaying turbulence for about six large-eddy turnover times ($\tau = {L_I}/{u^{{\rm{rms}}}}$), the first two of which are used for the adjoint-based optimization to determine the model coefficients of VOMM model (only the dissipation spectrum is used, stored once every 0.1$\tau$, twenty sets in total). 

\begin{figure}\centering
	\begin{subfigure}{0.33\textwidth}
		\centering
		{($a$)}
		\includegraphics[width=0.88\linewidth,valign=t]{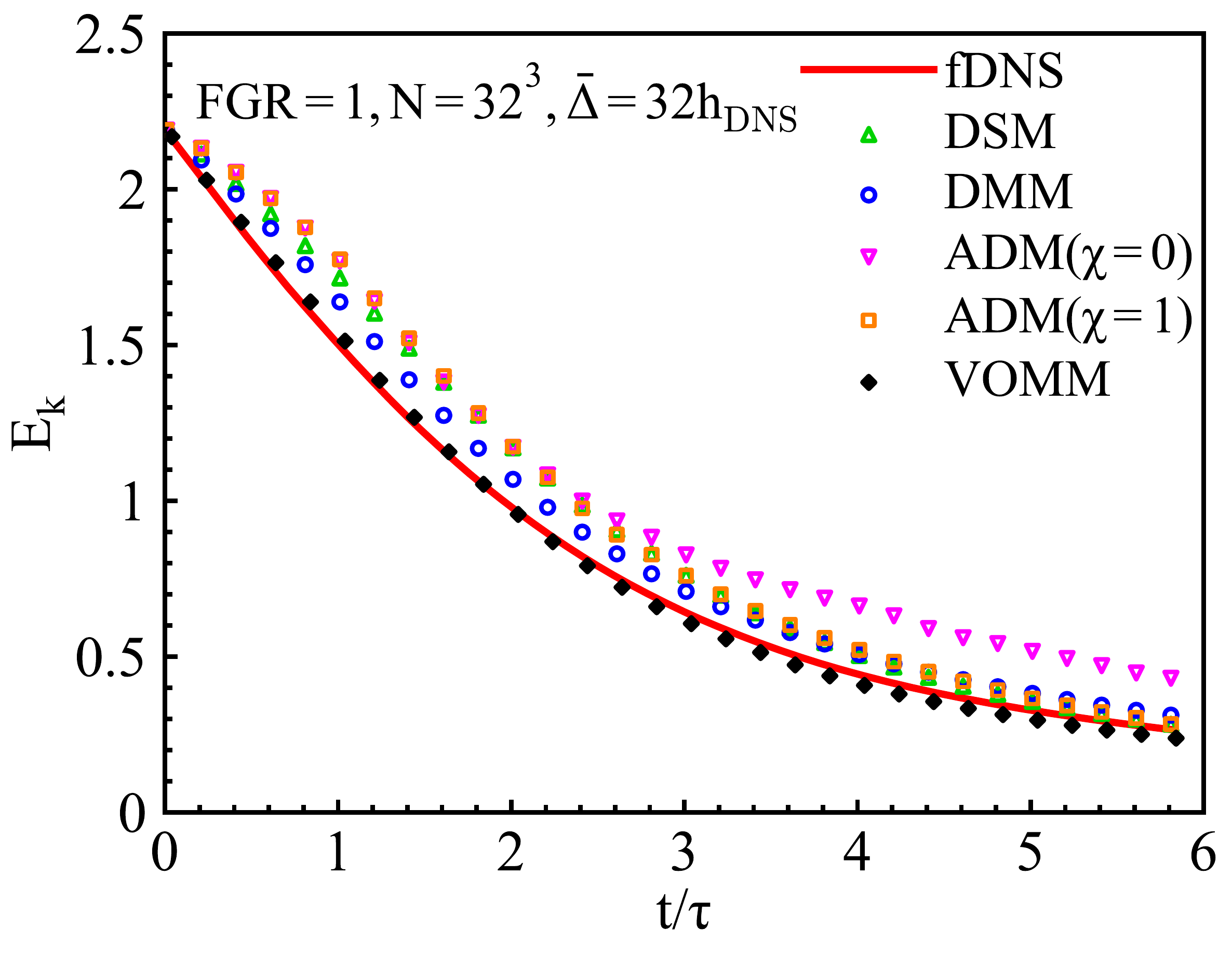}
	\end{subfigure}%
	\begin{subfigure}{0.33\textwidth}
		\centering
		{($b$)}
		\includegraphics[width=0.88\linewidth,valign=t]{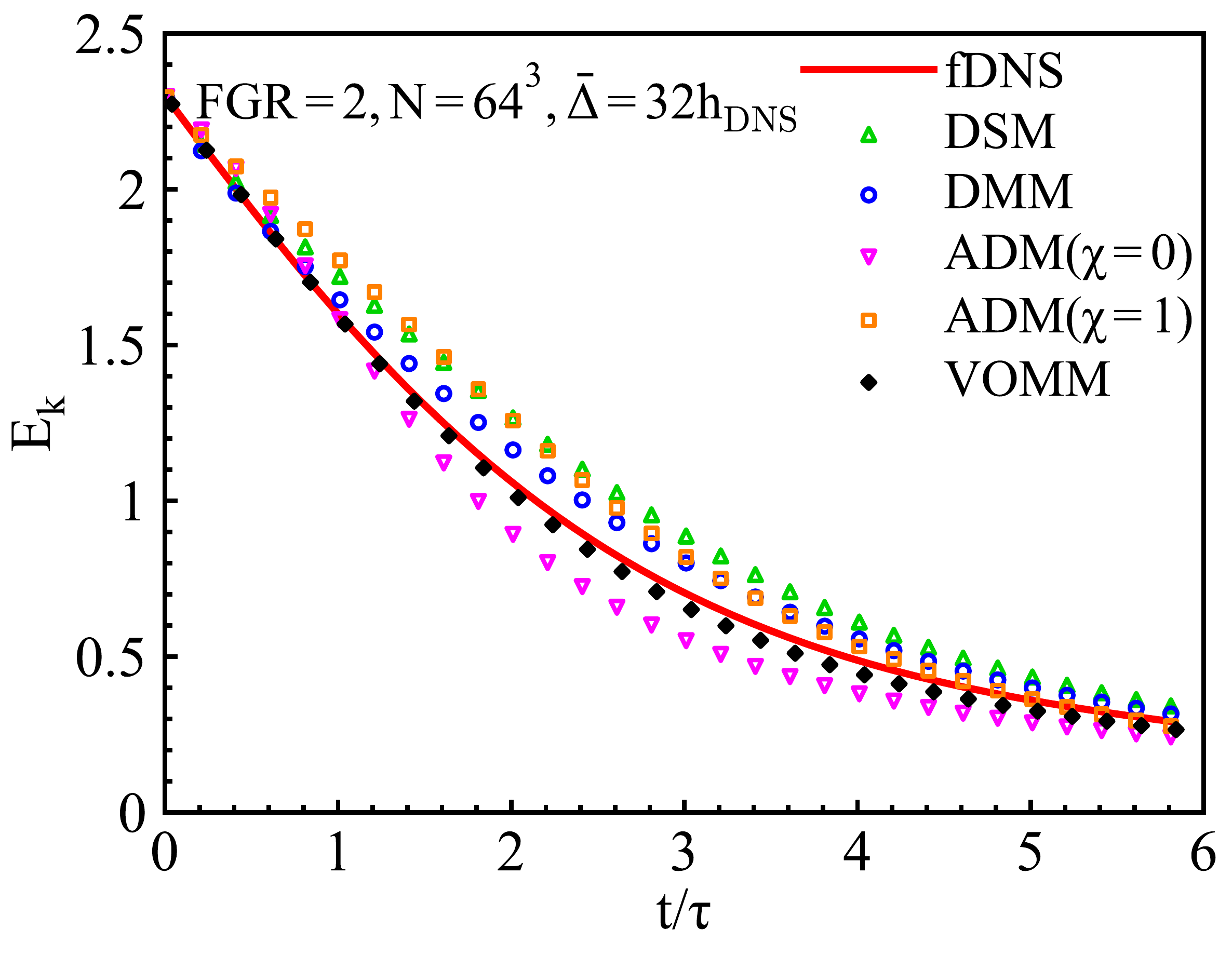}
	\end{subfigure}
	\begin{subfigure}{0.33\textwidth}
		\centering
		{($c$)}
		\includegraphics[width=0.88\linewidth,valign=t]{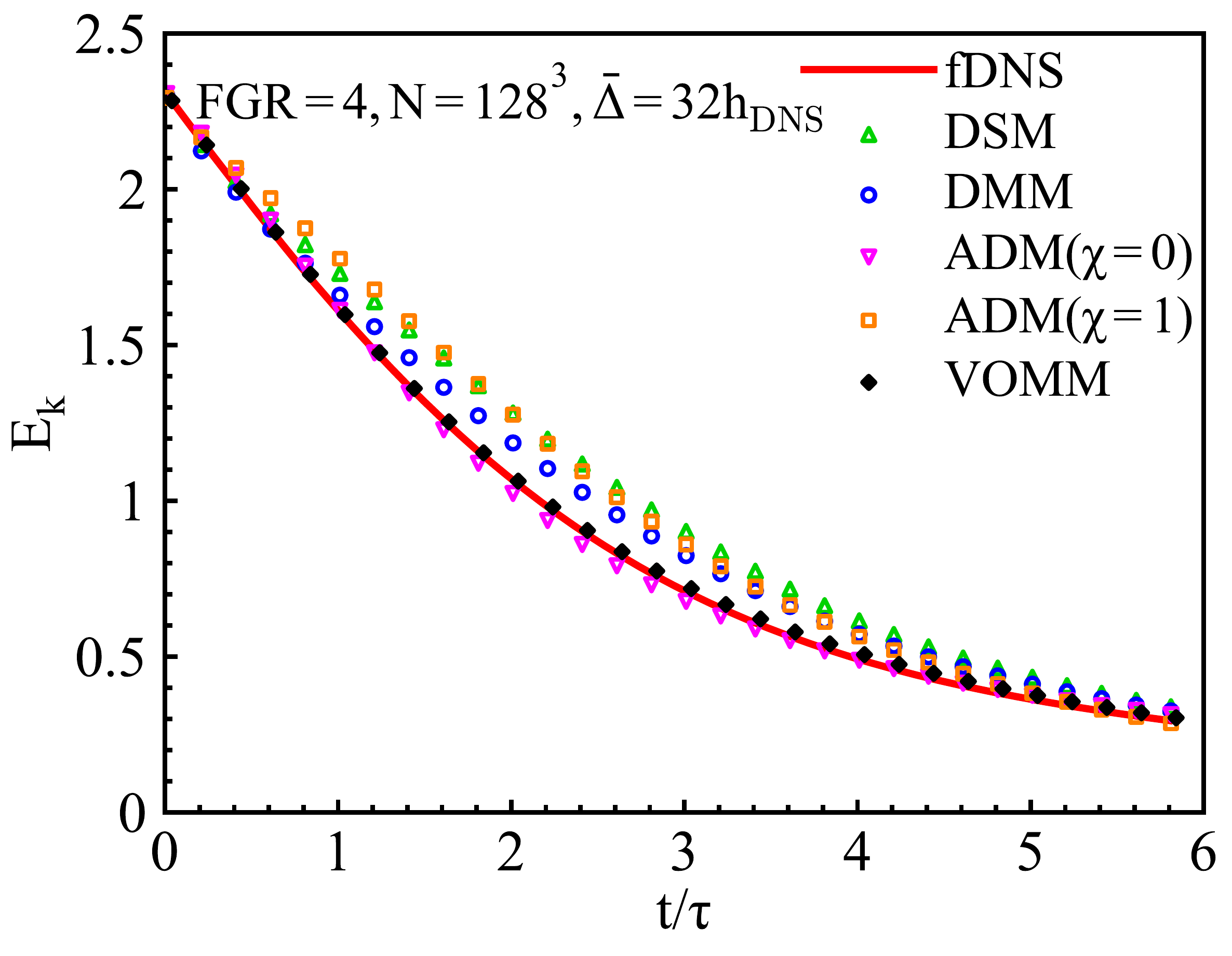}
	\end{subfigure}%
	\caption{Temporal evolutions of the turbulent kinetic energy $E_k$ for LES in the \emph{a posteriori} analysis of decaying homogeneous isotropic turbulence with the same filter scale $\bar \Delta  = 32{h_{{\rm{DNS}}}}$: (a) FGR=1, $N=32^3$; (b) FGR=2, $N=64^3$; and (c) FGR=4, $N=128^3$.}
	\label{fig:11}
\end{figure}

The \emph{a posteriori} studies of LES adopt the consistent kinematic viscosity ($\nu=0.001$) with the DNS. The Gaussian filter (Eq.~\ref{G}) is selected as the explicit filter with the given filter width $\bar \Delta =32h_{\rm{DNS}}$. Similar to the forced isotropic turbulence, three different filter-to-grid ratios FGR=$\bar \Delta/ h_{\rm{LES}}$=1,2 and 4 are chosen to investigate the impact of the spatial discretization on the SGS stress modeling with the corresponding grid resolutions of LES $N=32^3$, $64^3$ and $128^3$. The adjoint-based optimization of the VOMM model (cf. Fig.~\ref{fig:1}) is first performed to determine the optimal model coefficients using the dissipation spectra as the cost functional. The pure ADM model without the Smagorinsky part is used as the initial SGS model with parameters $C_1^{\left( 0 \right)}=0$ and  $C_2^{\left( 0 \right)}=1$. The adjoint-based gradients of the cost functional for the model coefficients are evaluated by successively forward solving the LES equations (Eqs.~\ref{fns1} and \ref{fns2}) and backward integrating the stabilized adjoint LES equations (Eqs.~\ref{adj_LES1} and \ref{adj_MLES2}). The gradient-based L-BFGS optimization algorithm (Eq.~\ref{opt_iterative}) is used for iteratively updating the SGS model parameters until reaching the stopping criteria. 

\begin{figure}\centering
	\begin{subfigure}{0.33\textwidth}
		\centering
		{($a$)}
		\includegraphics[width=0.88\linewidth,valign=t]{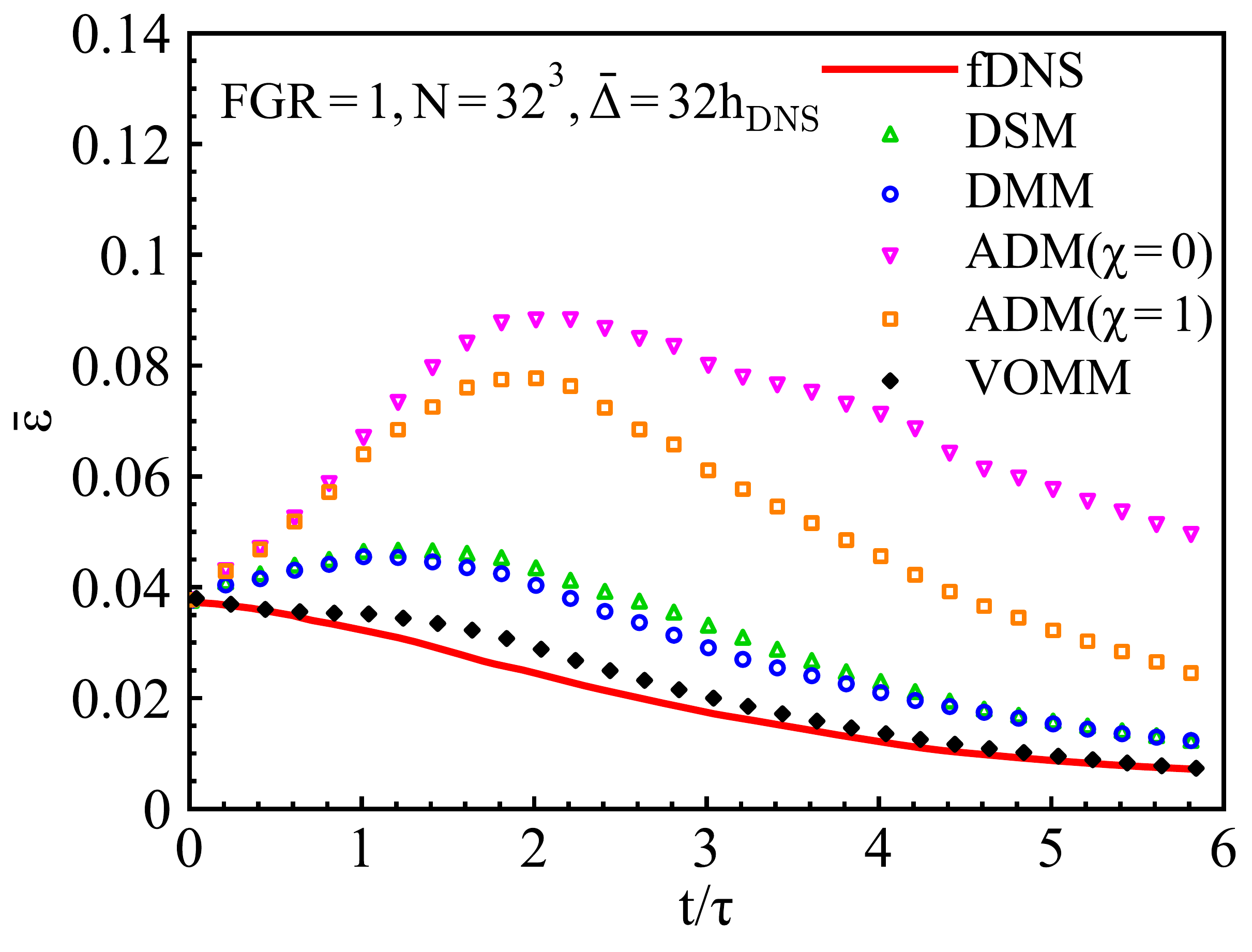}
	\end{subfigure}%
	\begin{subfigure}{0.33\textwidth}
		\centering
		{($b$)}
		\includegraphics[width=0.88\linewidth,valign=t]{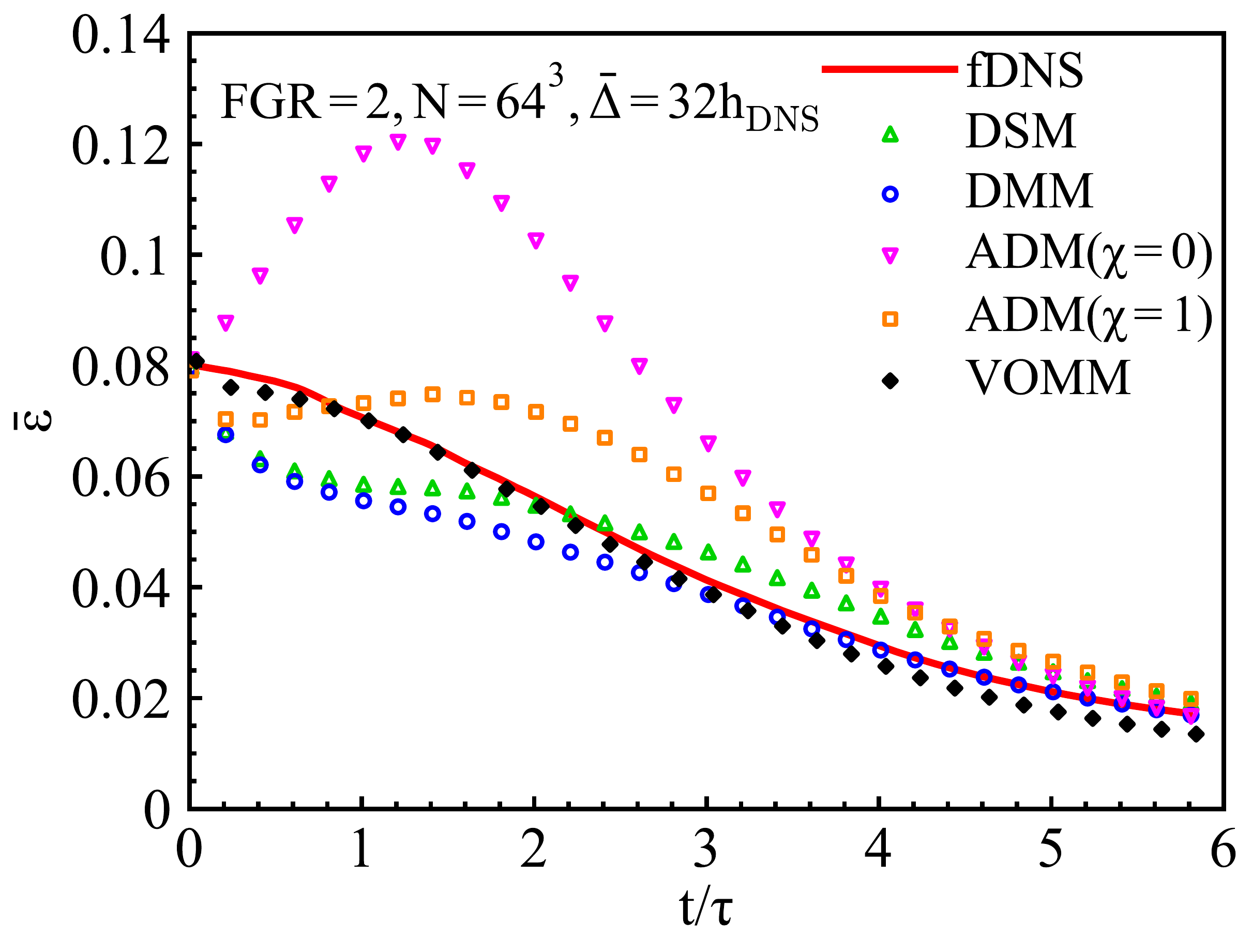}
	\end{subfigure}
	\begin{subfigure}{0.33\textwidth}
		\centering
		{($c$)}
		\includegraphics[width=0.88\linewidth,valign=t]{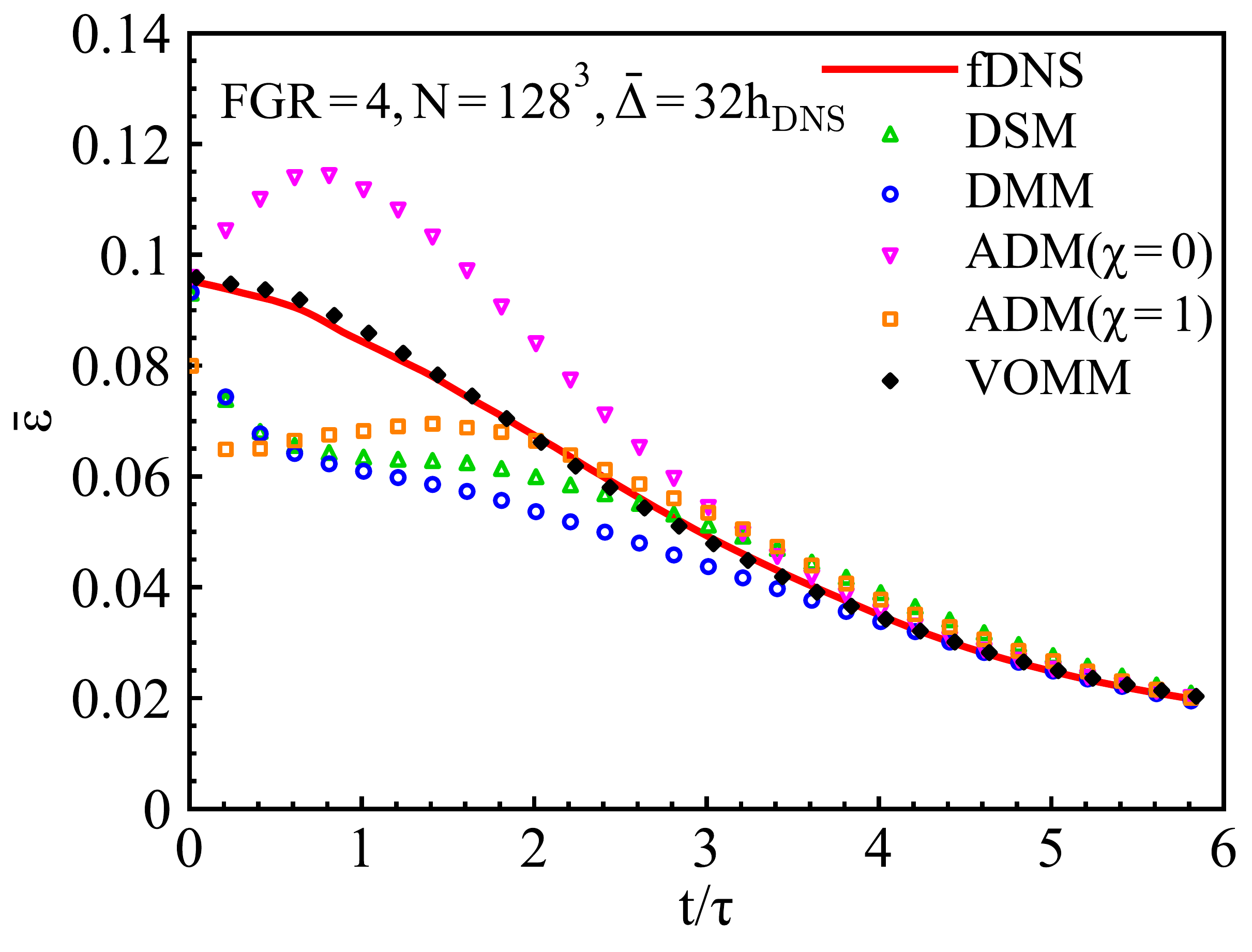}
	\end{subfigure}%
	\caption{Temporal evolutions of the average dissipation rate $\bar \varepsilon$ for LES in the \emph{a posteriori} analysis of decaying homogeneous isotropic turbulence with the same filter scale $\bar \Delta  = 32{h_{{\rm{DNS}}}}$: (a) FGR=1, $N=32^3$; (b) FGR=2, $N=64^3$; and (c) FGR=4, $N=128^3$.}
	\label{fig:12}
\end{figure}

The evolution of the cost function normalized by the initial loss during the adjoint-based optimization for the decaying isotropic turbulence is displayed in Fig.~\ref{fig:10}. The loss functions for all three cases of different grid resolutions (FGR=1,2 and 4) drop rapidly at the beginning and gradually reach a plateau within approximately twenty iterations.  The prediction errors of the optimization objective are considerably reduced to 10\% of the initial state for both FGR=1 and 2, and substantially decreased to about 20\% of the original value at FGR=4. The adjoint-based gradient optimization can quickly obtain the optimal model parameters within a limited number of iterations (less than 100 optimization iterations, namely, 200 LES evaluations). Table~\ref{tab:4} gives the optimal parameters of the VOMM model. The magnitude of the dissipative Smagorinsky coefficient ($\left| {C_1^{\rm{opt}}} \right|$) significantly drops from 0.0398 to 0.002 as the LES resolution increases, which is slightly lower than that in forced homogeneous isotropic turbulence. In contrast, the coefficient of the structural part ($C_2^{\rm{opt}}$)  is asymptotically close to unity as the grid spacing of LES becomes smaller, similar to the results of forced isotropic turbulence.

\begin{figure}\centering
	\begin{subfigure}{0.5\textwidth}
		\centering
		{($a$)}
		\includegraphics[width=0.9\linewidth,valign=t]{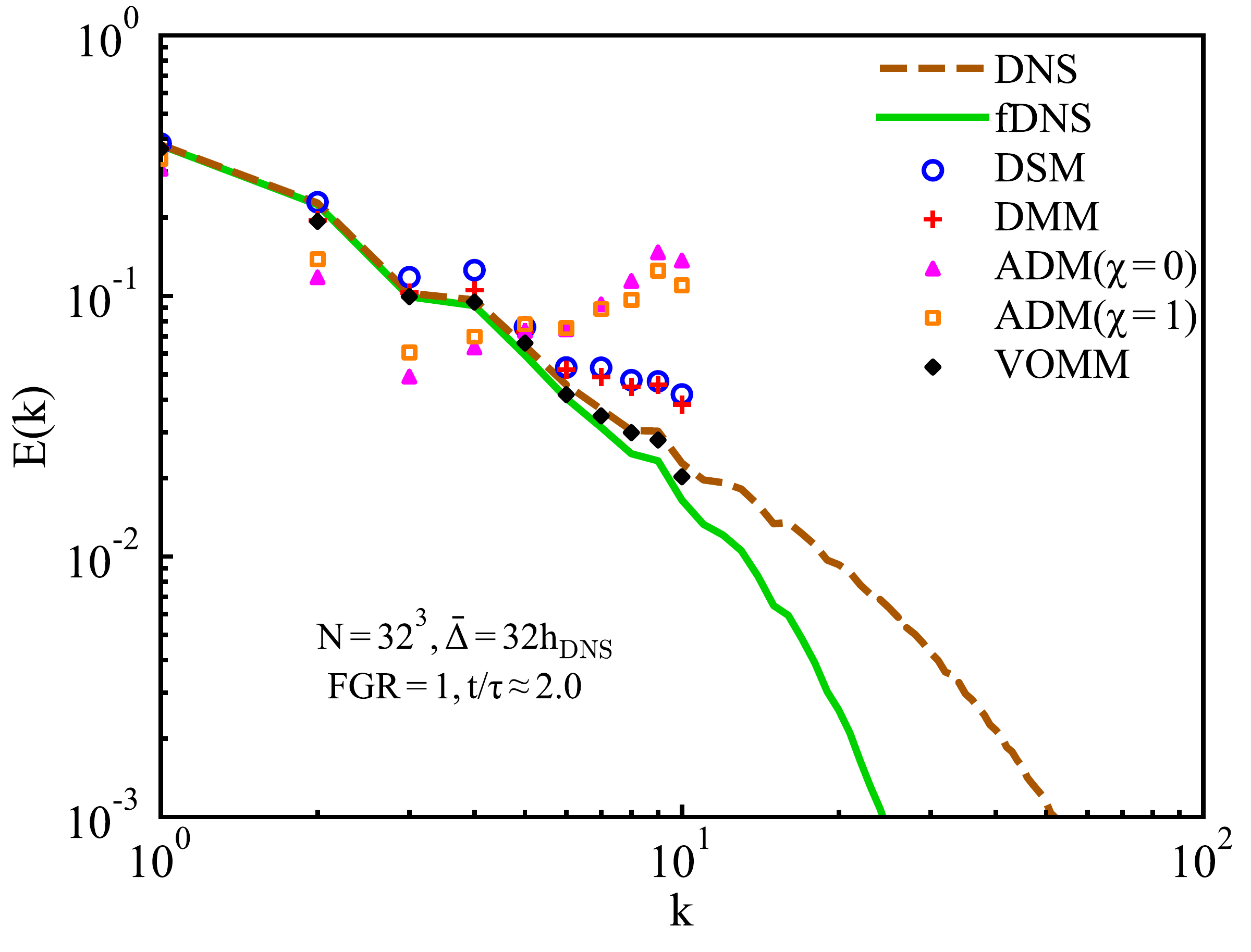}
	\end{subfigure}%
	\begin{subfigure}{0.5\textwidth}
		\centering
		{($b$)}
		\includegraphics[width=0.9\linewidth,valign=t]{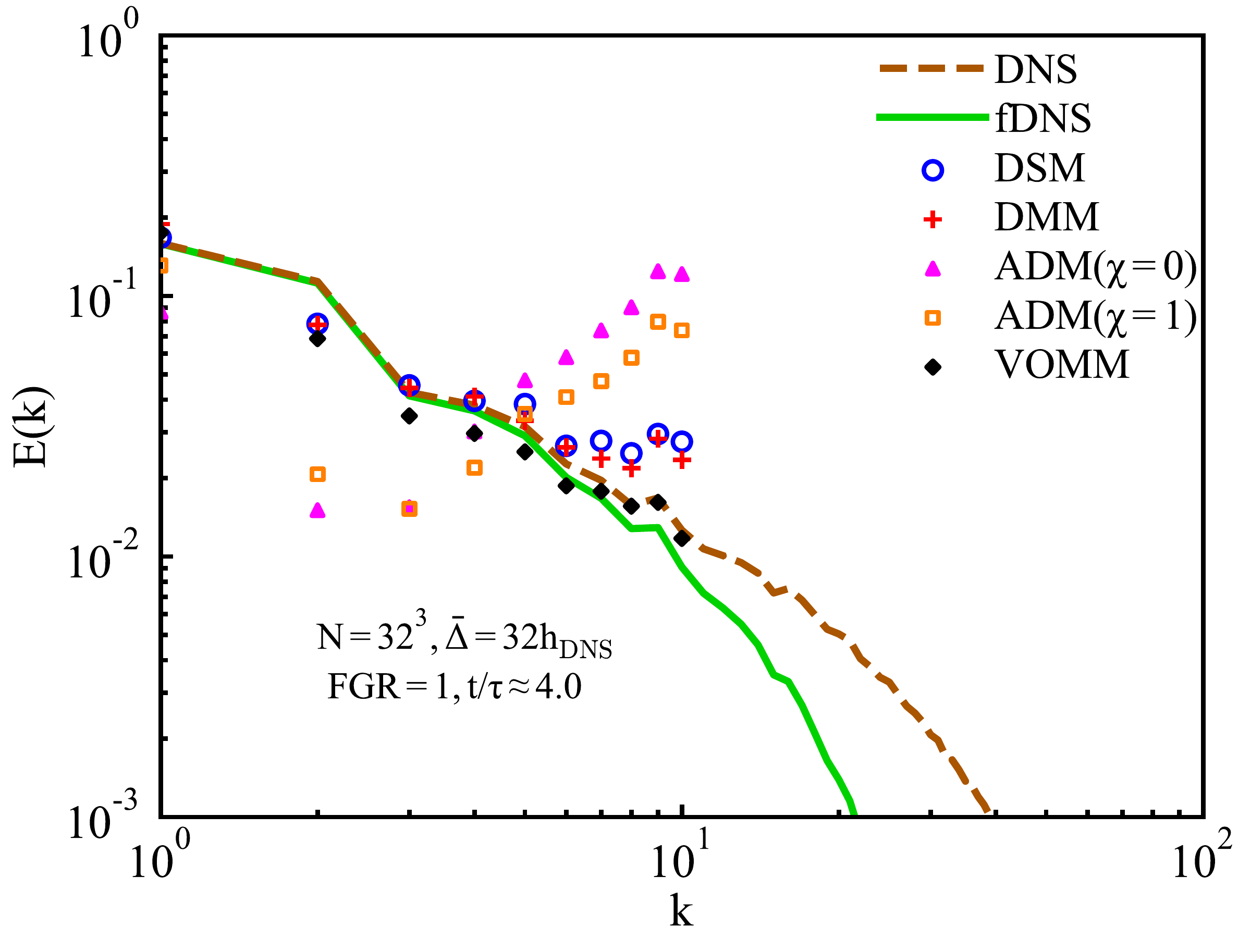}
	\end{subfigure}\\
	\begin{subfigure}{0.5\textwidth}
		\centering
		{($c$)}
		\includegraphics[width=0.9\linewidth,valign=t]{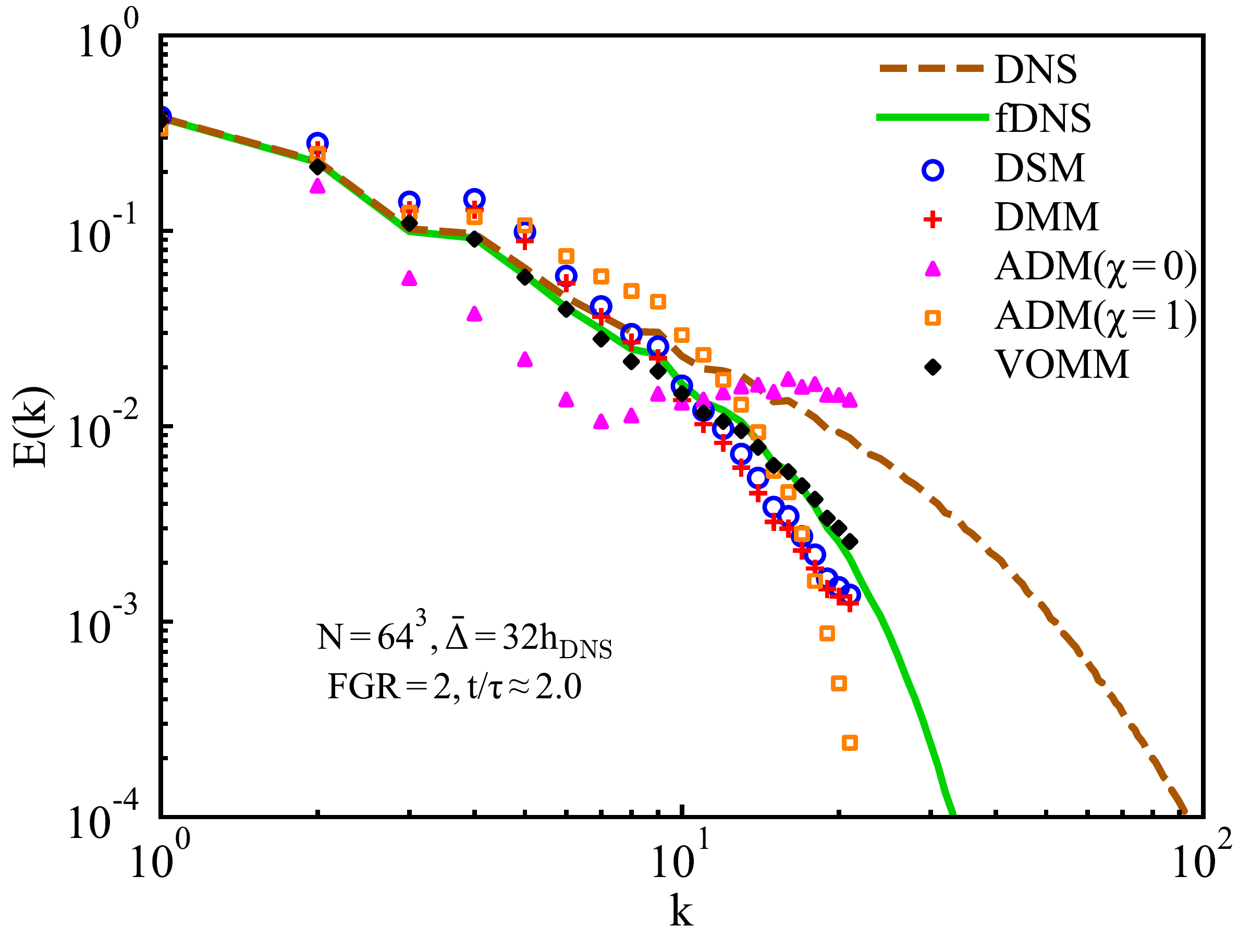}
	\end{subfigure}%
	\begin{subfigure}{0.5\textwidth}
		\centering
		{($d$)}
		\includegraphics[width=0.9\linewidth,valign=t]{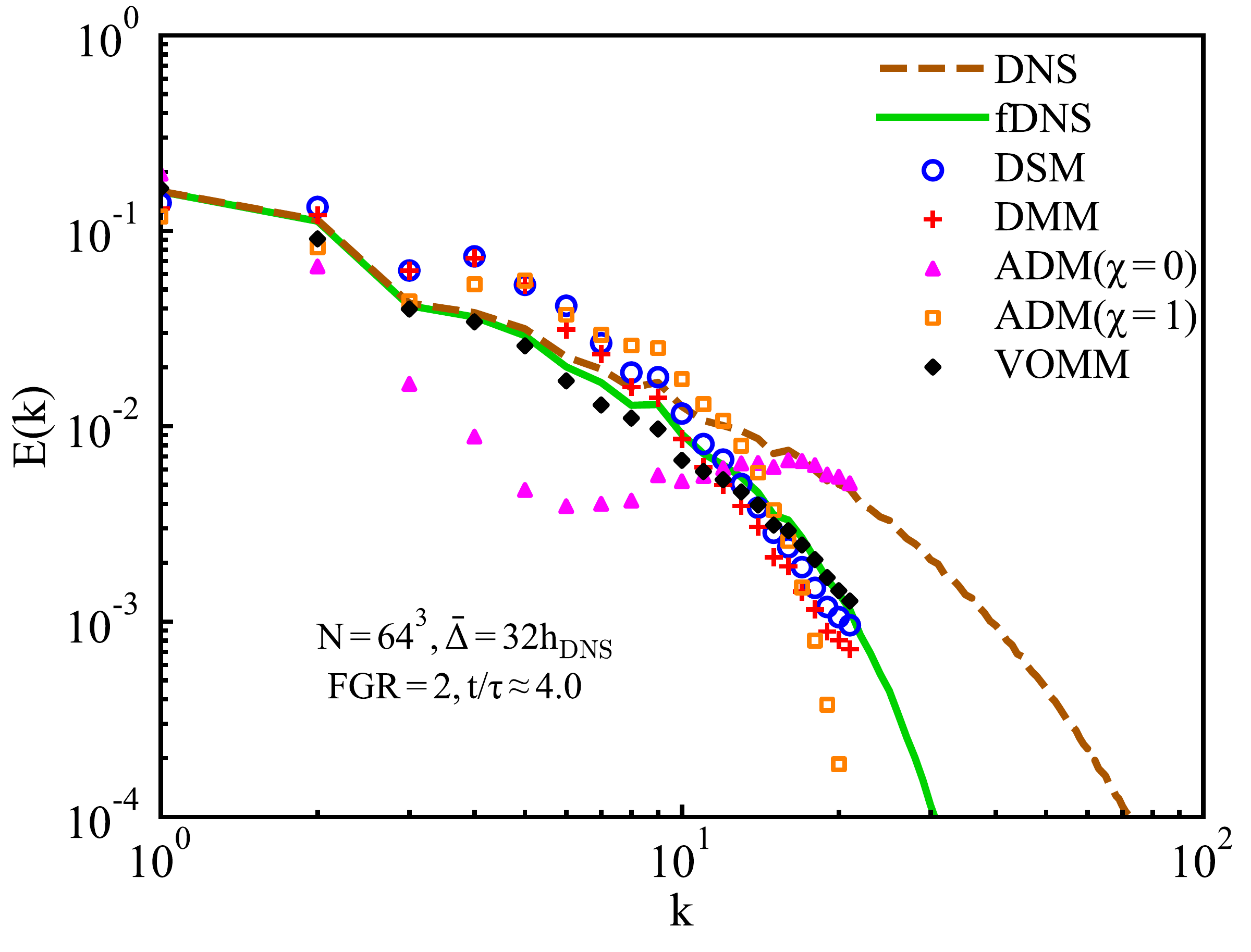}
	\end{subfigure}\\
	\begin{subfigure}{0.5\textwidth}
		\centering
		{($e$)}
		\includegraphics[width=0.9\linewidth,valign=t]{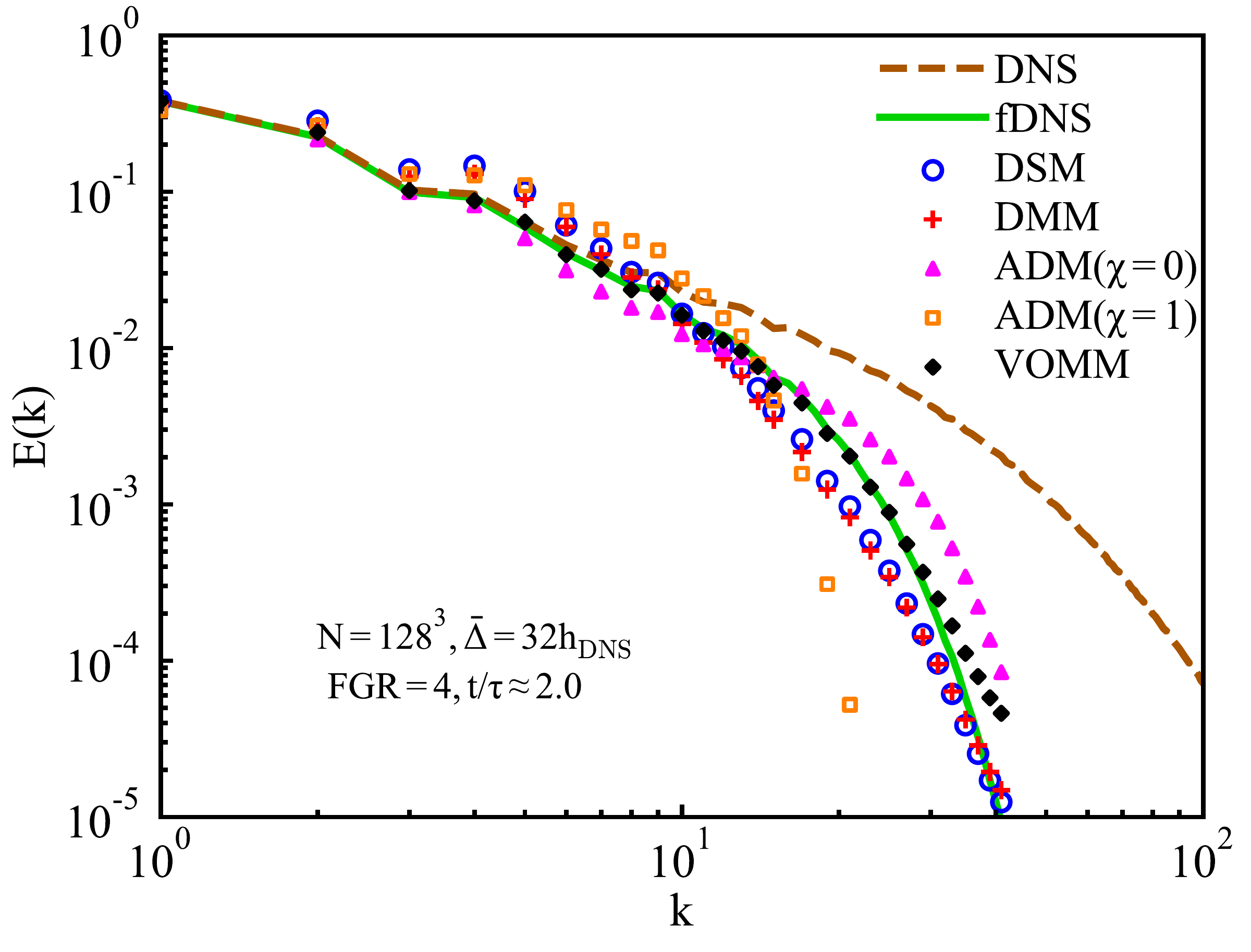}
	\end{subfigure}%
	\begin{subfigure}{0.5\textwidth}
		\centering
		{($f$)}
		\includegraphics[width=0.9\linewidth,valign=t]{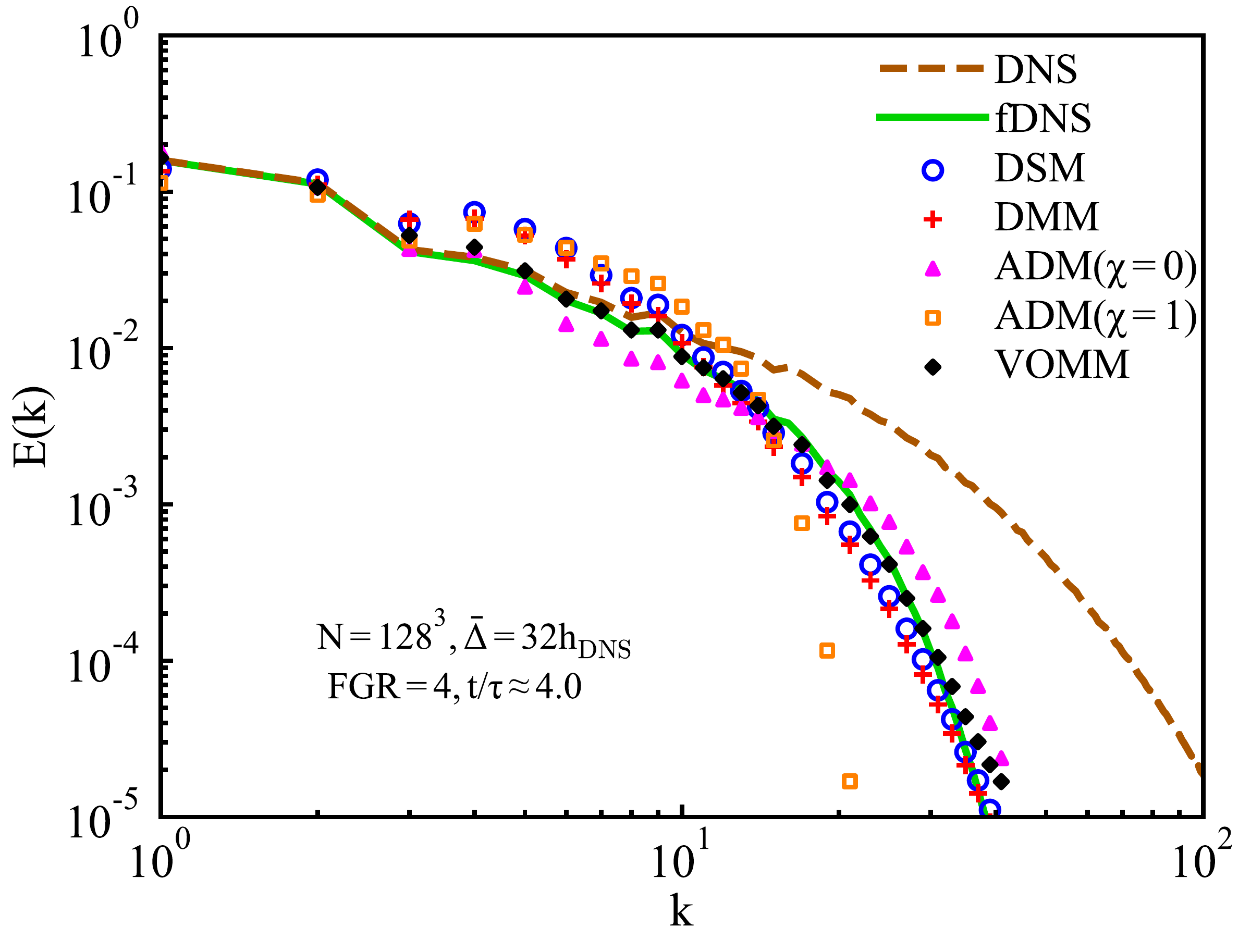}
	\end{subfigure}
	\caption{Velocity spectra for different SGS models in the \emph{a posteriori} analysis of decaying homogeneous isotropic turbulence with the same filter scale $\bar \Delta  = 32{h_{{\rm{DNS}}}}$ at $t /\tau \approx$ 2 and 4: (a) FGR=1, $N=32^3$ at $t/\tau \approx$ 2; (b) FGR=1, $N=32^3$ at $t/\tau \approx$ 4; (c) FGR=2, $N=64^3$ at $t/\tau \approx$ 2; (d) FGR=2, $N=64^3$ at $t/\tau \approx$ 4; (e) FGR=4, $N=128^3$ at $t/\tau \approx$ 2; and (f) FGR=4, $N=128^3$ at $t/\tau \approx$ 4.}
	\label{fig:13}
\end{figure}

The \emph{a posteriori} performance of the VOMM model is further validated after determining the optimal SGS model coefficients by the adjoint-based gradient optimization. We compare the proposed VOMM model (Eq.~\ref{VOMM}) with the classical SGS models including the DSM model (Eq.~\ref{tau_sm}), DMM model (Eq.~\ref{dmm1}) and the ADM model regularized by the standard  secondary-filtering technique (Eqs.~\ref{ADM} and \ref{ADM_relax}). The time steps of LES are given as $\Delta {t_{{\rm{LES}}}}/\Delta {t_{{\rm{DNS}}}} = \left\{ {10,10,5} \right\}$ for different grid resolutions (FGR=1, 2 and 4 with $N=32^3$, $64^3$ and $128^3$). The average computational costs for the SGS stress modeling with different grid resolutions using different SGS models at the same filter scale $\bar \Delta=32h_{\rm{DNS}}$ are summarized in Table~\ref{tab:5}. The computation time of the VOMM model only accounts for approximately 30\% of the time of DMM model and slightly increases in computational cost compared to the ADM models with $\chi = 0$ and 1. 

\begin{figure}\centering
	\begin{subfigure}{0.33\textwidth}
		\centering
		{($a$)}
		\includegraphics[width=0.88\linewidth,valign=t]{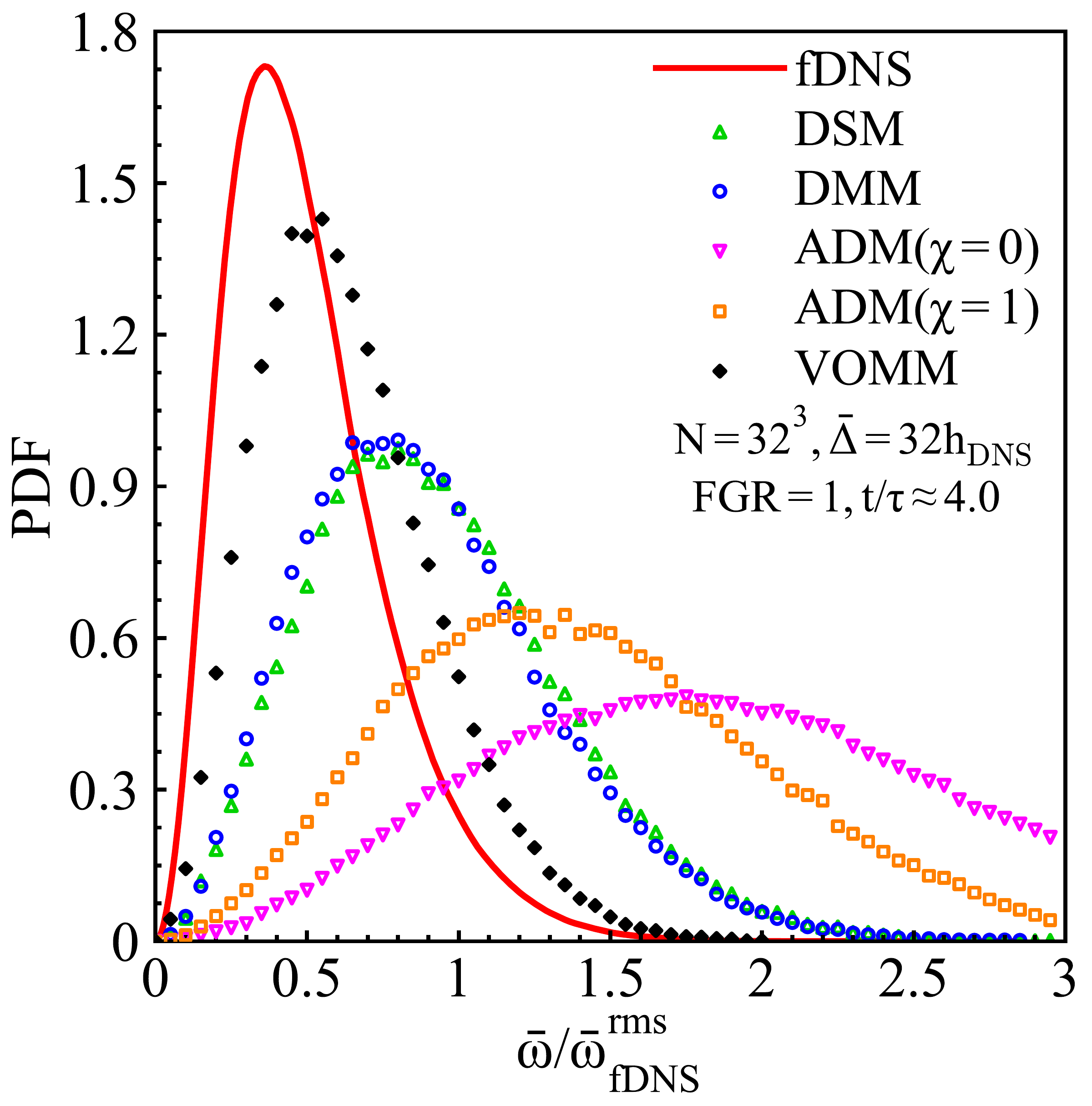}
	\end{subfigure}%
	\begin{subfigure}{0.33\textwidth}
		\centering
		{($b$)}
		\includegraphics[width=0.88\linewidth,valign=t]{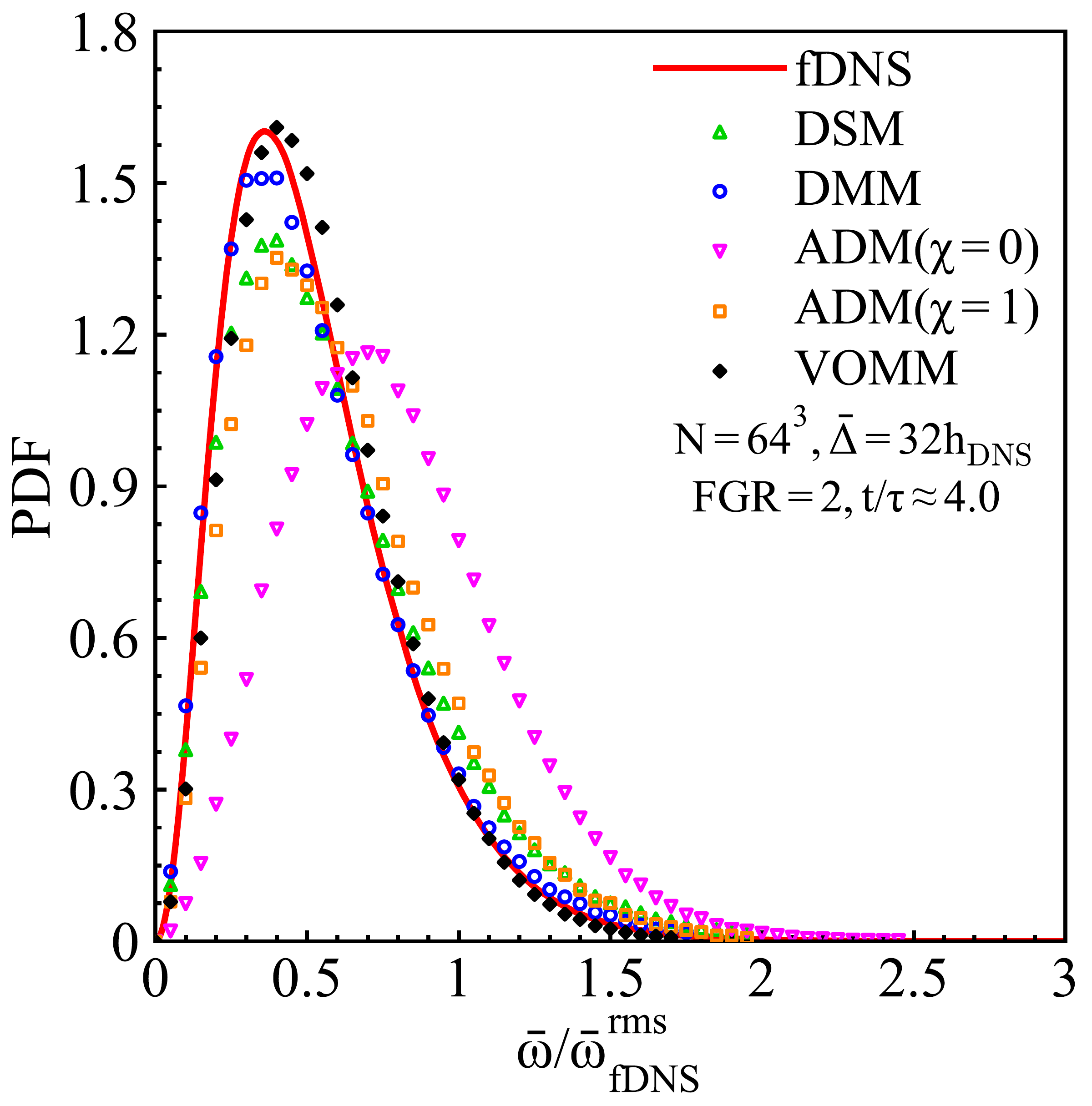}
	\end{subfigure}
	\begin{subfigure}{0.33\textwidth}
		\centering
		{($c$)}
		\includegraphics[width=0.88\linewidth,valign=t]{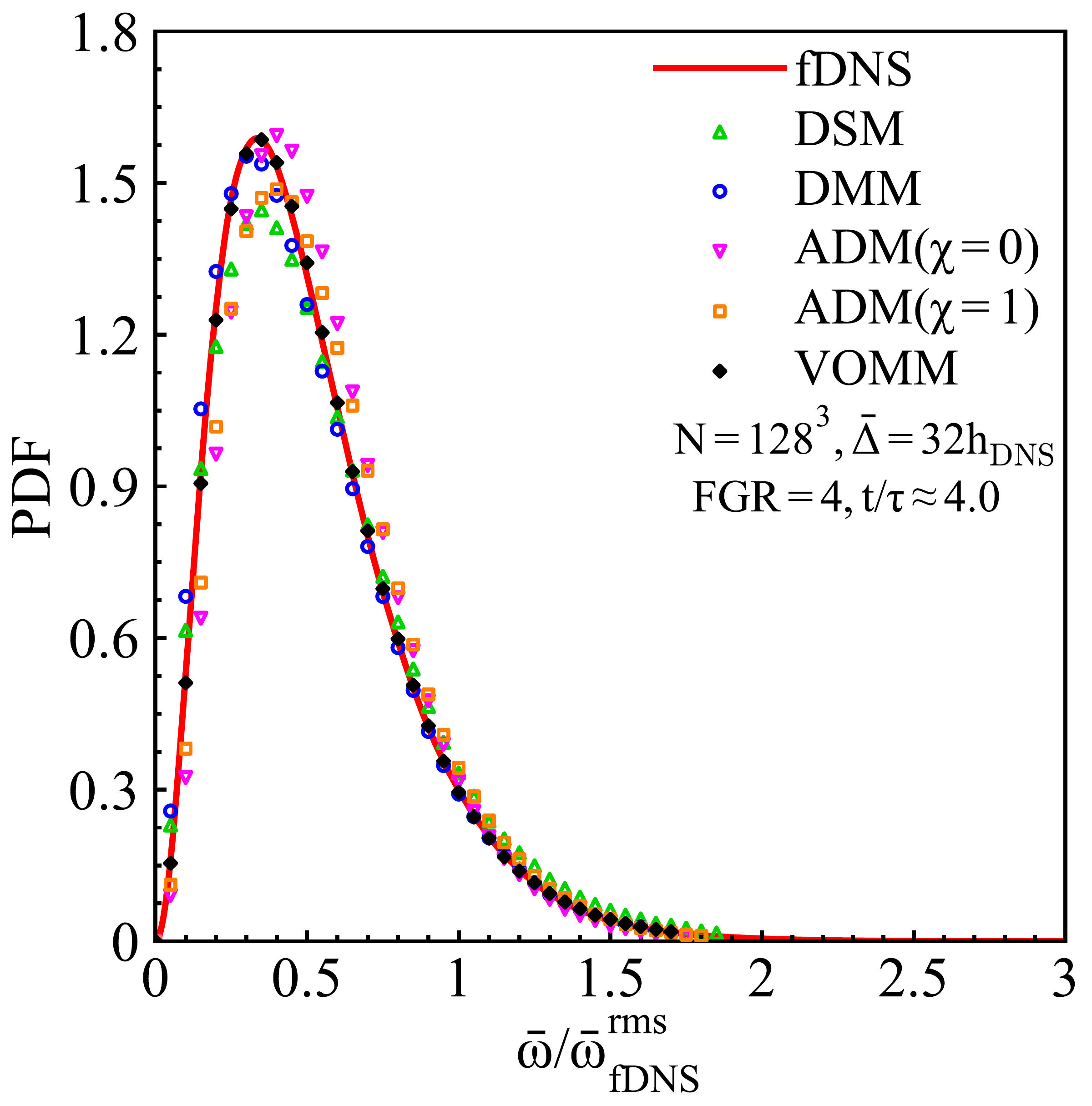}
	\end{subfigure}%
	\caption{PDFs of the normalized vorticity $\bar \omega /\bar \omega _{{\rm{fDNS}}}^{{\rm{rms}}}$ for LES in the \emph{a posteriori} analysis of decaying homogeneous isotropic turbulence with the same filter scale $\bar \Delta  = 32{h_{{\rm{DNS}}}}$ at $t/\tau \approx$ 4: (a) FGR=1, $N=32^3$; (b) FGR=2, $N=64^3$; and (c) FGR=4, $N=128^3$.}
	\label{fig:14}
\end{figure}

Figures~\ref{fig:11} and \ref{fig:12} respectively compare the temporal evolutions of the turbulent kinetic energy and the resolved dissipation rate ($\bar \varepsilon =2\nu \left\langle {{\bar S}_{ij}}{{\bar S}_{ij}} \right\rangle $) of different SGS models with the filtered DNS (fDNS) data. The turbulent kinetic energy gradually decays from the initial statistically steady state over time, since there are no additional forcing driving the dissipative turbulent system. All the classical SGS models (DSM, DMM and ADM models) clearly overestimate the kinetic energy throughout the time, which differs significantly from the benchmark fDNS data. In contrast, the VOMM model gives reasonable predictions of the turbulent kinetic energy, which is the closest to the fDNS data. The average dissipation rate displays a decline trend with time, similar to that of the turbulent kinetic energy. However, all conventional SGS models wrongly predict the non-monotonic tendency of the average dissipation rate over time. For the case of sufficiently coarse grid resolution of LES (FGR=1 with $N=32^3$), DSM, DMM and ADM models overpredict the dissipation rate with an erroneous temporal evolution that first increases and then decreases. When the grid resolution of LES becomes fine (FGR=2 and 4 with $N=64^3$  and $128^3$), DSM and DMM models obviously underestimate the dissipative rate at the early stage of decaying turbulence ($t/\tau \le 3$), then DMM model gradually becomes closer to the fDNS data while DSM model overestimates the dissipation rate with the decaying of turbulence. The pure ADM model ( $\chi = 0$ ) always gives the overestimations of the dissipation rate for all three different grid resolutions of LES, even though the pure ADM model can accurately predict the turbulent kinetic energy at a sufficiently high grid resolution (FGR=4). These results demonstrate that the pure structural ADM model without any dissipative terms might not accurately predict all physical quantities of LES (\emph{i.e.}, the average dissipation rate), even if the grid resolution is high enough compared to the filter scale (FGR=4). The ADM model with standard secondary-filtering regularization ($\chi=1$) provides excessive dissipation similar to the DSM model with mispredictions of first underestimating and then overestimating the average dissipation rate over time at FGR=2 and 4. In comparison to these classical SGS models, the VOMM model accurately predicts the temporal evolutions of average dissipation rate for all three different grid resolutions, which agrees fairly well with the benchmark filtered DNS data. 

\begin{figure}\centering
	\begin{subfigure}{0.40\textwidth}
		\centering
		{($a$)}
		%\caption{fDNS}
		\includegraphics[width=0.9\linewidth,valign=t]{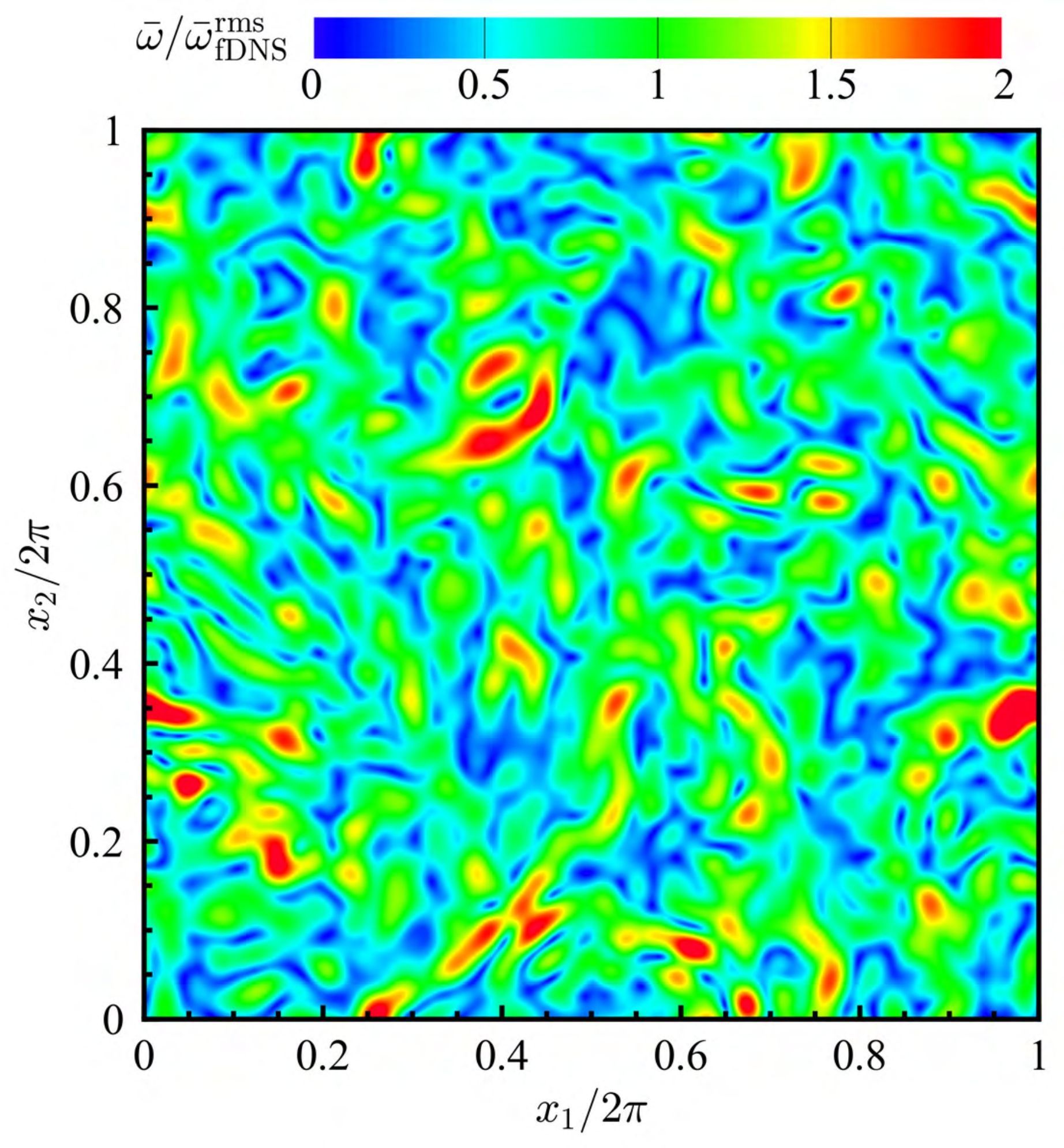}
		
	\end{subfigure}%
	\begin{subfigure}{0.40\textwidth}
		\centering
		{($b$)}
		%\caption{DMM}
		\includegraphics[width=0.9\linewidth,valign=t]{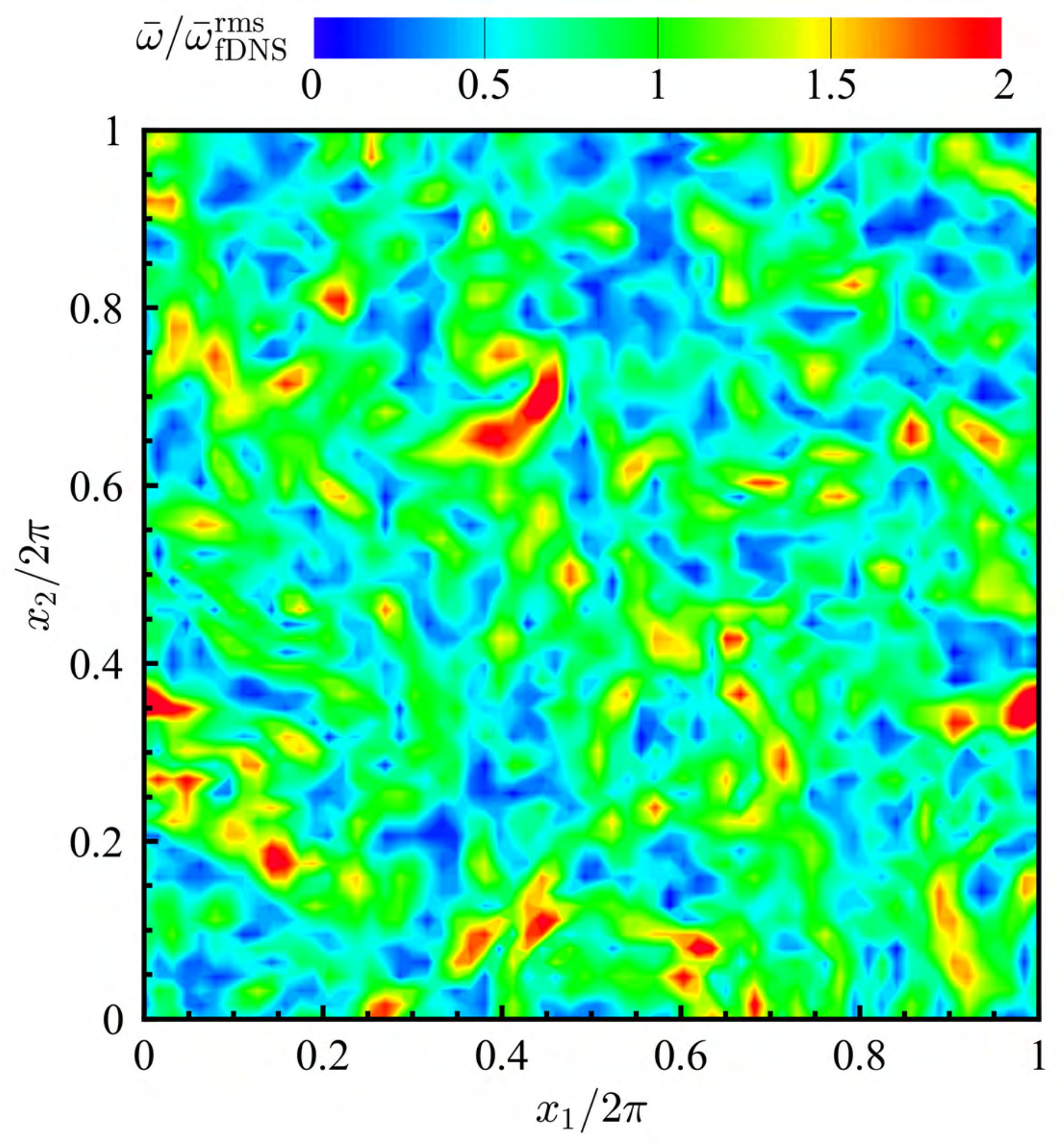}
	\end{subfigure}\\
	\begin{subfigure}{0.40\textwidth}
		\centering
		{($c$)}
		%\caption{ADM (${\chi \!=\! 0}$)}
		\includegraphics[width=0.9\linewidth,valign=t]{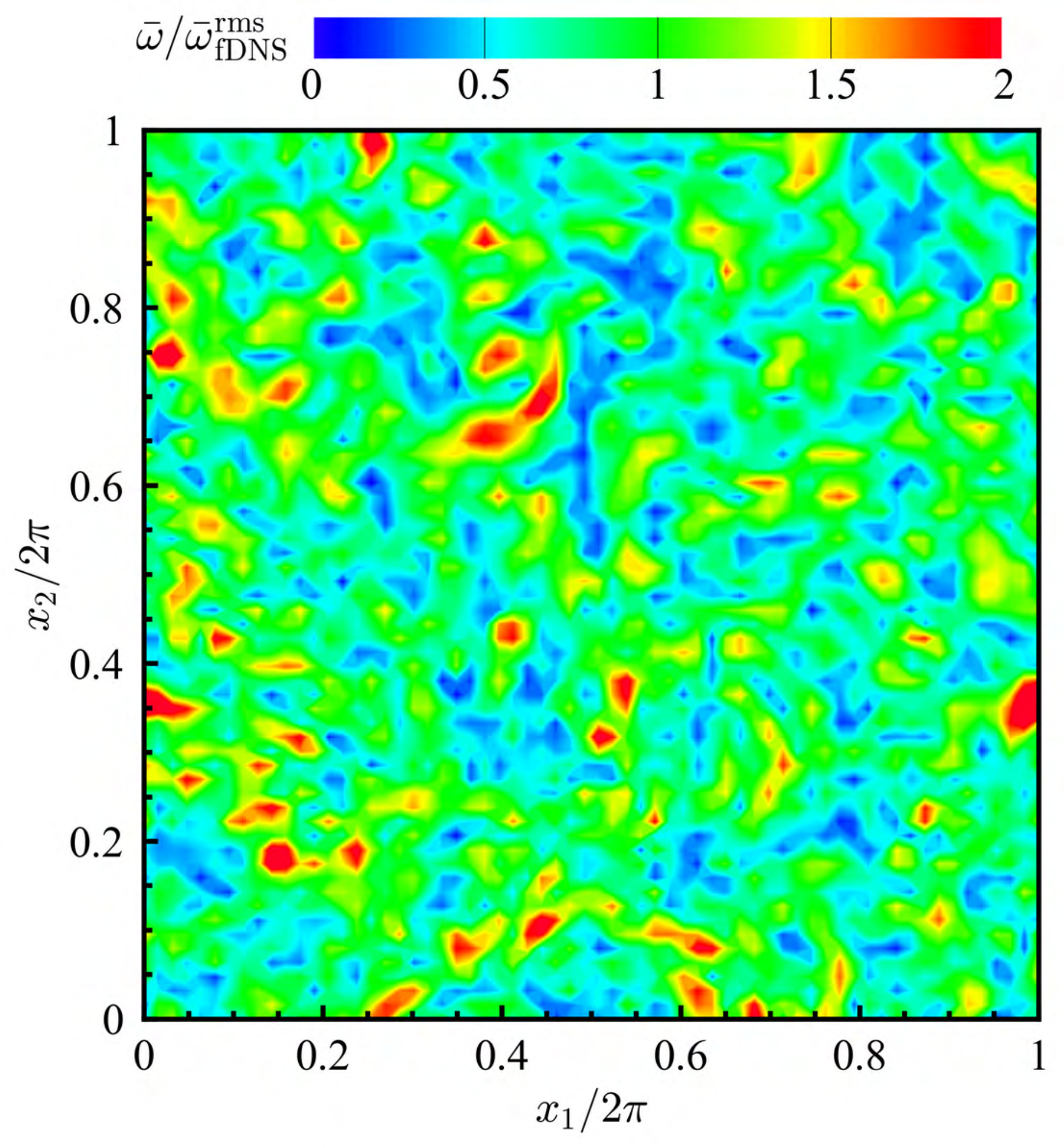}
	\end{subfigure}%
	\begin{subfigure}{0.40\textwidth}
		\centering
		{($d$)}
		%\caption{ADM (${\chi \!=\! 1}$)}
		\includegraphics[width=0.9\linewidth,valign=t]{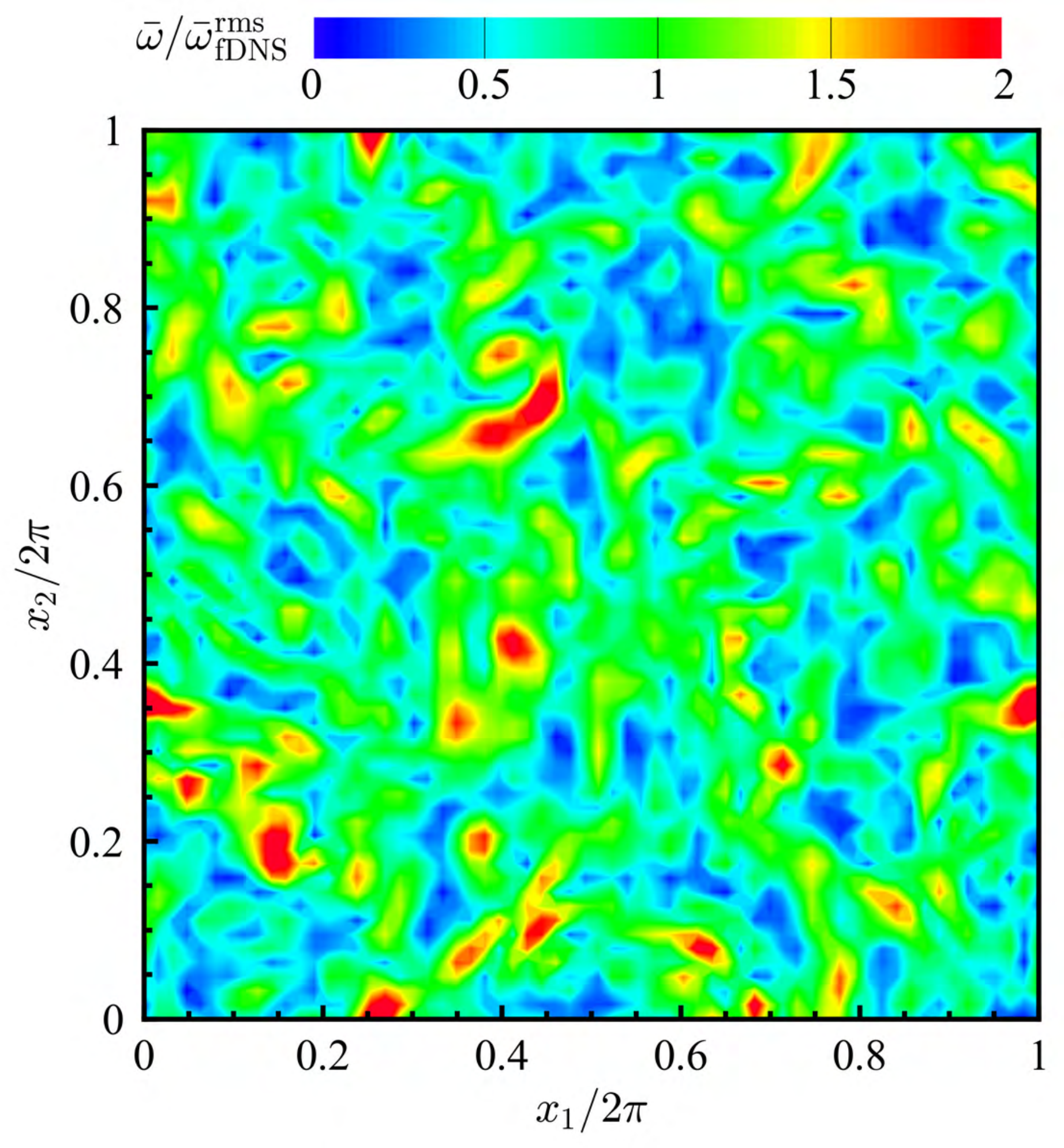}
	\end{subfigure}\\
	\begin{subfigure}{0.40\textwidth}
		\centering
		{($e$)}
		%\caption{DSM}
		\includegraphics[width=0.9\linewidth,valign=t]{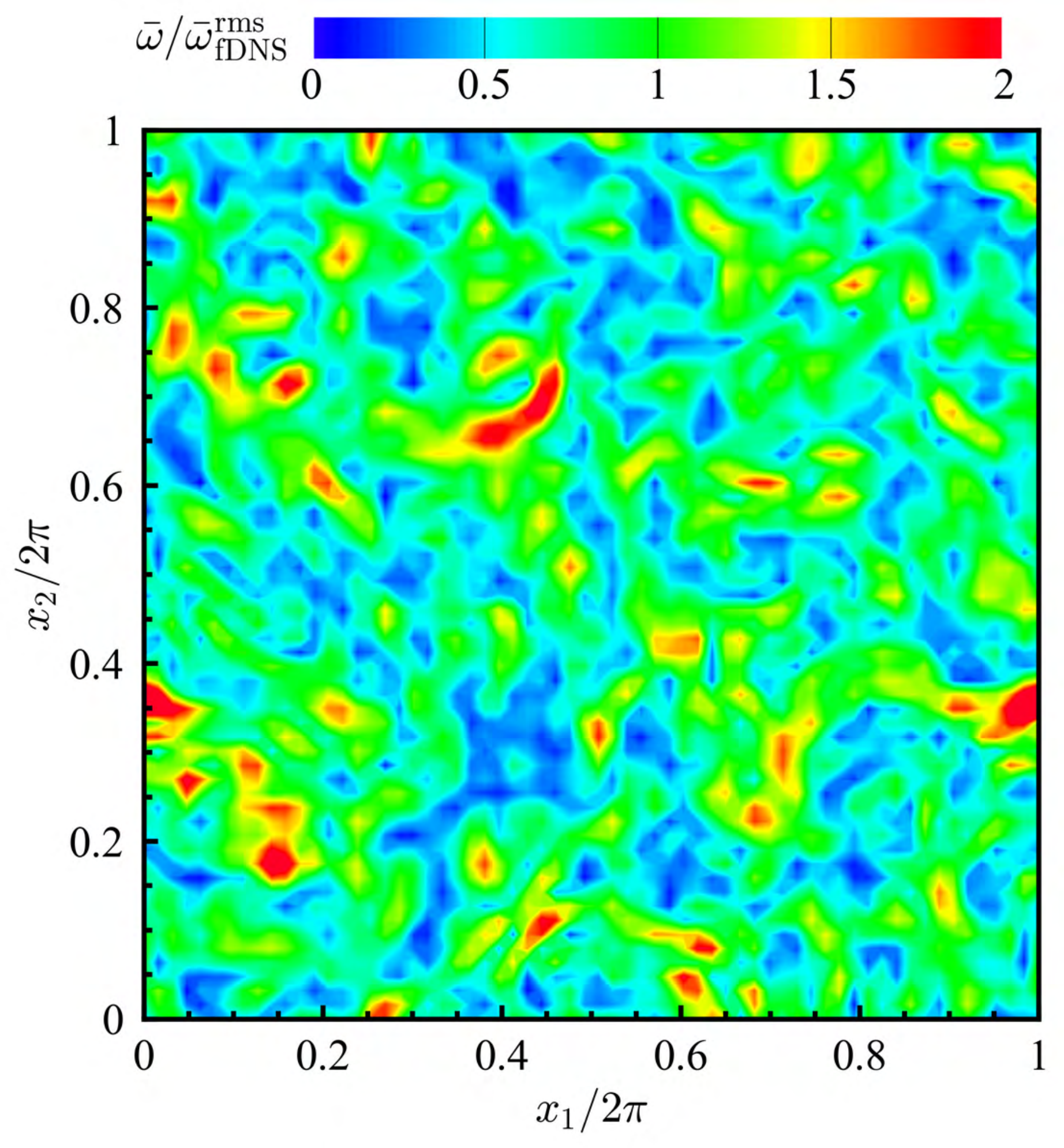}
	\end{subfigure}%
	\begin{subfigure}{0.40\textwidth}
		\centering
		{($f$)}
		%\caption{VOMM}
		\includegraphics[width=0.9\linewidth,valign=t]{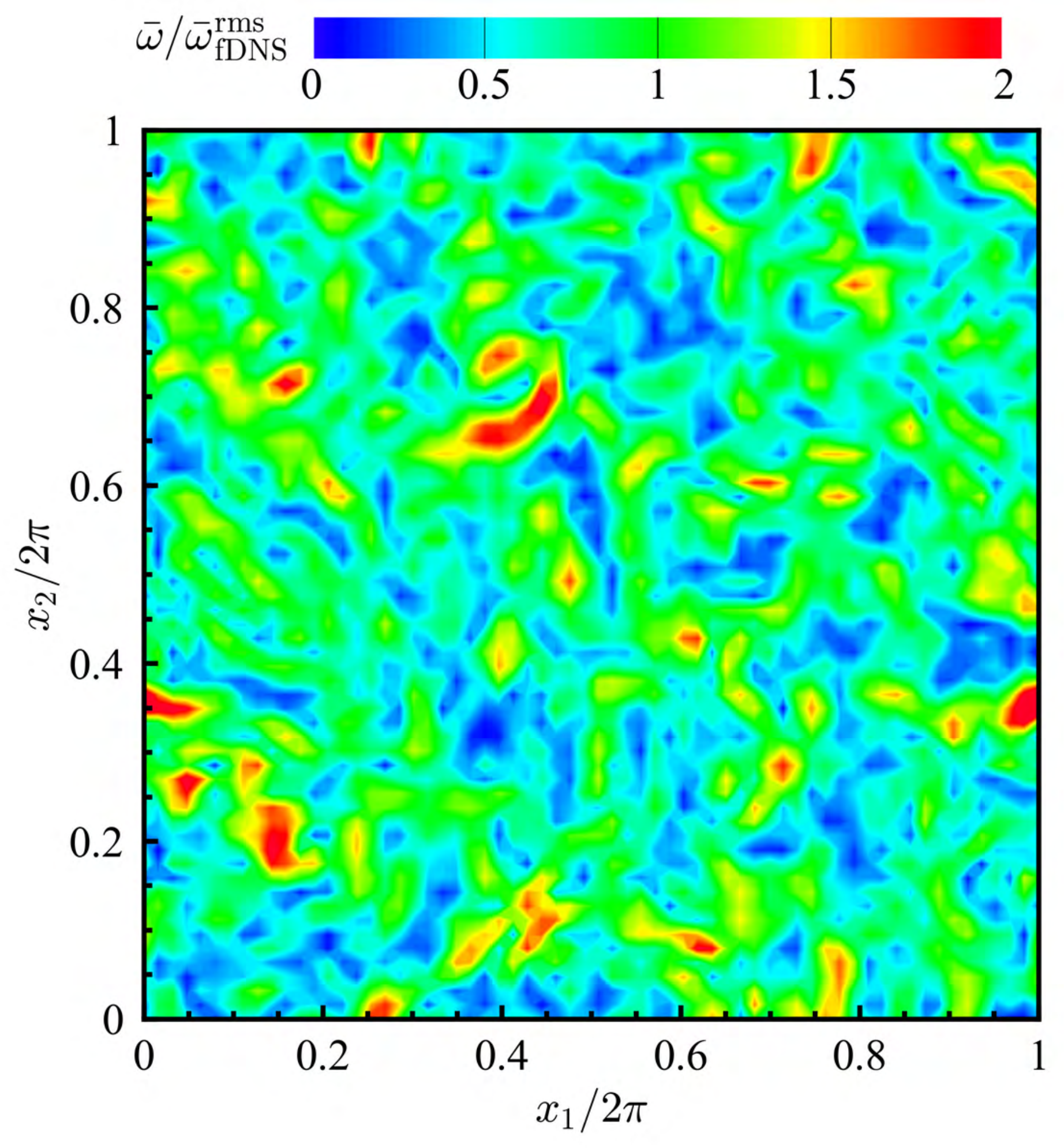}
	\end{subfigure}	
	\caption{Contours of the normalized vorticity $\bar \omega /\bar \omega _{{\rm{fDNS}}}^{{\rm{rms}}}$ at an arbitrary $x_1$-$x_2$ plane at $t/\tau \approx$ 4 for LES at a grid resolution of $64^3$ (FGR=2) in decaying homogeneous isotropic turbulence with the filter width $\bar \Delta  = 32{h_{{\rm{DNS}}}}$: (a) fDNS, (b) DMM, (c) ADM($\chi$=0), (d) ADM($\chi$=1), (e) DMM, and (f) VOMM.}
	\label{fig:15}
\end{figure}

The transient velocity spectra of different SGS models at the filter width $\bar \Delta=32 h_{\rm{DNS}}$ with two different time instants $t/\tau \approx 2$ and 4 are further illustrated in Fig.~\ref{fig:13}. The velocity spectra exhibit an overall decrease, and the kinetic energy at all wavenumbers declines with the decaying of turbulence. All the classical SGS models (DSM, DMM and ADM models) overpredict the kinetic energy at high wavenumbers for the coarse grid-resolution case (FGR=1 with $N=32^3$ ) and the excessive kinetic energy stacked at small scales leads to the numerical instability of LES, which gradually intensifies with the evolution of time. The conventional SGS models provide insufficient model dissipation to balance the discretization errors and the small-scale kinetic energy cannot be effectively dissipated in time at FGR=1. For the fine grid-resolution cases (FGR=2 and 4 with $N=64^3$  and $128^3$), the dissipation of the traditional SGS models (DSM, DMM models, and ADM model with $\chi = 1$) is too strong to diminish most small-scale flow structures near the truncated wavenumber, which hinders the normal transmission of turbulent kinetic energy cascades from large scales to small scales.  Therefore, the kinetic energy of classical SGS models accumulates in the region of intermediate wavenumbers, leading to the overestimations of the turbulent kinetic energy with time (Fig.~\ref{fig:11}) at FGR=2 and 4 with $N=64^3$  and $128^3$.  LES using the pure ADM model with $\chi = 0$ is always numerically unstable and lacks necessary SGS dissipation to drain out the small-scale kinetic energy for all different grid resolutions. Compared to these classical SGS models, the VOMM model can accurately reconstruct the kinetic energy cascade with the predictions that nearly coincide with those of fDNS at all three different grid resolutions.  

\begin{figure}\centering
	\begin{subfigure}{0.50\textwidth}
		\centering
		{($a$)}
		\includegraphics[width=0.92\linewidth,valign=t]{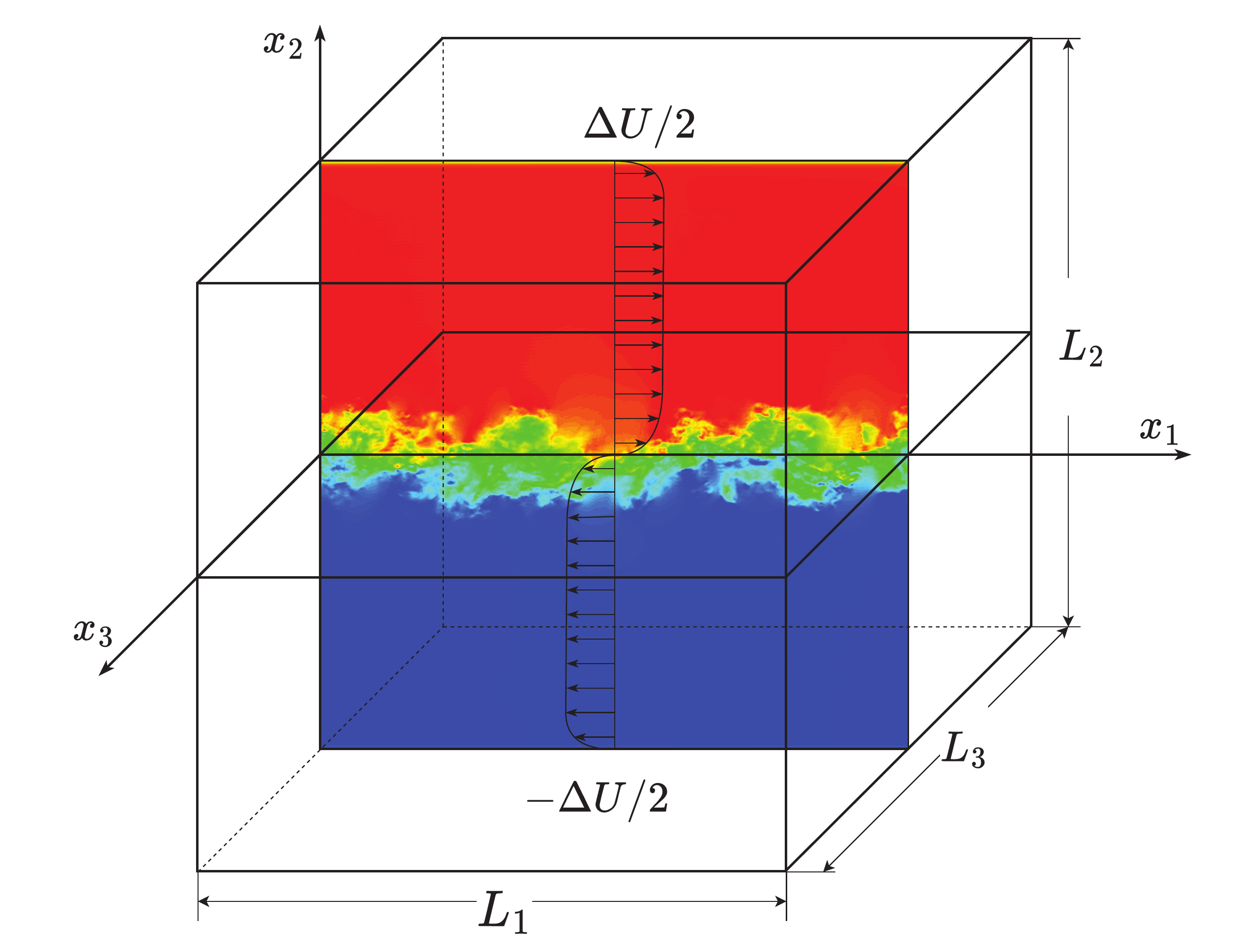}
	\end{subfigure}%
	\begin{subfigure}{0.50\textwidth}
		\centering
		{($b$)}
		\includegraphics[width=0.80\linewidth,valign=t]{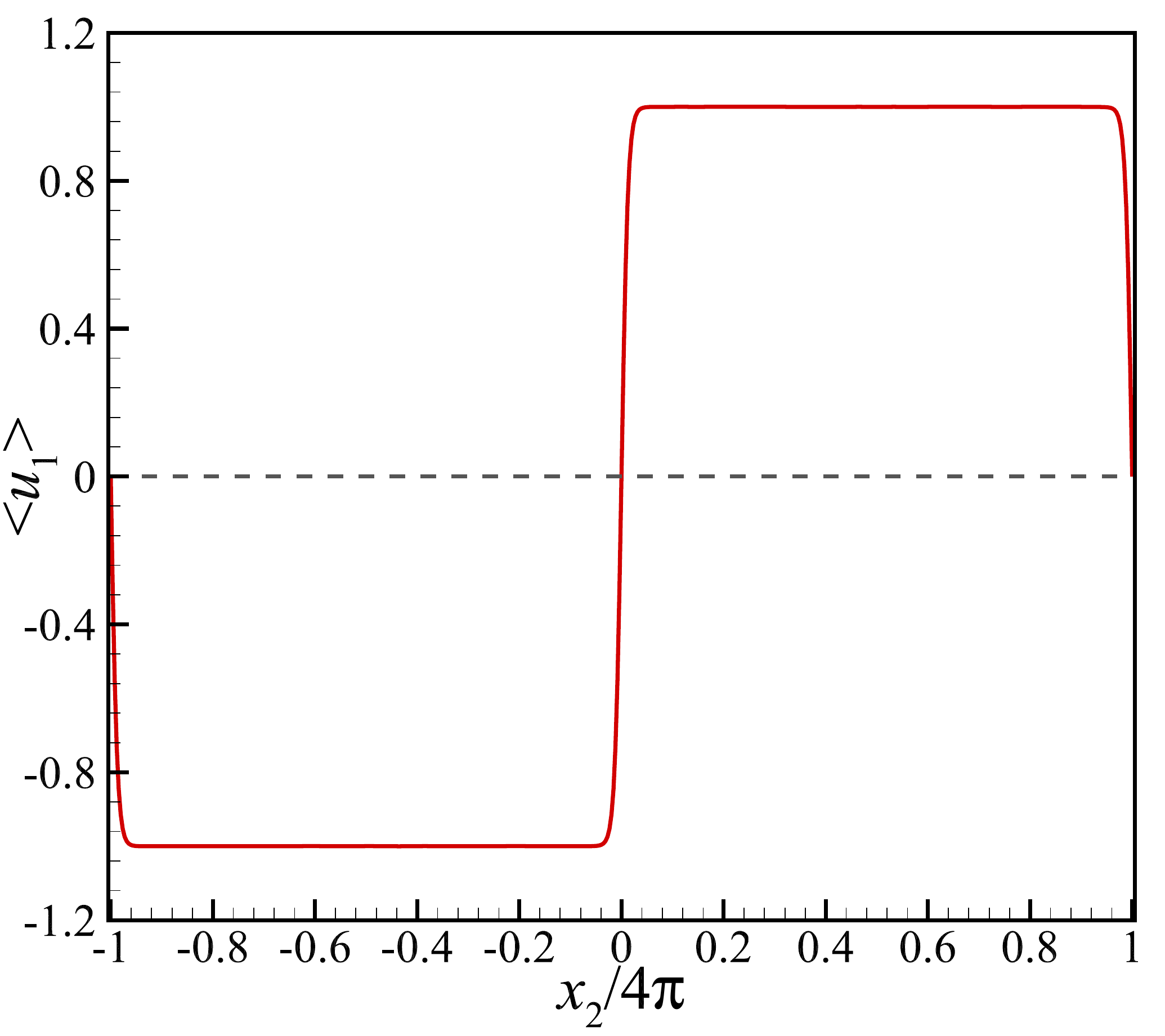}
	\end{subfigure}
	\caption{Diagram of the temporally evolving mixing layer with the mean velocity profile: (a) schematic of the mixing layer, (b) mean streamwise velocity  profile $\left\langle {{u_1}} \right\rangle$ along the normal ($x_2$) direction.}
	\label{fig:16}
\end{figure}

Furthermore, we compare the PDFs of the normalized vorticity magnitude at the dimensionless time $t/\tau \approx 4$ as shown in Fig.~\ref{fig:14}. The vorticity is normalized by the root-mean-square values of the vorticity calculated by the fDNS data for comparisons of different grid resolutions. The pure ADM models with $\chi = 0$ gives the worst prediction of the vorticity with erroneous peaks of PDFs significantly different from the fDNS data for all three grid resolutions. The secondary filtering technique ($\chi = 1$) of the ADM model  cannot improve the prediction of vorticity very well, whose estimations are still obviously different from the benchmark fDNS data. DSM and DMM models underestimate the PDF of vorticity and have wrong predictions of the PDF peak at the coarse grid-resolution case (FGR=1 with $N=32^3$ ), while greatly improving the predictions of PDFs with the increasing of the grid resolution (FGR=2 and 4 with $N=64^3$  and $128^3$). In contrast, the VOMM model outperforms these classical SGS models at all three different grid resolutions, which gives a reasonably good prediction for both the locations and the peaks of the PDFs of the vorticity.

The reconstruction of transient spatial vorticity structures are finally demonstrated by the contours of the normalized vorticity magnitude shown in Fig.~\ref{fig:15}. The instantaneous snapshots are selected on an arbitrary $x_1$-$x_2$ slice at the consistent time instant  $t/\tau \approx 4$. The pure ADM model predicts the excessive stochastic small-scale structures, which significantly differ from the fDNS data. The other SGS models can predict the large-scale vorticity structures quite well, but the VOMM model reconstructs the spatial vortex structures very similar to the benchmark fDNS data. The VOMM model can accurately recover more flow structures and the temporal evolution of the vortex with suitable SGS dissipation and accurate structural modeling. 

\subsection {Temporally evolving turbulent mixing layer}
The turbulent mixing layer is one of the cardinal flows in the fluid-mechanics community, which is widely applied to the investigation of turbulent combustion, chemical reaction mixing process, and fundamental studies of flow instabilities. The turbulent mixing layer involves the unsteady shear process of vortex shedding and transition from laminar to turbulent flows, which are remarkably suitable for investigating the impact of non-uniform turbulent shear and mixing on the SGS models. The temporally evolving turbulent mixing layer characterized by the Kelvin–Helmholtz instability induced by the initial velocity difference is considered in this paper \citep{vreman1996comparision,vreman1997,sharan2019,wang2022b}.  The free-shear mixing layer is governed by the same Navier-Stokes equations (Eqs.~\ref{ns1} and \ref{ns2}) without the forcing term. Figure~\ref{fig:16} illustrates the diagram of the flow configuration for the temporally evolving turbulent mixing layer with the initial hyperbolic tangent streamwise velocity profile. The numerical simulation of mixing layer is performed in a cuboid domain with lengths ${L_1} \times {L_2} \times {L_3} = 8\pi  \times 8\pi  \times 4\pi $ at the uniform grid resolution of ${N_1} \times {N_2} \times {N_3} = 512 \times 512 \times 256$ where $x_1 \in \left[ { - {L_1}/2,{L_1}/2} \right]$, $x_2 \in \left[ { - {L_2}/2,{L_2}/2} \right]$ and $x_3 \in \left[ { - {L_3}/2,{L_3}/2} \right]$ denote the streamwise, transverse and spanwise directions, respectively. To enable a periodic configuration in the normal direction, the initial mean streamwise velocity (cf. Fig.~\ref{fig:16}b) is given by \citep{sharan2019,wang2022b}
\begin{equation}
	\left\langle {{u_1}} \right\rangle  = \frac{{\Delta U}}{2}\left[ {\tanh \left( {\frac{{{x_2}}}{{2\delta _\theta ^0}}} \right) - \tanh \left( {\frac{{{x_2} + {L_2}/2}}{{2\delta _\theta ^0}}} \right) - \tanh \left( {\frac{{{x_2} - {L_2}/2}}{{2\delta _\theta ^0}}} \right)} \right],\;{\rm{for}}\; -\frac{L_2}{2} \le {x_2} \le \frac{L_2}{2}, 
	\label{meanU_TML}
\end{equation}
where ${\Delta U}=2$ is the velocity difference between two equal and opposite free streams across the shear layer,  ${\delta _\theta ^0}=0.08$ denotes the initial momentum thickness, and $\left\langle  \cdot  \right\rangle $ stands for a spatial average over all the homogeneous directions (\emph{i.e.}, $x_1$ and $x_3$ directions for the mixing layer). The initial mean transverse and spanwise velocities are both set to zero, namely, $\left\langle {{u_2}} \right\rangle  = \left\langle {{u_3}} \right\rangle  = 0$. Since the initial mean velocity field is periodic in all three directions, the triply periodic boundary conditions are adopted and the pseudo-spectral method with the two-thirds dealiasing rule is used for the spatial discretization. An explicit two-step Adam-Bashforth scheme is selected as the time-advancing scheme. To effectively suppress the influence of the top and bottom boundaries on the central mixing layer, two numerical diffusion buffer zones are applied near the vertical edges of domain \citep{wang2022b}. The thickness of the buffer layer is set to $15{\delta _\theta ^0}$ in the paper, which is sufficiently large and has a negligible effect on the calculations of mixing layer \citep{wang2022b}. 

The digital filter method is used to generate the spatially-correlated initial perturbation imposed on the mean velocities with the digital filter width $\Delta_d = \bar \Delta = 8h_{\rm{DNS}}$  consistent to the filter scale of LES \citep{klein2003,wang2022a}. The initial Reynolds stress distribution (${R_{ij}} = \left\langle {{u_i^\prime}{u_j^\prime}} \right\rangle $ where ${u_i^\prime} = {u_i} - \left\langle {{u_i}} \right\rangle $ represents the fluctuated velocity) of the digital filter method is assumed as a vertical distribution of ${R_{ij}} = A\left( {1 - {{\left\langle {{u_1}} \right\rangle }^2}} \right) I_{ij}$ with the identity ${I_{ij}}$ and peak amplitude $A=0.025 \Delta U$. The kinematic viscosity of shear layer is set to $\nu_{\infty}  = 5 \times {10^{ - 4}}$. The momentum thickness quantifies the range of turbulence region in the mixing layer, which is defined by \citep{rogers1994,sharan2019}
\begin{equation}
	{\delta _\theta } = \int\limits_{ - {L_2}/4}^{{L_2}/4} {\left[ {\frac{1}{4} - {{\left( {\frac{{\left\langle {{{\bar u}_1}} \right\rangle }}{{\Delta U}}} \right)}^2}} \right]d{x_2}}.
	\label{theta_TML}
\end{equation}
Correspondingly, the Reynolds number based on the momentum thickness ${\rm Re}_{\theta}$ is expressed as
\begin{equation}
	{{\rm Re}_{\theta}}=\frac{ \Delta U \delta_\theta}{\nu_{\infty}}.
	\label{Re_theta}
\end{equation}  
Here, the initial momentum thickness Reynolds number is ${{\rm Re}_{\theta}^{0}} = 320$. The detailed numerical parameters of DNS for the temporally evolving mixing layer is summarized in Table~\ref{tab:6}.

\begin{table}
	\begin{center}
		\caption{Numerical parameters for the DNS of the temporally evolving mixing layer.} 	\label{tab:6}%	
		\small 	
		\begin{tabular*}{0.95\textwidth}{@{\extracolsep{\fill}} lcccccccc}
			\hline\hline
			$N_1 \times N_2 \times N_3$ & $L_1 \times L_2 \times L_3$ & $\nu_{\infty}$ & ${\rm Re}_{\theta}$ & $\delta_\theta^0$ & $\Delta U$ & ${\Delta_d}/{h_{{\rm{DNS}}}}$ & ${h_{{\rm{DNS}}}}$ & $\Delta t_{\rm{DNS}}$ \\ \hline
			$512 \times 512 \times 256$ & $8 \pi \times 8\pi \times 4\pi$ & $5 \times 10^{-4}$ & 4000 & 0.08 & 2 & 8 & $\pi/64$ & 0.002 \\ \hline\hline
		\end{tabular*}%
	
	\end{center}
\end{table}% 

\begin{figure}\centering
	\includegraphics[width=0.7\textwidth]{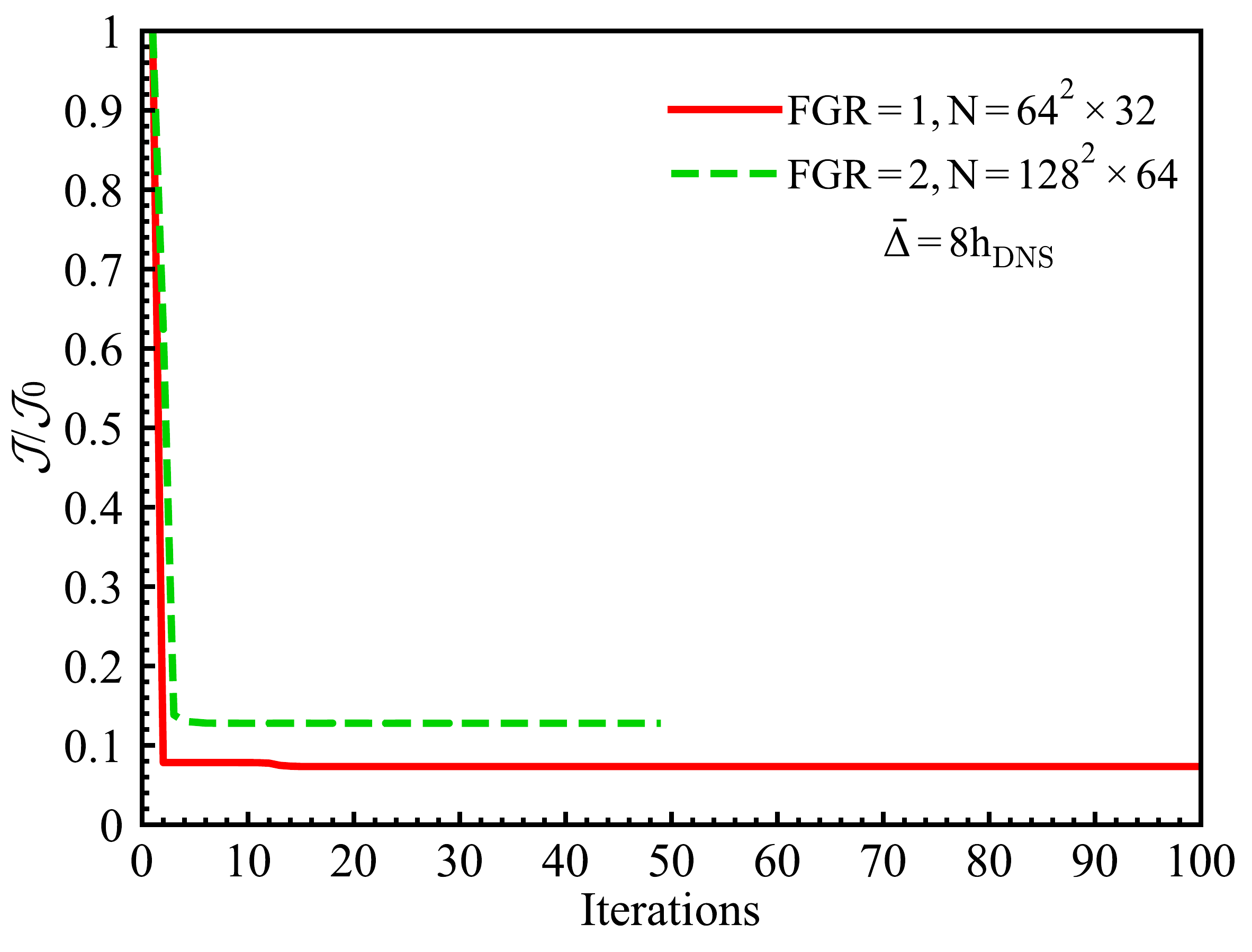}
	\caption{The evolution of the normalized cost function in temporally evolving turbulent mixing layer.}\label{fig:17}
\end{figure}	

We calculate the DNS of the mixing layer for total of eight hundred time units ($t/{\tau _\theta } = 800$) normalized by $\tau_\theta =\delta _\theta ^0/\Delta U$. To reduce the impact of initial random disturbances on the temporal development of the shear layer, six numerical experiments with different random initializations are performed, one of which is adopted for the parameter optimization of the VOMM model, while the remaining five are used to evaluate the ensemble-averaged physical quantities. The \emph{a posteriori} studies of LES are conducted using the explicit Gaussian filter (Eq.~\ref{G}) with the given filter scale $\bar \Delta = 8h_{\rm{DNS}}$ and initialized by the same instantaneous velocity field of the filtered DNS at $t/{\tau _\theta } = 50$. Two different filter-to-grid ratios FGR=$\bar \Delta/h_{\rm{LES}}$=1 and 2 are selected to study the influence of the spatial resolution or discretization error on the SGS stress modeling with the corresponding grid resolutions of LES: $N = {64^2} \times 32$ and ${128^2} \times 64$. The results from the previous two turbulence problems (forcing and decaying homogenous isotropic turbulence) indicate that the statistics of turbulence are very close and similar when the grid resolution is sufficiently fine (FGR=2 and 4) and the discretization error is considered negligible. However, the statistics of LES with a relatively coarse grid resolution (FGR=1) are distinctly different from those of LES with satisfactory grid resolutions (FGR=2 and 4), since the spatial discretization error of FGR=1 is considerably significant and dominates the SGS modelling error. Therefore, the \emph{a posteriori} testings of LES at both FGR=1 and 2 are essential for performance evaluations of the SGS model. 

\begin{table}
	\begin{center}
		\caption{The initial and optimal parameters of the VOMM model for LES computations with the filter width $\bar \Delta = 8 h_{\rm{DNS}}$ in temporally evolving mixing layer.} 	\label{tab:7}%	
		\small 	
		\begin{tabular*}{0.95\textwidth}{@{\extracolsep{\fill}} lccccc}	
			\hline\hline
			FGR & LES Resolution & $C_1^{\left( 0 \right)}$ & $C_2^{\left( 0 \right)}$ & $C_1^{\rm{opt}}$ & $C_2^{\rm{opt}}$ \\ \hline
			1 & $64^2 \times 32$   & 0 & 1 & -0.0637 & 1.188 \\ 
			2 & $128^2 \times 64$  & 0 & 1 & -0.0126 & 1.000  \\ \hline\hline
		\end{tabular*}%

	\end{center}
\end{table}% 

The dissipation spectrum of the filtered DNS is consistently used as the objective function to optimize the model parameters of the VOMM model during the period (assess every $t/{\tau _\theta } = 10$ with total thirty-six groups at $50 \le t/{\tau _\theta } \le 400$) of the adjoint-based optimization (cf. Fig.~\ref{fig:1}). The pure ADM model without the dissipative term is adopted as the initial SGS model with coefficients $C_1^{\left( 0 \right)}=0$ and  $C_2^{\left( 0 \right)}=1$. We calculate the adjoint-based gradients of the cost functional for the model parameters by backward integrating the stabilized adjoint LES equations (Eqs.~\ref{adj_LES1} and \ref{adj_MLES2}). The SGS model coefficients are iteratively updated by the L-BFGS optimization method (Eq.~\ref{opt_iterative}) until the stopping criterion is ultimately satisfied. Figure~\ref{fig:17} gives the optimization process of the cost function during the adjoint-based optimization for the temporally evolving mixing layer. The loss functions for both FGR=1 and 2 drop dramatically and reach a steady plateau within less than ten iterations. The cost function of FGR=1 shows a more distinct reduction with approximately 8\% of the initial level than that of FGR=2 decreasing to the 10\% of original value. The optimal parameters of VOMM model are quickly obtained by the effective gradient-based optimization within a limited number of iterations (around 10 optimization evaluations, namely, 20 LES calculations). Table~\ref{tab:7} summarizes the optimal parameters of the VOMM model. The parameter magnitude of the dissipative Smagorinsky term ($\left| {C_1^{\rm{opt}}} \right|$) obviously decreases from 0.0637 to 0.0126 when the FGR increases from 1 to 2, while the ADM coefficient ($C_2^{\rm{opt}}$) generally tends towards unity, similar to the cases of isotropic turbulence.

\begin{table}
	\begin{center}
		\caption{The average computational cost of SGS stress modeling $\tau_{ij}$ for LES computations with filter width $\bar \Delta=8 h_{\rm{DNS}}$ in temporally evolving turbulent mixing layer.}\label{tab:8}%	
		\small 				
		\begin{tabular*}{0.95\textwidth}{@{\extracolsep{\fill}} lccccc}
			\hline\hline
			Model(FGR=1,$N=64^2 \times 32$)       & DSM   & DMM    & ADM($\chi$=0) & ADM($\chi$=1) & VOMM \\ \hline
			t(CPU$\cdot$s) & 0.646 & 1.096 & 0.254 & 0.247 & 0.362 \\
			t/t$_{\rm{DMM}}$ & 0.590 & 1 & 0.232 & 0.225 & 0.330 \\ 
			\hline
			Model(FGR=2,$N=128^2 \times 64$)       & DSM   & DMM    & ADM($\chi$=0) & ADM($\chi$=1) & VOMM \\ 
			t(CPU$\cdot$s) & 3.756 & 6.370 & 1.465 & 1.460 & 1.908 \\
			t/t$_{\rm{DMM}}$ & 0.590 & 1 & 0.230 & 0.229 & 0.300 \\ \hline\hline
		\end{tabular*}%

	\end{center}
\end{table}% 

We then examine the \emph{a posteriori} performance of the proposed VOMM model once the SGS model coefficients are determined by the adjoint-based gradient optimization strategy. To demonstrate the generality of the optimal model parameters that are insensitive to the initial perturbations, ensemble-averaged quantities are evaluated by five numerical experiments with different initial random disturbances from the optimization process. The time steps of LES are set as $\Delta {t_{{\rm{LES}}}}/\Delta {t_{{\rm{DNS}}}} = \left\{ {10,5} \right\}$ to guarantee the consistent CFL number for different grid resolutions (FGR=1 and 2 with $N = {64^2} \times 32$ and ${128^2} \times 64$). The VOMM model is compared with the conventional SGS models (DSM, DMM and ADM models), and the average modeling costs for different SGS models are listed in Table~\ref{tab:8}. The VOMM model evaluates efficiently with about 30\% computational cost of the DMM model which is similar to those of the ADM models. 

\begin{figure}\centering
	\begin{subfigure}{0.50\textwidth}
		\centering
		{($a$)}
		\includegraphics[width=0.90\linewidth,valign=t]{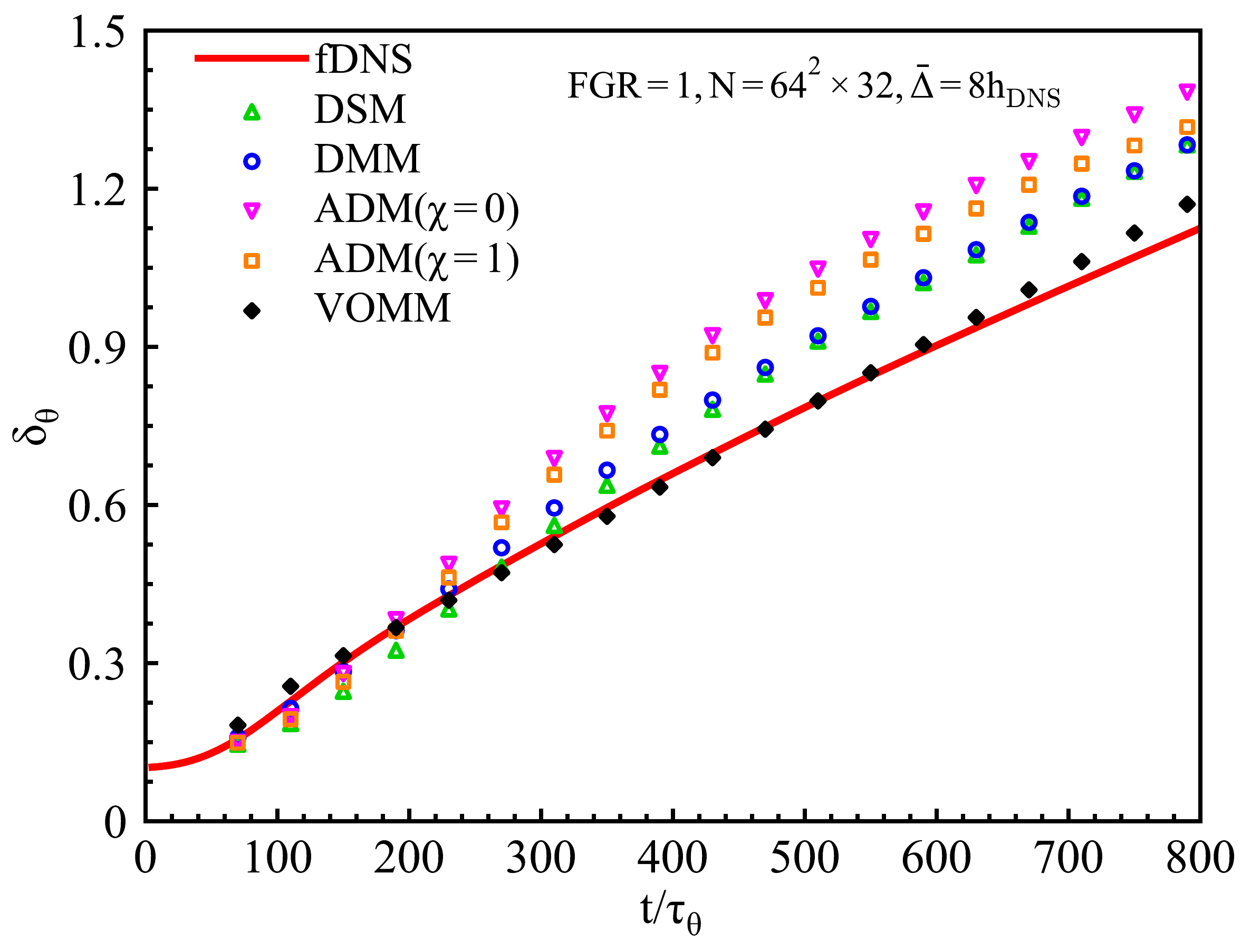}
	\end{subfigure}%
	\begin{subfigure}{0.50\textwidth}
		\centering
		{($b$)}
		\includegraphics[width=0.90\linewidth,valign=t]{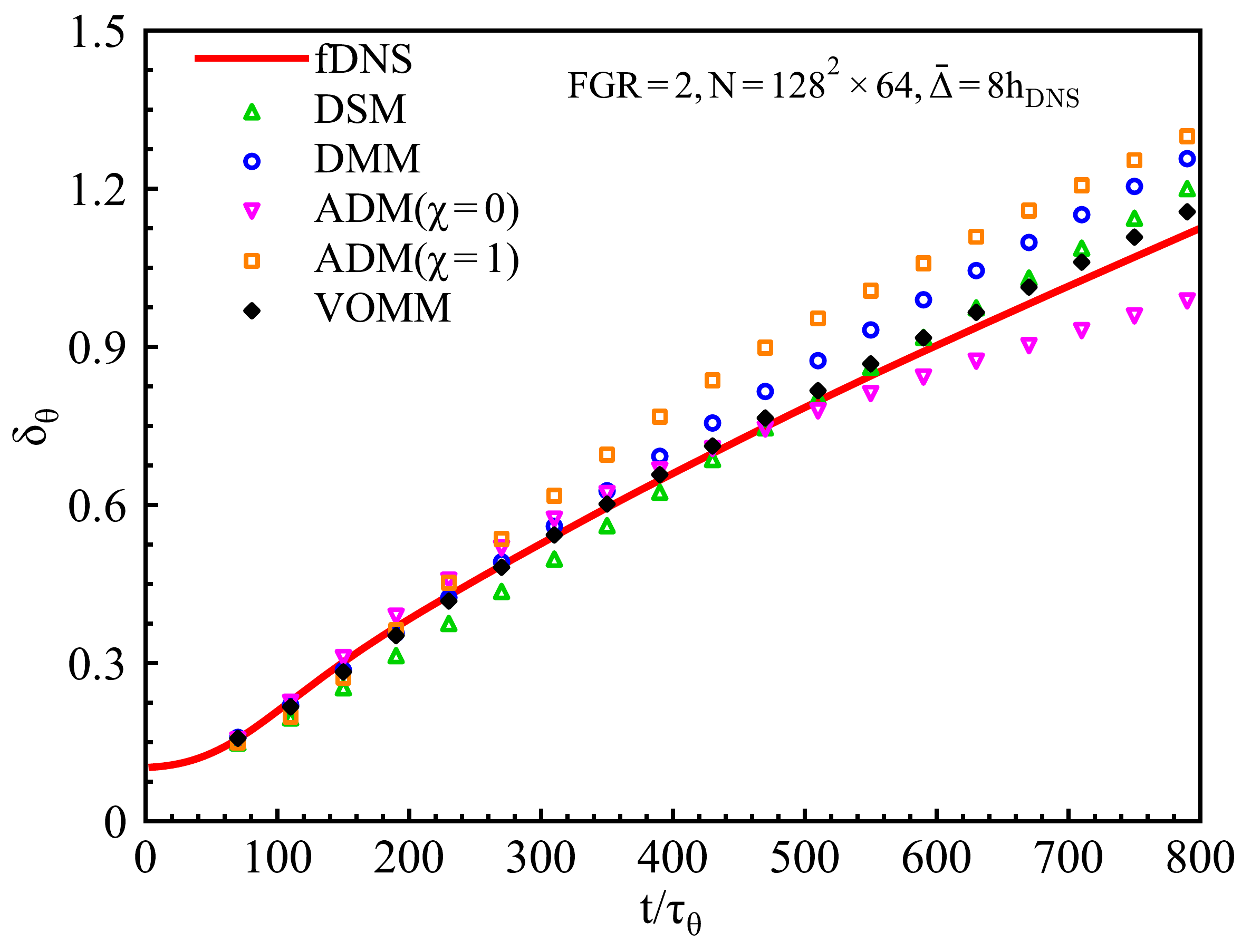}
	\end{subfigure}
	\caption{Temporal evolutions of the momentum thickness $\delta_\theta$ for LES in the \emph{a posteriori} analysis of temporally evolving turbulent mixing layer with the same filter scale $\bar \Delta  = 8{h_{{\rm{DNS}}}}$: (a) FGR=1, $N=64^2 \times 32$; (b) FGR=2, $N=128^2 \times 64$.}
	\label{fig:18}
\end{figure}

Figure~\ref{fig:18} illustrates the temporal evolutions of the momentum thickness $\delta_\theta$ in LES calculations of different SGS models compared to the benchmark fDNS data.
At the case of coarse grid resolution (FGR=1 with $N = {64^2} \times 32$),  all conventional SGS models underpredict the momentum thickness at the early stage of shear layer development ($t/{\tau _\theta } \le 300$) but give obvious overestimations in the linear growth region. For the fine-grid-resolution case (FGR=2 with $N = {128^2} \times 64$), DMM and ADM ($\chi$=1) models can capture the growth rate of momentum thickness well at the beginning of temporal development, but still overpredict the thickness with the developing of shear layer. The prediction of the pure ADM model with $\chi=0$ is irregular and nonlinear all the time without an apparent linear self-similar region. The DSM model at different grid resolutions gives the clearly tilted temporal evolutions, where the momentum thickness is underestimated at the beginning of transition region and overpredicted in the region of linear growth. In contrast, the predictions of the VOMM model always coincide well with those of fDNS, and they accurately capture the temporal growth rate in the linear region at both grid resolutions. 

\begin{figure}\centering
	\begin{subfigure}{0.50\textwidth}
		\centering
		{($a$)}
		\includegraphics[width=0.90\linewidth,valign=t]{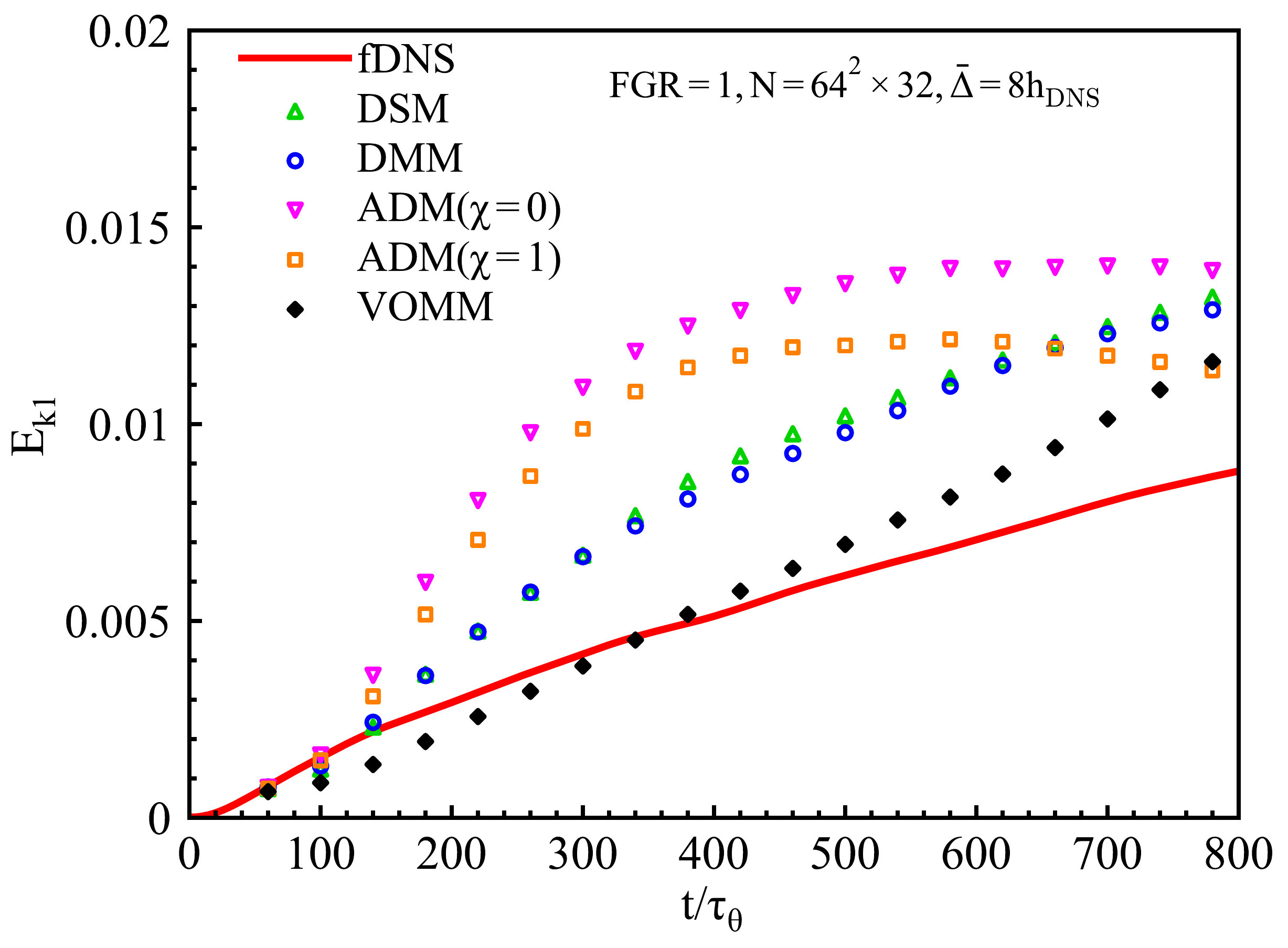}
	\end{subfigure}%
	\begin{subfigure}{0.50\textwidth}
		\centering
		{($b$)}
		\includegraphics[width=0.90\linewidth,valign=t]{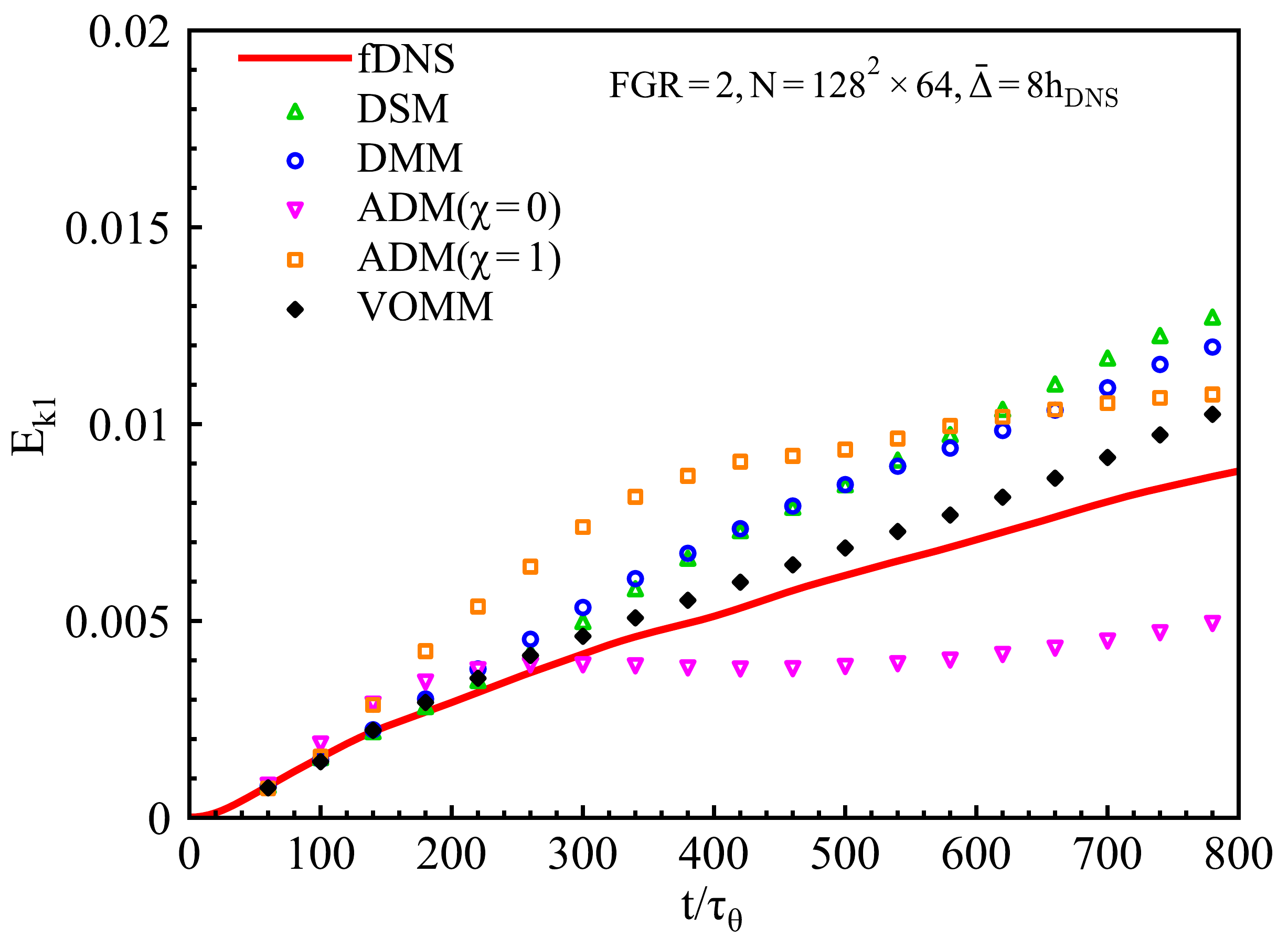}
	\end{subfigure}
	\caption{Temporal evolutions of the streamwise turbulent kinetic energy $E_{k1}$ for LES in the \emph{a posteriori} analysis of temporally evolving turbulent mixing layer with the same filter scale $\bar \Delta  = 8{h_{{\rm{DNS}}}}$: (a) FGR=1, $N=64^2 \times 32$; (b) FGR=2, $N=128^2 \times 64$.}
	\label{fig:19}
\end{figure}

\begin{figure}\centering
	\begin{subfigure}{0.50\textwidth}
		\centering
		{($a$)}
		\includegraphics[width=0.90\linewidth,valign=t]{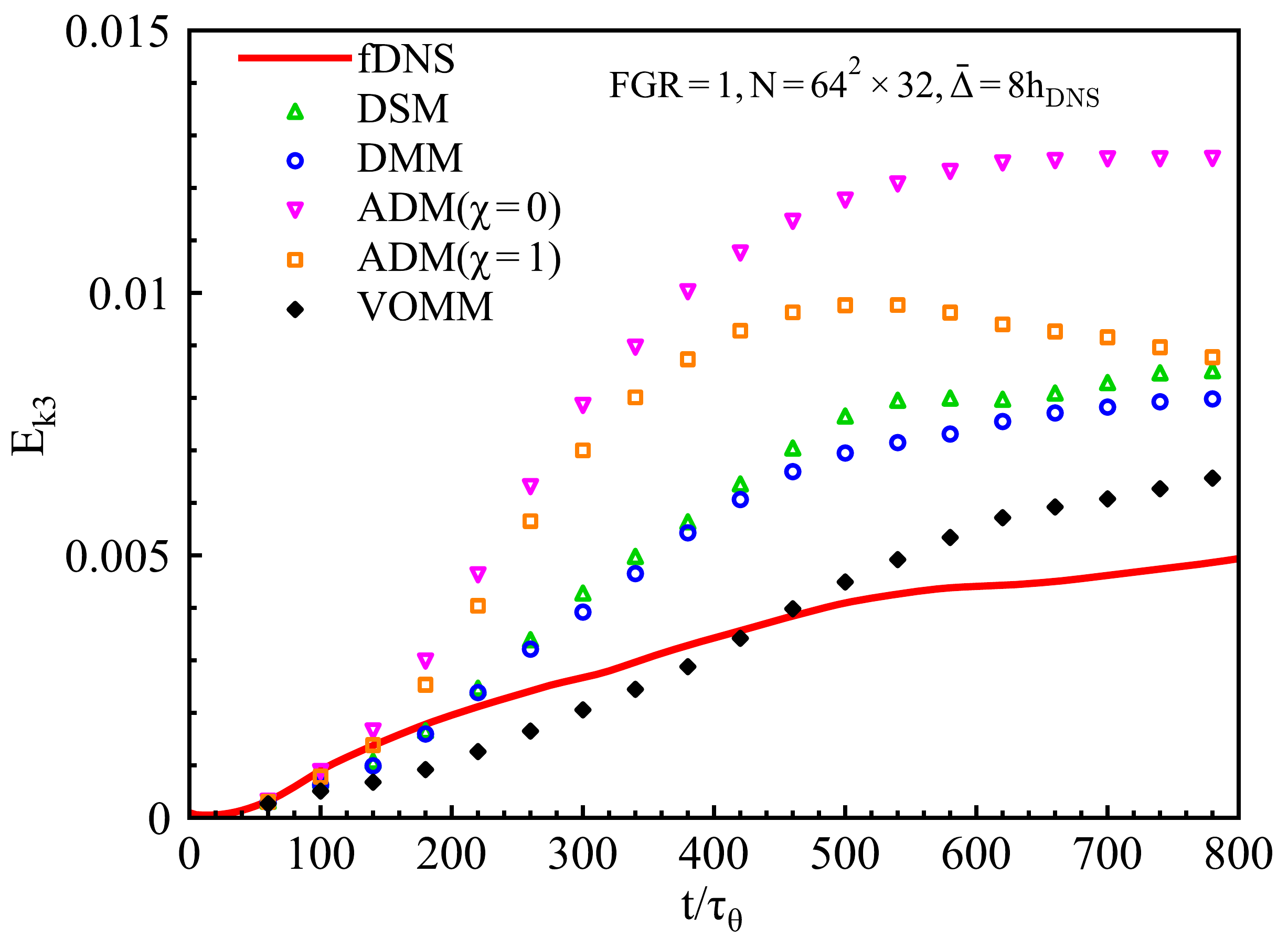}
	\end{subfigure}%
	\begin{subfigure}{0.50\textwidth}
		\centering
		{($b$)}
		\includegraphics[width=0.90\linewidth,valign=t]{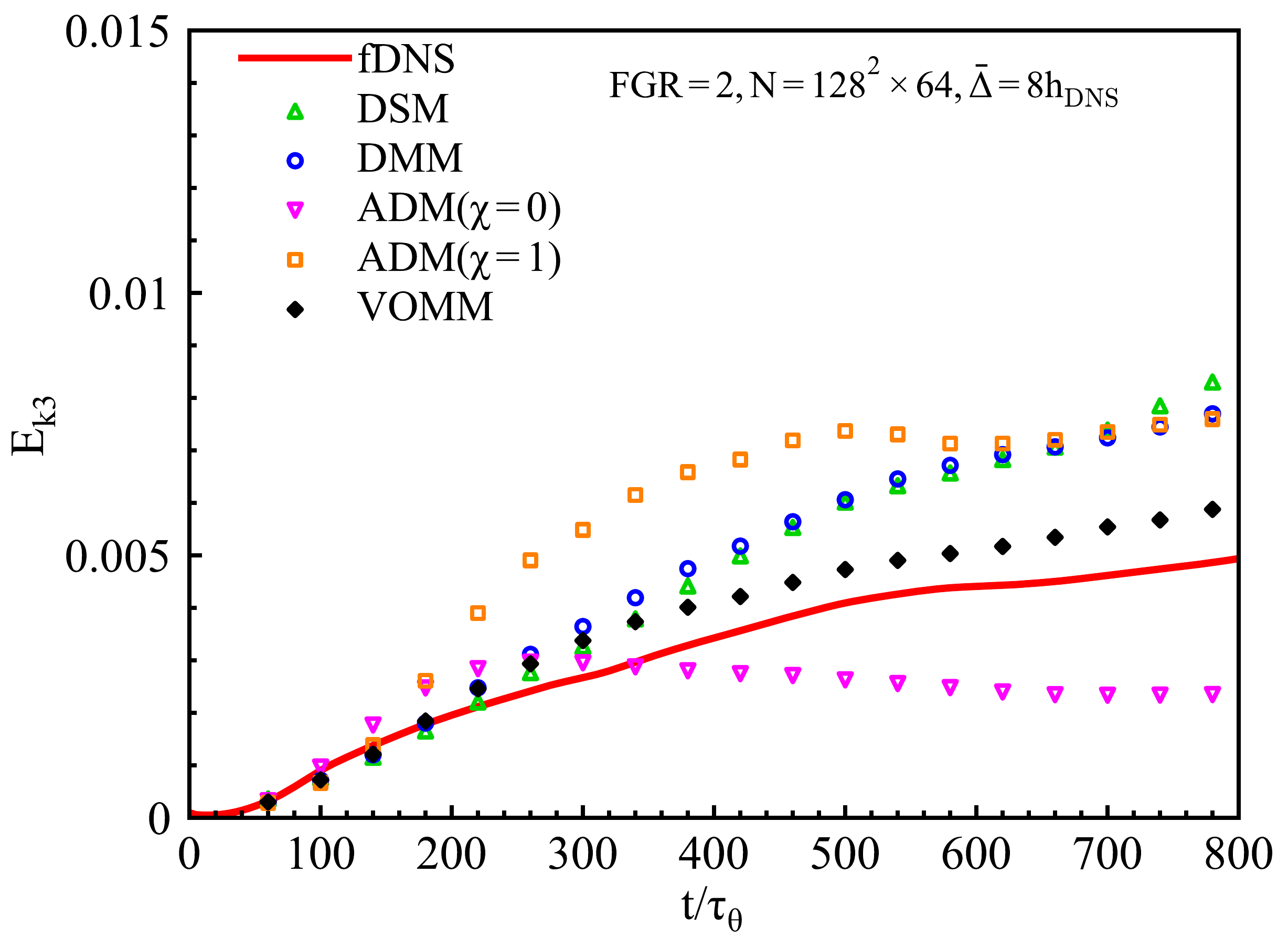}
	\end{subfigure}
	\caption{Temporal evolutions of the spanwise turbulent kinetic energy $E_{k3}$ for LES in the \emph{a posteriori} analysis of temporally evolving turbulent mixing layer with the same filter scale $\bar \Delta  = 8{h_{{\rm{DNS}}}}$: (a) FGR=1, $N=64^2 \times 32$; (b) FGR=2, $N=128^2 \times 64$.}
	\label{fig:20}
\end{figure}

Furthermore, the evolutions of the turbulent kinetic energy in the streamwise and spanwise directions are displayed in Figs.~\ref{fig:19} and \ref{fig:20}, respectively. The comparisons of transverse turbulent kinetic energy for different SGS models are very similar to those in the spanwise direction, not shown in the paper. The turbulent kinetic energy of DNS in different directions gradually increases with the developing of the shear layer, since the initial perturbated velocity field is approximately laminar and steadily transitions to turbulence. The temporal development of the streamwise kinetic energy can be approximately regarded as a linear growth with time, which is distinctly different from that of spanwise kinetic energy. All classical SGS models predict both streamwise and spanwise kinetic energy much larger than the benchmark fDNS results at both grid resolutions of LES, except that the pure ADM model gives underestimations of kinetic energy in the fine-grid-resolution case (FGR=2). Compared to these traditional models, the VOMM model accurately predicts the kinetic energy at different grid resolutions in both streamwise and spanwise directions, and is the closest to the fDNS data. 

\begin{figure}\centering
	\begin{subfigure}{0.50\textwidth}
		\centering
		{($a$)}
		\includegraphics[width=0.90\linewidth,valign=t]{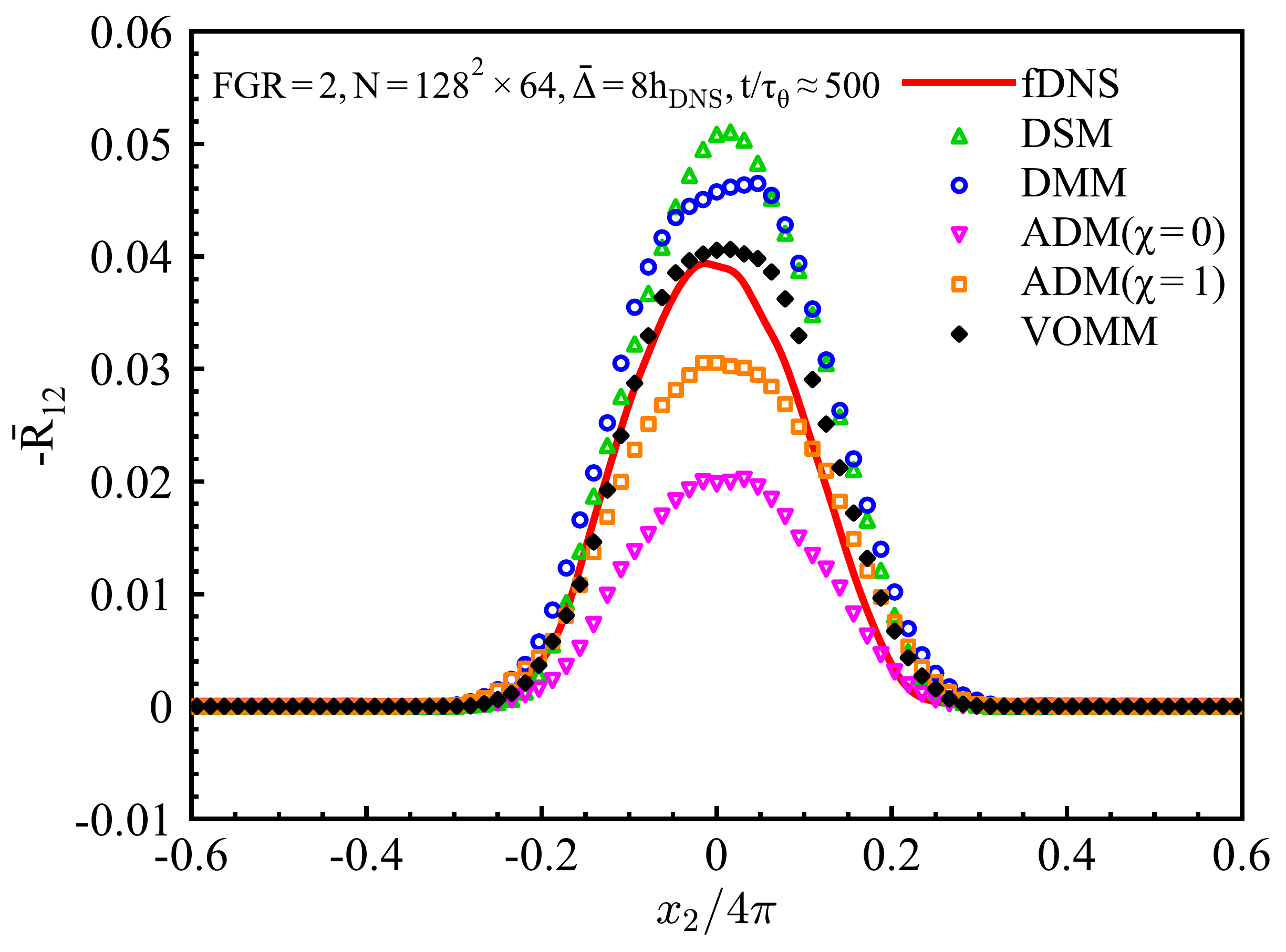}
	\end{subfigure}%
	\begin{subfigure}{0.50\textwidth}
		\centering
		{($b$)}
		\includegraphics[width=0.90\linewidth,valign=t]{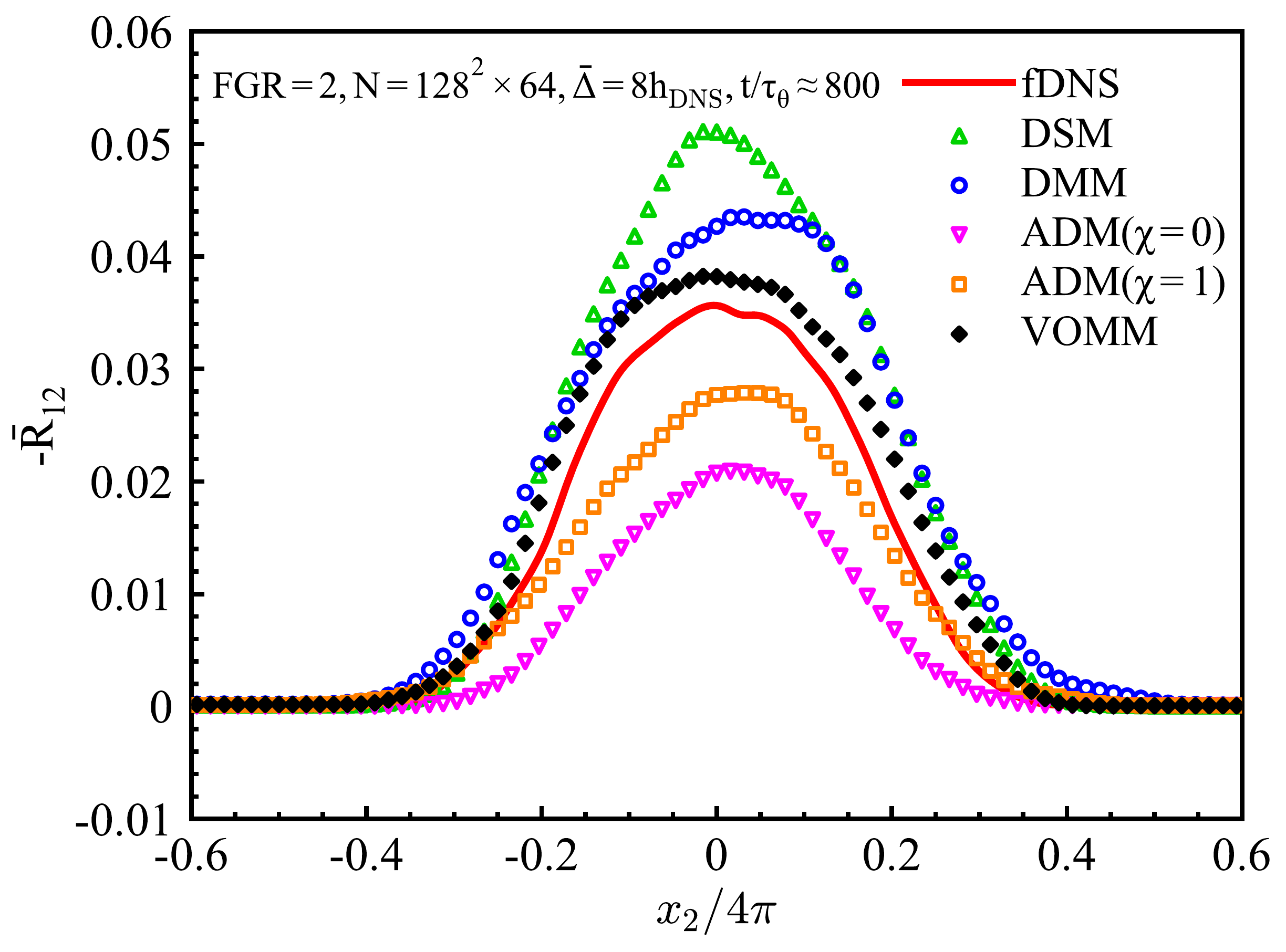}
	\end{subfigure}
	\caption{The transient profile of the resolved Reynolds shear stress ${{\bar R}_{12}} = \left\langle {{{\bar u}_1^\prime}{{\bar u}_2^\prime}} \right\rangle$ along the cross-stream direction for LES in the \emph{a posteriori} analysis of temporally evolving turbulent mixing layer with filter scale $\bar \Delta  = 8{h_{{\rm{DNS}}}}$ at grid resolution of $N=128^2 \times 64$: (a) $t/\tau_\theta \approx$ 500; (b) $t/\tau_\theta \approx$ 800.}
	\label{fig:21}
\end{figure}

The profiles of the resolved Reynolds shear stress component ${\bar R_{12}} = \left\langle {{\bar u_1^\prime}{\bar u_2^\prime}} \right\rangle$ at time instants $t/\tau_\theta \approx$ 500 and 800 are illustrated in Fig.~\ref{fig:21}, which is the dominant Reynolds stress term due to the intense mixing along the streamwise and normal directions \citep{vreman1997,sharan2019}. The normal distribution of the Reynolds stress is a second-order statistic of turbulence which has high requirements for the accuracy of SGS modeling of LES. The ADM models underpredict the Reynolds stress, while DSM and DMM models give obvious overestimations at different times. Compared to these classical SGS models, the VOMM model gives the prediction closest to the fDNS results, and accurately recovers the transient profiles of  Reynolds stress. 

\begin{figure}\centering
	\begin{subfigure}{0.5\textwidth}
		\centering
		{($a$)}
		\includegraphics[width=0.9\linewidth,valign=t]{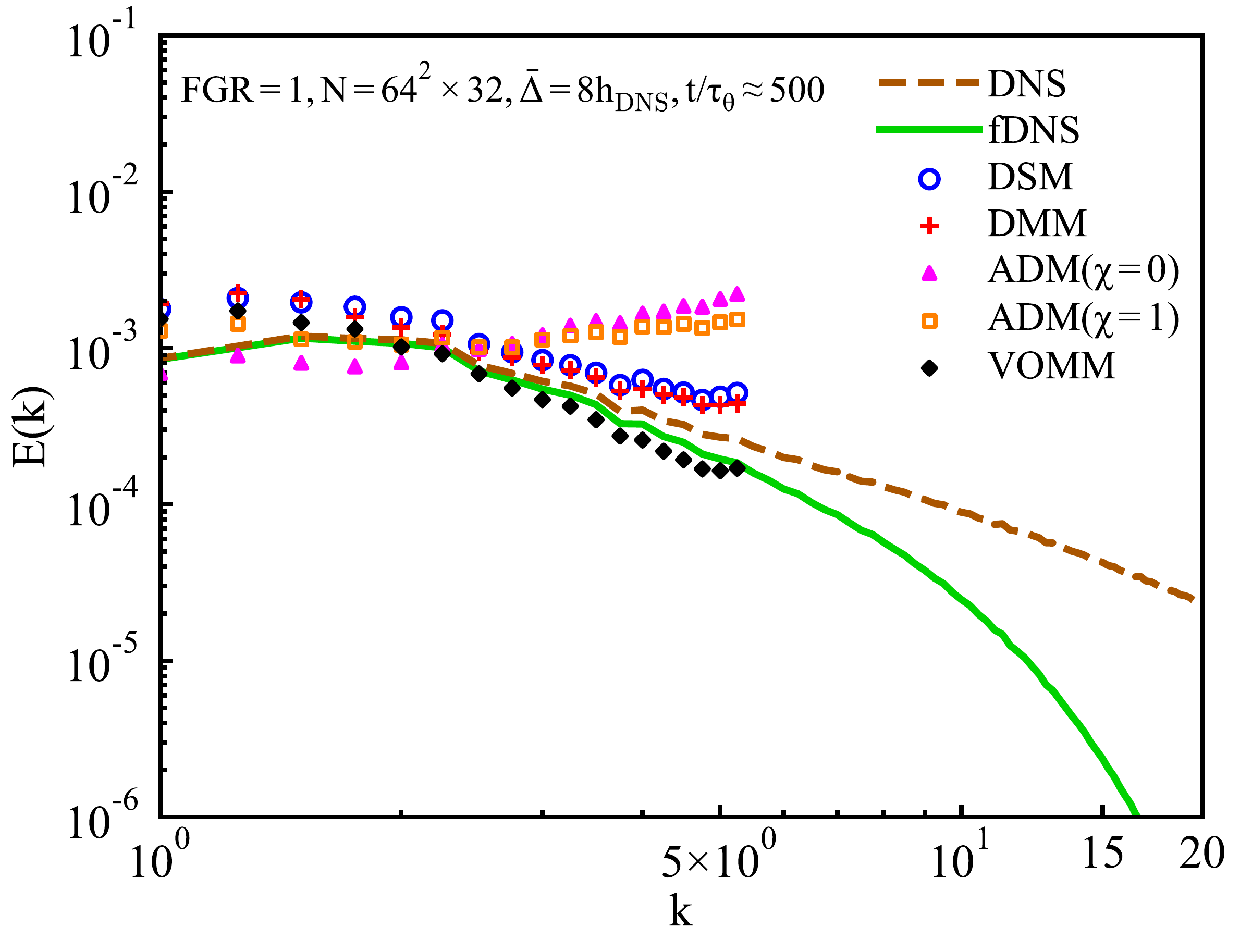}
	\end{subfigure}%
	\begin{subfigure}{0.5\textwidth}
		\centering
		{($b$)}
		\includegraphics[width=0.9\linewidth,valign=t]{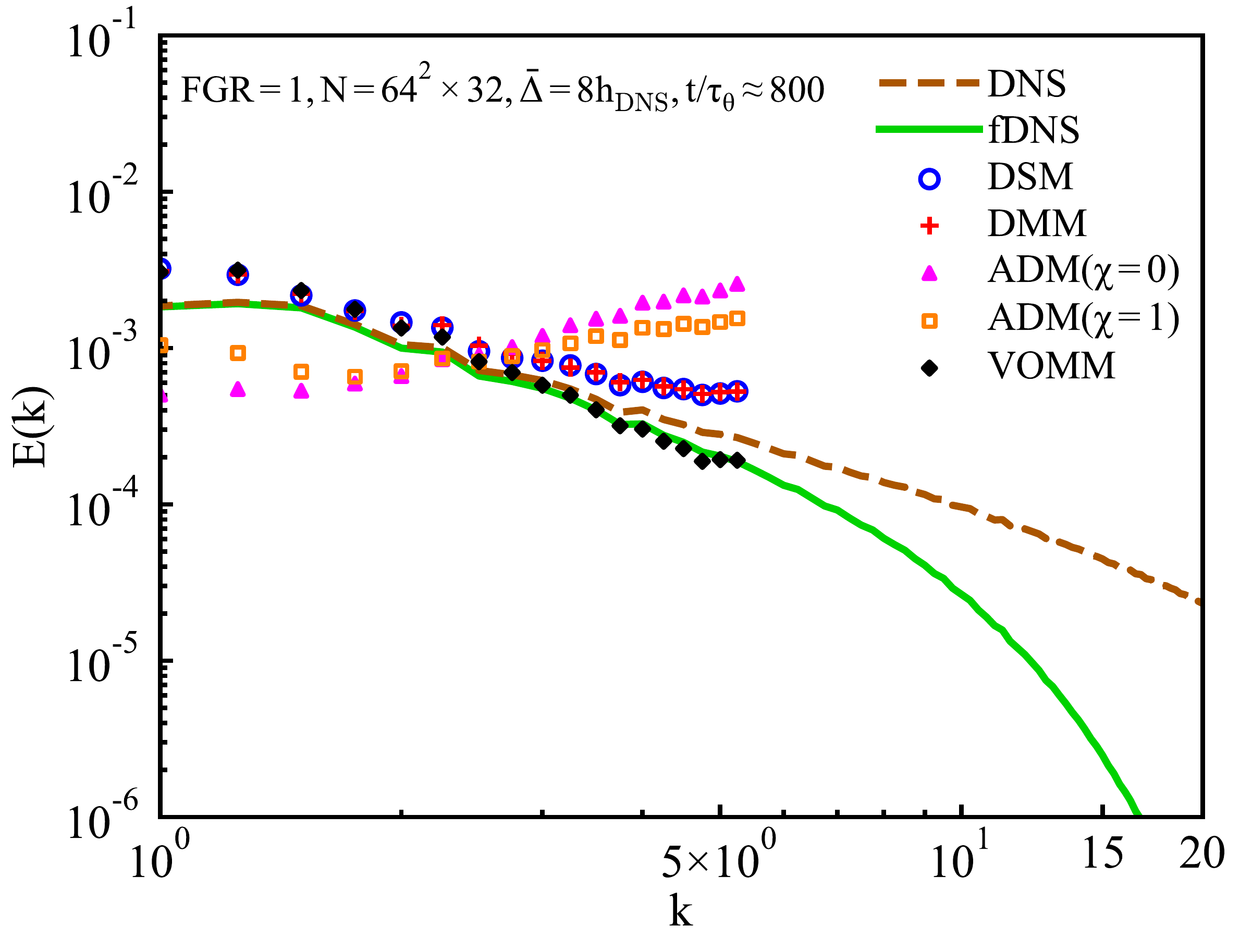}
	\end{subfigure}\\
	\begin{subfigure}{0.5\textwidth}
		\centering
		{($c$)}
		\includegraphics[width=0.9\linewidth,valign=t]{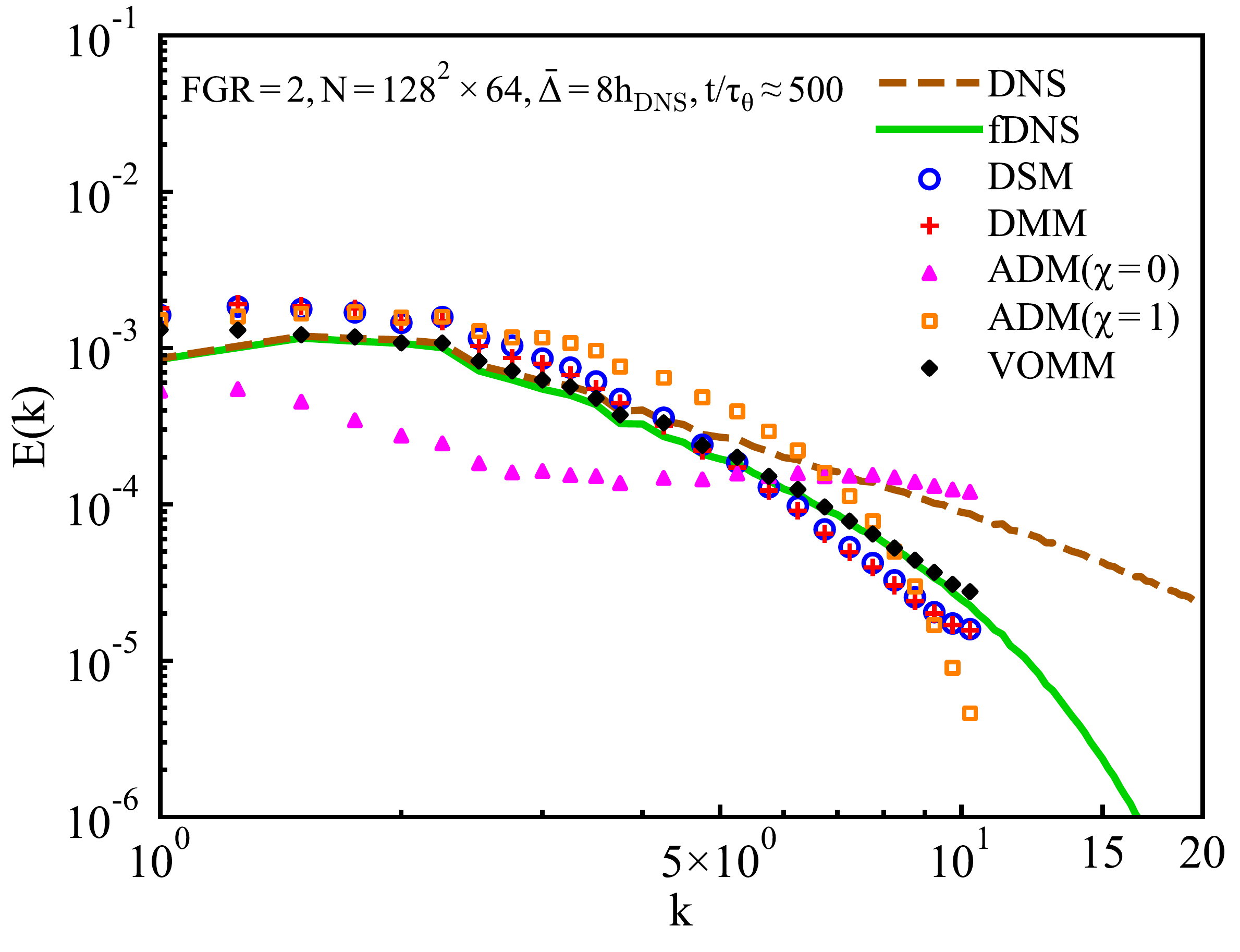}
	\end{subfigure}%
	\begin{subfigure}{0.5\textwidth}
		\centering
		{($d$)}
		\includegraphics[width=0.9\linewidth,valign=t]{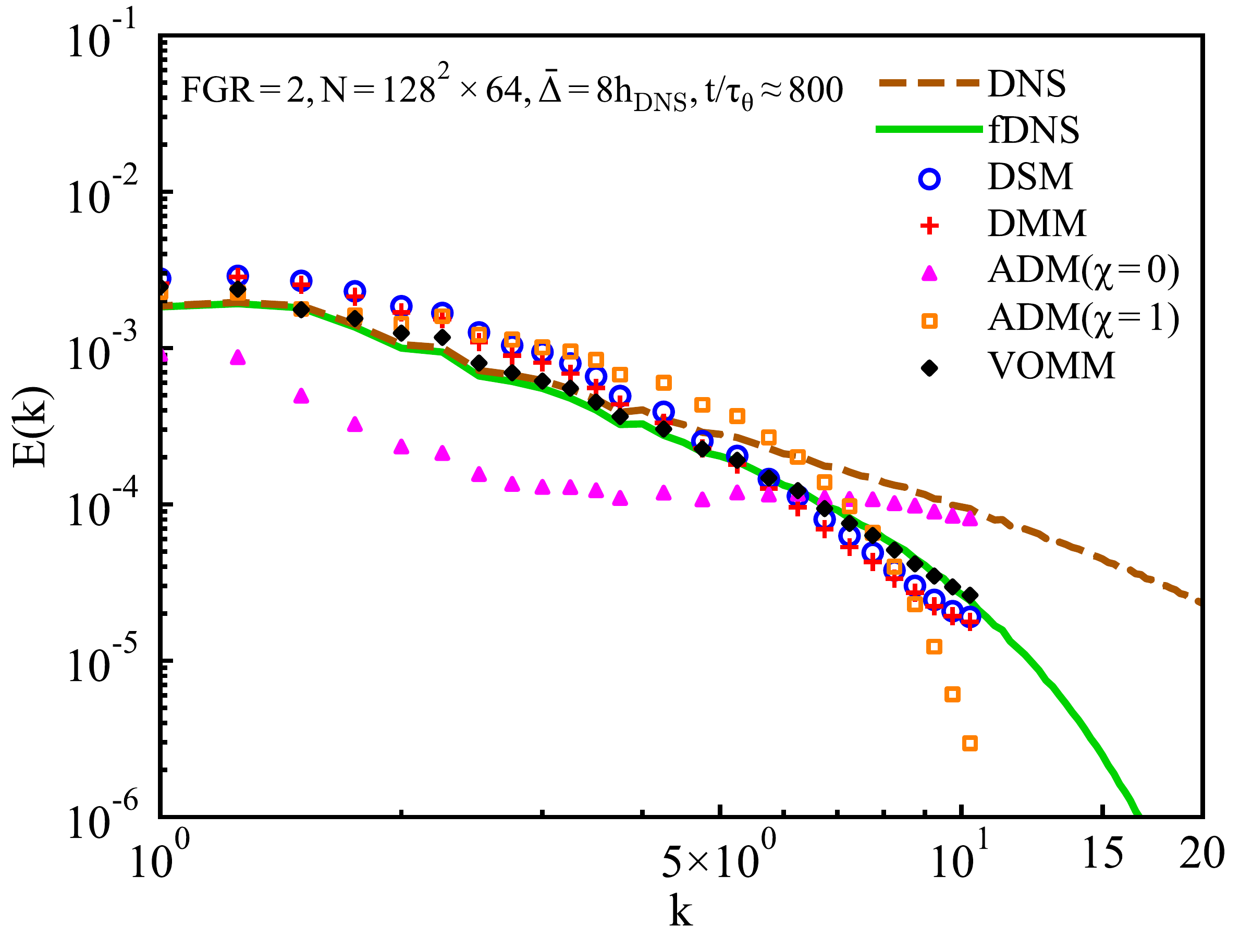}
	\end{subfigure}
	\caption{Velocity spectra for different SGS models in the \emph{a posteriori} analysis of temporally evolving turbulent mixing layer with the same filter scale $\bar \Delta  = 8{h_{{\rm{DNS}}}}$ at $t /\tau_\theta \approx$ 500 and 800: (a) FGR=1, $N=64^2 \times 32$ at $t/\tau_\theta \approx$ 500; (b) FGR=1, $N=64^2 \times 32$ at $t/\tau_\theta \approx$ 800; (c) FGR=2, $N=128^2 \times 64$ at $t/\tau_\theta \approx$ 500; (d) FGR=2, $N=128^2 \times 64$ at $t/\tau_\theta \approx$ 800.}
	\label{fig:22}
\end{figure}

\begin{figure}\centering
	\begin{subfigure}{0.45\textwidth}
		\centering
		{($a$)}
		%\caption{fDNS}
		\includegraphics[width=0.9\linewidth,valign=t]{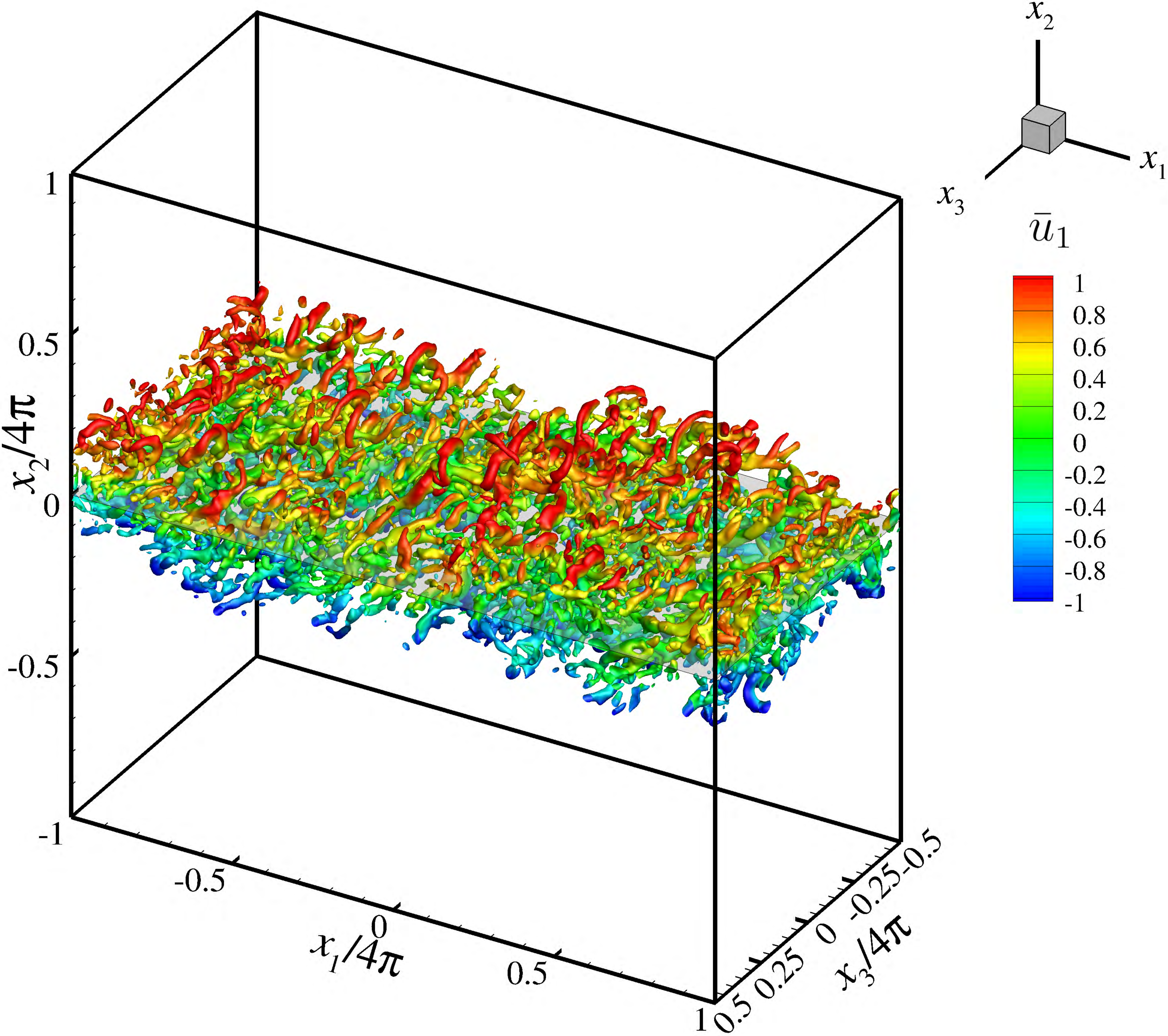}
		
	\end{subfigure}%
	\begin{subfigure}{0.45\textwidth}
		\centering
		{($b$)}
		%\caption{DMM}
		\includegraphics[width=0.9\linewidth,valign=t]{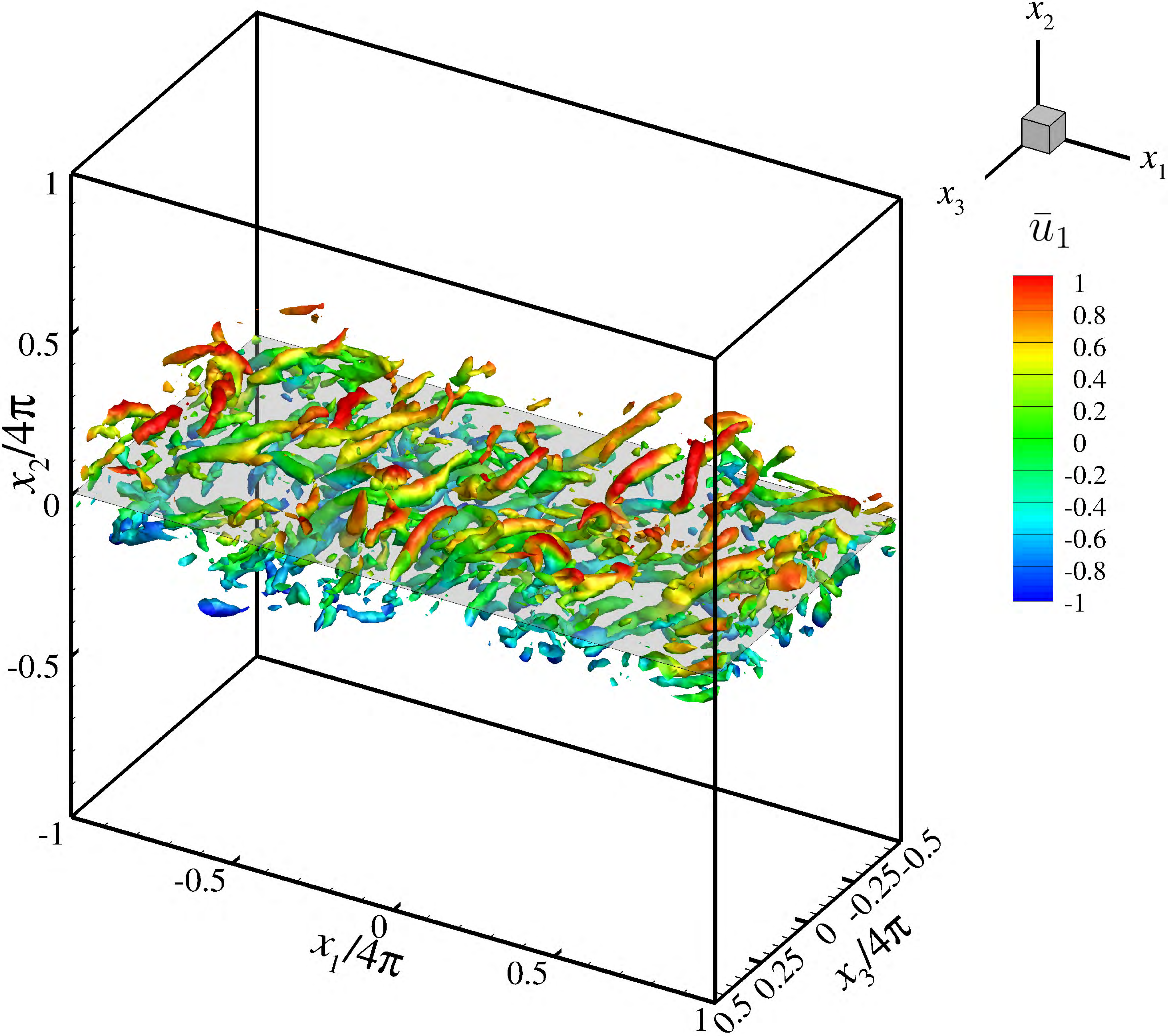}
	\end{subfigure}\\
	\begin{subfigure}{0.45\textwidth}
		\centering
		{($c$)}
		%\caption{ADM (${\chi \!=\! 0}$)}
		\includegraphics[width=0.9\linewidth,valign=t]{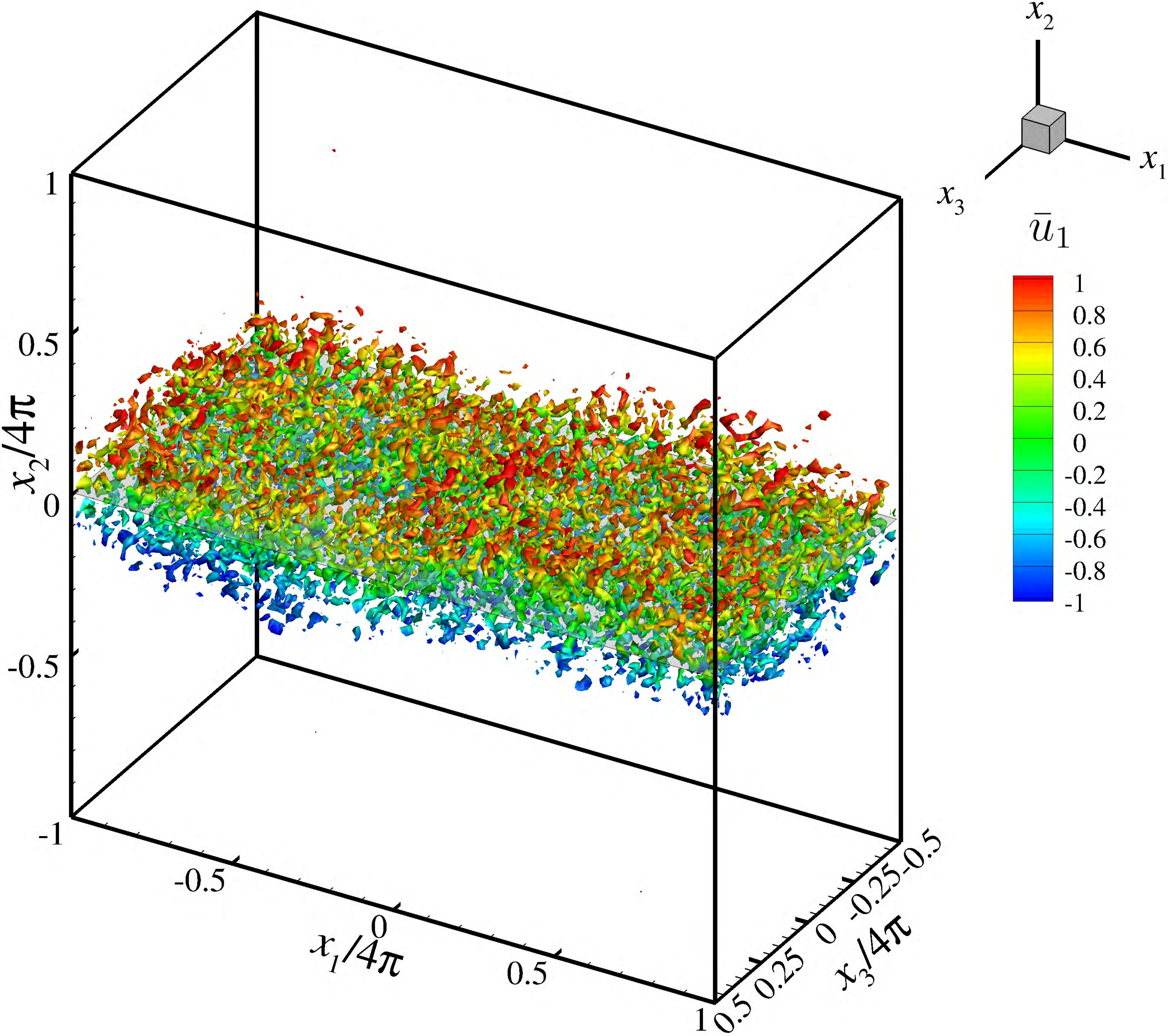}
	\end{subfigure}%
	\begin{subfigure}{0.45\textwidth}
		\centering
		{($d$)}
		%\caption{ADM (${\chi \!=\! 1}$)}
		\includegraphics[width=0.9\linewidth,valign=t]{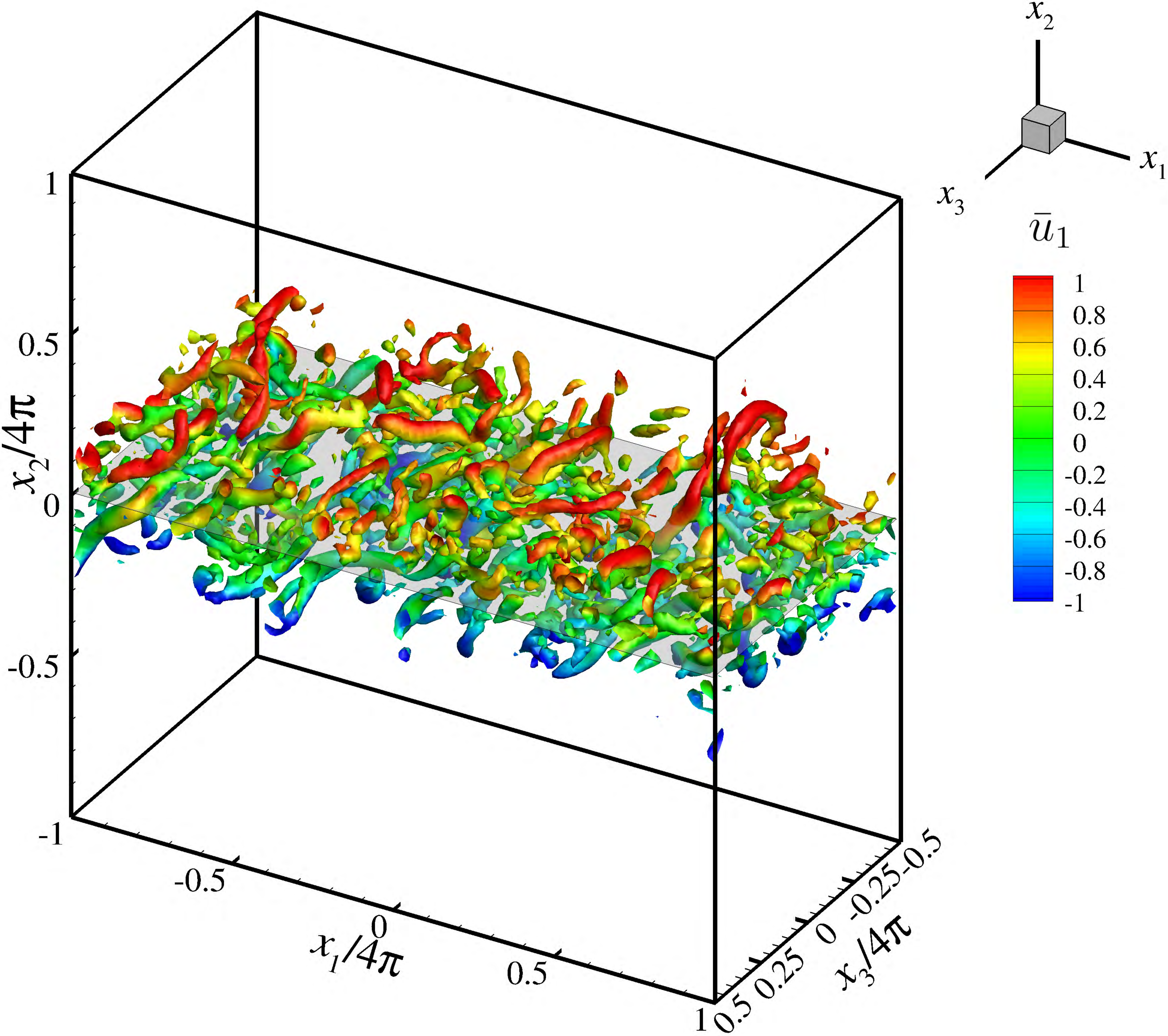}
	\end{subfigure}\\
	\begin{subfigure}{0.45\textwidth}
		\centering
		{($e$)}
		%\caption{DSM}
		\includegraphics[width=0.9\linewidth,valign=t]{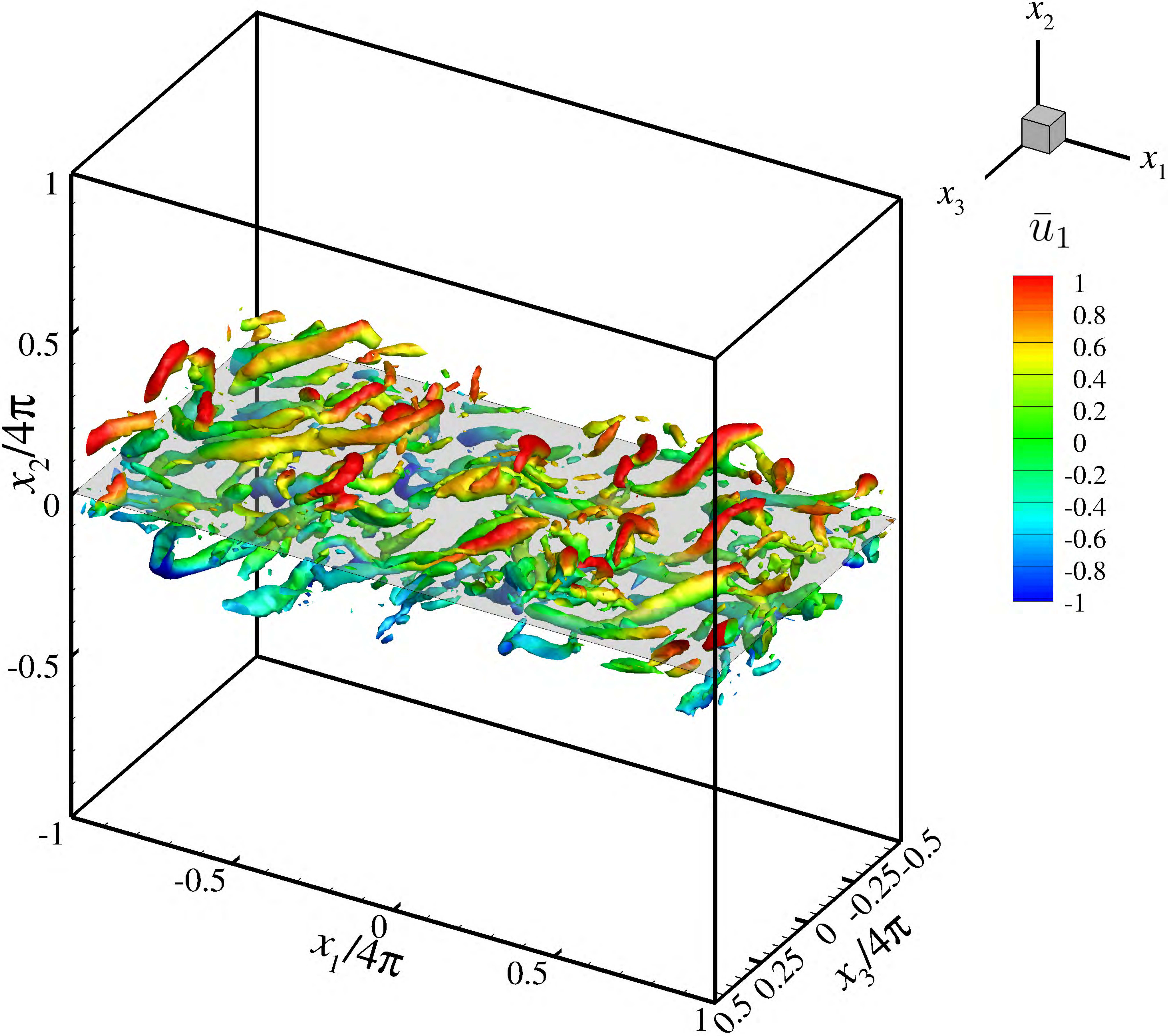}
	\end{subfigure}%
	\begin{subfigure}{0.45\textwidth}
		\centering
		{($f$)}
		%\caption{VOMM}
		\includegraphics[width=0.9\linewidth,valign=t]{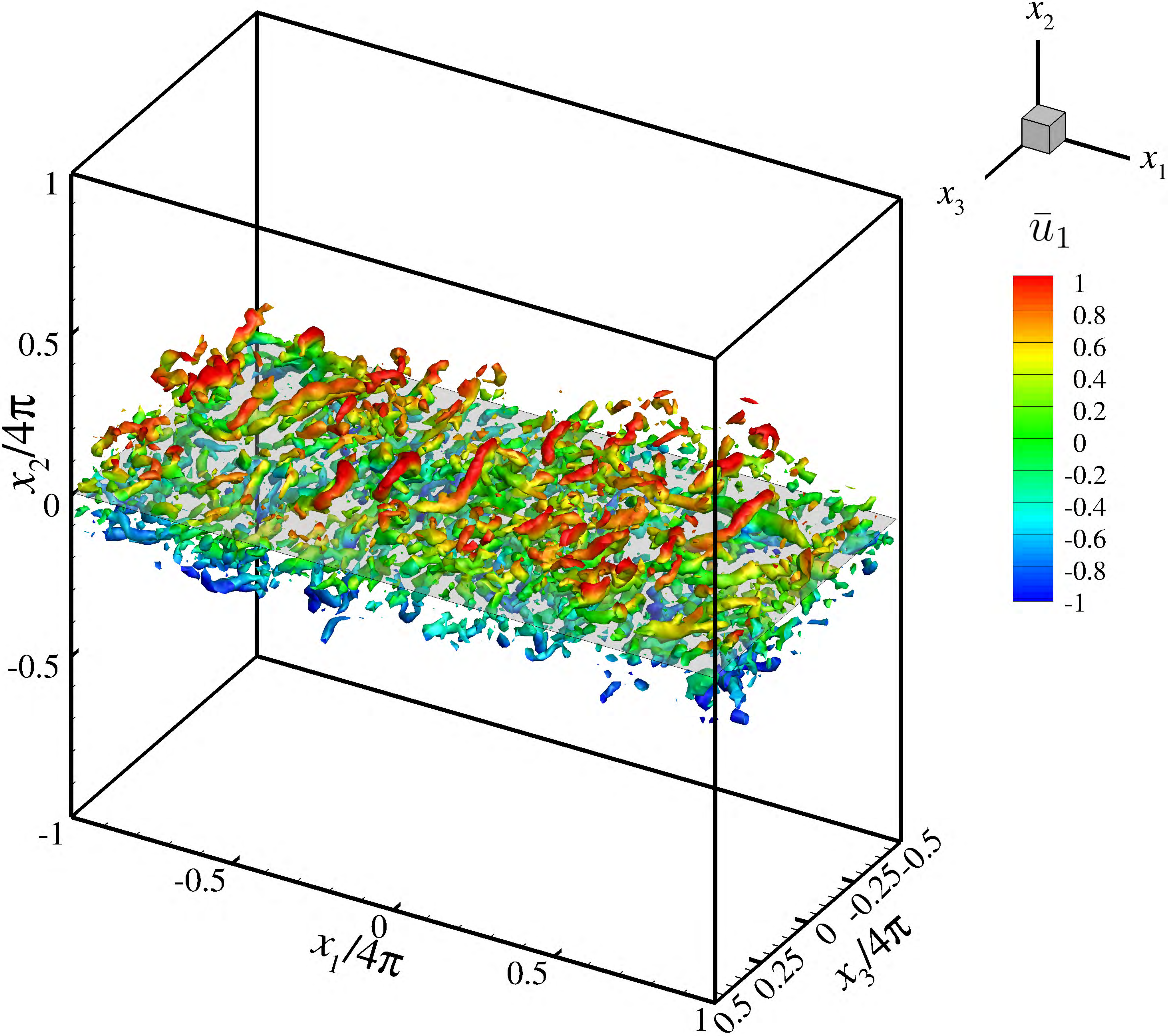}
	\end{subfigure}	
	\caption{The iso-surface of the Q-criterion at $Q$=0.2 colored by the streamwise velocity at $t/\tau_\theta \approx$ 500 in the \emph{a posteriori} analysis of temporally evolving turbulent mixing layer with filter scale $\bar \Delta  = 8{h_{{\rm{DNS}}}}$ at grid resolution of $N=128^2 \times 64$: (a) fDNS, (b) DMM, (c) ADM($\chi$=0), (d) ADM($\chi$=1), (e) DMM, and (f) VOMM.}
	\label{fig:23}
\end{figure}

We further compare the velocity spectra of different SGS models with the DNS and filtered DNS data at time instants $t/\tau_\theta \approx$ 500 and 800, as shown in Fig.~\ref{fig:22}. The spectra of DNS at $t/\tau_\theta \approx$ 500 and 800 are very similar since the instantaneous velocity fields at different moments are both at the self-similar stage of mixing layer. For the coarse grid-resolution case at FGR=1 with $N = {64^2} \times 32$, the conventional SGS models (DSM, DMM and ADM models) always give the overestimations of the small-scale kinetic energy at high wavenumbers, and the excess kinetic energy accumulates at small scales and exacerbates the numerical instability of LES over time. The SGS dissipation provided by these conventional SGS models is insufficient to stabilize the numerical perturbations induced by the spatial discretization errors, which cannot effectively drain out the small-scale kinetic energy in time at FGR=1. For the case of fine grid resolution at FGR=2 with  $N = {128^2} \times 64$, the pure ADM model is still numerically unstable, whose prediction distinctly deviates from the fDNS data. And the velocity spectra predicted by the other conventional SGS models (DSM, DMM and ADM with $\chi$=0) diminish at high-wavenumber regions and accumulate in the region of intermediate wavenumbers, since these traditional SGS models are too dissipative at the fine grid-resolution case to recover the effect of small-scale flow structures near the cutoff wavenumber, giving rise to the blockage of the kinetic energy cascade from large scales to small scales. In contrast, the kinetic energy cascade can be correctly constructed with high accuracy by the VOMM model, and the predictions are always in reasonable agreement with those of fDNS at different grid resolutions and time instants. 

The reconstruction of vortex structures is finally compared with different SGS models by displaying the iso-surface of the Q-criterion. The Q-criterion is a useful visualization tool for observing vortex structures in turbulent flows, and is the second invariant of velocity gradient tensor, namely \citep{hunt1988,dubief2000,zhan2019}
\begin{equation}
	Q = \frac{1}{2}\left( {{{\bar \Omega }_{ij}}{{\bar \Omega }_{ij}} - {{\bar S}_{ij}}{{\bar S}_{ij}}} \right),
	\label{Q_criterion}
\end{equation}
where ${{\bar \Omega }_{ij}} = \frac{1}{2}\left( {\partial {{\bar u}_i}/\partial {x_j} - \partial {{\bar u}_j}/\partial {x_i}} \right)$ represents the rotation-rate tensor. The instantaneous iso-surface of Q at $t/\tau_\theta \approx 500$ is illustrated in Fig.~\ref{fig:23} during the self-similar stage of the mixing layer for Q=0.2 colored by the streamwise velocity. The Q iso-surface of fDNS contains a large number of elaborate vortex structures near the middle $x_1$-$x_3$ plane of the shear layer, including the rib-like vortices, hairpin vortices and complex helical vortices. DSM, DMM and ADM ($\chi$=1) models exhibit an excessive dissipation that only large-scale rib-like vortex structures remain, while the pure ADM model with $\chi$=0 suffers from numerical instability of LES and overpredicts many nonphysical small-scale structures caused by numerical noise. In contrast, the VOMM model can accurately reconstruct much more vortex structures, highlighting its advantage in improving the accuracy of LES.

\section {Generalization ability of the VOMM model}\label{sec:level6}

To examine the general performance of the proposed VOMM model, we further study the impact of unsteady evolutions and large-scale forcing, the impact of initial random disturbances on the temporally evolving turbulent mixing layer, and the generalization to higher Reynolds numbers.

\begin{figure}[h!]\centering
	\begin{subfigure}{0.33\textwidth}
		\centering
		{($a$)}
		\includegraphics[width=0.88\linewidth,valign=t]{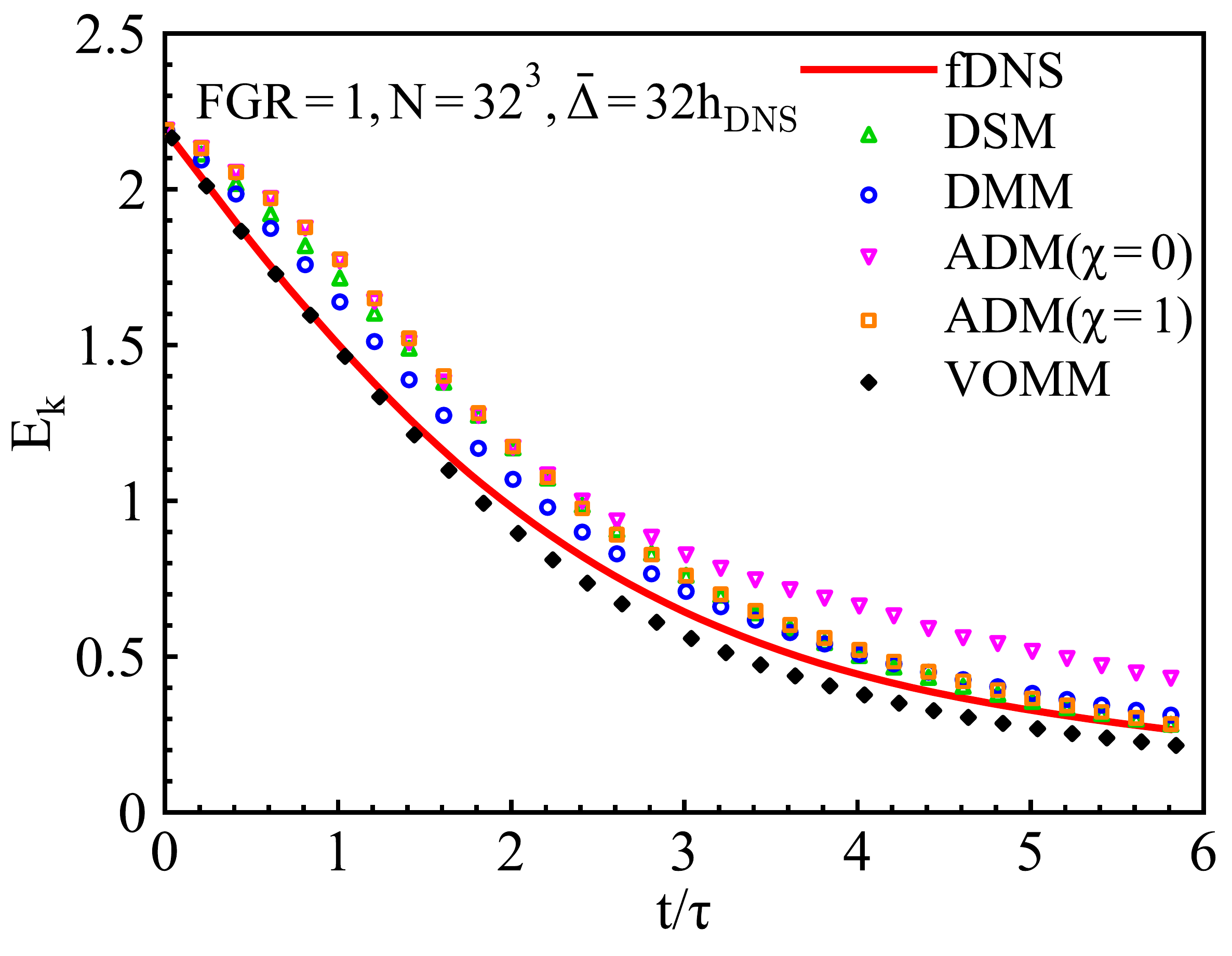}
	\end{subfigure}%
	\begin{subfigure}{0.33\textwidth}
		\centering
		{($b$)}
		\includegraphics[width=0.88\linewidth,valign=t]{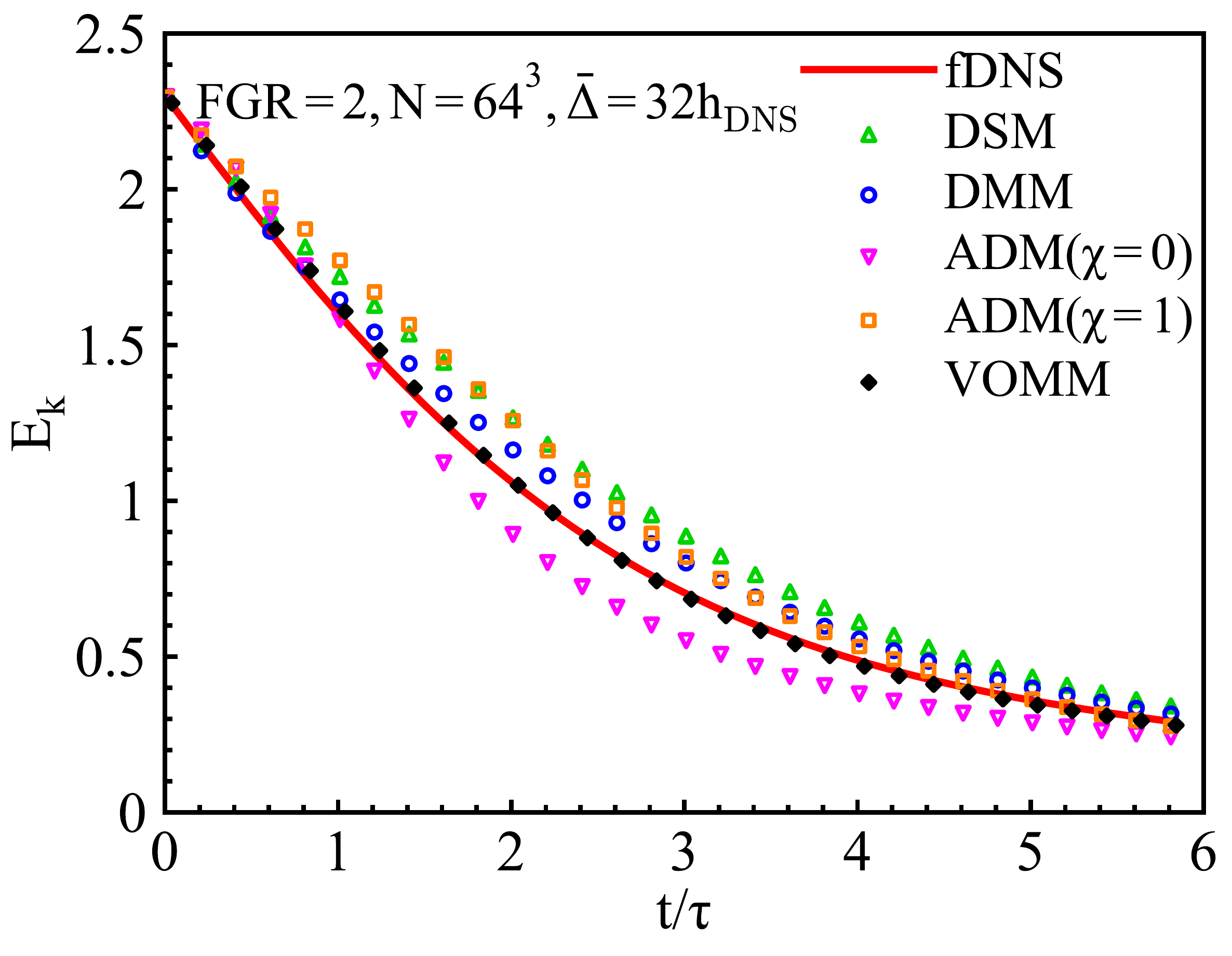}
	\end{subfigure}
	\begin{subfigure}{0.33\textwidth}
		\centering
		{($c$)}
		\includegraphics[width=0.88\linewidth,valign=t]{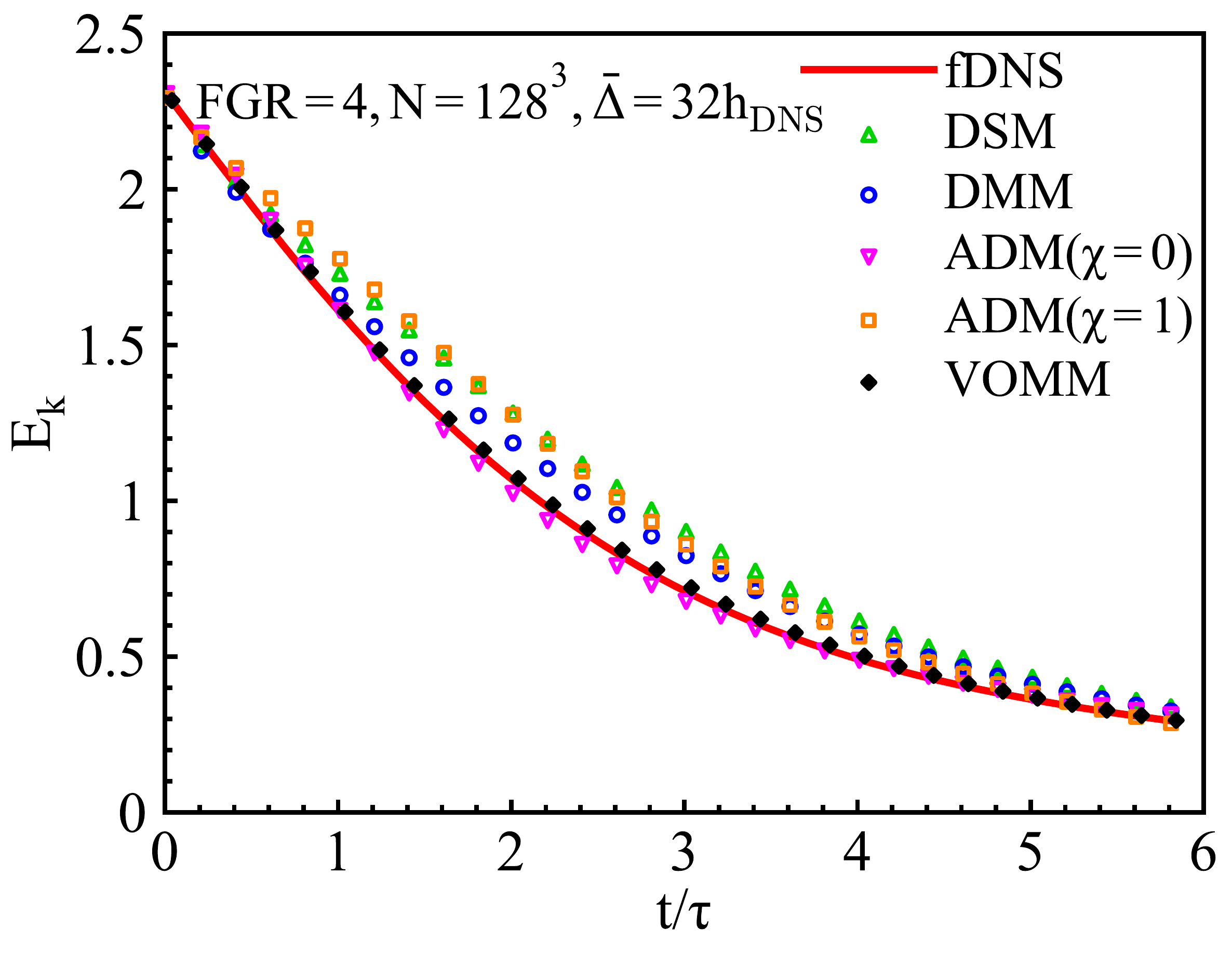}
	\end{subfigure}%
	\caption{Temporal evolutions of the turbulent kinetic energy $E_k$ for VOMM models with parameters optimized by the forced homogeneous isotropic turbulence in the \emph{a posteriori} analysis of decaying homogeneous isotropic turbulence with the same filter scale $\bar \Delta  = 32{h_{{\rm{DNS}}}}$: (a) FGR=1, $N=32^3$; (b) FGR=2, $N=64^3$; and (c) FGR=4, $N=128^3$.}
	\label{fig:24}
\end{figure}

\begin{figure}[h!]\centering
	\begin{subfigure}{0.33\textwidth}
		\centering
		{($a$)}
		\includegraphics[width=0.88\linewidth,valign=t]{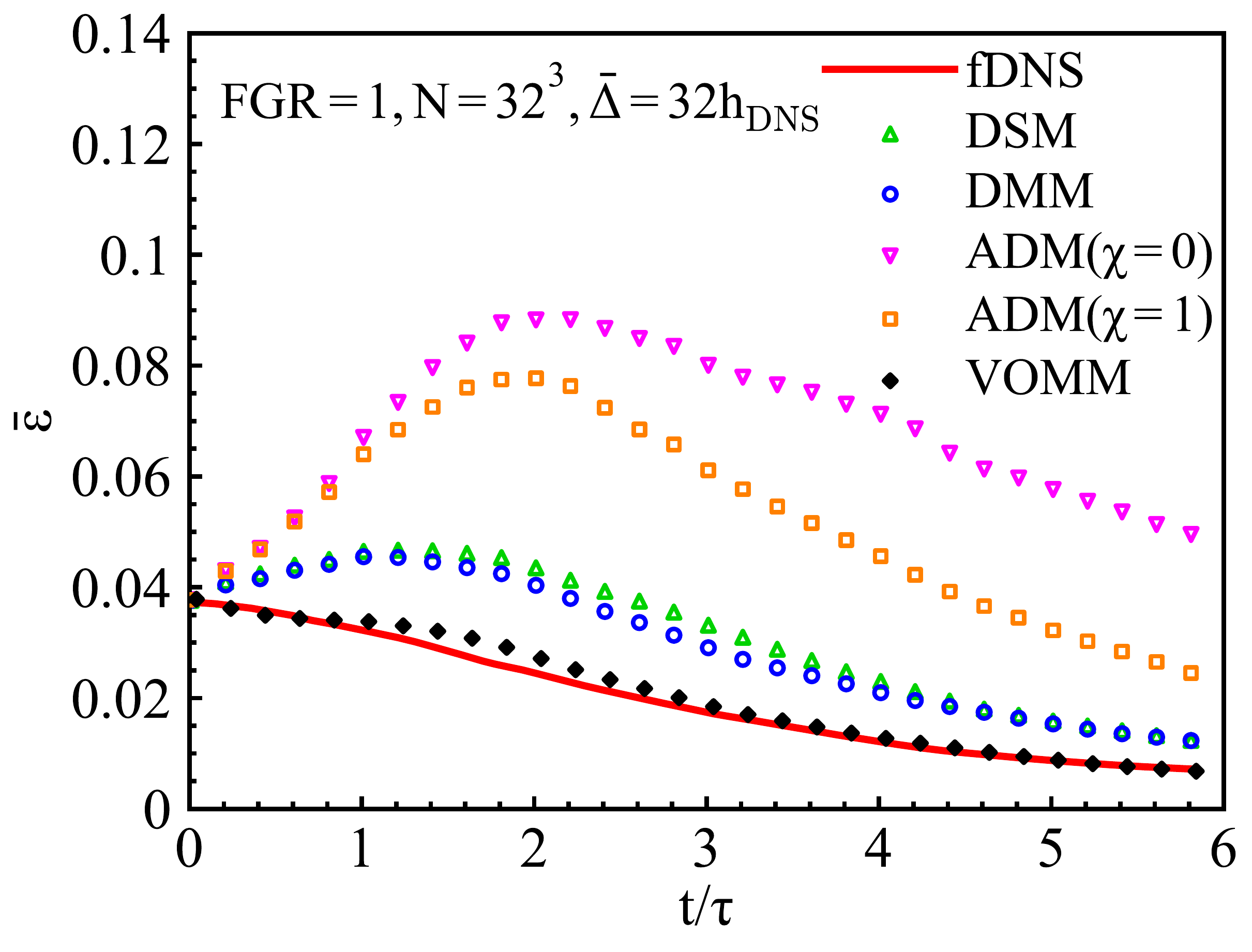}
	\end{subfigure}%
	\begin{subfigure}{0.33\textwidth}
		\centering
		{($b$)}
		\includegraphics[width=0.88\linewidth,valign=t]{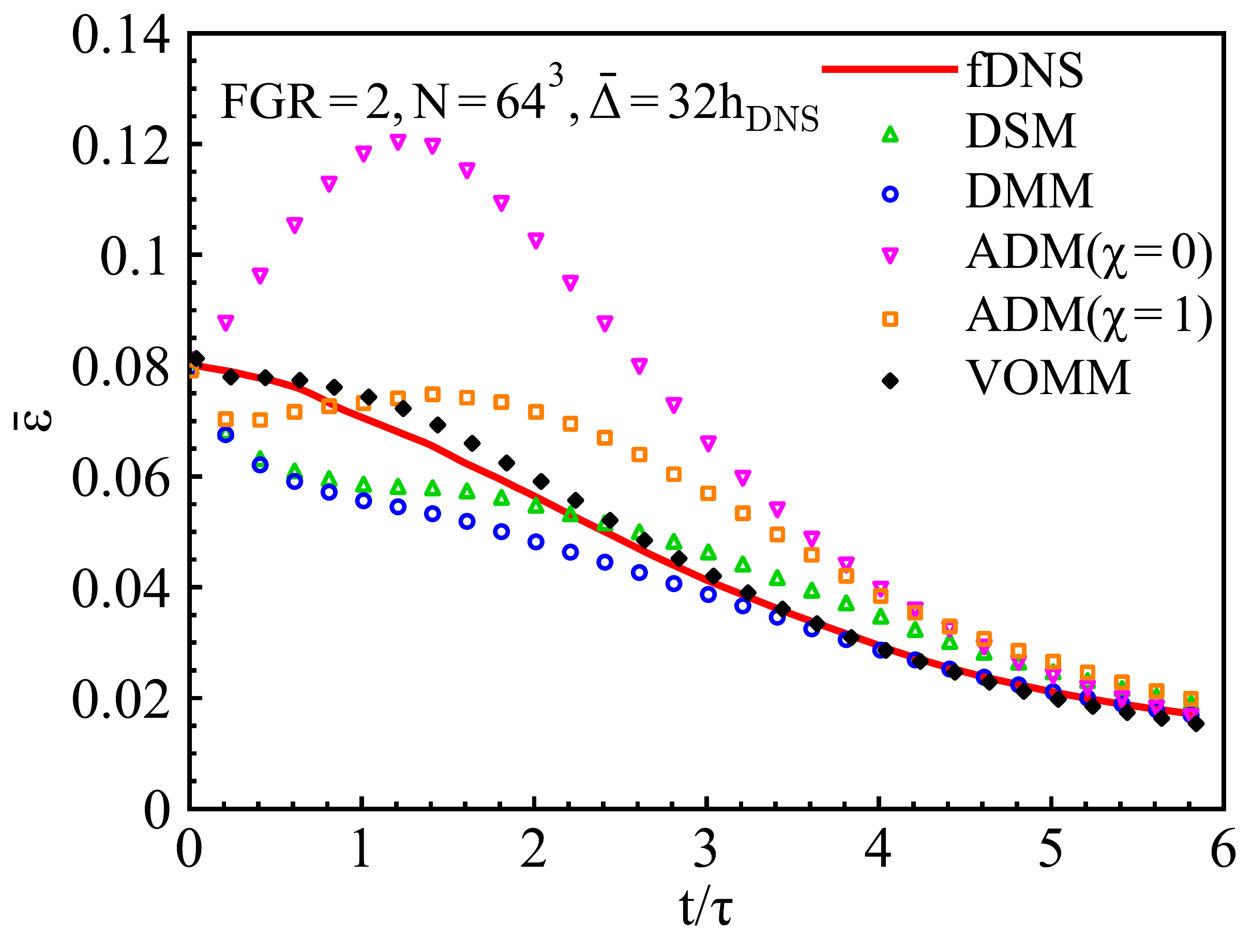}
	\end{subfigure}
	\begin{subfigure}{0.33\textwidth}
		\centering
		{($c$)}
		\includegraphics[width=0.88\linewidth,valign=t]{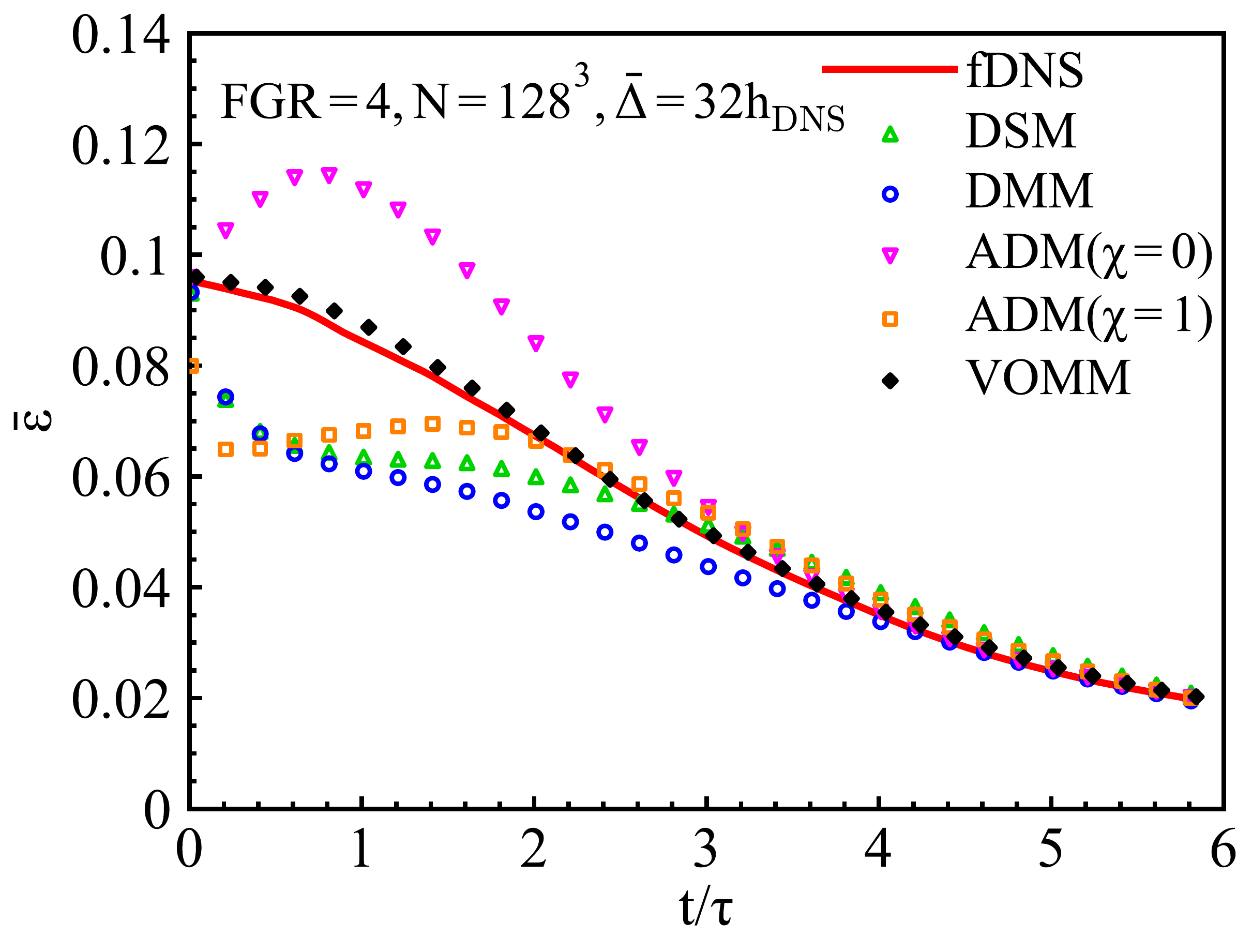}
	\end{subfigure}%
	\caption{Temporal evolutions of the average dissipation rate $\bar \varepsilon$ for VOMM models with parameters optimized by the forced homogeneous isotropic turbulence in the \emph{a posteriori} analysis of decaying homogeneous isotropic turbulence with the same filter scale $\bar \Delta  = 32{h_{{\rm{DNS}}}}$: (a) FGR=1, $N=32^3$; (b) FGR=2, $N=64^3$; and (c) FGR=4, $N=128^3$.}
	\label{fig:25}
\end{figure}

\subsection{Impact of unsteady evolutions and large-scale forcing}
\begin{figure}[h!]\centering
	\begin{subfigure}{0.5\textwidth}
		\centering
		{($a$)}
		\includegraphics[width=0.9\linewidth,valign=t]{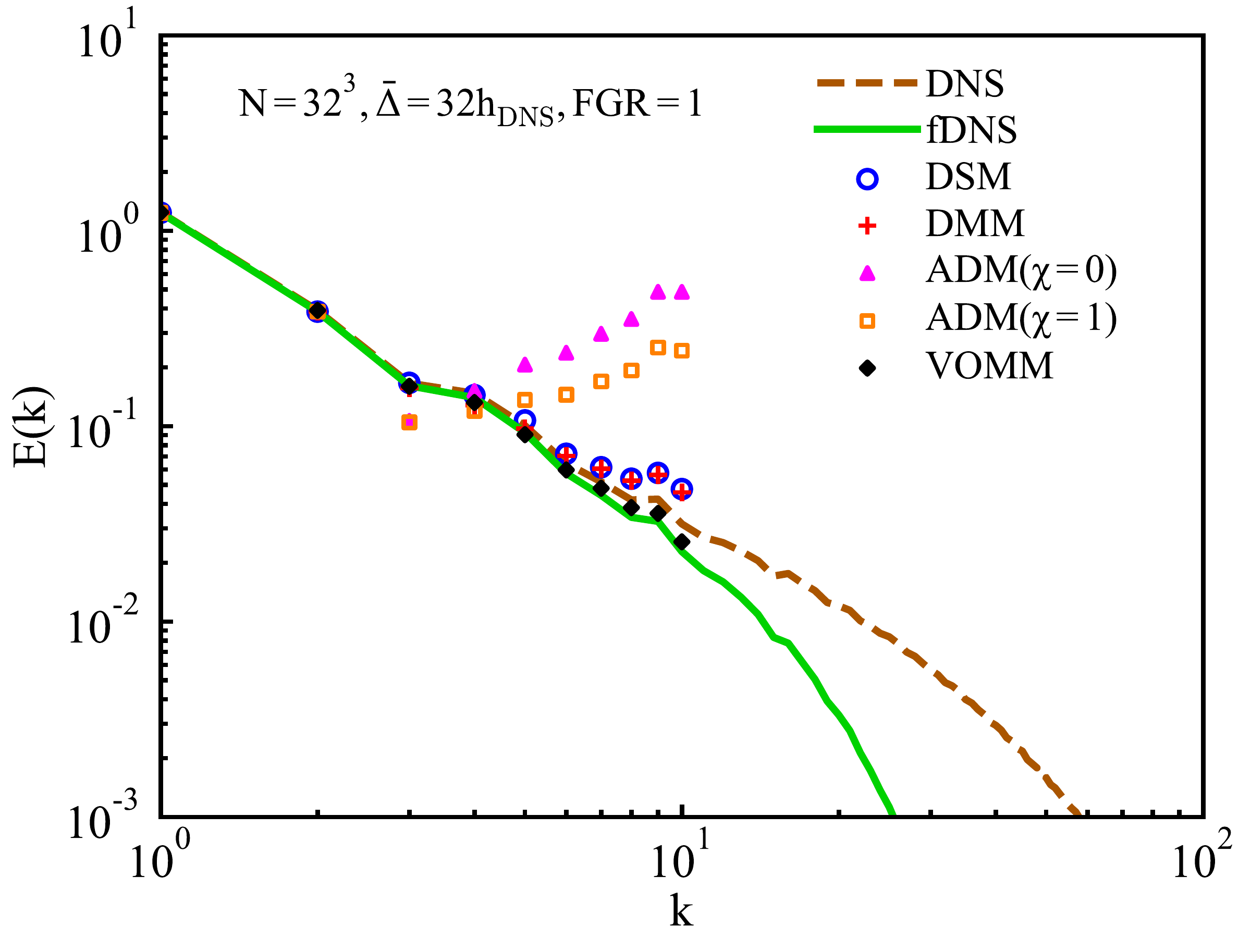}
	\end{subfigure}%
	\begin{subfigure}{0.5\textwidth}
		\centering
		{($b$)}
		\includegraphics[width=0.9\linewidth,valign=t]{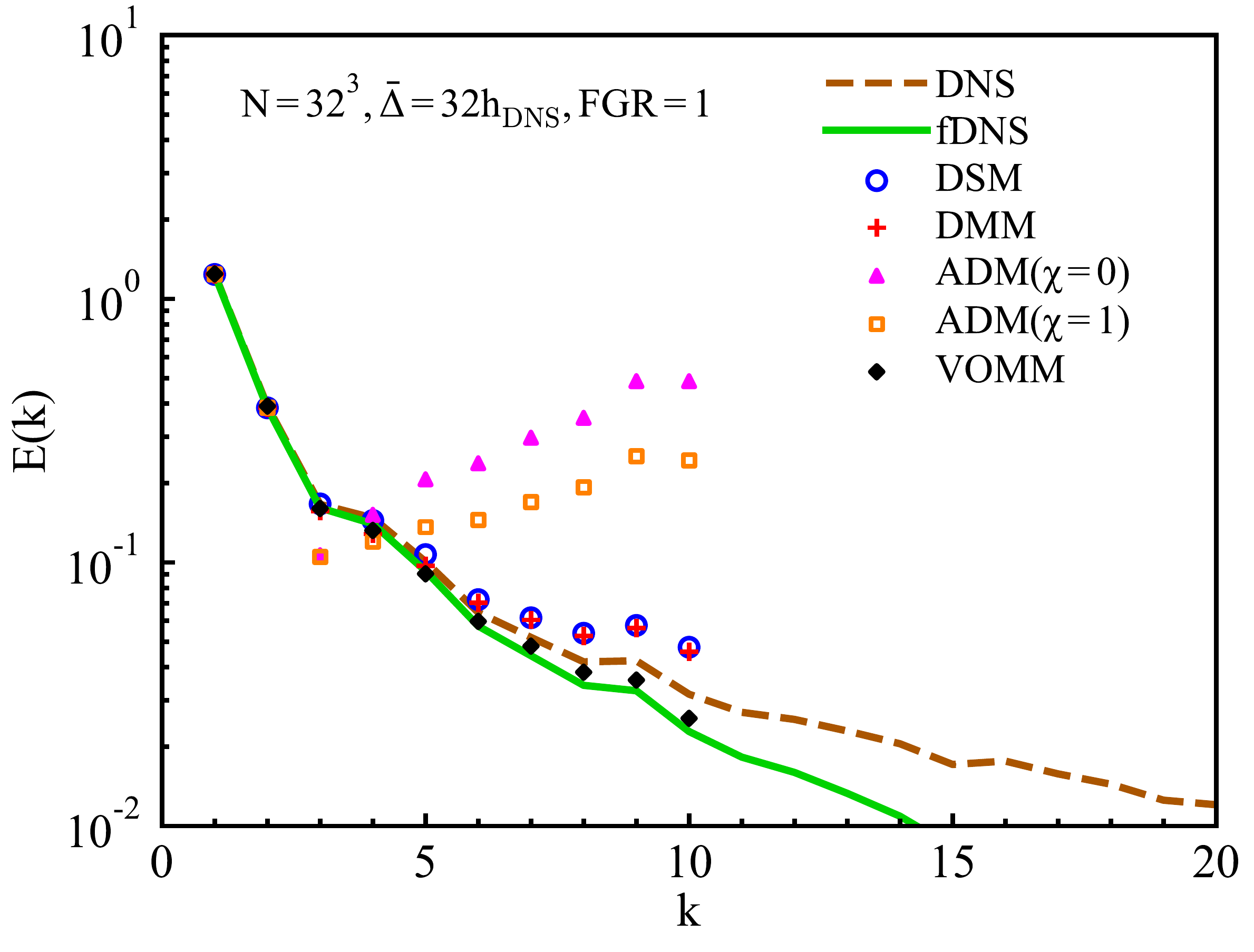}
	\end{subfigure}\\
	\begin{subfigure}{0.5\textwidth}
		\centering
		{($c$)}
		\includegraphics[width=0.9\linewidth,valign=t]{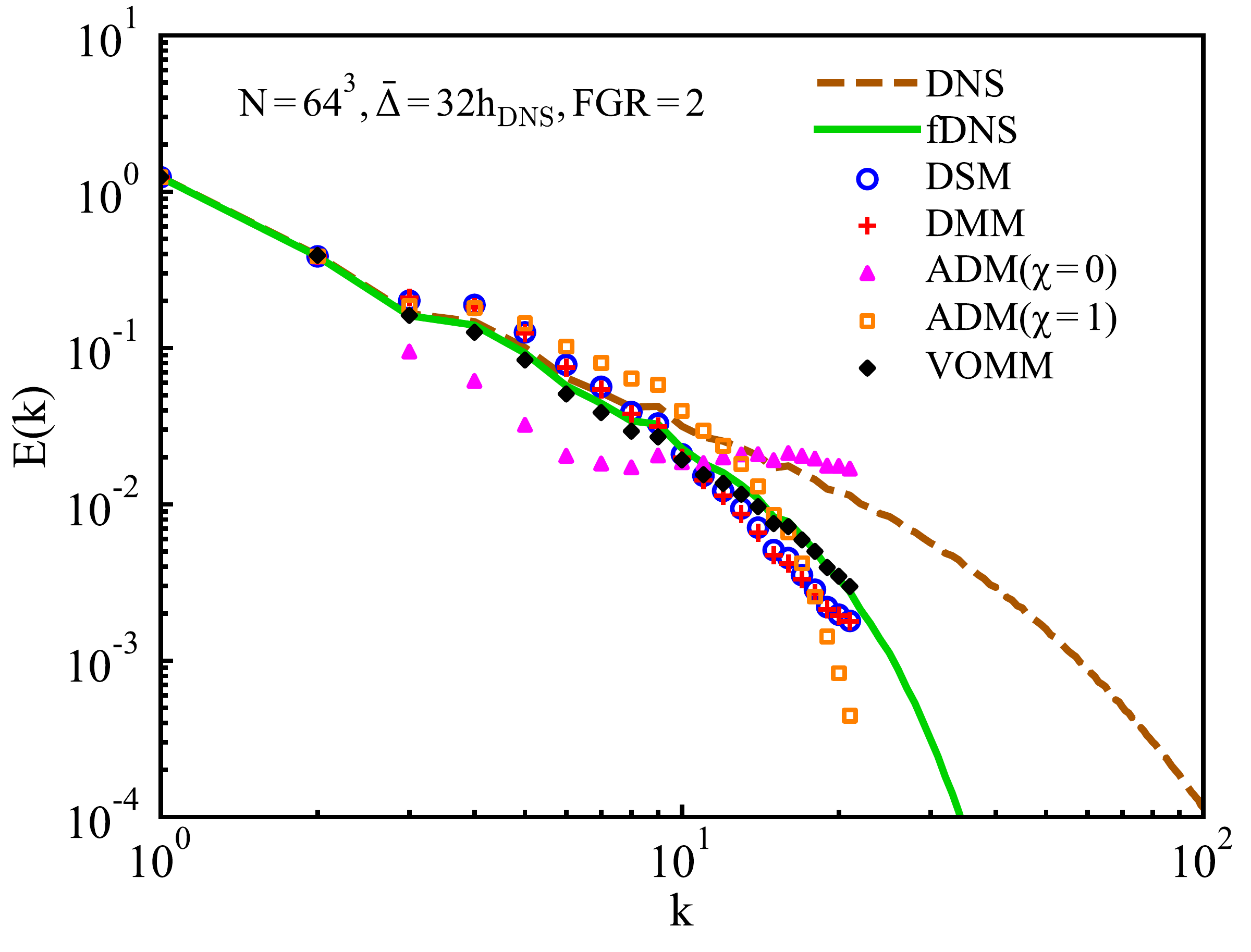}
	\end{subfigure}%
	\begin{subfigure}{0.5\textwidth}
		\centering
		{($d$)}
		\includegraphics[width=0.9\linewidth,valign=t]{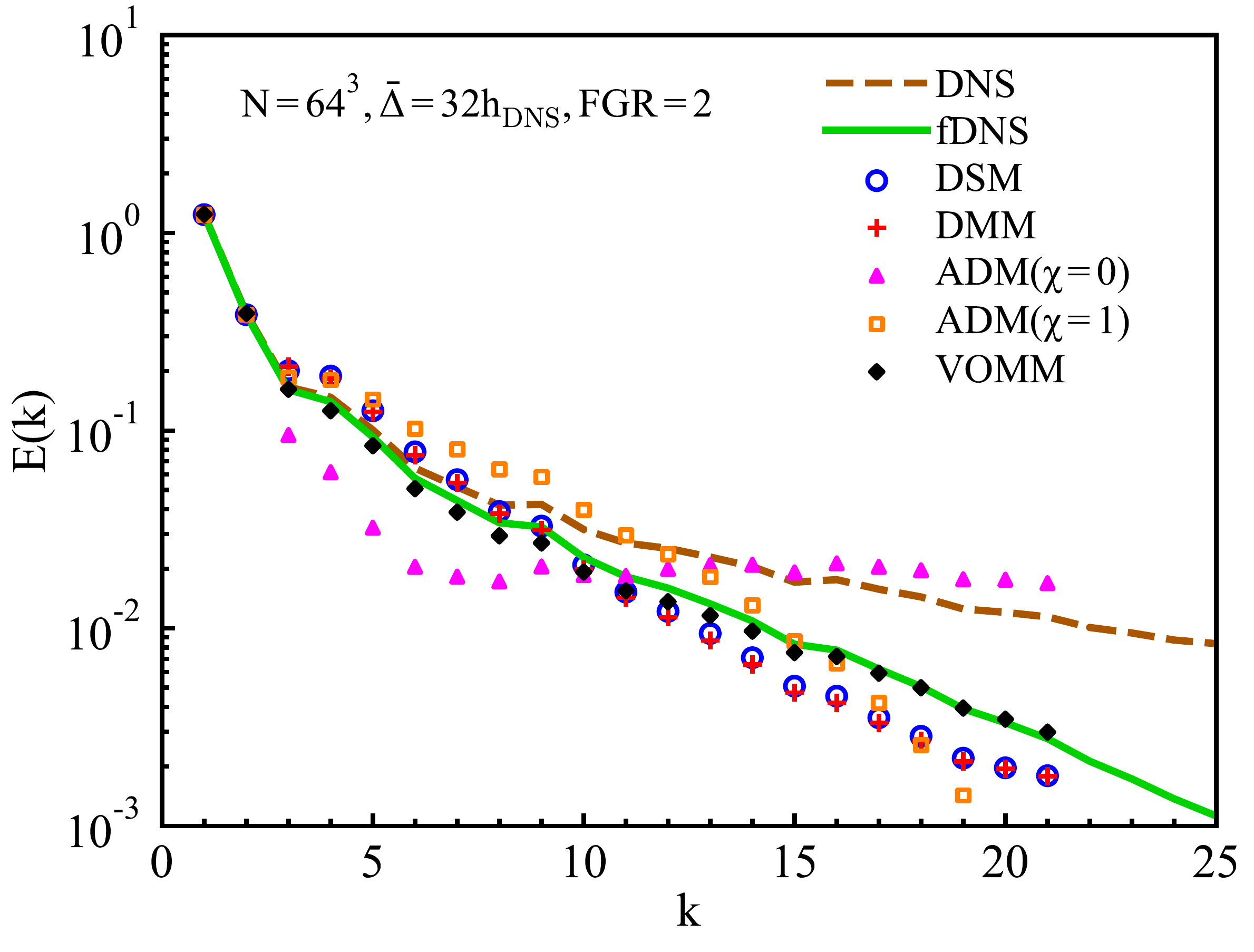}
	\end{subfigure}\\
	\begin{subfigure}{0.5\textwidth}
		\centering
		{($e$)}
		\includegraphics[width=0.9\linewidth,valign=t]{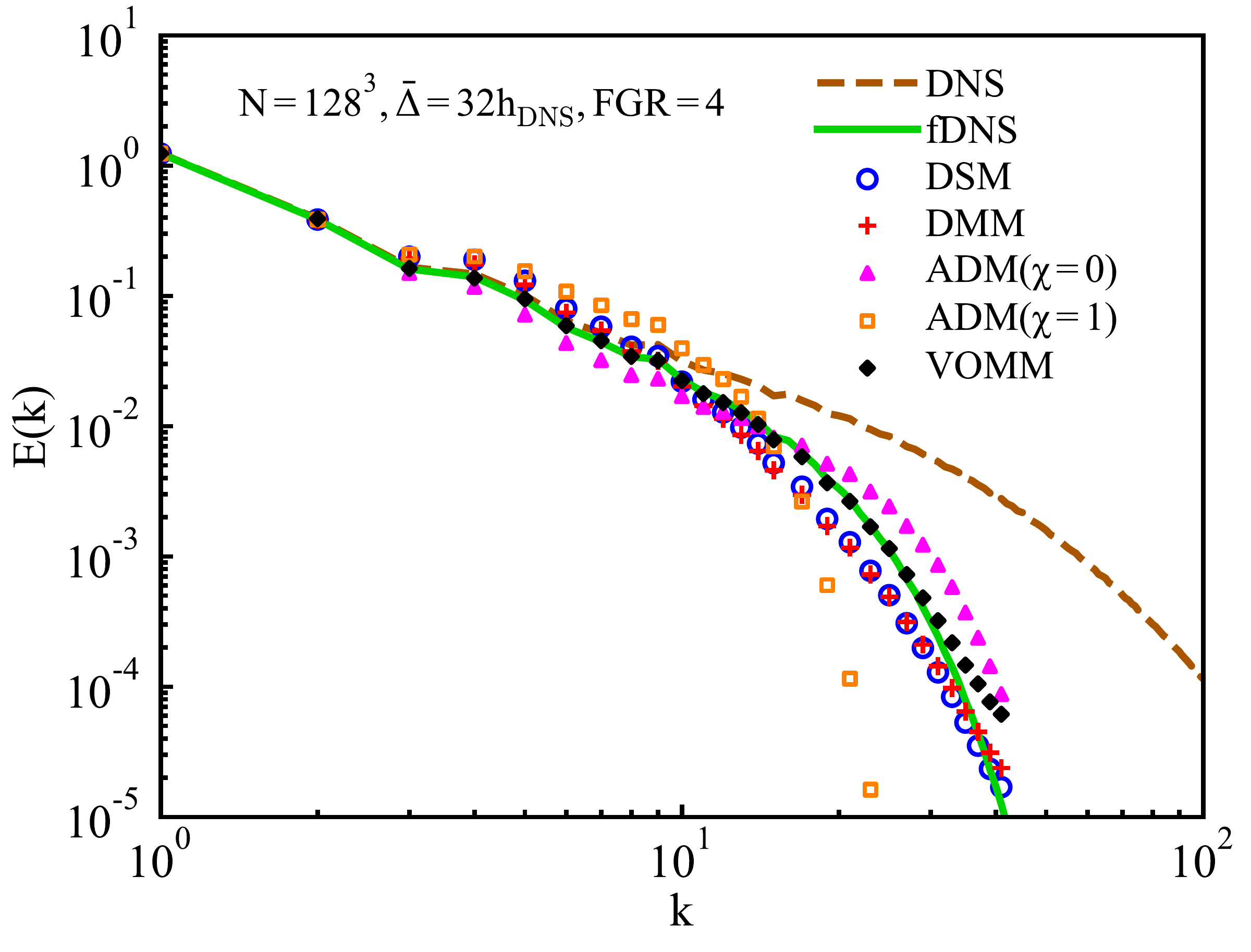}
	\end{subfigure}%
	\begin{subfigure}{0.5\textwidth}
		\centering
		{($f$)}
		\includegraphics[width=0.9\linewidth,valign=t]{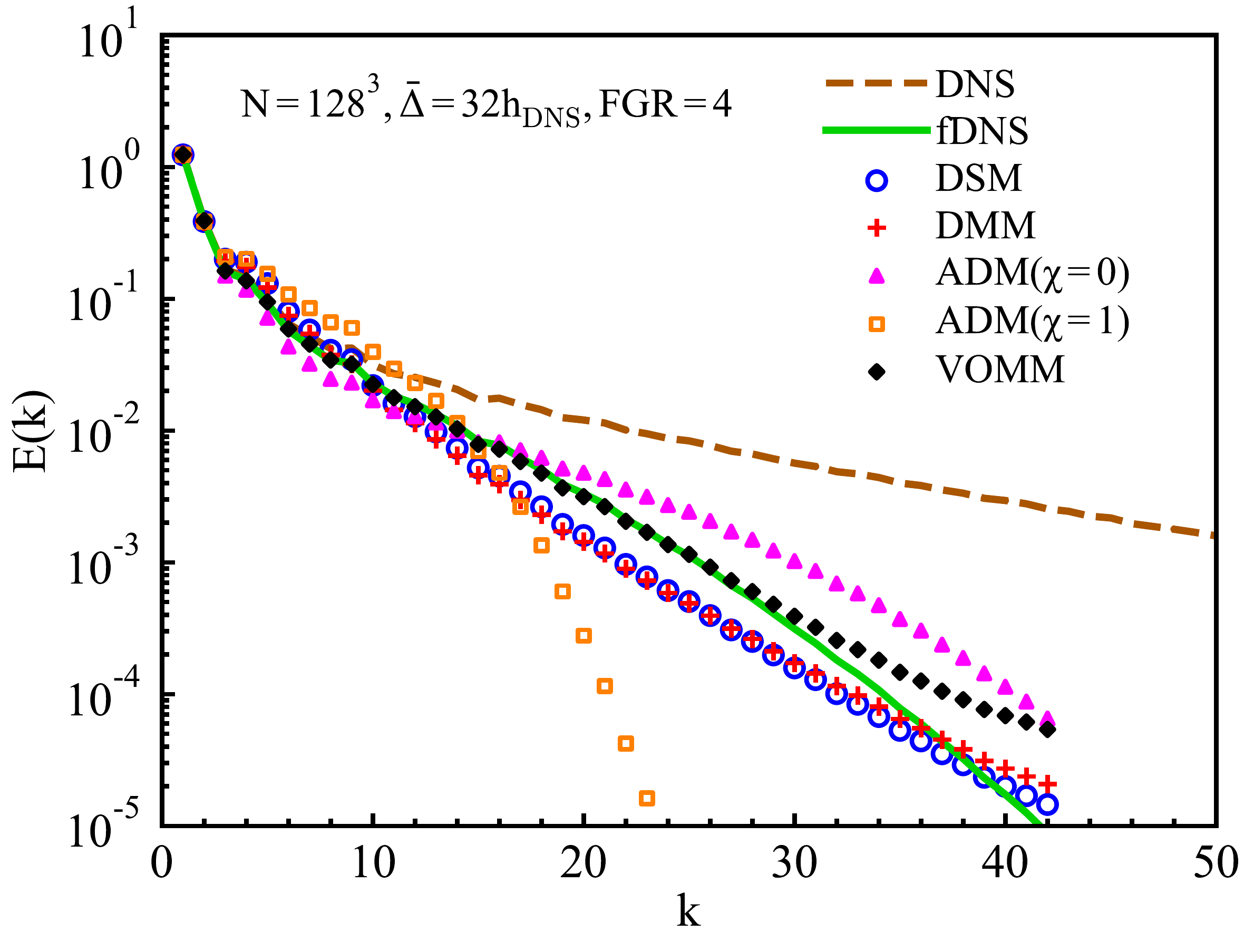}
	\end{subfigure}
	\caption{Velocity spectra of VOMM models with parameters optimized by the decaying homogeneous isotropic turbulence in the \emph{a posteriori} analysis of forced homogeneous isotropic turbulence with filter scale $\bar \Delta  = 32{h_{{\rm{DNS}}}}$: (a) log-log for FGR=1, $N=32^3$; (b) semi-log for FGR=1, $N=32^3$; (c) log-log for FGR=2, $N=64^3$; (d) semi-log for FGR=2, $N=64^3$; (e) log-log for FGR=4, $N=128^3$; and (f) semi-log for FGR=4, $N=128^3$.}
	\label{fig:26}
\end{figure}

We respectively perform the \emph{a posteriori} studies of VOMM model in the decaying homogeneous isotropic turbulence (DHIT) using the optimal model parameters optimized by the statistics of the forced homogeneous isotropic turbulence (FHIT), and \emph{a posteriori} simulations for the FHIT using model coefficients optimized by the DHIT data to investigate the impact of unsteady evolutions and large-scale forcing on the \emph{a posteriori} accuracy of the VOMM model. The optimal parameters of the VOMM model for LES calculations with the filter width $\bar \Delta=32h_{\rm{DNS}}$ in  FHIT and DHIT are respectively summarized in Tables~\ref{tab:2} and \ref{tab:4}. The temporal evolutions of the turbulent kinetic energy and the resolved dissipation rate for LES of DHIT at different grid resolutions (FGR=1,2 and 4 with $N=32^3, 64^3$ and $128^3$) are respectively shown in Figs.~\ref{fig:24} and \ref{fig:25}. The VOMM model with parameters optimized using the stationary FHIT data still accurately predicts the temporal decay of turbulence without any \emph{a priori} knowledge and outperforms other classical SGS models at different grid resolutions. Moreover, we evaluate the \emph{a posteriori} performance of the VOMM model in FHIT using the optimal coefficients of DHIT data. Figure~\ref{fig:26} illustrates the velocity spectra of VOMM models with parameters optimized using the unsteady DHIT data in the \emph{a posteriori} studies of the stationary FHIT. Compared to the classical SGS models (DSM, DMM and ADM), the velocity spectra predicted by the VOMM model is very close to the benchmark fDNS results, indicating that the \emph{a posteriori} accuracy of VOMM models is almost unaffected by the large-scale forcing and temporal turbulence evolutions. 

\subsection{Impact of initial random disturbances on the turbulent mixing layer}

\begin{figure}[h!]\centering
	\begin{subfigure}{0.5\textwidth}
		\centering
		{($a$)}
		\includegraphics[width=0.9\linewidth,valign=t]{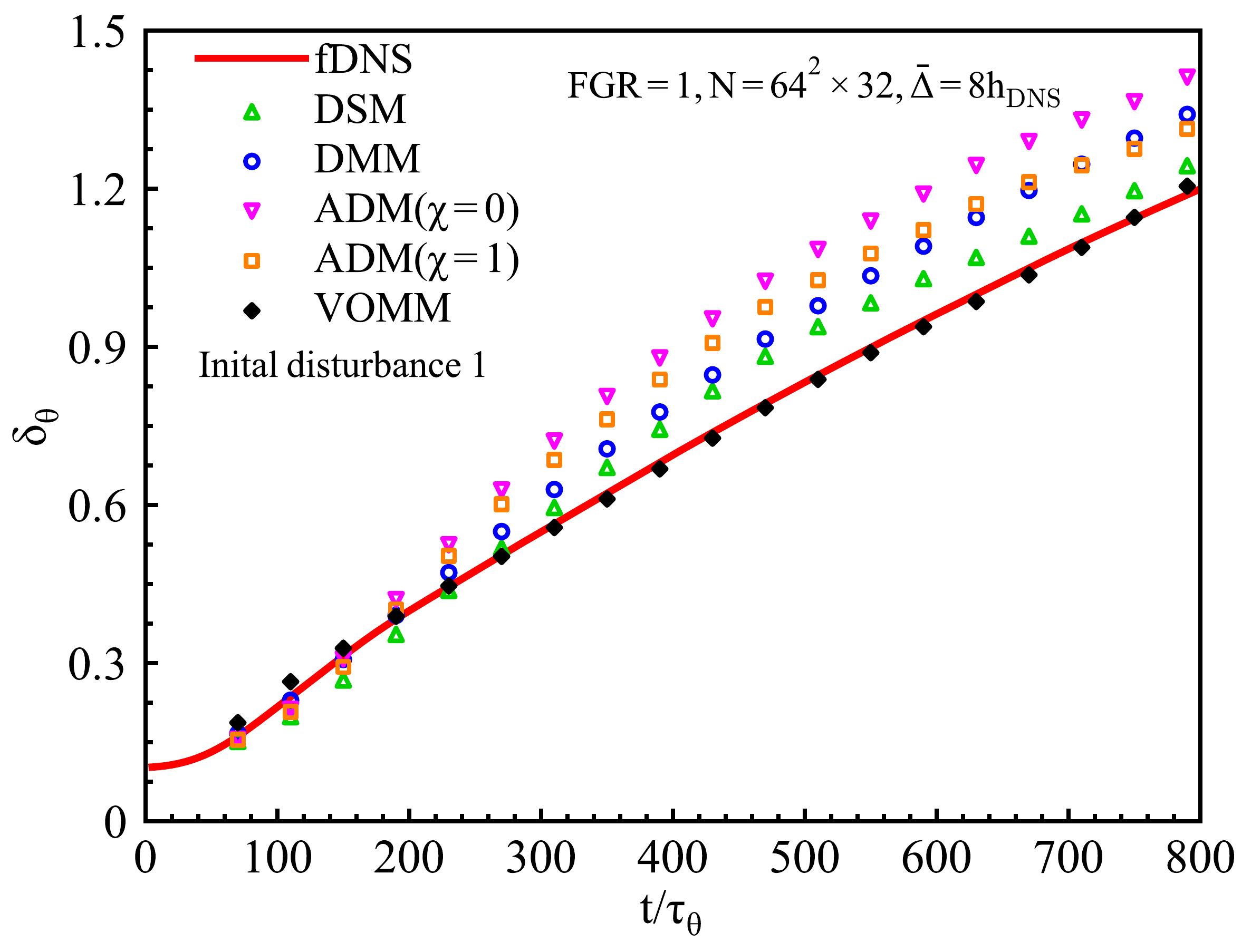}
	\end{subfigure}%
	\begin{subfigure}{0.5\textwidth}
		\centering
		{($b$)}
		\includegraphics[width=0.9\linewidth,valign=t]{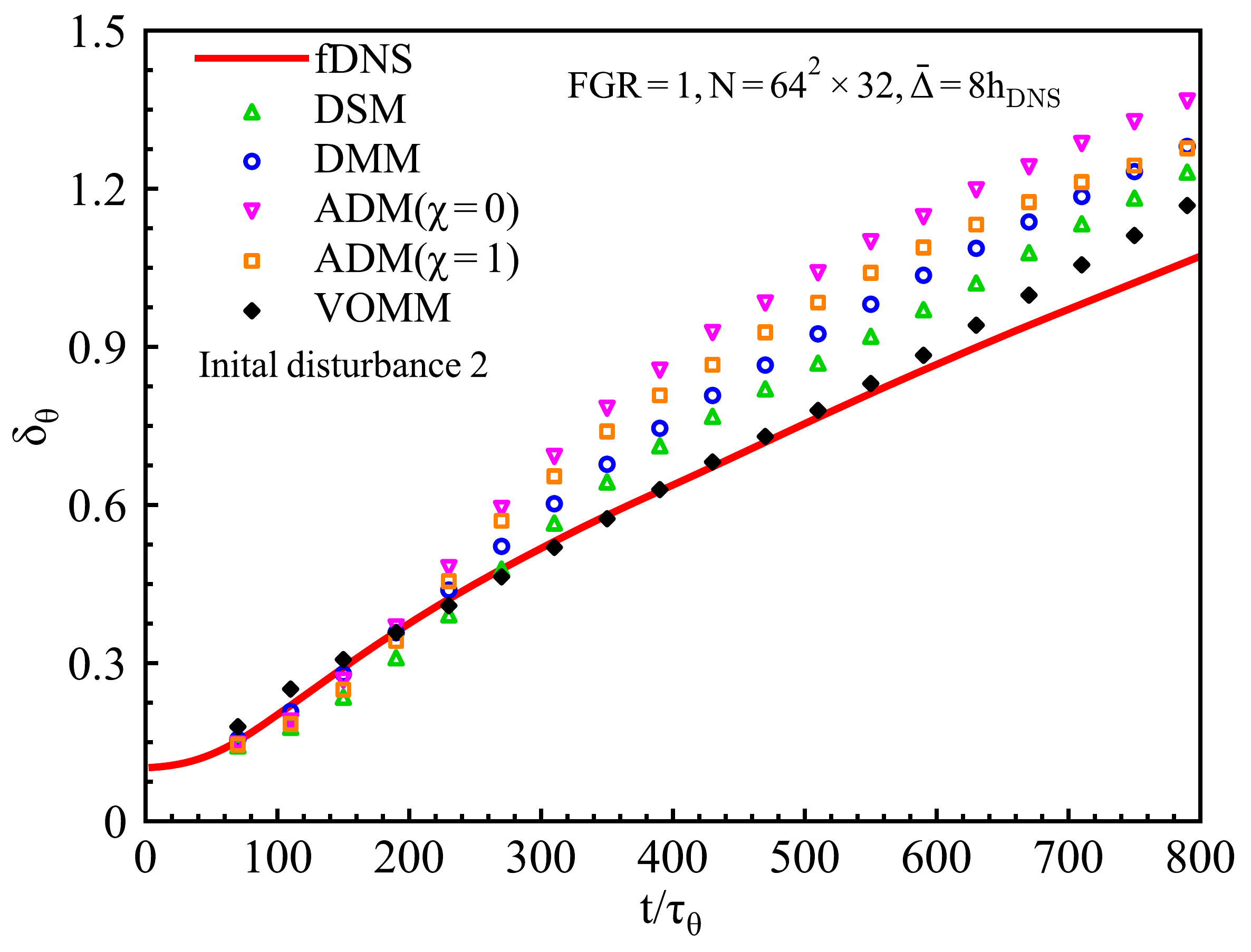}
	\end{subfigure}\\
	\begin{subfigure}{0.5\textwidth}
		\centering
		{($c$)}
		\includegraphics[width=0.9\linewidth,valign=t]{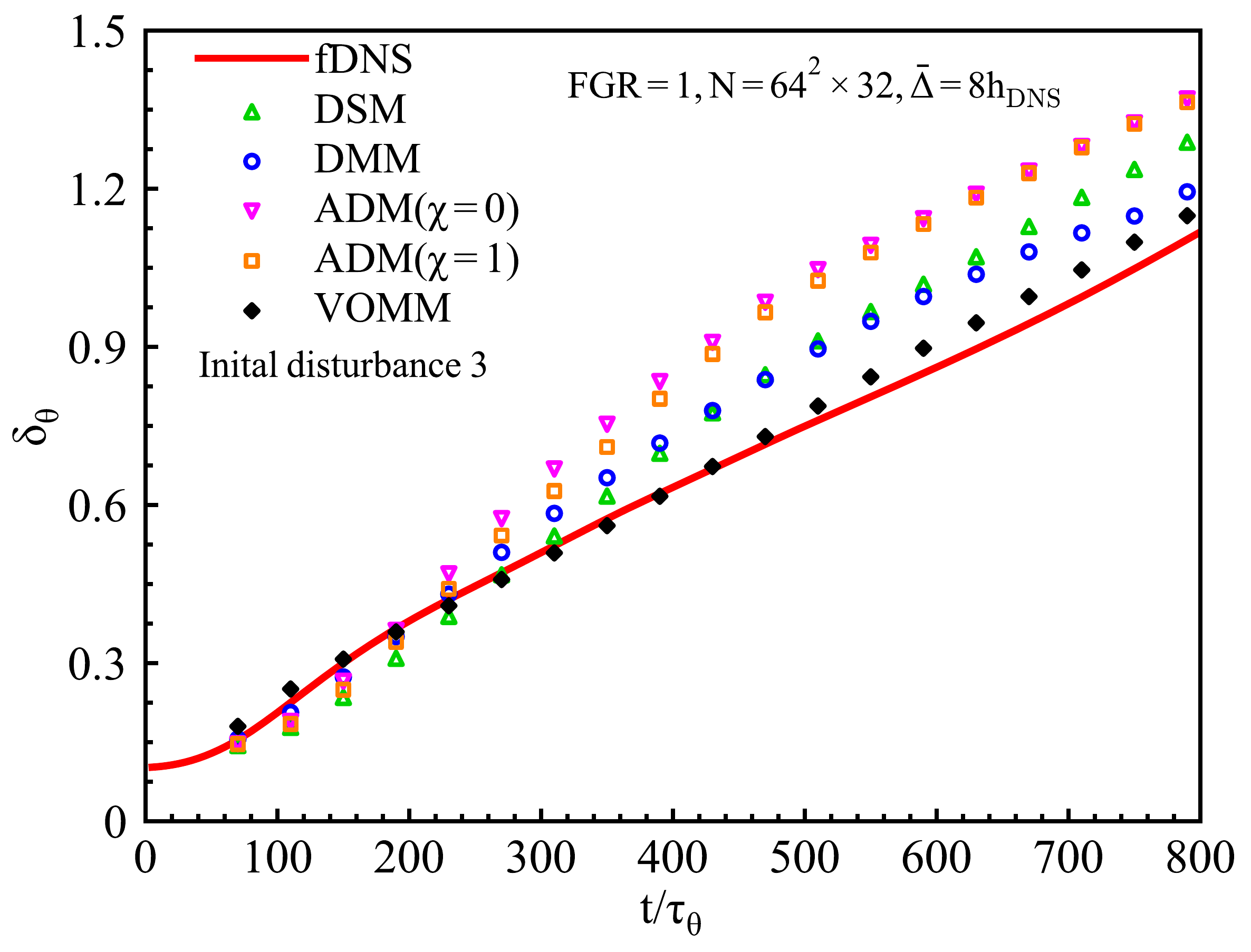}
	\end{subfigure}%
	\begin{subfigure}{0.5\textwidth}
		\centering
		{($d$)}
		\includegraphics[width=0.9\linewidth,valign=t]{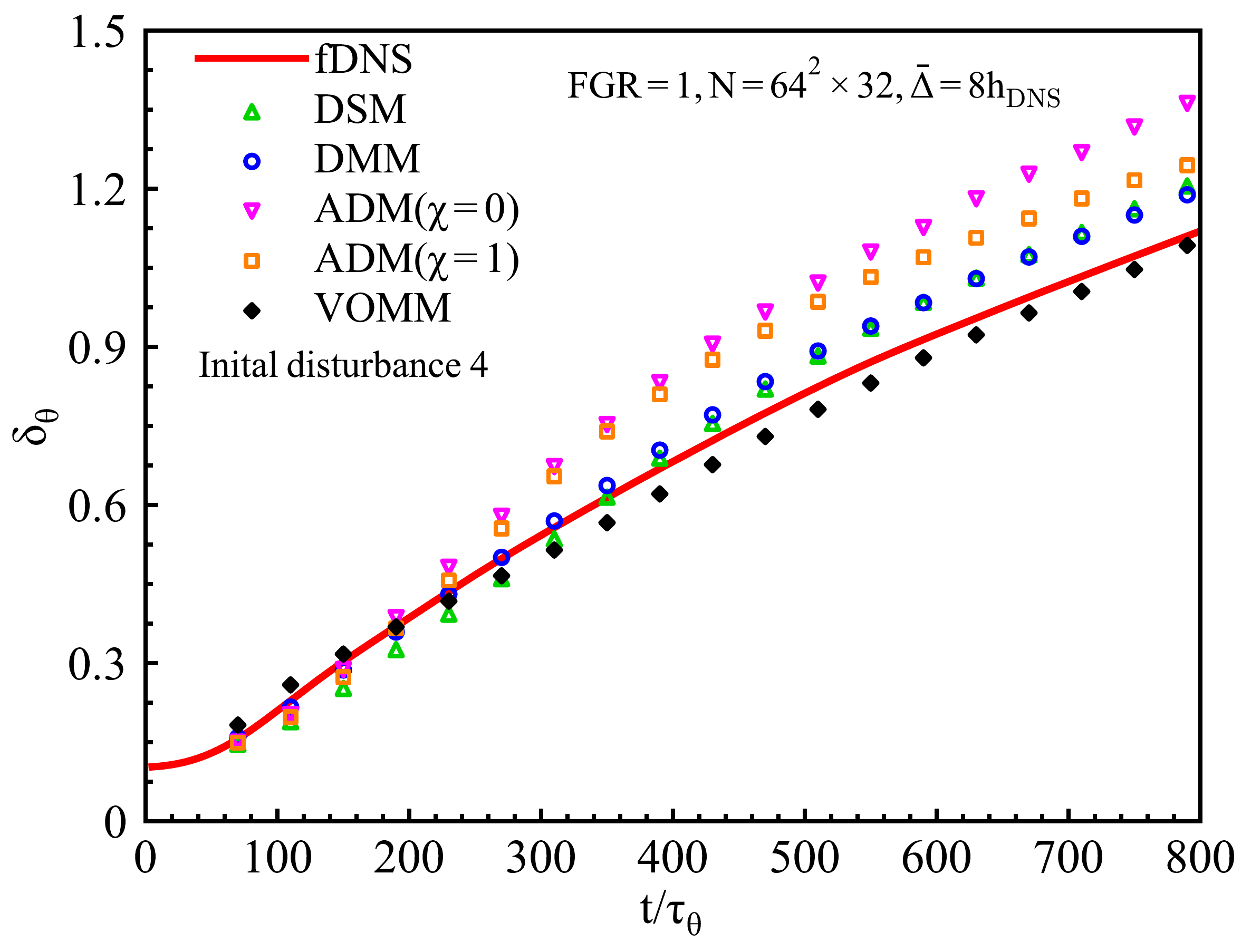}
	\end{subfigure}\\
	\begin{subfigure}{0.5\textwidth}
		\centering
		{($e$)}
		\includegraphics[width=0.9\linewidth,valign=t]{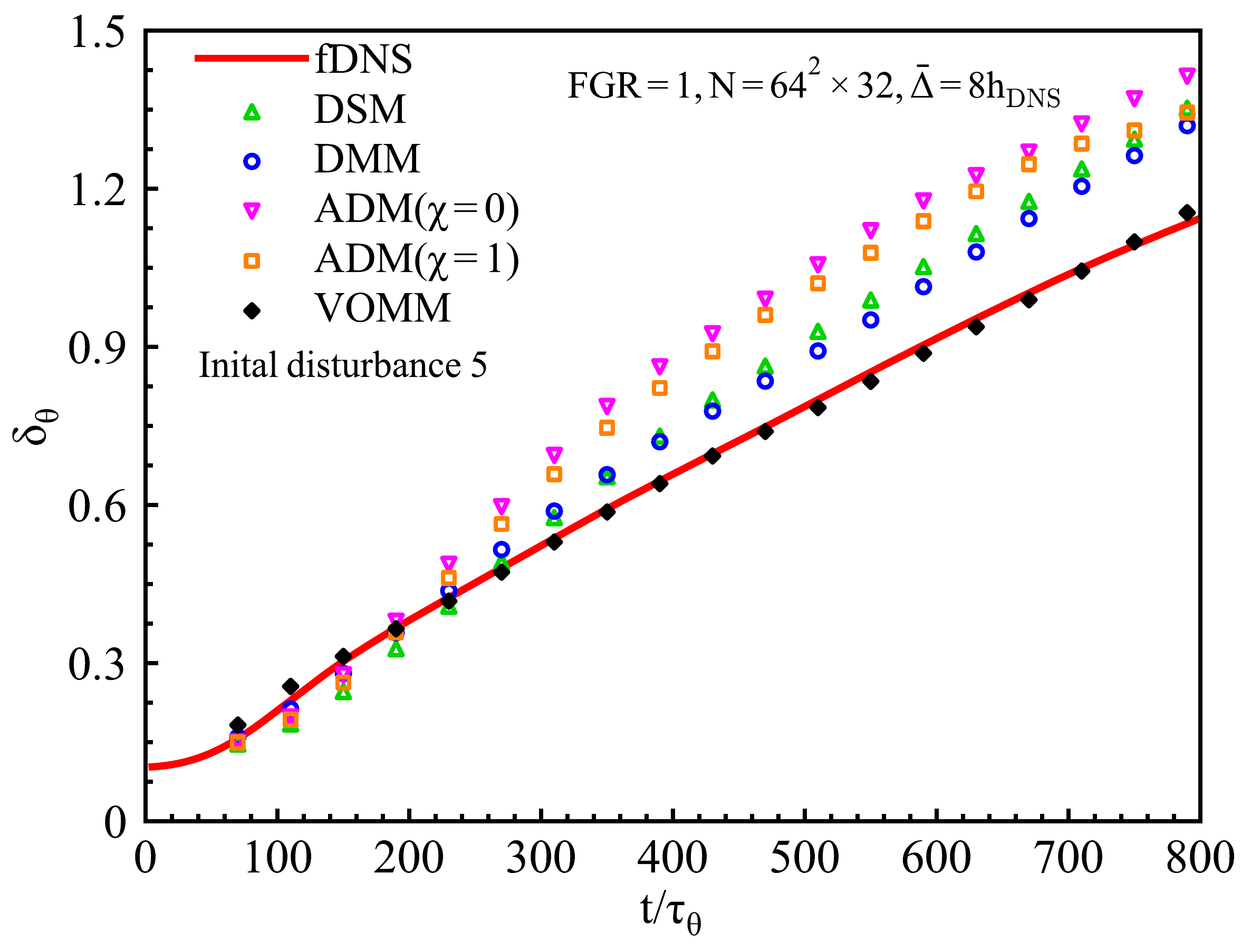}
	\end{subfigure}%
	\begin{subfigure}{0.5\textwidth}
		\centering
		{($f$)}
		\includegraphics[width=0.9\linewidth,valign=t]{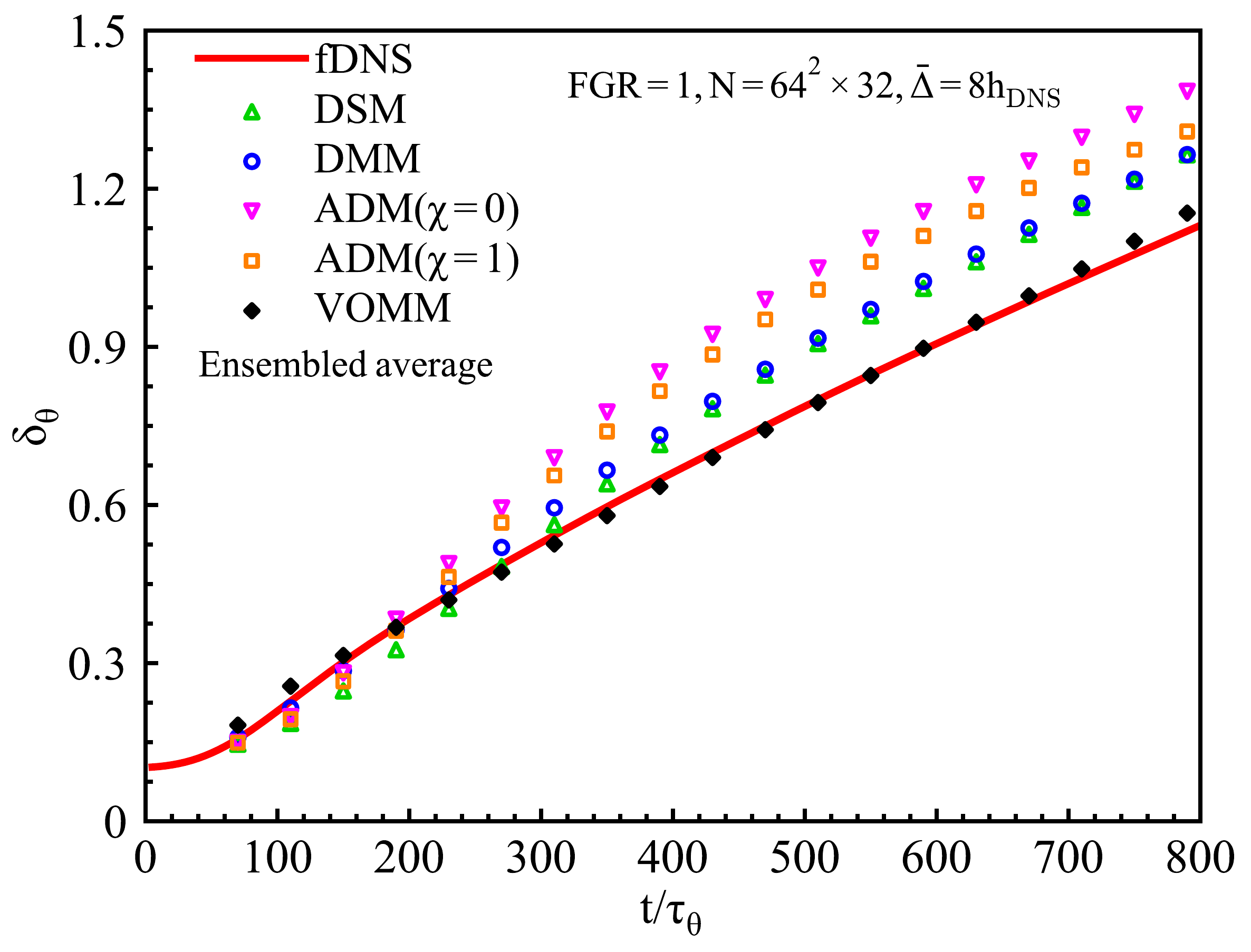}
	\end{subfigure}
	\caption{Temporal evolutions of the momentum thickness  for LES in the \emph{a posteriori} analysis of turbulent mixing layer with filter scale $\bar \Delta=8h_{\rm{DNS}}$ at FGR=1 and $N = 64^2 \times 32$  using different initial random disturbances. (a)$\sim$(e) initialization with random disturbances 1$\sim$5; and (f) ensembled average of five different results.}
	\label{fig:27}
\end{figure}

We conduct six numerical experiments with different random initializations in the case of the temporally evolving mixing layer, one of which is used for the parameter optimization of the VOMM model and the rest five are used to calculate the ensemble-averaged turbulence statistics to reduce the impact of initial random disturbances on the development of mixing layer. As it turns out, the VOMM model has an excellent generalization ability to the initial disturbance of turbulence. Figure~\ref{fig:27} displays the temporal evolutions of the momentum thickness for LES in the \emph{a posteriori} studies of turbulent mixing layer with filter scale $\bar \Delta=8 h_{\rm{DNS}}$ at FGR=1 and $N=64^2 \times 32$ using different initial random disturbances. The predictions of the VOMM model at fine grid resolution (FGR=2 and $N=128^2 \times 64$ ) using different random initializations (ensembled-average quantities shown in Fig.~\ref{fig:18}) are very similar to those at FGR=1 and $N=64^2 \times 32$, and not repeated here. VOMM model can accurately reconstruct the entire temporal development of the shear layer including both the early transition stage and the self-similar region with the linear growth of momentum thickness. In contrast, other classical SGS models (DSM, DMM and ADM)  have distinct deviations from the fDNS data. The predictions of the VOMM model using different initial random disturbances are consistently in reasonable agreement with the benchmark fDNS results, which demonstrates that the accuracy of the VOMM model is nearly insensitive to the initial perturbations of turbulence.

\begin{table}
	\begin{center}	
		\normalsize
		\caption{Parameters and statistics of the coarse-grid simulations in forced homogeneous isotropic turbulence with grid resolution of $512^3$ at different Reynolds numbers.}\label{tab:9}
		\small 
		\begin{tabular*}{0.95\textwidth}{@{\extracolsep{\fill}}lcccccc}
				\hline\hline
				${ {\rm Re}}$ & ${ \nu}$ & ${ {\rm Re}_{\lambda }}$ & $k_{\rm{max}} \eta$ &  $\eta /{{h }_{\rm{DNS}}}$ & ${{L}_{I}}/\eta $ & $\lambda /\eta$   \\  \hline
				$1\times 10^3$   & $1\times 10^{-3}$  & $2.55\times 10^{2}$  & 1.042 & 0.497 & $2.34\times 10^{2}$ & 31.40  \\
				$2\times 10^3$   & $5\times 10^{-4}$  & $3.95\times 10^{2}$  & 0.653 & 0.312 & $3.76\times 10^{2}$ & 39.11  \\
				$5\times 10^3$   & $2\times 10^{-4}$  & $7.92\times 10^{2}$  & 0.365 & 0.174 & $6.59\times 10^{2}$ & 55.45  \\
				$1\times 10^4$   & $1\times 10^{-4}$  & $1.46\times 10^{3}$  & 0.249 & 0.119 & $9.76\times 10^{2}$ & 75.21  \\
				$2\times 10^4$   & $5\times 10^{-5}$  & $2.76\times 10^{3}$  & 0.172 & 0.082 & $1.42\times 10^{3}$ & 103.6  \\
				$5\times 10^4$   & $2\times 10^{-5}$  & $6.72\times 10^{3}$  & 0.106 & 0.051 & $2.27\times 10^{3}$ & 161.6  \\
				$1\times 10^5$   & $1\times 10^{-5}$  & $1.33\times 10^{4}$  & 0.075 & 0.036 & $3.23\times 10^{3}$ & 227.6  \\
				\hline\hline
			\end{tabular*}
	\end{center}
\end{table}

\subsection{Generalization to higher Reynolds numbers}

\begin{figure}[h!]\centering
	\begin{subfigure}{0.5\textwidth}
		\centering
		{($a$)}
		\includegraphics[width=0.9\linewidth,valign=t]{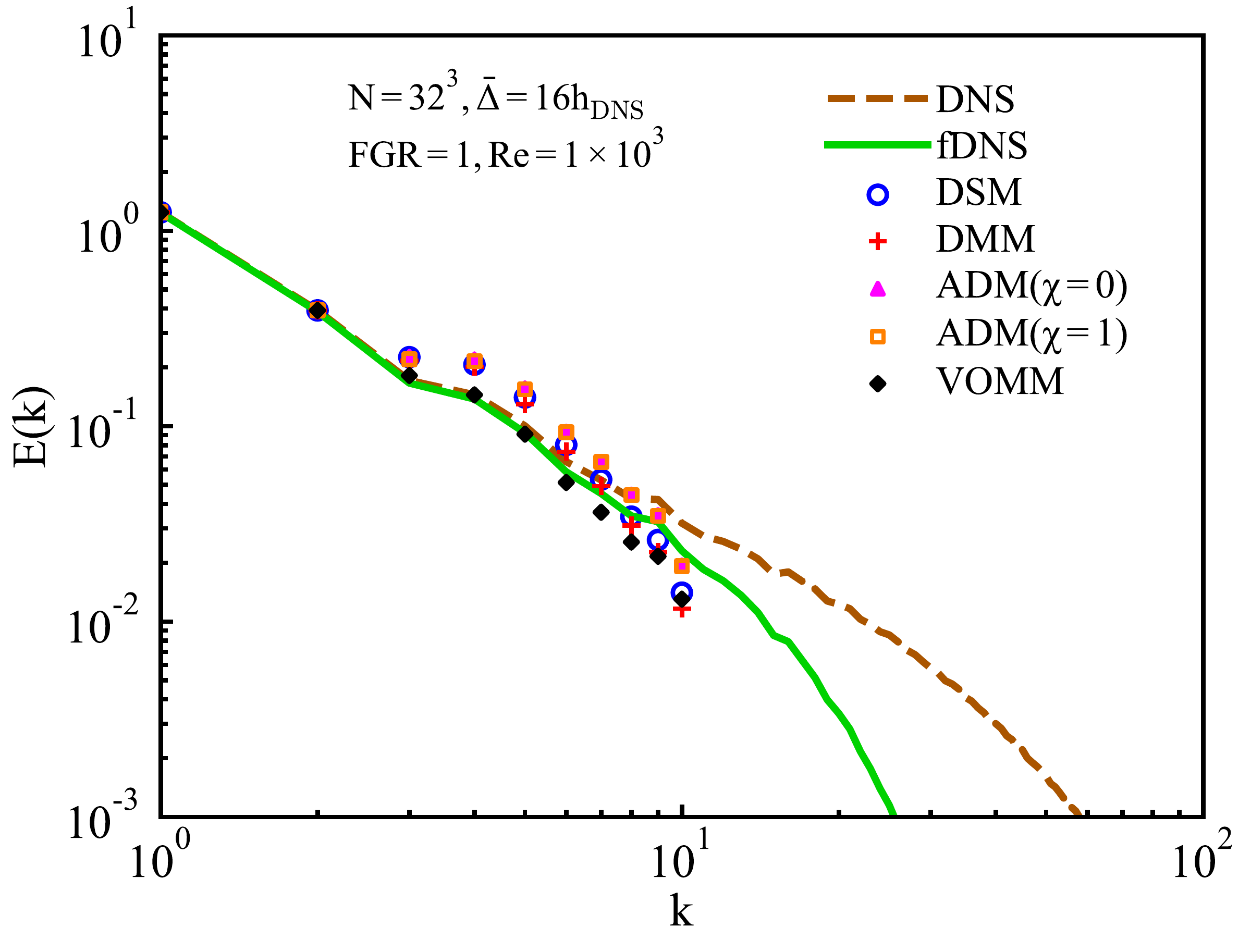}
	\end{subfigure}%
	\begin{subfigure}{0.5\textwidth}
		\centering
		{($b$)}
		\includegraphics[width=0.9\linewidth,valign=t]{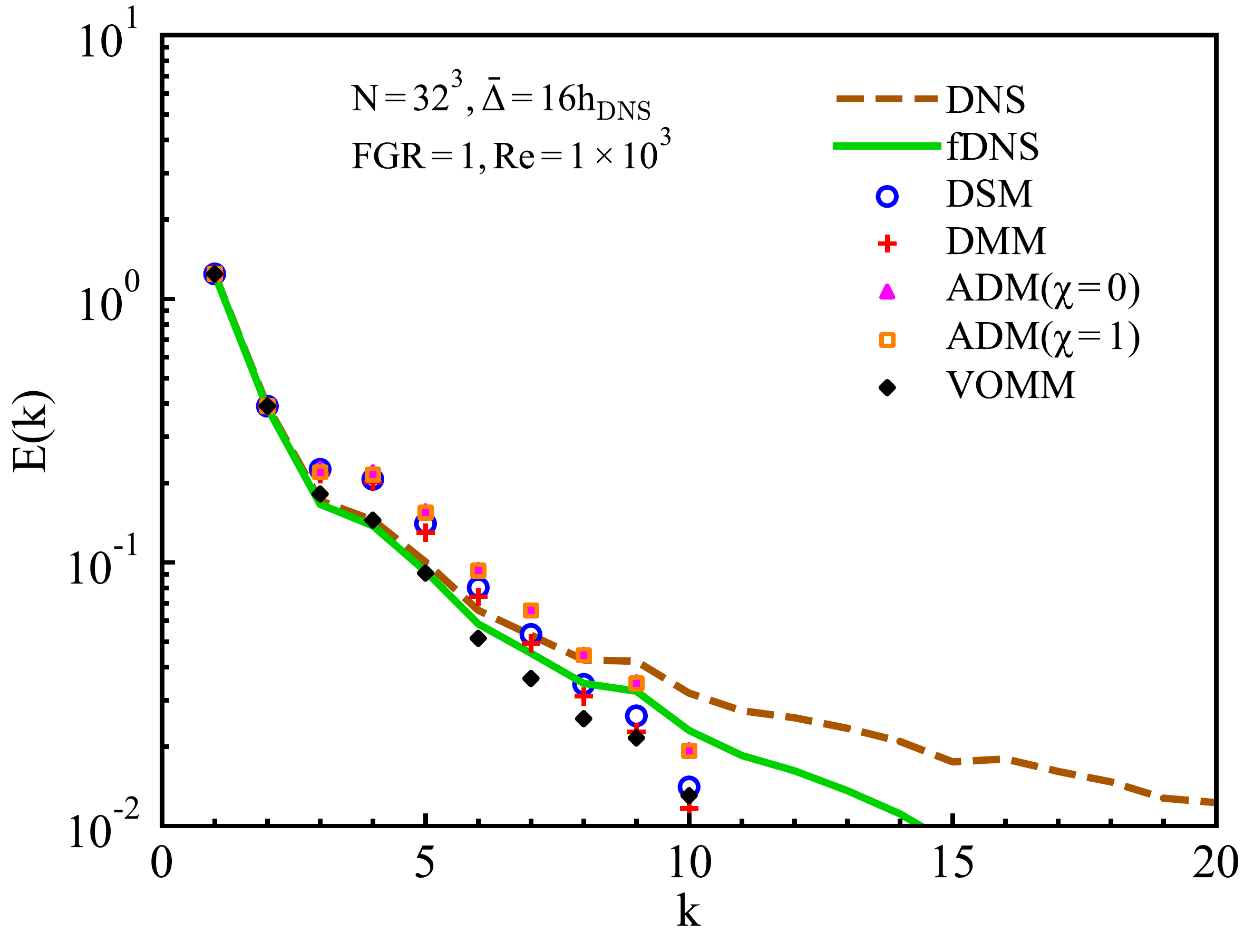}
	\end{subfigure}%
	\caption{Velocity spectra for different SGS models in the \emph{a posteriori} analysis of forced homogeneous isotropic turbulence with filter scale $\bar \Delta  = 16{h_{{\rm{DNS}}}}$ and ${\rm{Re}}=10^3$ ($\nu=0.001$): (a) log-log for FGR=1, $N=32^3$; (b) semi-log for FGR=1, $N=32^3$.}
	\label{fig:28}
\end{figure}

\begin{figure}[h!]\centering
	\begin{subfigure}{0.5\textwidth}
		\centering
		{($a$)}
		\includegraphics[width=0.9\linewidth,valign=t]{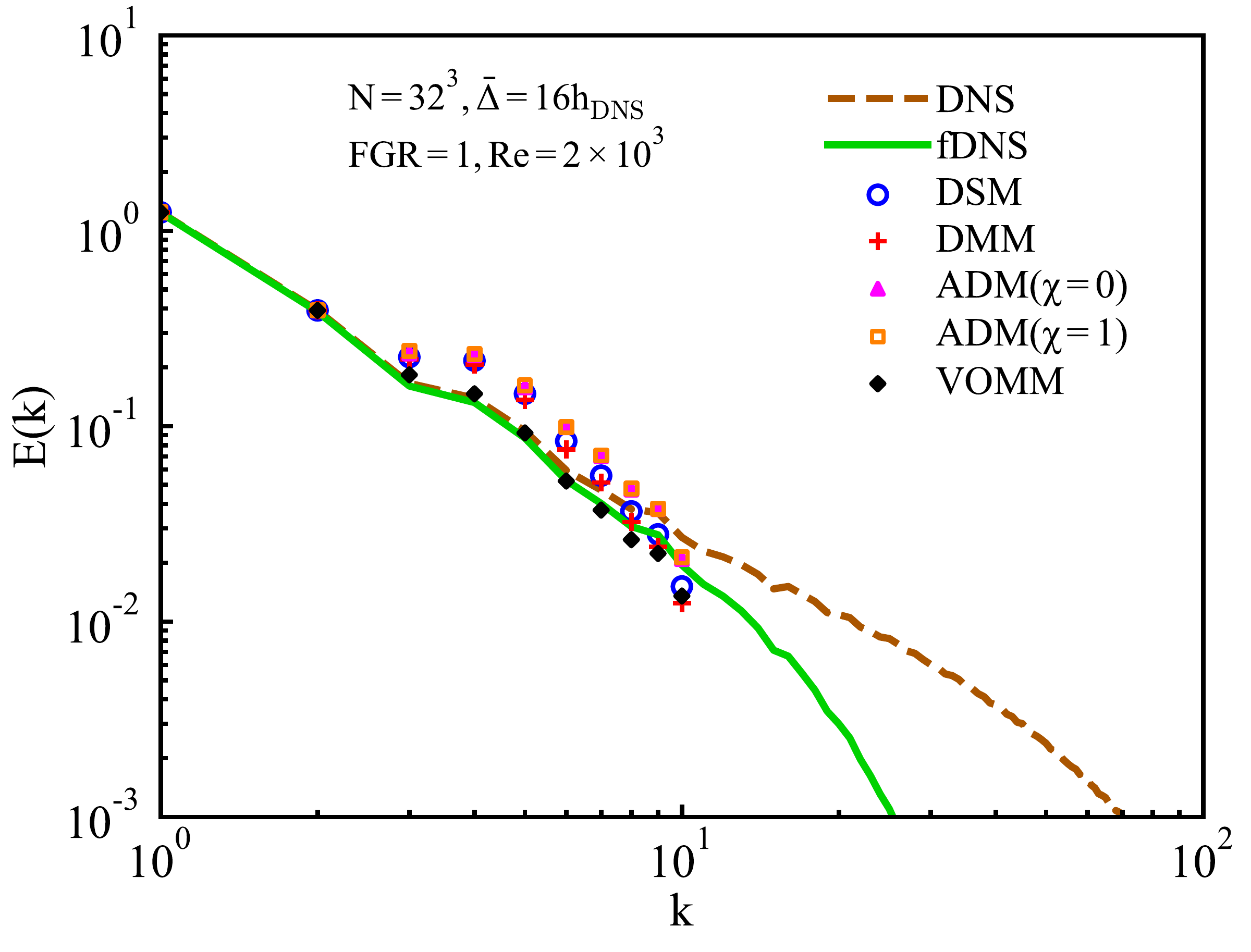}
	\end{subfigure}%
	\begin{subfigure}{0.5\textwidth}
		\centering
		{($b$)}
		\includegraphics[width=0.9\linewidth,valign=t]{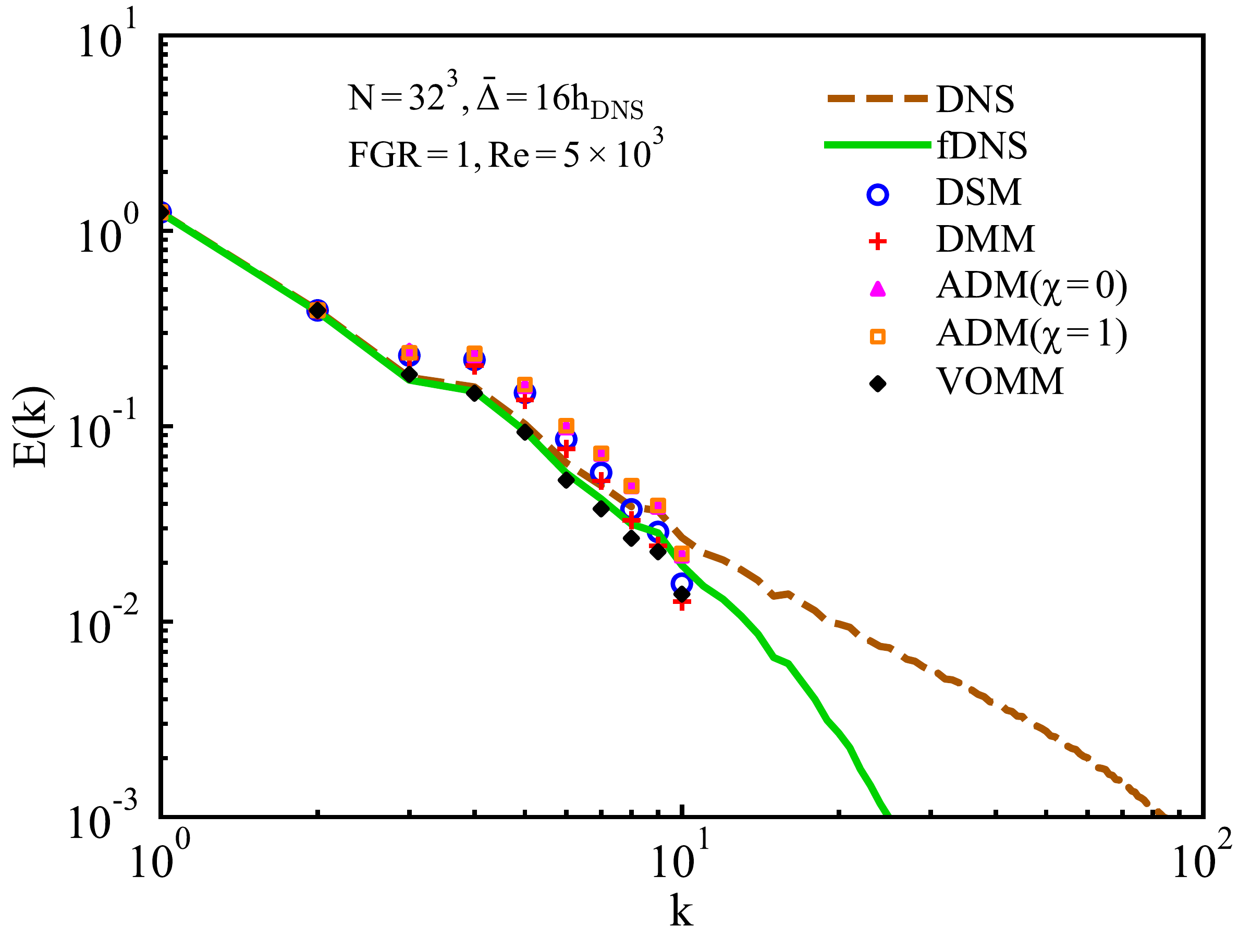}
	\end{subfigure}%
	
	\begin{subfigure}{0.5\textwidth}
		\centering
		{($c$)}
		\includegraphics[width=0.9\linewidth,valign=t]{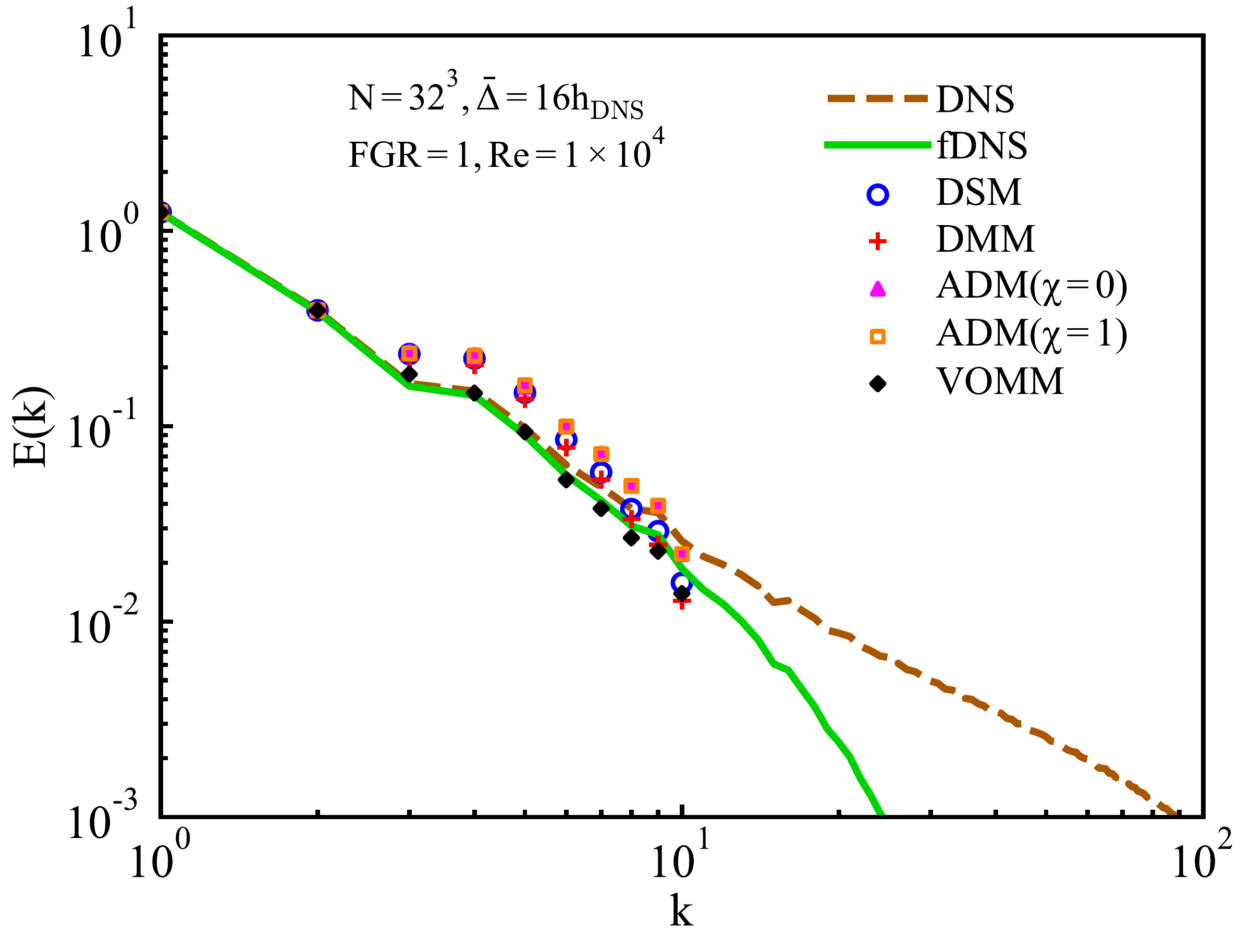}
	\end{subfigure}%
	\begin{subfigure}{0.5\textwidth}
		\centering
		{($d$)}
		\includegraphics[width=0.9\linewidth,valign=t]{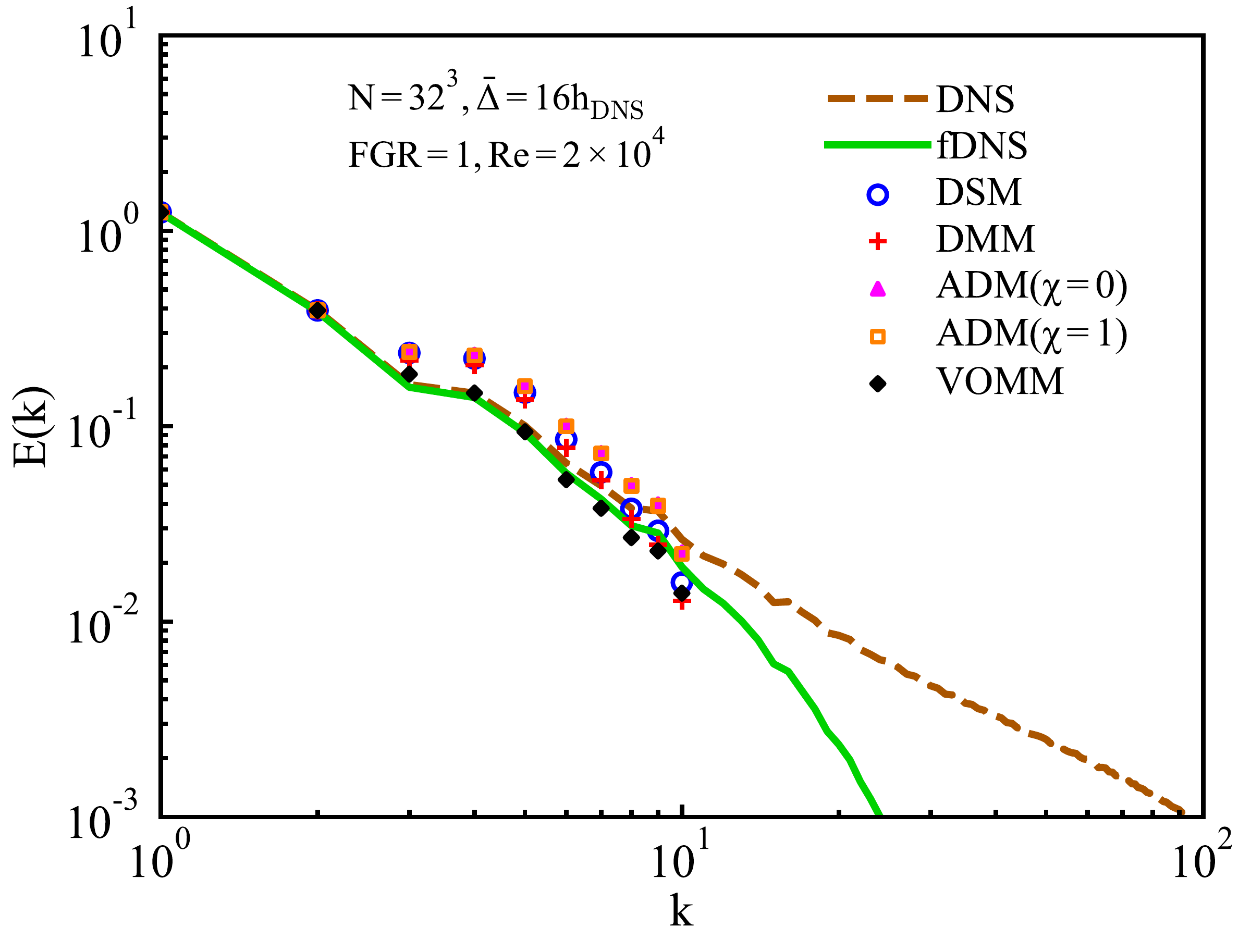}
	\end{subfigure}%
	
	\begin{subfigure}{0.5\textwidth}
		\centering
		{($e$)}
		\includegraphics[width=0.9\linewidth,valign=t]{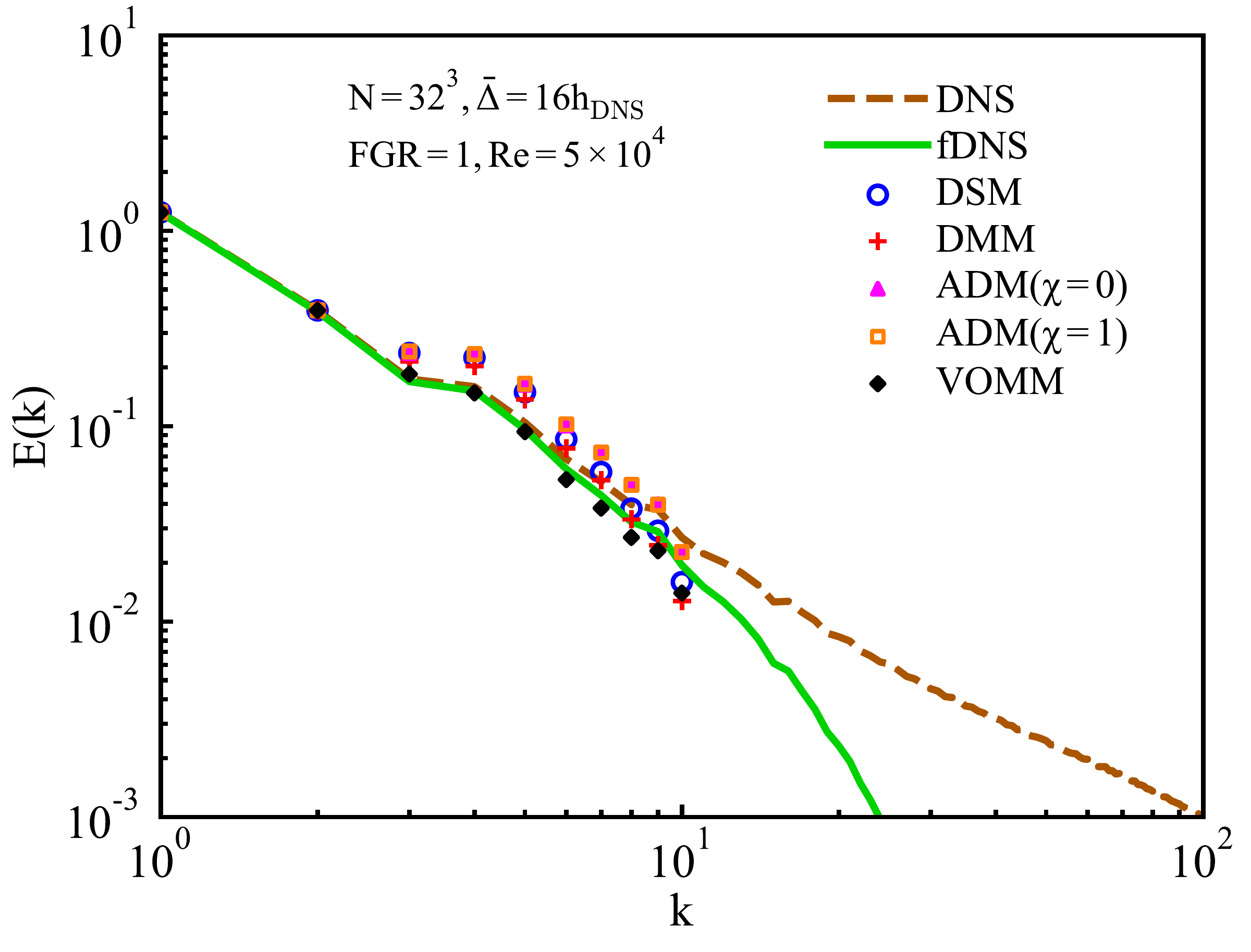}
	\end{subfigure}%
	\begin{subfigure}{0.5\textwidth}
		\centering
		{($f$)}
		\includegraphics[width=0.9\linewidth,valign=t]{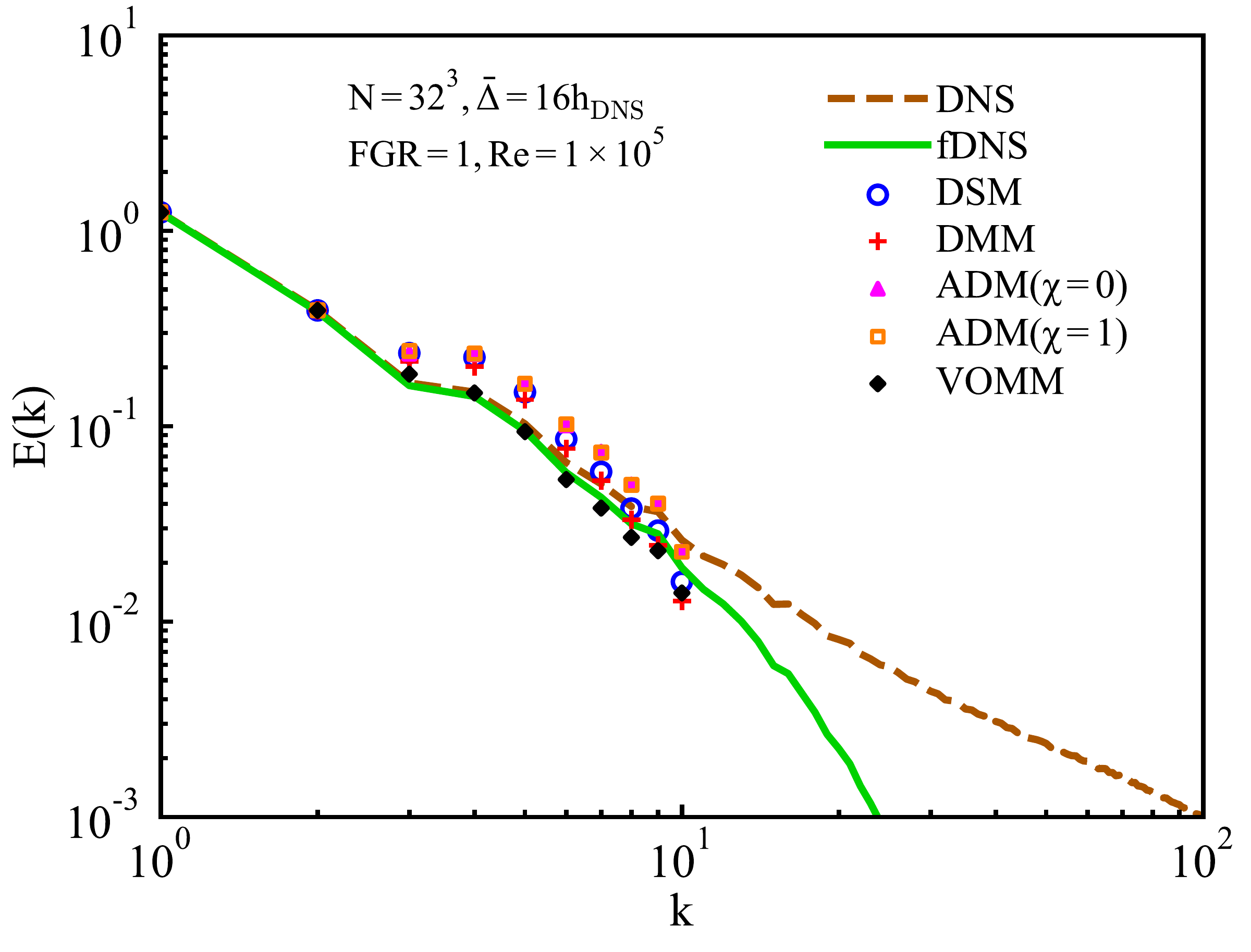}
	\end{subfigure}%
	
	\caption{Velocity spectra for different SGS models in the \emph{a posteriori} analysis of forced homogeneous isotropic turbulence with filter scale $\bar \Delta  = 16{h_{{\rm{DNS}}}}$ at higher Reynolds numbers (VOMM models with parameters optimized at ${\rm{Re}}=10^3$): (a) ${\rm{Re}}=2 \times 10^3$; (b) ${\rm{Re}}=5 \times 10^3$; (c) ${\rm{Re}}=1 \times 10^4$; (d) ${\rm{Re}}=2 \times 10^4$; (e) ${\rm{Re}}=5 \times 10^4$; (f) ${\rm{Re}}=1 \times 10^5$.}
	\label{fig:29}
\end{figure}

In this section, we perform the numerical simulations of forced homogeneous isotropic turbulence  using a fixed uniform grid resolution $N=512^3$ in a cubic box of $(2\pi)^3$ at seven different Reynolds numbers ${\rm{Re}}=1/\nu=\{1 \times 10^3,2 \times 10^3, 5 \times 10^3, 1 \times 10^4, 2 \times 10^4, 5 \times 10^4, 1 \times 10^5\}$. The numerical simulation approach is consistent with the FHIT at $N=1024^3$, which is detailedly introduced in Sec.~\ref{sec:level5}. The six-order compact-difference filtering scheme is adopted to provide necessary artificial dissipation for the numerical stability of coarse-grained numerical simulations\citep{visbal2002,visbal2002a,yuan2022}. Table~\ref{tab:9} summarizes the detailed simulation parameters and statistics of the coarse-grid simulations at different Reynolds numbers. The Gaussian filter (Eq.~\ref{G}) with the filter width $\bar \Delta=16 h_{\rm{DNS}}$ is selected as the explicit filter, ensuring that the filter type and filter scale are the same with the case of FHIT at $N=1024^3$. The corresponding \emph{a posteriori} studies of LES adopt the consistent kinematic viscosity (cf. Table~\ref{tab:9} ) and simulation settings. We select the VOMM model at FGR=1 and $N = 32^3$ as the example to examine the generalization ability of VOMM model for higher-Reynolds-number turbulent flows. The dissipation spectra of fDNS  with the filter width $\bar \Delta=16 h_{\rm{DNS}}$ at ${\rm{Re}}=1\times 10^3$ are used for the parameter optimization of VOMM model, and the optimal model coefficients are $C_{1}^{\rm{opt}} = -0.1276$ and $C_{2}^{\rm{opt}} = -6.645$. The comparisons of velocity spectra for different SGS models (DSM, DMM, ADM and VOMM) in the \emph{a posteriori} testings at  FGR=1 and $N = 32^3$ are illustrated in Fig.~\ref{fig:28}.  The classical SGS models (DSM, DMM and ADM) exhibit the obviously tilde distribution due to the excessive dissipation. The kinetic energy near the truncated wavenumbers are diminished by the numerical dissipation, resulting in the blockage of the kinetic energy cascade from large scales to small scales. Hence, the kinetic energy accumulates in the region of medium wavenumbers. In contrast, the VOMM model can flexibly adjust the dissipation induced by the numerical scheme, and accurately mimic the turbulent kinetic energy cascades. Furthermore, we directly apply the VOMM model optimized at ${\rm{Re}}=1\times 10^3$ to the LES calculations at higher Reynolds numbers ${{\rm Re}}=\{2 \times 10^3, 5 \times 10^3, 1 \times 10^4, 2 \times 10^4, 5 \times 10^4, 1 \times 10^5\}$ to evaluate the performance at unseen higher Reynolds numbers.  Fig.~\ref{fig:29} illustrates velocity spectra of different SGS models at different Reynolds numbers. The VOMM models are fixed with parameters optimized at  ${\rm{Re}}=1\times 10^3$. Compared to these classical SGS models (DSM, DMM and ADM), the velocity spectra predicted by the VOMM model almost overlap with the benchmark fDNS results, demonstrating that the proposed VOMM model has excellent generalization performance at different Reynolds numbers.

\section{Conclusion}\label{sec:level7}

In this work, an adjoint-based variational optimal mixed model (VOMM) is developed for the large-eddy simulation of turbulence. We first derive the original adjoint LES equations with the general SGS model, and then carry out the energy budget analysis of adjoint equations. These detailed derivations demonstrate that the quadratic term with negative eigenvalues of the shear strain rate is responsible for the exponential temporal growth of the adjoint-based gradients, giving rise to the numerical divergence in a long time horizon for the chaotic turbulent flows. This issue might greatly limits the application of the adjoint-based variational methods and optimal control strategy in turbulence problems. An additional stabilization term is introduced to maintain the numerical stability of the adjoint LES equations and is efficiently calculated by the sequential quadratic programming (SQP) approach, without degrading the accuracy of gradient evaluations for the SGS model parameters. Subsequently, the stabilized adjoint LES equations are correspondingly formulated. 

The approximate deconvolution model (ADM) in the scale-similarity form and the dissipative Smagorinsky term are selected as the basis tensors of the proposed VOMM model. The parameters of the VOMM model are optimally identified by minimizing the statistical discrepancies between dissipation spectra of the LES and those of the benchmark filtered DNS data. The adjoint-based gradients of cost functional for model coefficients are  efficiently evaluated by successively forward solving the LES equations and backward integrating the stabilized adjoint LES equations. The gradient-based L-BFGS optimization algorithm is adopted for iteratively updating the VOMM model parameters until the optimal values are obtained. 

Three turbulent flow scenarios including the forced homogeneous isotropic turbulence, decaying homogeneous isotropic turbulence and temporally evolving turbulent mixing layer are investigated to examine the \emph{a posteriori} performance of the VOMM model. The pure structural ADM model without the dissipative Smagorinsky term is selected as the initial SGS model for the parameter optimization. The loss functions of the dissipation spectra can dramatically converge and reach the optimal state of only about 10\% of the initial value within less than twenty iterations (about forty LES evaluations) during the adjoint-based gradient optimization at different grid resolutions for these three types of turbulence. These results indicate that the adjoint-based gradient optimization is an effective tool to obtain the optimal parameters of VOMM model within only a few iterations. Meanwhile, the computational efficiency of the proposed method is independent of the number of parameters.

Once the optimal SGS model coefficients are determined by the adjoint-based gradient optimization, the \emph{a posteriori} accuracy of the VOMM model is further tested in comparison with the classical SGS models, including the dynamic Smagorinsky model (DSM), dynamic mixed model (DMM), the pure ADM model and ADM model with the standard secondary-filtering regularization, respectively. The various statistics of turbulence and the instantaneous flow structures are comprehensively compared for LES calculations of different SGS models with the benchmark filtered DNS data at different grid resolutions of three turbulent flow scenarios.

In the cases of forced and decaying homogeneous isotropic turbulence, the filter scale is fixed to $\bar \Delta  = 32{h_{{\rm{DNS}}}}$ and the impact of the spatial discretization errors on the SGS modeling is studied by changing the grid resolution of LES with three different filter-to-grid ratios FGR=1, 2 and 4. The \emph{a posteriori} performance of the proposed VOMM model is systematically evaluated by comparison to the conventional SGS models (DSM, DMM and ADM models) in terms of the velocity spectra, structure functions with different orders, PDFs of the velocity increments and vorticity, temporal evolutions of the turbulent kinetic energy and average dissipation rate, as well as the instantaneous vorticity contours at different grid resolutions. The pure ADM model always exhibits numerical instability due to the insufficient sufficient SGS dissipation for all grid-resolution cases. The dynamic models and standard regularized ADM model underpredict the model dissipation in the case of coarse grid resolution (FGR=1), with the excess kinetic energy accumulated at small scales leading to the numerical instability of LES. The SGS dissipation imposed by these classical SGS models is insufficient to suppress the numerical perturbations dominated by the spatial discretization, and it cannot effectively drain out the small-scale kinetic energy in time at FGR=1. However, the traditional SGS models are too dissipative that most small-scale flow structures near the truncated wavenumber are diminished, giving rise to the blockage of the kinetic energy cascade from large scales to small scales at situations of satisfactory grid resolutions. In contrast, the VOMM model can correctly reconstruct the kinetic energy cascade and the evolution of dissipation rate with high accuracy, which is essential for the isotropic turbulence. In addition, the VOMM model accurately predicts various flow statistics and transient spatial flow structures, which are always in reasonable agreement with the benchmark filtered DNS results at different grid resolutions and times. 

In the context of the temporally evolving turbulent mixing layer, the unsteady evolution of the shear layer from the initial perturbed velocity field gradually transitions to fully developed turbulence is challenging for the SGS modeling of LES. The VOMM model can accurately reconstruct the temporal evolutions of characteristic physical quantities of the mixing layer, including the momentum thickness, turbulent kinetic energy in different directions and transient velocity spectra at different times. The corresponding predictions of VOMM are closest to the filtered DNS results and superior to these conventional SGS models (DSM, DMM and ADM models). The profiles of Reynolds shear stress at the self-similar stage of the shear layer are critical for the development of mixing layer, and all conventional SGS models are not able to accurately predict the vertical distributions with significant deviations from the benchmark fDNS result. In contrast, the VOMM model predicts the Reynolds stress fairly well at different time instants. Besides, it can be clearly observed from the iso-surface of Q-criterion that the VOMM model accurately recovers the diverse spatial vortex structures very similar to the benchmark fDNS data in comparison to the classical SGS models. 

Furthermore, for the cases of three turbulent flow scenarios with different grid resolutions, the computational cost of the proposed VOMM model is only about 30\% the time of the DMM model, which is very efficient and competitive compared to the classical SGS models. We also examine the generalization ability of the VOMM model by studying the impact of unsteady evolutions and large-scale forcing, the impact of initial random disturbances on the temporally evolving turbulent mixing layer and the generalization of higher Reynolds numbers. The VOMM model can accurately reconstruct the spectra statistics accurately in these general cases with satisfactory performance. These results suggest that the proposed VOMM model has high \emph{a posteriori} accuracy and computational efficiency by assimilating the \emph{a priori} knowledge of turbulence statistics, and can be a promising tool to develop advanced SGS models in the LES of turbulence.

Eventually, fine-tuning a small number of model parameters of some traditional SGS models can significantly improve the \emph{a posteriori} accuracy of LES using the proposed adjoint-based optimization framework. In addition, the predictions of LES in complex turbulent flows using the VOMM model might be dramatically accurate as the number of model coefficients increases, while the computational cost of the adjoint-based approach hardly varies with to the number of parameters. Although the high-fidelity turbulence statistics is provided by DNS data in the current study, the experimental measurements can also be assimilated using the same optimization procedure to increase the accuracy of LES modeling for a particular type of complex turbulent flow. In the current research, the VOMM model has the spatially constant parameters that cannot vary with the spatial position over the entire computational domain. We would further apply the VOMM model with the spatially varying  parameters to more realistic wall turbulence in the follow-up studies.  

\acknowledgments{This work was supported by the National Natural Science Foundation of China (NSFC Grants No. 91952104, No. 92052301, No. 12172161, and No. 91752201), by the NSFC Basic Science Center Program (Grant No. 11988102), by the Shenzhen Science and Technology Program (Grants No. KQTD20180411143441009), by Key Special Project for Introduced Talents Team of Southern Marine Science and Engineering Guangdong Laboratory (Guangzhou) (Grant No. GML2019ZD0103), and by Department of Science and Technology of Guangdong Province (No. 2020B1212030001). This work was also supported by Center for Computational Science and Engineering of Southern University of Science and Technology.}

\section{DATA AVAILABILITY} \label{sec:level8}
The data that support the findings of this study are available from the corresponding
author upon reasonable request.

\appendix
\section{Derivation of the adjoint large-eddy simulation equations}\label{AppendixA}
The large-eddy simulation (LES) equations are expressed as \citep{pope2000,sagaut2006}
\begin{equation}
	{R_0}\left( {{{\bar u}_i}} \right) = \frac{{\partial {{\bar u}_i}}}{{\partial {x_i}}} = 0,
	\label{LES_eq1}
\end{equation}
\begin{equation}
	{R_i}\left( {{{\bar u}_i},\bar p} \right) = \frac{{\partial {{\bar u}_i}}}{{\partial t}} + \frac{{\partial \left( {{{\bar u}_i}{{\bar u}_j}} \right)}}{{\partial {x_j}}} + \frac{{\partial \bar p}}{{\partial {x_i}}} - \nu \frac{{{\partial ^2}{{\bar u}_i}}}{{\partial {x_j}\partial {x_j}}} - {\overline {\mathcal F} _i} + \frac{{\partial {\tau _{ij}}}}{{\partial {x_j}}} = 0,
	\label{LES_eq2}
\end{equation}
where an overbar denotes the filtered variables with filter scale $\bar \Delta$, ${{{\bar u}_i}}$ and $\bar p$ denote the filtered velocity and pressure, respectively. Here, $\nu$ is the kinematic viscosity, and $\bar {\mathcal {F}_i}$ represents the large-scale forcing. The unclosed SGS stress ${\tau _{ij}} = \overline {{u_i}{u_j}}  - {{\bar u}_i}{{\bar u}_j}$ is modeled by the $N$-parameter mixed model ${\tau _{ij}} = \sum\limits_{n = 1}^N {{C_n}T_{ij}^{\left( n \right)}\left( {{{\bar u}_i};\bar \Delta } \right)} $ with the basis stress tensors ${T_{ij}^{\left( n \right)}}$ and model coefficients ${C_n}\;\left( {n = 1,2,...,N} \right)$.
The sensitivities of the governing equations for the LES variables ${{{\bf{\bar v}}} } = {\left[ {\bar p,\bar u_1,\bar u_2,\bar u_3 } \right]^T}$ are given by
\begin{equation}
	\delta R_k = \frac{{\partial {R_k}}}{{\partial {\bf{\bar v}}}} \cdot \delta {\bf{\bar v}} = \left[ \begin{array}{l}
		\frac{{\partial \delta {{\bar u}_i}}}{{\partial {x_i}}}\\
		\frac{{\partial \delta {{\bar u}_i}}}{{\partial t}} + \frac{{\partial \left( {{{\bar u}_j}\delta {{\bar u}_i}} \right)}}{{\partial {x_j}}} + \frac{{\partial \left( {{{\bar u}_i}\delta {{\bar u}_j}} \right)}}{{\partial {x_j}}} + \frac{{\partial \delta \bar p}}{{\partial {x_i}}} - \nu \frac{{{\partial ^2}\delta {{\bar u}_i}}}{{\partial {x_j}\partial {x_j}}} + \frac{\partial \delta {\tau _{ij}}}{{\partial {x_j}}}
	\end{array} \right] = 0.
	\label{sen_eqnA}
\end{equation}
The adjoint LES equations are derived by the adjoint identity acting on the adjoint variables ${{{\bf{\bar v}^\dag}} } = {\left[ {\bar p^\dag,\bar u_1^\dag,\bar u_2^\dag,\bar u_3^\dag } \right]^T}$, namely
\begin{equation}
	{\left\langle {\frac{{\partial {R_k}}}{{\partial {\bf{\bar v}}}} \cdot \delta {\bf{\bar v}},{{{\bf{\bar v}}}^\dag }} \right\rangle _{{\bf{x}},t}} = {\left\langle {\delta {\bf{\bar v}}, {{\left( {\frac{{\partial {R_k}}}{{\partial {\bf{\bar v}}}}} \right)}^\dag } \cdot {{{\bf{\bar v}}}^\dag }} \right\rangle _{{\bf{x}},t}} + BT,
	\label{adj_identityA}
\end{equation}
where $BT$ denotes the boundary and temporal integral terms, and $BT=0$ can identify the boundary and terminal conditions of the adjoint equations. The corresponding adjoint LES equations can be expressed as
\begin{equation}
	\sum\limits_{k = 0}^3 {{{\left( {\frac{{\partial {R_k}}}{{\partial {\bf{\bar v}}}}} \right)}^{\dag}} \cdot {{{\bf{\bar v}}}^{\dag}}} -\frac{{\partial J}}{{\partial {\bf{\bar v}}}} = 0,
	\label{adj_EqnA}
\end{equation}
where ${\partial J}/{\partial {\bf{\bar v}}}= {\left[ {0,\frac{{\partial J}}{{\partial {{\bar u}_1}}},\frac{{\partial J}}{{\partial {{\bar u}_2}}},\frac{{\partial J}}{{\partial {{\bar u}_3}}}} \right]^T}$ denotes the sensitivity of the cost functional ${J\left( {{{\bar u}_i},\bar u_i^{{\rm{ref}}};{C_n},{\bf{x}},t} \right)}$ which quantifies the discrepancy between ${{{\bar u}_i}}$ and the reference data ${\bar u_i^{{\rm{ref}}}}$ in the LES calculations under the given parameters $C_n \left( {n = 1,2,...,N} \right)$ at a certain space-time state $\left( {{\bf{x}},t} \right)$.
Here, the terms ${{{\left( {{\partial {R_k}}/{{\partial {\bf{\bar v}}}}} \right)}^{\dag}} \cdot {{{\bf{\bar v}}}^{\dag}}} \left( {k = 0,1,2,3} \right)$ are derived by multiplying the perturbation LES equations (Eq.~\ref{sen_eqnA}) with the adjoint LES variables ${\bf{\bar v}}^{\dag}$, and then integrating by parts to rearrange all of the differential operators without $\delta \bf{\bar v}$  onto the adjoint variables  ${\bf{\bar v}}^{\dag}$ , yielding
\begin{equation}
	\begin{array}{*{20}{l}}
		{\frac{{\partial \delta {{\bar u}_i}}}{{\partial {x_i}}}{{\bar p}^{\dag}} + \left[ {\frac{{\partial \delta {{\bar u}_i}}}{{\partial t}} + \frac{{\partial \left( {{{\bar u}_j}\delta {{\bar u}_i}} \right)}}{{\partial {x_j}}} + \frac{{\partial \left( {{{\bar u}_i}\delta {{\bar u}_j}} \right)}}{{\partial {x_j}}} + \frac{{\partial \delta \bar p}}{{\partial {x_i}}} - \nu \frac{{{\partial ^2}\delta {{\bar u}_i}}}{{\partial {x_j}\partial {x_j}}} + \frac{{\partial \delta {\tau _{ij}}}}{{\partial {x_j}}}} \right]\bar u_i^{\dag} = }\\
		{ - \left( {\frac{{\partial \bar u_i^{\dag}}}{{\partial {x_i}}}} \right)\delta \bar p - \left[ {\frac{{\partial \bar u_i^{\dag}}}{{\partial t}} + \left( {\frac{{\partial \bar u_i^{\dag}}}{{\partial {x_j}}} + \frac{{\partial \bar u_j^{\dag}}}{{\partial {x_i}}}} \right){{\bar u}_j} + \frac{{\partial {{\bar p}^{\dag}}}}{{\partial {x_i}}}+ \nu \frac{{{\partial ^2}\bar u_i^{\dag}}}{{\partial {x_j}\partial {x_j}}} + \frac{\partial }{{\partial {x_j}}}\left( {\bar u_k^{\dag}\frac{{\partial {\tau _{jk}}}}{{\partial {{\bar u}_i}}}} \right) - \bar u_k^{\dag}\frac{{{\partial ^2}{\tau _{jk}}}}{{\partial {{\bar u}_i}\partial {x_j}}}} \right]\delta {{\bar u}_i} + }\\
		{ + \underbrace {\frac{{\partial \left( {\bar u_i^{\dag}\delta {{\bar u}_i}} \right)}}{{\partial t}}}_{{\rm{terminal}}\;{\rm{condition}}}{\rm{ + }}\underbrace {\frac{\partial }{{\partial {x_j}}}\left[ {\left( {\bar u_i^{\dag}{{\bar u}_j} + \nu \frac{{\partial \bar u_i^{\dag}}}{{\partial {x_j}}} + \bar u_k^{\dag}\frac{{\partial {\tau _{jk}}}}{{\partial {{\bar u}_i}}}} \right)\delta {{\bar u}_i} - \nu \bar u_i^{\dag}\frac{{\partial \delta {{\bar u}_i}}}{{\partial {x_j}}}} \right] + \frac{\partial }{{\partial {x_i}}}\left[ {\bar u_i^{\dag}\delta \bar p + \left( {{{\bar p}^{\dag}} + {{\bar u}_j}\bar u_j^{\dag}} \right)\delta {{\bar u}_i}} \right]}_{{\rm{boundary}}\;{\rm{condition}}}.}
	\end{array}
	\label{adj_EqnA1}
\end{equation}
The adjoint LES equations are written in detail as
\begin{equation}
	{\frac{{\partial \bar u_i^\dag }}{{\partial {x_i}}}}=0,
	\label{adj_LESA1}
\end{equation}
\begin{equation}
	\frac{{\partial \bar u_i^{\dag}}}{{\partial t}} + \left( {\frac{{\partial \bar u_i^{\dag}}}{{\partial {x_j}}} + \frac{{\partial \bar u_j^{\dag}}}{{\partial {x_i}}}} \right){{\bar u}_j} + \frac{{\partial {{\bar p}^{\dag}}}}{{\partial {x_i}}} + \nu \frac{{{\partial ^2}\bar u_i^{\dag}}}{{\partial {x_j}\partial {x_j}}} + \frac{\partial }{{\partial {x_j}}}\left( {\bar u_k^{\dag}\frac{{\partial {\tau _{jk}}}}{{\partial {{\bar u}_i}}}} \right) - \bar u_k^{\dag}\frac{{{\partial ^2}{\tau _{jk}}}}{{\partial {{\bar u}_i}\partial {x_j}}} + \frac{{\partial J}}{{\partial {{\bar u}_i}}} = 0.
	\label{adj_LESA2}
\end{equation}
It is worth noting that the adjoint SGS term ${\bar u_k^\dag \frac{{{\partial ^2}{\tau _{jk}}}}{{\partial {{\bar u}_i}\partial {x_j}}}}$ can lead to the non-conservation of the adjoint momentum and deteriorate the evaluation of the adjoint-based gradients. To our knowledge, few previous studies have addressed this critical issues that make the LES adjoint field prone to numerical instability and eventual divergence.
To maintain the momentum conservation in the adjoint equations, we remove ${\bar u_k^\dag \frac{{{\partial ^2}{\tau _{jk}}}}{{\partial {{\bar u}_i}\partial {x_j}}}}$ from Eq.~\ref{adj_LESA2}, and the conservative adjoint LES equations are obtained as
\begin{equation}
	{\frac{{\partial \bar u_i^\dag }}{{\partial {x_i}}}}=0,
	\label{adj_LESA3}
\end{equation}
\begin{equation}
	\frac{{\partial \bar u_i^{\dag}}}{{\partial t}} + \left( {\frac{{\partial \bar u_i^{\dag}}}{{\partial {x_j}}} + \frac{{\partial \bar u_j^{\dag}}}{{\partial {x_i}}}} \right){{\bar u}_j} + \frac{{\partial {{\bar p}^{\dag}}}}{{\partial {x_i}}} + \nu \frac{{{\partial ^2}\bar u_i^{\dag}}}{{\partial {x_j}\partial {x_j}}} + \frac{{\partial \tau _{ij}^{\dag}}}{{\partial {x_j}}} + \frac{{\partial J}}{{\partial {{\bar u}_i}}} = 0,
	\label{adj_LESA4}
\end{equation}
where $\tau _{ij}^\dag  = \bar u_k^\dag \frac{{\partial {\tau _{jk}}}}{{\partial {{\bar u}_i}}}$ is the adjoint SGS stress. If the unclosed SGS terms is modeled by the $N$-parameter mixed model ${\tau _{ij}} = \sum\limits_{n = 1}^N {{C_n}T_{ij}^{\left( n \right)}\left( {{{\bar u}_i};\bar \Delta } \right)} $ with the basis stress tensors ${T_{ij}^{\left( n \right)}}$ and model coefficients ${C_n}$, the adjoint SGS stresses are correspondingly represented as $\tau _{ij}^\dag  = \sum\limits_{n = 1}^N {{C_n}T_{ij}^{\left( n \right),\dag }}$ with the associated adjoint basis stress tensors ${T_{ij}^{\left( n \right),\dag }}\; \left( {n = 1,2,...,N} \right)$.
\section{Derivation of the adjoint SGS stress for the VOMM model}\label{AppendixB}
The present variational optimal mixed model (VOMM) combines the approximate deconvolution model (ADM) in the scale-similarity form with the dissipative Smagorinsky part, expressed as
\begin{equation}
	{\tau _{ij}} = {C_1}T_{ij}^{\left( 1 \right)} + {C_2}T_{ij}^{\left( 2 \right)},\;\;{\rm{with}}\;\;T_{ij}^{\left( 1 \right)} = {{\bar \Delta }^2}|\bar S|{{\bar S}_{ij}},\;\;T_{ij}^{\left( 2 \right)} = \overline {u_i^*u_j^*}  - \overline {u_i^*} \;\overline {u_j^*} ,
	\label{VOMM_model}
\end{equation}
where ${u_i^*}=\sum\limits_{n = 1}^N {{{(I - G)}^{n - 1}}} \otimes  {{{\bar u}_i}}$ stands for the $i$-th approximate unfiltered velocity component recovered by the iterative van Cittert procedure, $N$ is the number of iterations for the AD procedure, $I$ is the identity, and the symbol ``$\otimes$'' is the spatial convolution operator. Here, $C_1$ and $C_2$ are SGS model coefficients. 
The variation of the first basis SGS tensor $T_{ij}^{\left( 1 \right)}$ with respect to the velocity, is derived by
\begin{equation}
	\delta T_{ij}^{\left( 1 \right)} = {{\bar \Delta }^2}\left[ {|\bar S|\delta {{\bar S}_{ij}} + \left( {\delta |\bar S|} \right){{\bar S}_{ij}}} \right] = {{\bar \Delta }^2}\left( {|\bar S|\frac{{\partial {{\bar S}_{ij}}}}{{\partial {{\bar u}_k}}} + \frac{{\partial |\bar S|}}{{\partial {{\bar u}_k}}}{{\bar S}_{ij}}} \right)\delta {{\bar u}_k},
	\label{delta_Tij1}
\end{equation}
where the derivatives of the shear strain-rate tensor and characteristic strain rate for the velocity are further written as
\begin{equation}
	\frac{{\partial {{\bar S}_{ij}}}}{{\partial {{\bar u}_k}}} = \frac{1}{2}\frac{\partial }{{\partial {{\bar u}_k}}}\left( {\frac{{\partial {{\bar u}_i}}}{{\partial {x_j}}} + \frac{{\partial {{\bar u}_j}}}{{\partial {x_i}}}} \right) = \frac{1}{2}\left( {\frac{{\partial {\delta _{ik}}}}{{\partial {x_j}}} + \frac{{\partial {\delta _{jk}}}}{{\partial {x_i}}}} \right),
	\label{dSij_duj}
\end{equation}
and 
\begin{equation}
	\frac{{\partial |\bar S|}}{{\partial {{\bar u}_k}}} = \frac{{\partial |\bar S|}}{{\partial {{\bar S}_{ij}}}}\frac{{\partial {{\bar S}_{ij}}}}{{\partial {{\bar u}_k}}} = \frac{{{{\bar S}_{ij}}}}{{|\bar S|}}\left( {\frac{{\partial {\delta _{ik}}}}{{\partial {x_j}}} + \frac{{\partial {\delta _{jk}}}}{{\partial {x_i}}}} \right).
	\label{dS_uj}
\end{equation}
The inner product between the variation of the first basis SGS force and the adjoint velocity is derived by
\begin{equation}
	\begin{array}{l}
		\frac{{\partial \delta T_{ij}^{\left( 1 \right)}}}{{\partial {x_j}}}\bar u_i^\dag  =  - \frac{{\partial \bar u_i^\dag }}{{\partial {x_j}}}\delta T_{ij}^{\left( 1 \right)} + \frac{\partial }{{\partial {x_j}}}\left[ {\bar u_i^\dag \delta T_{ij}^{\left( 1 \right)}} \right]\\
		=  - \frac{{{{\bar \Delta }^2}}}{2}\left[ {\left( {|\bar S|\frac{{\partial \bar u_i^\dag }}{{\partial {x_j}}}} \right)\left( {\frac{{\partial {\delta _{ik}}}}{{\partial {x_j}}} + \frac{{\partial {\delta _{jk}}}}{{\partial {x_i}}}} \right) + \left( {\frac{{\partial {\delta _{mk}}}}{{\partial {x_n}}} + \frac{{\partial {\delta _{nk}}}}{{\partial {x_m}}}} \right)\left( {\frac{{2{{\bar S}_{mn}}}}{{|\bar S|}}{{\bar S}_{ij}}\frac{{\partial \bar u_i^\dag }}{{\partial {x_j}}}} \right)} \right]\delta {{\bar u}_k} + \frac{\partial }{{\partial {x_j}}}\left[ {\bar u_i^\dag \delta T_{ij}^{\left( 1 \right)}} \right]\\
		=  - \frac{{{{\bar \Delta }^2}}}{2}\left\{ {\frac{\partial }{{\partial {x_j}}}\left[ {|\bar S|\left( {\frac{{\partial \bar u_k^\dag }}{{\partial {x_j}}} + \frac{{\partial \bar u_j^\dag }}{{\partial {x_k}}}} \right)} \right] + \frac{\partial }{{\partial {x_j}}}\left[ {\frac{{2{{\bar S}_{jk}}}}{{|\bar S|}}{{\bar S}_{mn}}\left( {\frac{{\partial \bar u_m^\dag }}{{\partial {x_n}}} + \frac{{\partial \bar u_n^\dag }}{{\partial {x_m}}}} \right)} \right]} \right\}\delta {{\bar u}_k} + \frac{\partial }{{\partial {x_j}}}\left[ {\bar u_i^\dag \delta T_{ij}^{\left( 1 \right)}} \right],
	\end{array}
	\label{dT1dx_u}
\end{equation}
Here, the adjoint strain-rate tensor $\bar S_{ij}^\dag  = \left( {\partial \bar u_i^\dag /\partial {x_j} + \partial \bar u_j^\dag /\partial {x_i}} \right)/2$, and the inner product term can be further expressed as 
\begin{equation}
	\bar u_i^\dag \frac{{\partial \delta T_{ij}^{\left( 1 \right)}}}{{\partial {x_j}}} = \left\{ {\frac{\partial }{{\partial {x_j}}}\left[ { - {{\bar \Delta }^2}\left( {|\bar S|\bar S_{ij}^\dag  + \frac{{2{{\bar S}_{kl}}\bar S_{kl}^\dag }}{{|\bar S|}}{{\bar S}_{ij}}} \right)} \right]} \right\}\delta {{\bar u}_i} + \frac{\partial }{{\partial {x_j}}}\left[ {\bar u_i^\dag \delta T_{ij}^{\left( 1 \right)}} \right].
	\label{dT1dx_u1}
\end{equation}
Thus, the adjoint basis stress tensor $T_{ij}^{\left( 1 \right),\dag }$ is given by
\begin{equation}
	T_{ij}^{\left( 1 \right),\dag } =  - {{\bar \Delta }^2}\left( {|\bar S|\bar S_{ij}^\dag  + \frac{{2{{\bar S}_{kl}}\bar S_{kl}^\dag }}{{|\bar S|}}{{\bar S}_{ij}}} \right).
	\label{Tij*1}
\end{equation}
The common filter function $G$ (\emph{e.g.} top-hat, Gaussian and spectral filters) is symmetric spatial filter, and is self-adjoint, namely \citep{vreman2004a}
\begin{equation}
	{\left\langle {G \otimes f,g} \right\rangle _{\bf{x}}} = {\left\langle {f,G \otimes g} \right\rangle _{\bf{x}}},
	\label{self_adjG}
\end{equation}
where $f\left( {\bf{x}} \right)$ and $g\left( {\bf{x}} \right)$ are arbitrary variables. The ${G^n}$ filter with spatially filtering $n$ times (${G^n} = G \otimes G \otimes  \cdots  \otimes G$)  also satisfies the self-adjoint property proved by the mathematical induction method, expressed as
\begin{equation}
	{\left\langle {{G^n} \otimes f,g} \right\rangle _{\bf{x}}} = {\left\langle {G \otimes {G^{n - 1}} \otimes f,g} \right\rangle _{\bf{x}}} = {\left\langle {{G^{n - 1}} \otimes f,G \otimes g} \right\rangle _{\bf{x}}} =  \cdots  = {\left\langle {f,{G^n} \otimes g} \right\rangle _{\bf{x}}}.
	\label{self_adjGn}
\end{equation}
The $\left( I-G \right)$ filter is also a symmetric filter, and the approximate deconvolution procedure $H = \sum\limits_{n = 1}^N {{{\left( {I - G} \right)}^{n - 1}}} $ is thus the self-adjoint filter. The second basis SGS tensor $T_{ij}^{\left( 2\right)}$can be described using the AD abbreviated notation, namely
\begin{equation}
	T_{ij}^{\left( 2 \right)} = \overline {u_i^*u_j^*}  - \overline {u_i^*} \;\overline {u_j^*}  = G \otimes \left[ {\left( {H \otimes {{\bar u}_i}} \right)\left( {H \otimes {{\bar u}_j}} \right)} \right] - \left[ {G \otimes \left( {H \otimes {{\bar u}_i}} \right)} \right]\;\left[ {G \otimes \left( {H \otimes {{\bar u}_j}} \right)} \right].
	\label{Tij2_B}
\end{equation}
The variation of the second basis SGS tensor $T_{ij}^{\left( 2\right)}$ with respect to the velocity, expressed as
\begin{equation}
	\delta T_{ij}^{\left( 2 \right)} = G \otimes \left[ {\left( {H \otimes \delta {{\bar u}_i}} \right)u_j^*} \right] + G \otimes \left[ {u_i^*\left( {H \otimes \delta {{\bar u}_j}} \right)} \right] - \left[ {G \otimes \left( {H \otimes \delta {{\bar u}_i}} \right)} \right]\;\overline {u_j^*}  - \overline {u_i^*} \;\left[ {G \otimes \left( {H \otimes \delta {{\bar u}_j}} \right)} \right].
	\label{delta_Tij2}
\end{equation}
The inner product between the variation of the second basis SGS force and the adjoint velocity is given by
\begin{equation}
	\begin{array}{l}
		\frac{{\partial \delta T_{ij}^{\left( 2 \right)}}}{{\partial {x_j}}}\bar u_i^\dag  =  - \frac{{\partial \bar u_i^\dag }}{{\partial {x_j}}}\delta T_{ij}^{\left( 2 \right)} + \frac{\partial }{{\partial {x_j}}}\left[ {\bar u_i^\dag \delta T_{ij}^{\left( 2 \right)}} \right]\\
		=  - 2\bar S_{ij}^\dag \left\{ {G \otimes \left[ {\left( {H \otimes \delta {{\bar u}_i}} \right)u_j^*} \right]} \right\} + 2\bar S_{ij}^\dag \left[ {G \otimes \left( {H \otimes \delta {{\bar u}_i}} \right)} \right]\;\overline {u_j^*}  + \frac{\partial }{{\partial {x_j}}}\left[ {\bar u_i^\dag \delta T_{ij}^{\left( 2 \right)}} \right].
	\end{array}
	\label{dT2dx_u}
\end{equation}
The inner product term can be further simplified by the self-adjoint property, such that
\begin{equation}
	\begin{array}{*{20}{l}}
		{\frac{{\partial \delta T_{ij}^{\left( 2 \right)}}}{{\partial {x_j}}}\bar u_i^\dag  =  - 2\left( {G \otimes \bar S_{ij}^\dag } \right)\left[ {\left( {H \otimes \delta {{\bar u}_i}} \right)u_j^*} \right] + 2\left[ {G \otimes \left( {\bar S_{ij}^\dag \overline {u_j^*} } \right)} \right]\left( {H \otimes \delta {{\bar u}_i}} \right)\; + \frac{\partial }{{\partial {x_j}}}\left[ {\bar u_i^\dag \delta T_{ij}^{\left( 2 \right)}} \right]}\\
		{ = H \otimes \left( { - 2\overline {\bar S_{ij}^\dag } u_j^* + 2\overline {\bar S_{ij}^\dag \overline {u_j^*} } } \right)\delta {{\bar u}_i} + \frac{\partial }{{\partial {x_j}}}\left[ {\bar u_i^\dag \delta T_{ij}^{\left( 2 \right)}} \right]}\\
		{ = \left\{ {\frac{\partial }{{\partial {x_j}}}\left[ {H \otimes \left( {\overline {\bar u_i^\dag \overline {u_j^*} }  - \overline {\bar u_i^\dag } u_j^*} \right)} \right] + H \otimes \left( {\overline {\frac{{\partial \bar u_j^\dag }}{{\partial {x_i}}}\overline {u_j^*} }  - \frac{{\partial \overline {\bar u_j^\dag } }}{{\partial {x_i}}}u_j^*} \right)} \right\}\delta {{\bar u}_i} + \frac{\partial }{{\partial {x_j}}}\left[ {\bar u_i^\dag \delta T_{ij}^{\left( 2 \right)}} \right].}
	\end{array}
	\label{dT2dx_u1}
\end{equation}
It is quite notable that the second adjoint SGS term makes the non-conservation of the adjoint momentum, therefore we discard the second adjoint SGS term. Thus, the second adjoint basis stress tensor $T_{ij}^{\left( 2 \right),\dag }$ can be written as
\begin{equation}
	T_{ij}^{\left( 2 \right),\dag } = H \otimes \left( {\overline {\bar u_i^\dag \overline {u_j^*} }  - \overline {\bar u_i^\dag } u_j^*} \right) = \sum\limits_{n = 1}^N {{{\left( {I - G} \right)}^{n - 1}} \otimes \left( {\overline {\bar u_i^\dag \overline {u_j^*} }  - \overline {\bar u_i^\dag } u_j^*} \right)} .
	\label{Tij*2}
\end{equation}
In summary, the adjoint SGS stress of the proposed VOMM model is represented by
\begin{equation}
	\tau _{ij}^\dag  = {C_1}T_{ij}^{\left( 1 \right),\dag } + {C_2}T_{ij}^{\left( 2 \right),\dag }, 
	\label{adj_SGS_B}
\end{equation}  
where the adjoint basis stress tensors are $T_{ij}^{\left( 1 \right),\dag } =  - {{\bar \Delta }^2}\left( {|\bar S|\bar S_{ij}^\dag  + \frac{{2{{\bar S}_{kl}}\bar S_{kl}^\dag }}{{|\bar S|}}{{\bar S}_{ij}}} \right)$ and $T_{ij}^{\left( 2 \right),\dag } = \sum\limits_{n = 1}^N {{{\left( {I - G} \right)}^{n - 1}} \otimes \left( {\overline {\bar u_i^\dag \overline {u_j^*} }  - \overline {\bar u_i^\dag } u_j^*} \right)} $.

%\nocite{*}
\bibliography{reference}
\end{document}